\documentclass[useAMS,usenatbib]{mn2e}
\usepackage{natbib}
\usepackage{graphicx}
\usepackage[T1]{fontenc}
\usepackage{aecompl}
\usepackage{xr}
\usepackage{pdfpages}
\usepackage{grffile}

\def\ga{\mathrel{\mathchoice {\vcenter{\offinterlineskip\halign{\hfil
$\displaystyle##$\hfil\cr>\cr\sim\cr}}}
{\vcenter{\offinterlineskip\halign{\hfil$\textstyle##$\hfil\cr>\cr\sim\cr}}}
{\vcenter{\offinterlineskip\halign{\hfil$\scriptstyle##$\hfil\cr>\cr\sim\cr}}}
{\vcenter{\offinterlineskip\halign{\hfil$\scriptscriptstyle##$\hfil
\cr>\cr\sim\cr}}}}}
\def\la{\mathrel{\mathchoice {\vcenter{\offinterlineskip\halign{\hfil
$\displaystyle##$\hfil\cr<\cr\sim\cr}}}
{\vcenter{\offinterlineskip\halign{\hfil$\textstyle##$\hfil\cr<\cr\sim\cr}}}
{\vcenter{\offinterlineskip\halign{\hfil$\scriptstyle##$\hfil\cr<\cr\sim\cr}}}
{\vcenter{\offinterlineskip\halign{\hfil$\scriptscriptstyle##$\hfil
\cr<\cr\sim\cr}}}}}

\title{Vlasov versus $N$-body: the H\'enon sphere}

\author[S. Colombi, T. Sousbie, S. Peirani, G. Plum \& Y. Suto]{S. Colombi,$^1$\thanks{E-mail: colombi@iap.fr} T. Sousbie,$^{1,2,3}$ S. Peirani,$^1$ G. Plum$^1$ 
and Y. Suto$^{2,3}$
\\
\\
$^1$Institut d'Astrophysique de Paris, CNRS UMR 7095 and UPMC, 98bis, bd Arago, F-75014 Paris, France\\
$^2$Department of Physics, The University of Tokyo, Tokyo 113-0033, Japan,\\
$^3$Research Center for the Early Universe, School of Science, The
University of Tokyo, Tokyo 113-0033, Japan}
\begin{document}
\voffset -1cm
\date{\today}
\pagerange{\pageref{firstpage}--\pageref{lastpage}} \pubyear{2014}
\maketitle
\label{firstpage}
\begin{abstract}
We perform a detailed comparison of the phase-space density traced by
the particle distribution in {\tt Gadget} simulations to the result
obtained with a spherical Vlasov solver using the splitting
algorithm. The systems considered are apodized H\'enon spheres with
two values of the virial ratio, $R \simeq 0.1$ and $0.5$. After
checking that spherical symmetry is well preserved by the $N$-body
simulations, visual and quantitative comparisons are performed. In
particular we introduce new statistics, correlators and entropic
estimators, based on the likelihood of whether $N$-body simulations
actually trace randomly the Vlasov phase-space density. When taking
into account the limits of both the $N$-body and the Vlasov codes,
namely collective effects due to the particle shot noise in the first
case and diffusion and possible nonlinear instabilities due to finite
resolution of the phase-space grid in the second case, we find a
spectacular agreement between both methods, even in regions of
phase-space where nontrivial physical instabilities develop. However,
in the colder case, $R=0.1$, it was not possible to prove actual
numerical convergence of the $N$-body results after a number of
dynamical times, even with $N=10^8$ particles.
\end{abstract}

\section{Introduction}

Stars in galaxies and dark matter in the Universe can be modeled in
phase-space as self-gravitating collisionless fluids obeying
the Vlasov-Poisson equations:
\begin{eqnarray}
\frac{\partial f}{\partial t} 
+ \bmath{u}.\nabla_{\bmath{r}}f - \nabla_{\bmath{r}}\phi.\nabla_{\bmath{u}}f=0, \\
\Delta_{\bmath{r}} \phi=4\pi G\rho=4\pi G 
\int f(\bmath{r},\bmath{u},t)\ {\rm d} \bmath{u},
\end{eqnarray}
where $f(\bmath{r},\bmath{u},t)$ represents the phase-space density at
position $\bmath{r}$ and velocity $\bmath{u}$, $\phi$ is the
gravitational potential, and $G$ is the gravitational constant. 

In general, these equations do not have simple analytical
solutions. They are therefore often solved numerically. The most
widely used numerical scheme is the $N$-body approach and there exist
many different implementations, which mainly differ from each other in
the way Poisson equation is solved \citep[see, e.g.,][for reviews on
  the subject]{Bertschinger1998,Colombi2001,Dolag2008,Dehnen2011}. The
$N$-body method attempts to sample the phase-space density by
an ensemble of Dirac functions that represent particles
interacting with each other through gravitational
force. In order to avoid numerical artefacts due to the $1/r^2$
divergence of the force at small distances, the gravitational
potential is usually replaced by an effective one so that the force is
smoothed at scales smaller than a softening parameter $\epsilon$. This
procedure corresponds to assuming that the particles are
clouds of size ${\cal \epsilon}$ interacting with each other.

Approximating the phase-space density with macro-particles, however,
has its own limitation. In particular, the close $N$-body encounter is
one of the most notable sources of numerical artefacts, in
addition to more subtle collective effects induced by the discrete
nature of the distribution of the particles \citep[see,
  e.g.][]{Aarseth1988,Splinter1998,Boily2002,Binney2004,Joyce2009}.
Of course, the time integration scheme and the way to solve the
Poisson equation numerically are well-known sources of errors, even
though not particular to the $N$-body method.

There are several previous studies that discussed the limitations of
the $N$-body results, including underestimating strong numerical
artefacts, particularly in the cold case where the initial velocity
dispersion is null \citep[see, e.g.,][]{Melott1997,Melott2007}, and
long-term nonlinear resonant modes induced by the discrete nature of
the particles \citep[see,
  e.g.,][]{2005MNRAS.359..123A,ColombiTouma2014}.  We also note that
it is not yet obvious that the fine inner structure of dark matter
halos is completely understood from physical and even numerical points
of view, despite numerous intensive convergences studies of the
$N$-body approach \citep[see,
  e.g.,][]{Moore1998,JingSuto2000,JingSuto2002,Power2003,Springel2008,
  Stadel2009}.

It is therefore highly desired to develop alternative numerical methods to
the traditional $N$-body approach so that one can understand better
its validity and fundamental limitations.

In the cold case, relevant to the current paradigm of cold dark matter
scenario, the phase-space distribution function is supported by a
three-dimensional sheet evolving in six-dimensional phase-space, which
can be partitioned in a continuous way with an ensemble of tetrahedra
as proposed in recent works \citep[see,
  e.g.,][]{Shandarin2012,Hahn2013}. Unfortunately, the increasing
complexity of the structure of the system during evolution requires
more and more sampling elements, and the computational cost
becomes prohibitive after several dynamical time-scales.

In this article, we consider the warm case, in which the system presents
a non-negligible initial local velocity dispersion component relative
to  gravitational potential energy. In this case, 
the phase-space distribution
function has to be sampled on a 6-dimensional mesh, which makes again
the computational cost very high. Therefore, we shall restrict to spherical
systems, hence reducing the actual number of dimensions of
the dynamical setup to three. 

There exist many methods to solve the Vlasov-Poisson equations in the
warm case, mainly developed in plasma physics. One of the most famous
solvers is the splitting algorithm of \citet{CK} and its numerous
extensions \cite[see, e.g.][but this list is far from
  complete]{Shoucri1978,Sonnen1999,Filbet2001,Besse2003,2005MNRAS.359..123A,Umeda2008,Besse2008,Crouseilles2010,Campospinto2011,Rossmanith2011,Guclu2014}. This
algorithm, that we shall adopt below, exploits directly the Liouville
theorem: the phase-space density $f(\bmath{r},\bmath{v},t)$ is
conserved along motion. Then the equations of the dynamics during each
time step are divided into ``drift'' and ``kick'' parts according to
Hamiltonian dynamics and are solved backwards:
\begin{eqnarray}
f^*(\bmath{r},\bmath{u}) 
&= &f(\bmath{r}-\bmath{u} \Delta t/2,\bmath{u},t), 
\quad {\rm Drift}, \label{eq:ck1} \\
f^{**}(\bmath{r},\bmath{u}) 
&= &f^*(\bmath{r},\bmath{u}+\nabla_{\bmath{r}}\phi \Delta t), 
\quad {\rm Kick}, \label{eq:ck2} \\
f(\bmath{r},\bmath{u},t+\Delta t) 
&= & f^{**}(\bmath{r}-\bmath{u}\Delta t/2,\bmath{u}), 
\quad {\rm Drift}, \label{eq:ck3}
\end{eqnarray}
where $\nabla_{\bmath{r}}\phi$ is computed from $f^*$. In practice the
phase-space distribution function is sampled on a mesh, and each step
is performed by using tracer particles located at mesh sites and
following the equations of motion split as above. Resampling of $f^*$,
$f^{**}$ and finally the phase-space distribution function at the next
time step is performed by using an interpolation, e.g. based on
the spline method.

The splitting scheme was applied for the first time in astronomy in early
1980's, to one dimensional systems \citep[][]{Fujiwara1981}, galactic
disks \citep[][]{Watanabe1981,Nishida1981} and spherical systems
\citep{Fuji83}. Nevertheless, it has been almost forgotten since then 
except for a few contributions \citep[e.g.,][]{Hozumi1996,Hozumi2000}
that include a recent preliminary investigation of the algorithm in
full 6-dimensional phase-space
\citep[][]{Yoshikawa2013}.

As mentioned above, however, solving fully six-dimensional phase-space
problems with sufficient accuracy is still very unrealistic now. In
this article, therefore, we focus on spherical systems, where
phase-space is only three dimensional: the three coordinates of
interest are the radial position $r$, the radial velocity $v$ and the
angular momentum $j$. Following earlier works performed in the
framework of one dimensional gravity \citep[see, e.g.,][]{Mineau1990},
we carry out a detailed comparison between an $N$-body code, {\tt
  Gadget} \citep{Springel2001,Springel2005}, and an improved version
of the splitting algorithm implementation by \citet{Fuji83}, 
{\tt VlaSolve}.\footnote{{\tt VlaSolve} can be downloaded
from the following web page: {\tt www.vlasix.org}.}

 Our goal is to check how well the particle distribution in {\tt
   Gadget} traces the phase-space density obtained from {\tt
   VlaSolve}, and to see how the results depend on various parameters
 of the simulations, in particular the number of particles in the
 $N$-body simulations and the spatial resolution in the Vlasov code.
We would however like to emphasize here that the purpose of this article is not to compare the
performance of the two codes from the view-point of computational cost.

While a fairly good physical insight is obtained through visual
inspection of the resulting phase-space density plots, we also present
a more quantitative comparison. To do so, we introduce correlators and
entropic estimators based on a likelihood approach, ans ask whether
the $N$-body simulations can be considered as local Poisson
realizations of the Vlasov code phase-space density.  

Because of our restrictive choice of the geometry of the system, it is
important to simulate spherical configurations that are known to be
stable against small anisotropic perturbations induced by the shot
noise of the particles. Indeed, we shall use the public treecode {\tt
  Gadget} without any specific modification to enforce spherical
dynamics.  Although an alternative approach consisting in enforcing
pure radial dynamics in {\tt Gadget} \cite[see, e.g.,][]{Huss1999}
may facilitate comparisons with the Vlasov code, we do not adopt this approach
in order to avoid any possible subtle biases in the analyses.

In this respect, the H\'enon sphere \citep{Henon1964} is particularly
suited for our purpose since it is known to preserve well its
spherical nature during the course of dynamics even when being
simulated with a $N$-body technique and, in particular, it is not
prone to radial orbit instability \citep[see,
  e.g.,][]{VanAlbada1982,Hozumi1996,RoyPerez2004,Barnes2009}. In this
configuration, the initial phase-space distribution function is 
isotropic and Gaussian distributed in velocity space and given by
\begin{eqnarray}
f_{\rm H}(r,v,j)&=&\frac{\rho_0}{(2\pi \sigma_v^2)^{3/2}}
\exp\left( -\frac{1}{2} \frac{v^2+j^2/r^2}{\sigma_v^2} \right),
                      \nonumber \\
& & \quad \quad \quad \quad \quad \quad \quad \quad \quad \quad \quad \quad \quad r \leq
R_{\rm H},
\label{eq:henons}
\end{eqnarray}
with $(4\pi/3) \rho_0 R_{\rm H}^3=M$, the total mass of the system. In
the simulations discussed in this article, we work in units where $G=1$, and 
the initial radius of the
H\'enon sphere and its total mass are chosen to be
\begin{equation}
M=1, \quad R_{\rm H}=2,
\end{equation}
which fixes $\sigma_v$ in equation (\ref{eq:henons}) once the virial ratio is given. 

We shall consider ``warm'' and ``cold'' settings, which correspond to
the initial virial ratio $R=|2T/W|=5 R_{\rm H} \sigma_v^2/M$ of
$\approx 0.5$ and $\approx 0.1$, respectively, where $T$ and $W$ are
the total kinetic and potential energy of the system.  The two classes
of initial conditions exhibit distinct features, in particular
concerning the metastable state to which the system relaxes through
phase mixing. The warm system builds a core-halo structure, with the
halo displaying a power-law profile $\rho(r) \sim r^{-4}$ \citep[see,
  e.g.,][]{Henon1964,Gott73,VanAlbada1982}. In contrast, the cold
system develops a more concentrated smaller core \cite[see,
  e.g.,][]{VanAlbada1982,SylosLabini2012}, but never reaches a
strictly stationary regime because a significant fraction of the mass
acquires positive energy and escapes from the system
\citep[see, e.g.,][]{VanAlbada1982,Joyce2009,SylosLabini2012}.

This article is organized as follows. In \S~\ref{sec:thesimus} we
describe our Vlasov solver, {\tt VlaSolve}. Section \ref{sec:gadpres}
provides information about the $N$-body runs and the parameters used
in {\tt Gadget}. In \S~\ref{sec:sphericity}, we check that the
$N$-body simulations stay indeed spherical during evolution. Section
\ref{sec:visu} presents a visual inspection of the phase-space
density, which is followed by a quantitative statistical analysis in
\S~\ref{sec:statana}. Finally, \S~\ref{sec:discussion} summarizes and
discusses our present results.

\section{The Vlasov code: VlaSolve}
\label{sec:thesimus}

Under spherical symmetry, the Vlasov equation reads
\begin{equation}
\frac{\partial f}{\partial t}+v\frac{\partial f}{\partial r}
+\left(\frac{j^2}{r^3}-\frac{GM_r}{r^2}\right)\frac{\partial f}{\partial v}=0,
\label{eq:vl}
\end{equation}
where $v$ is the radial component of the velocity, $j$ is the angular
momentum, $M_r=M\left(<r\right)$ is the mass inside a sphere of
radius $r$.

Our code {\tt VlaSolve} solves equation (\ref{eq:vl}) numerically with the
splitting algorithm, following closely \cite{Fuji83}. 

Phase space is discretized into a rectangular mesh of size
$(N_r,N_v,N_j)$ for $R_{\rm min} \leq r \leq R_{\rm max}$, $-v_{\rm
  max} \leq v \leq v_{\rm max}$, and $0\leq j \leq J_{\rm max}$.  More
specifically, we use a logarithmically equal interval for $r$, a
linearly equal interval for $v$. The $k^{\rm th}$-bin of the angular
momentum slice corresponds to the interval $[J_{\rm max}
  (k-1)^2/N_j^2, J_{\rm max} k^2/N_j^2]$ and is represented by
$j_k=J_{\rm max}(k-1/2)^2/N_j^2$.

We modify the splitting algorithm using the fact that the angular
momentum is an invariant of the Hamiltonian system. Hence, 
one may treat each slice with a different value of $j$ in 
phase-space independently, except for gravitational coupling via the
Poisson equation. We include the inertial component of the force,
$j^2/r^3$, in the ``drift'' step (equations \ref{eq:ck1} and
\ref{eq:ck3}), while the ``kick'' step (equation \ref{eq:ck2})
corresponds solely to gravitational force:
\begin{eqnarray}
f^*(r,v,j) &=& f[r^*(-\Delta t/2),v^*(-\Delta t/2), j,t], \\
f^{**}(r,v,j)&=&f^*(r,v+ G M_r/r^2 \Delta t,j), \\
f(r,v,j,t+\Delta t)&=&f^{**}[r^*(-\Delta t/2),v^*(-\Delta t/2),j],
\end{eqnarray}
where $r^*$ and $v^*$ solve analytically the
motion in absence of gravity starting from coordinates $(r,v,j)$ in phase-space \citep[see, e.g.,][]{ColombiTouma2008}:
\begin{eqnarray}
{r}^*(h) &=& \sqrt{ \frac{ \left[ \sqrt{2 r^2 H_{\rm
  K}-j^2}+2\ {\rm sgn}(v) H_{\rm K} h \right]^2 + j^2 }{2
  H_{\rm K}}},\\
{v}^*(h) &=&{\rm sgn}(v) \sqrt{2 H_{\rm K}-
  \frac{j^2}{{r^*(h)}^2}},
\end{eqnarray}
with $H_{\rm K} \equiv v^2/2+j^2/(2 r^2)$ (when $v^* < 0$, these equations are
valid until $v^*=0$). 

Because a non-zero angular momentum bends the trajectories in
$(r,v)$ space, the drift step requires a
two-dimensional interpolation of the phase-space distribution function
in $(r,v)$ space, while the kick step, which only modifies the
velocities, can be completed with a one-dimensional interpolation.  We
follow \cite{Fuji83}, and carry out the interpolations using 
third-order splines. In this interpolation scheme, however, the
positivity of the phase-space distribution function is not warranted,
and numerical aliasing and diffusion effects are expected when the
phase-space distribution function varies over scales of the order
of, or smaller than, the mesh element size.

In order to reduce such numerical artefacts, we modify equation 
(\ref{eq:henons}) as follows:
\begin{eqnarray}
f_{\rm H}(r,v,j) &=&\frac{\rho_0}{(2\pi \sigma_v^2)^{3/2}}
\exp\left( -\frac{1}{2} \frac{v^2+j^2/r^2}{\sigma_v^2} \right) \times
\nonumber \\
&  & \frac{1}{2} 
\left[ 1+{\rm erf}\left(\frac{R_{\rm H}-r}{\Delta} \right) \right],
\quad r \leq R_{\rm H},
\label{eq:apodization}
\end{eqnarray}
with $\Delta=1/2$. Then we recompute $\rho_0$ in equation
(\ref{eq:henons}) so that the total mass remains unity. This
apodization slightly changes the actual values of the virial ratio to
$R\simeq 0.55$ and $0.11$, although we shall still denote them by
$0.5$ and $0.1$ just for simplicity. It may also
modify the long-term dynamical properties of the original H\'enon
sphere relative to what is expected. This is why we check again the
extent to which the spherical nature of the system is retained in
the $N$-body simulations (\S~\ref{sec:sphericity}).

Adopting a logarithmic binning for $r$ is well suited for
tracing small-scale features around the center of the system. This
implies, however, that radii smaller than a finite minimum value
$R_{\rm min}$ are missing from the computing domain.  A
conventional trick to overcome the problem is to assume a reflecting
boundary at $r=R_{\rm min}$ \cite[see, e.g..][]{Gott73,Fuji83}.
Usually, a systematic time-lag between orbits in this method is
neglected: particles reaching the reflective kernel boundary instantly
travel the $2R_{\rm min}$ distance through the central region, while
they should actually take a finite time depending on their radial
velocity and angular momentum.  In {\tt VlaSolve}, we improve the
reflecting sphere method by taking into account the actual time spent
by particles travelling inside the region $r \leq R_{\rm min}$, which
is made easily possible by neglecting the gravitational
force. Technical details about the implementation are provided in
Appendix~\ref{sec:monR0}.

To complete algorithmic details, Appendix~\ref{sec:para} discusses the hybrid
parallelization of {\tt VlaSolve} with OpenMP and MPI libraries.

In this paper, we perform 4 simulation runs with different resolutions,
each for $R=0.1$ and $0.5$ (Table~\ref{tab:resovla}).  To cover the
dynamical range of interest, the computing mesh uses
$R_{\rm min}=0.01$, $R_{\rm max}=25$ and $J_{\rm max}=1.6$. The
maximum amplitude of the velocity is $v_{\rm max}=2$ and
4 for $R=0.5$ and $0.1$, respectively. With this
choice of the parameters, the computational domains are sufficiently
large to contain all the system up to the end of the simulations, which
corresponds to $t=100$ for $R=0.5$ 
and $t=35$ for $R=0.1$. As will be illustrated later in phase-space density plots, these final
epochs are sufficient for the system to have relaxed at the coarse
level to a meta-stable state through mixing. Strictly speaking, this
is not the case in the $R=0.1$ case because a fraction of the mass
escapes from the system \citep[see, e.g.,][]{VanAlbada1982,SylosLabini2012}, as already
mentioned in the Introduction.
\begin{table}
\centerline{\hbox{
\begin{tabular}{cccc}
\hline 
$N_r$ & $N_v$ & $N_j$ & $\Delta t$ \\
\hline
1024 & 1024 & 512 & $5\times 10^{-4}$ \\
512  & 512 & 512 & $10^{-3}$ \\
2048 & 2048 & 32 & $2.5\times 10^{-4}$ \\
1024 & 1024 & 32 & $5\times 10^{-4}$\\
\hline
\end{tabular}}}
\caption[]{The parameters used for the {\tt VlaSolve} simulations.}
\label{tab:resovla}
\end{table}

We adopt a constant time step $\Delta t$ throughout each simulation.
Just to stay on the conservative side, we choose a resolutely small value of
$\Delta t$, despite the increased computational cost. Note however
that excessively small time step might artificially increase
diffusion effects related to successive 
interpolations of the phase-space distribution function
\citep[][]{Halle2015}.

In Appendix~\ref{sec:resoeffects}, a comparison among all the
simulations is performed for $R=0.1$. It indicates that diffusion and
aliasing effects discussed earlier are indeed significant, despite the
apodization of initial conditions, but do not seem to affect
the dynamical properties of the system. Note that 
is tempting to undersample angular momentum space since $j$ is an invariant of the
dynamics. However, we show in this appendix that it is not wise to do
so, because it can provoke nonlinear instabilities after a few
dynamical times.

\section{$N$-body simulation with Gadget}
\label{sec:gadpres}
We perform the $N$-body simulations using the latest version
of the {\tt Gadget}-2 code \citep{Springel2005}. Only the
treecode part of this ``treePM'' algorithm is employed.  The
particle number is varied from $N=10^4$ to $10^7$ for $R=0.1$ 
and $0.5$. We also run an additional simulation with $N=10^8$ for $R=0.1$. 

We choose the parameters for {\tt Gadget} runs as follows:
\begin{itemize}
\item The softening length of the gravitational force is set as
  $\epsilon=0.2 N^{-1/3}$, that is about $1/16$ of the
  initial mean interparticle distance $(4\pi/3N)^{1/3}R_{\rm H}$ (this
  estimate neglects the effects of the apodization \ref{eq:apodization}).
\item In {\tt Gadget}, each particle has its individual time step
  bounded by ${\rm d} t=\min[{\rm d} t_{\rm max}, (2 \eta
    \epsilon/|\bmath{a}|)^{1/2} ]$, where $\bmath{a}$ is the
  acceleration of the particle and $\eta$ is a control parameter. We choose
 $\eta=0.025$ and $\Delta t_{\rm max}=0.01$.
\item The tolerance parameter controlling the accuracy of the relative
  cell-opening criterion \citep[parameter designed by {\tt
      ErrTolForceAcc} in the documentation of {\tt Gadget}, see
    equation 18 of][]{Springel2005} is set as $\alpha_{\rm
    F}=0.005$.
\end{itemize}

Appendix~\ref{app:instab} presents the effects of changing these
parameters on the phase-space distribution function for simulations
with $N=10^6$ particles and a virial ratio of $R=0.1$.  These
analyses, performed at $t=15$, confirm that the parameters used for
the simulations of this paper are reasonable. Interestingly, changing
the softening length by large factors does not influence much the
results, as already noticed previously in the literature \citep[see,
  e.g.][]{Barnes2009}, as long as it is kept small enough.

\section{Consistency check: sphericity of the $N$-body results}
\label{sec:sphericity}

\begin{figure*}
\centerline{\hbox{
\includegraphics[height=6cm]{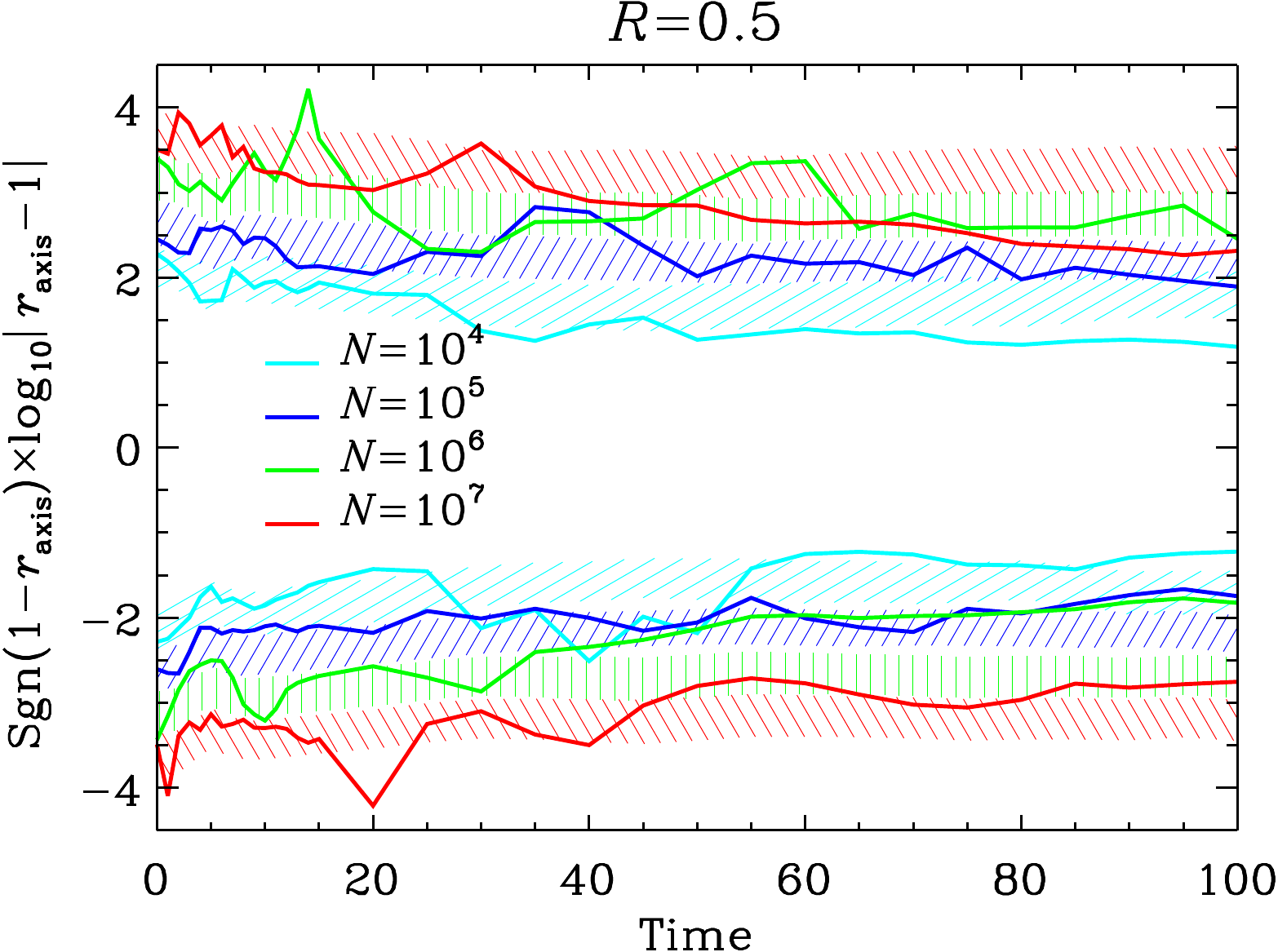}
\includegraphics[height=6cm]{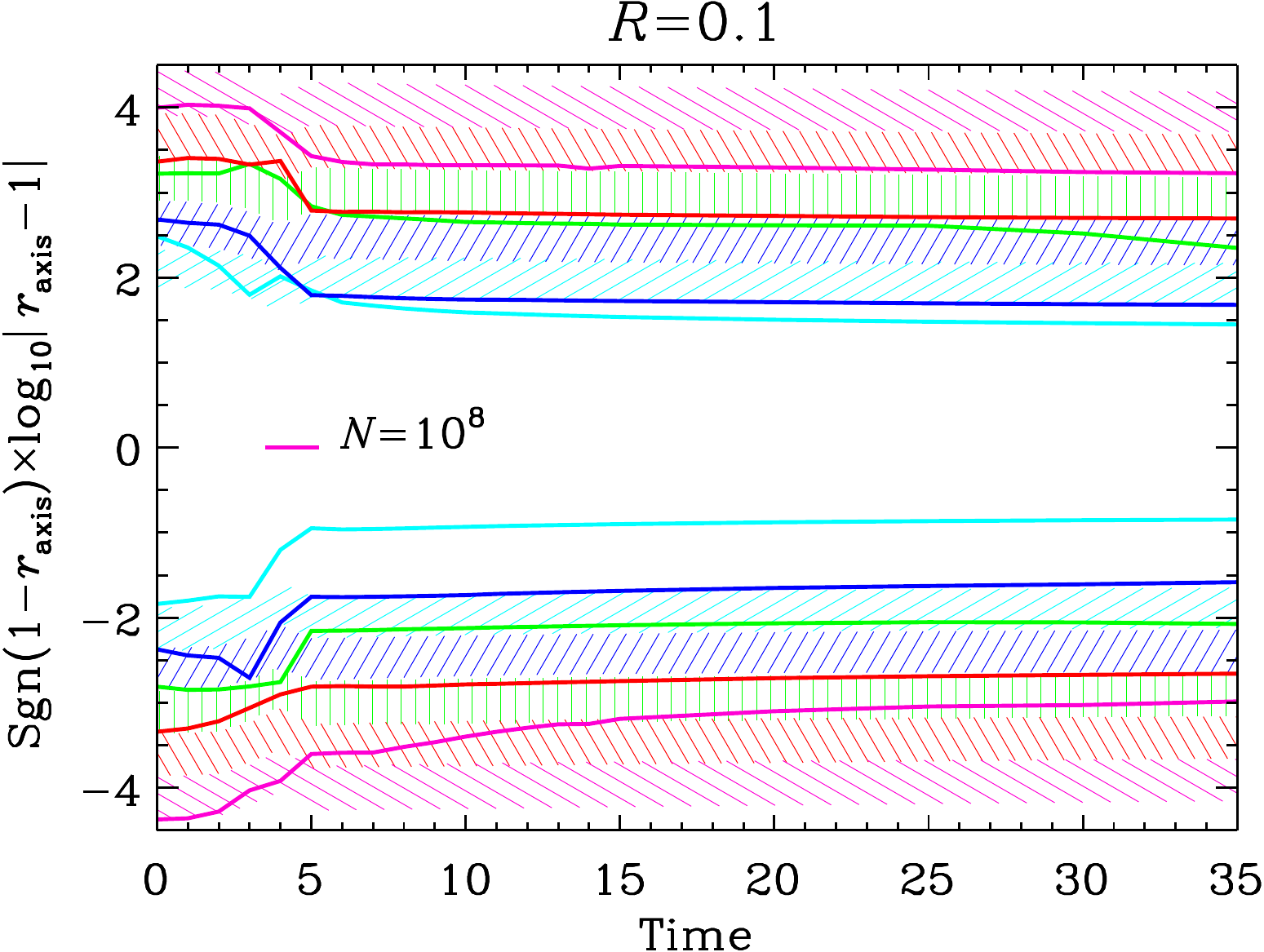}
}}
\caption[]{Evolution of the departure from spherical symmetry:
  ratios of the eigenvalues of the inertia tensor of the system as
  functions of time in the {\tt Gadget} simulations. To emphasize the
  small differences from unity, the quantity ${\rm sgn}(1-r_{\rm
    axis}) \log_{10}|1-r_{\rm axis}|$ is plotted as a function of
  time, where $r_{\rm axis}=b/a$ (upper curves on each panel) or $b/c$
  (lower curves) and $a \leq b \leq c$ are the eigenvalues of the
  inertia tensor of the {\tt Gadget} particle distribution. Each color
  corresponds to a given value of the number $N$ of particles as
  indicated in the panels. Dashed regions correspond to the one
  sigma confidence level zone expected for a particle distribution
  locally Poisson sampling the spherically symmetrical projected
  density profile $\rho(r,t)$, where $\rho(r,t)$ is estimated from
  interpolation of the {\tt Gadget} particle distribution in spherical
  shells. To calculate the average of $r_{\rm axis}$ and the
  associated one sigma error contours, 100 local Poisson realizations
  have been performed for each snapshot and value of $N$ considered,
  except for $N=10^7$ and $N=10^8$ (on right panel only for the
  latter). In the last cases, the dashed regions correspond to an
  extrapolation of the results obtained from $N=10^6$.}
\label{fig:sphe}
\end{figure*}

Before presenting comparisons between {\tt Gadget} and {\tt VlaSolve},
it is necessary to make sure that the sphericity of the system is
preserved in the {\tt Gadget} simulations because our Vlasov runs are
performed assuming exact spherical symmetry. Figure \ref{fig:sphe}
shows, for different values of the number of particles $N$, the
evolution with time of the ratios $b/a$ and $b/c$, where $a \leq b
\leq c$ are the eigenvalues of the inertia tensor of the particle
distribution.

The dashed regions correspond to the one sigma zone obtained from an
ensemble of 100 local Poisson realizations of the spherical density
$\rho(r)$, which is estimated from interpolation over spherical shells
from the {\tt Gadget} particles.  From the measurements in
Fig.~\ref{fig:sphe}, deviations from spherical symmetry
due to the particle shot noise can be roughly scaled to
\begin{equation}
\left\langle \frac{b}{a} \right\rangle-1 
\simeq 1-\left\langle \frac{b}{c} \right\rangle 
\simeq 2\sigma_{b/a} \simeq 2 \sigma_{b/c} \sim \frac{1}{\sqrt{N}},
\label{eq:spherr}
\end{equation}
where $\sigma_{b/a}^2$ and $\sigma^2_{b/c}$ are the variances of $b/a$
and $b/c$ obtained from the dispersion over the 100 realizations.
Note that equation (\ref{eq:spherr}) is not intended to be accurate.
The asphericity due to discreteness should depend on details of the
density profile, as shown in Fig.~\ref{fig:sphe}. While it would be
possible to compute in a perturbative way the quantities in equation
(\ref{eq:spherr}) from statistical analysis of the inertia tensor
assuming $N \gg 1$ and using error propagation formulae, this is a
cumbersome exercise far beyond the scope of this paper.

We also note that another possible source of errors comes from the
position of the center of the system. Indeed, an inaccurate
determination of the center obviously
worsens the apparent agreement with spherical symmetry.  In the
measurements presented in Fig.~\ref{fig:sphe}, the inertia matrix is
not computed with respect to the center of gravity of the particle
distribution, which can be affected by the fact that some particles
can get far away from the system through $N$-body relaxation. Instead,
we determine the center of the system using an iterative procedure
trying to optimize the match of the phase-space distribution function
with that of the Vlasov code, as detailed in \S~\ref{sec:cordef}. This
procedure is not free from errors either, and may contribute to the
fluctuations observed in the curves of Fig.~\ref{fig:sphe}.

Inspection of Fig.~\ref{fig:sphe} shows that the measured ratios $b/a$
and $b/c$ behave differently in the $R=0.5$ and $R=0.1$ simulations.
In the $R=0.5$ case, the agreement of the measurements with the
Poisson prediction is in general good, with a slight trend to
ellipticity, except for the top red curve and the bottom green curve
where the deviation from spherical symmetry is larger than the Poisson
expectation. Still, in the case of $R=0.5$, the system remains to a
very good approximation spherical for all values of $N$, given
the expected deviations due to pure statistical noise.

The curves representing the eigenvalue ratios are more steady for $R=0.1$
than for $R=0.5$, which might be slightly puzzling at first
sight. However, a very plausible explanation of this difference is
that the initial velocity dispersion is larger for $R=0.5$ than for
$R=0.1$, hence adding a more
prominent random component to the time behavior of the deviation from
sphericity.  

Regarding $R=0.1$, deviations from spherical symmetry are 
clearly more significant compared to local Poisson expectations 
after $t \approx 3$, roughly the collapse time of the sphere.
While the $N=10^4$ run exhibits a deviation larger than 10
percent, spherical symmetry is confirmed to be a good approximation 
for $N\geq 10^5$.

Finally, we also check deviations from spherical symmetry for subsets
of particles in excursions corresponding to $f \geq f_{\rm th}$, where
$f$ is the phase-space distribution function measured in the
$1024\times 1024 \times 512$ {\tt VlaSolve} runs. For each value of
the virial ratio, two thresholds $f_{\rm th}$ are chosen such that the
excursions contained initially about 90 and 60 percent of the total
mass (see bottom panels of Fig.~\ref{fig:entro} below). Given the
uncertainties in the measurements, the above conclusions still hold:
the properties of the deviations from spherical symmetry, that we do
not show here for simplicity, do not indeed depend significantly on
radius. We only notice a slight improvement in the $R=0.5$ case when
considering particles in the excursions.

\section{Phase-space density: visual inspection}
\label{sec:visu}

Now we are ready to perform direct comparisons between the Vlasov and $N$-body
simulation results. For this purpose, we consider the phase-space
density at different epochs (Figs.~ \ref{fig:0v5_12} to
\ref{fig:0v1_ALL} below). To be more specific, we plot 
the constant angular momentum slice of $f(r,v,j)$ at $j=0.244$, 
and its integral over the angular momentum:
\begin{equation}
f_{\rm summed}(r,v)=\int f(r,v,j)\ 2\pi j {\rm d}j.
\end{equation}

Figures~\ref{fig:0v5_12} and \ref{fig:0v5_ALLJ} plot $f(r,v,j\simeq
0.244)$ and $f_{\rm summed}(r,v)$, respectively, for the {\tt
  VlaSolve} and {\tt Gadget} simulations of the warm case, $R=0.5$.
In both figures, snapshots at $t=10$, 50, 80 and 100 are plotted from
left to right.  The panels correspond to the {\tt VlaSolve} runs with
$(N_r,N_v,N_j)=(2048,2048,32)$ and $(1024,1024,512)$, the {\tt Gadget} runs
with $N=10^7$, $10^6$ and $10^5$, from top to bottom.

\begin{figure*}
\hbox{
\includegraphics[width=4.25cm]{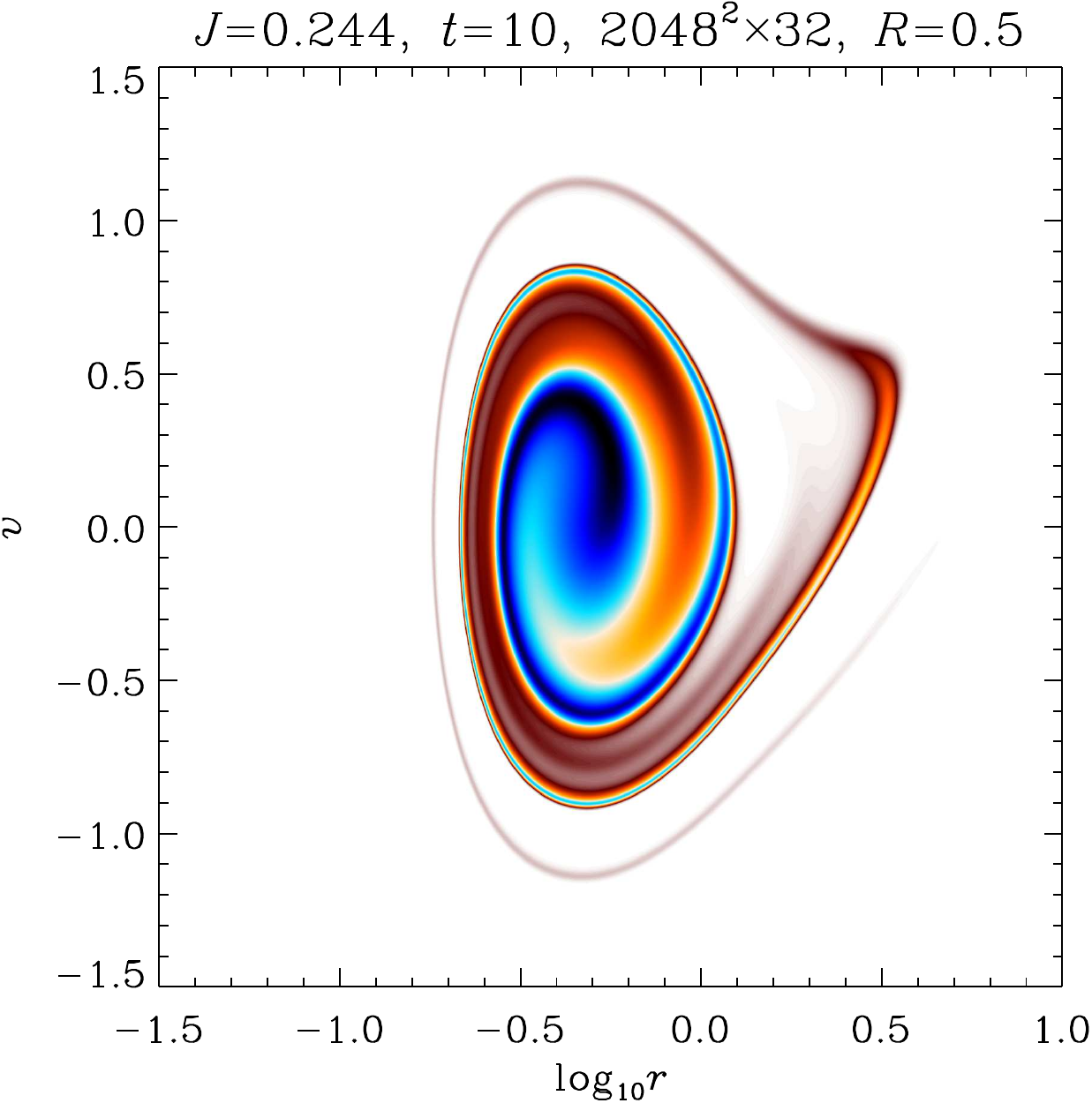}
\includegraphics[width=4.25cm]{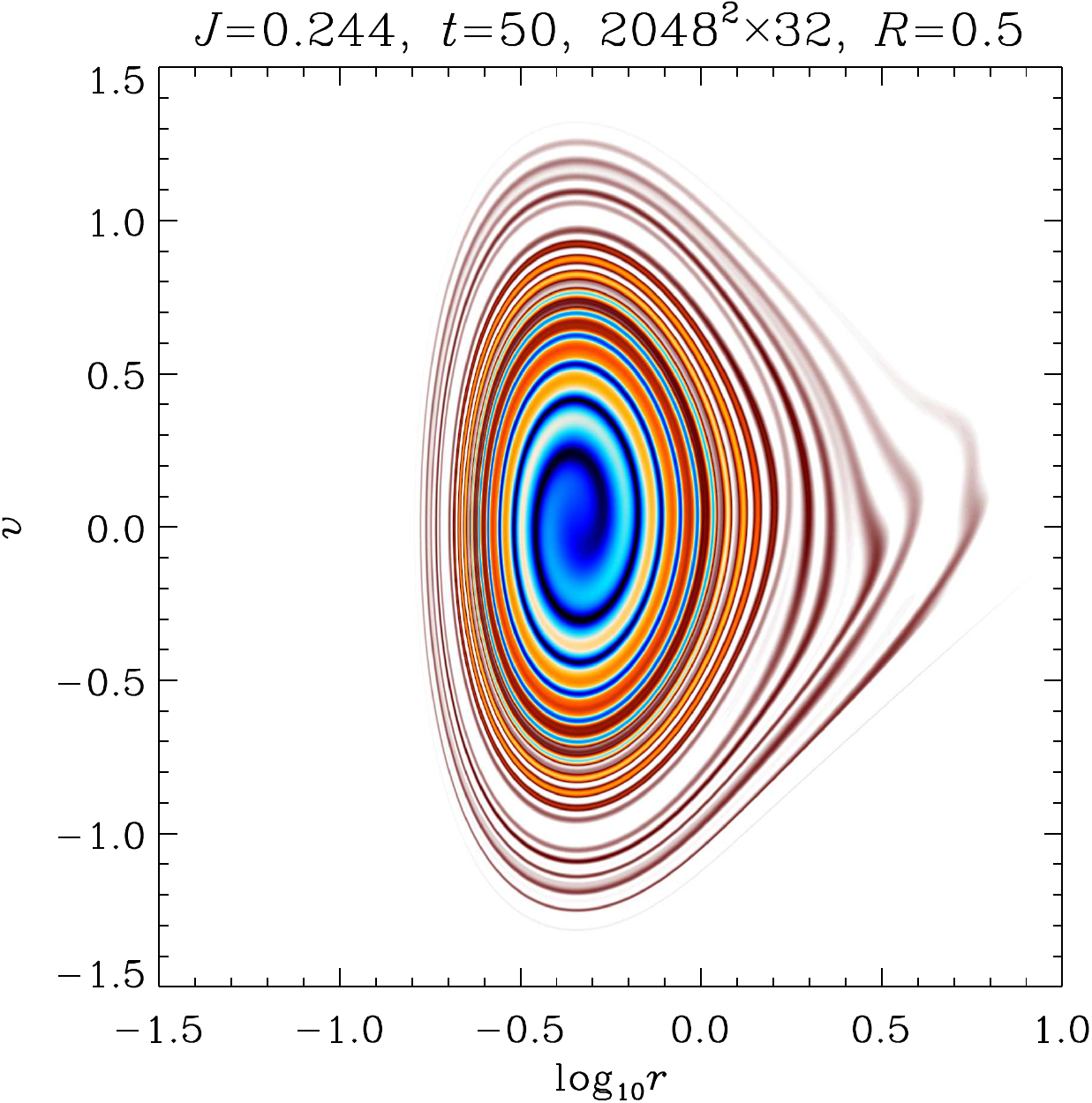}
\includegraphics[width=4.25cm]{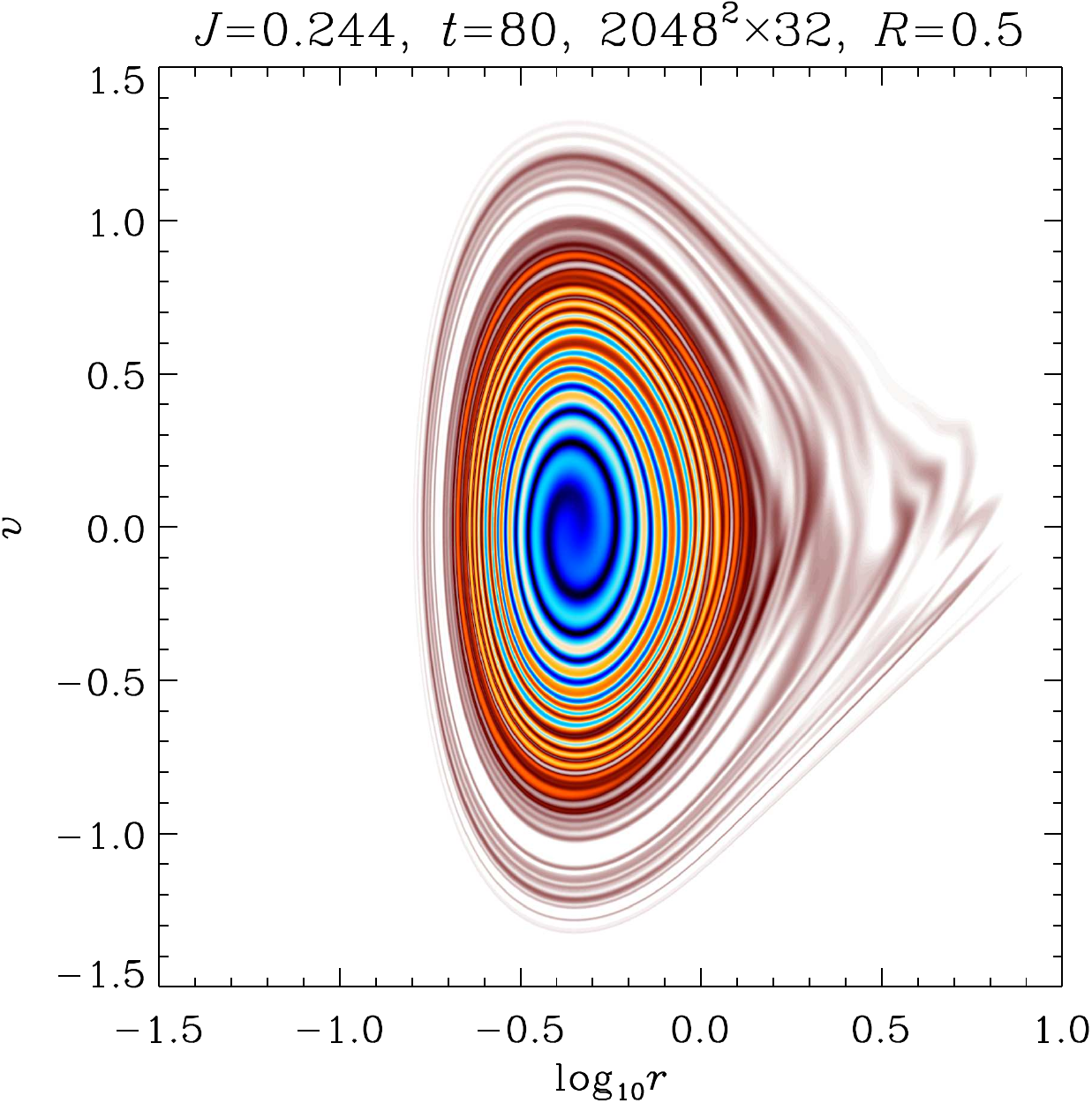}
\includegraphics[width=4.25cm]{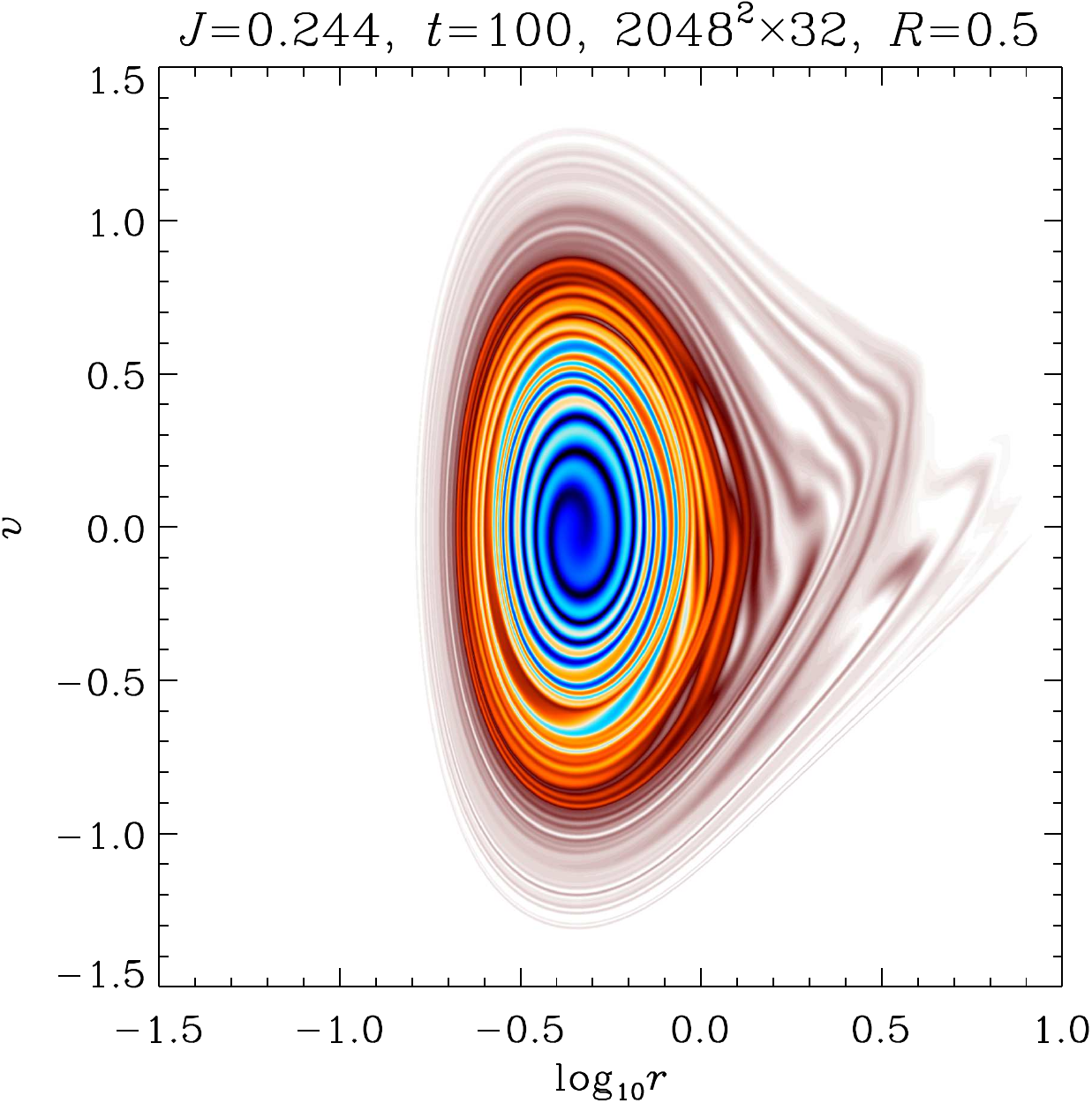}
}
\hbox{
\includegraphics[width=4.25cm]{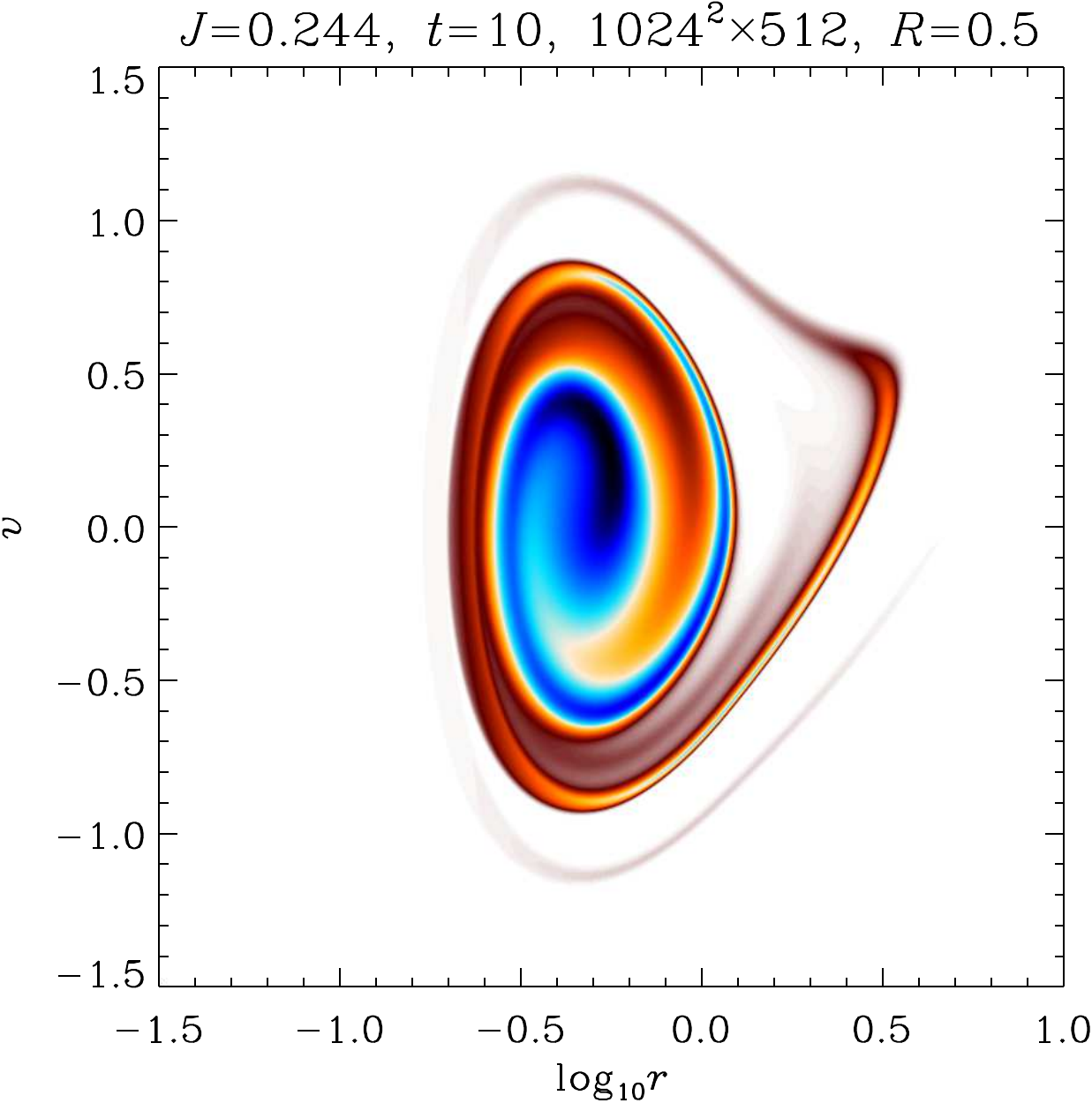}
\includegraphics[width=4.25cm]{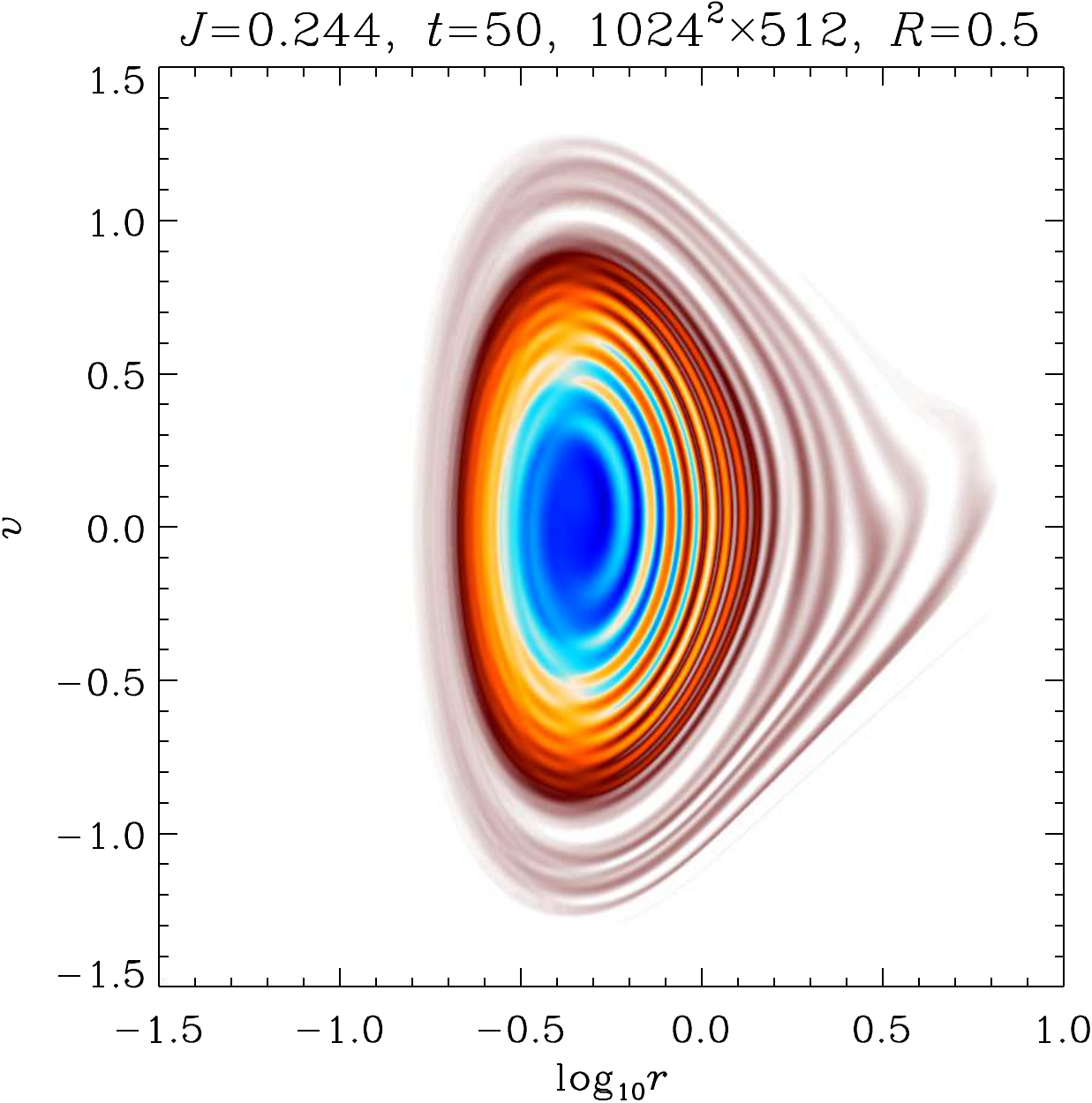}
\includegraphics[width=4.25cm]{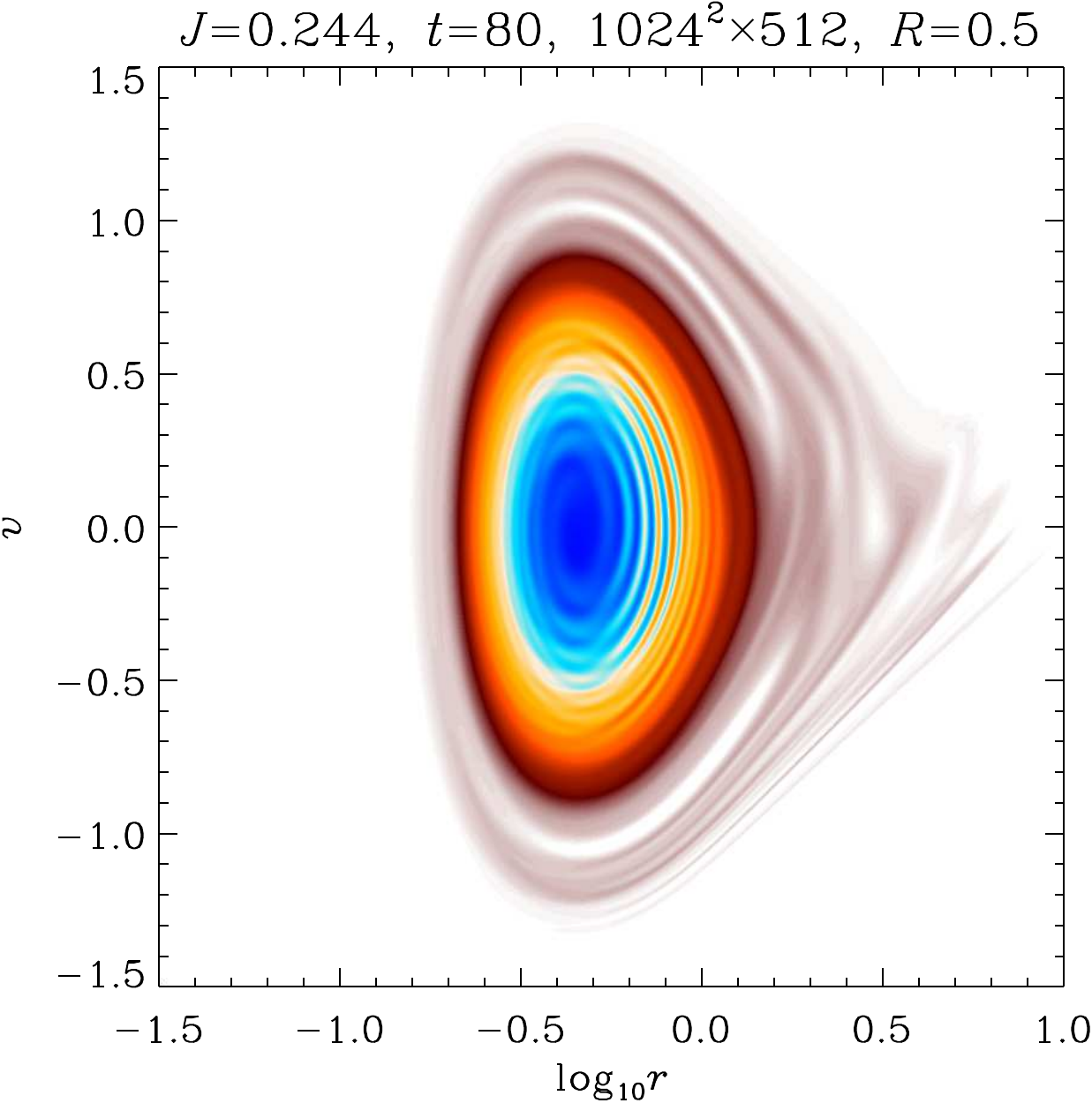}
\includegraphics[width=4.25cm]{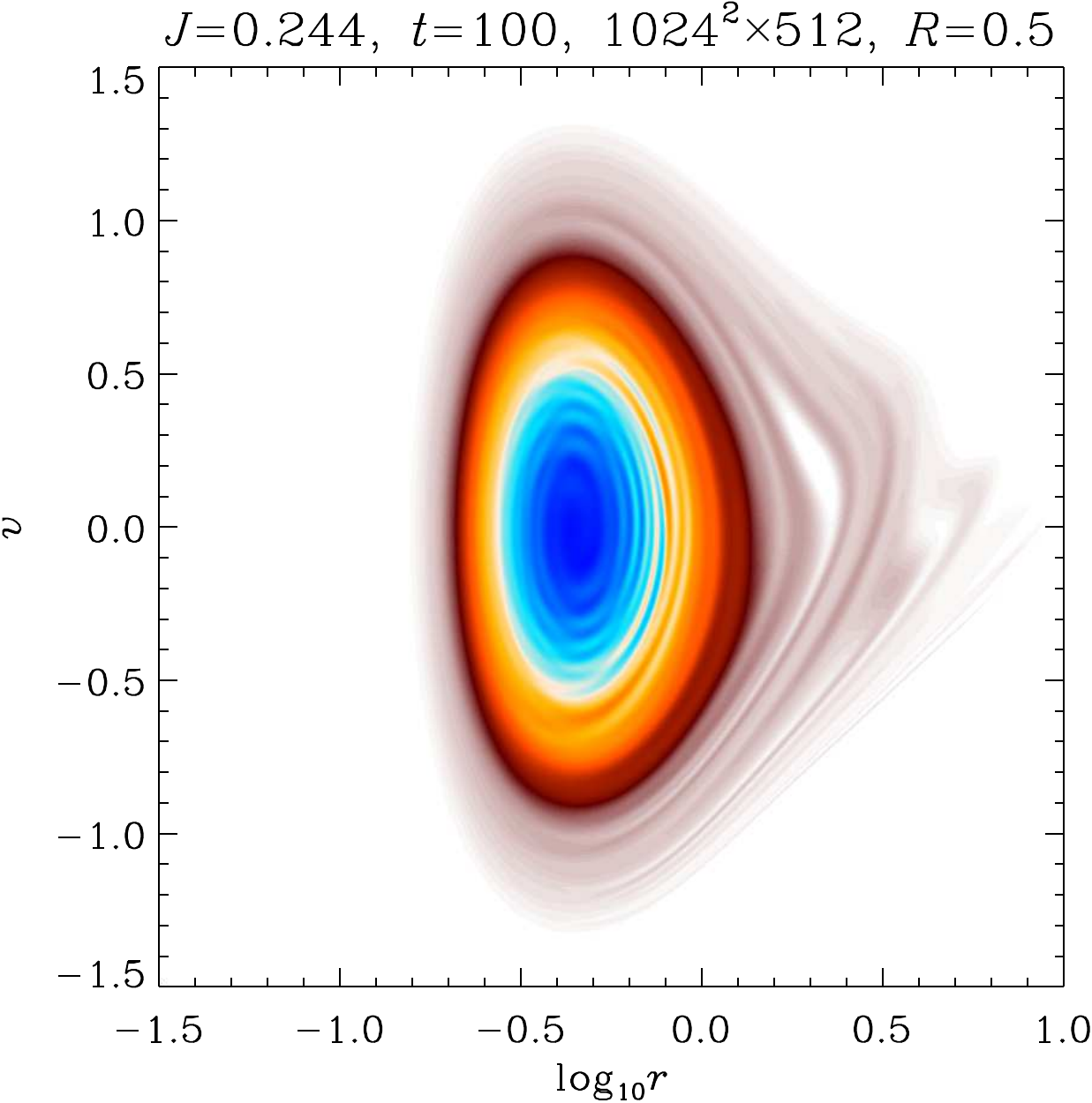}
}
\hbox{
\includegraphics[width=4.25cm]{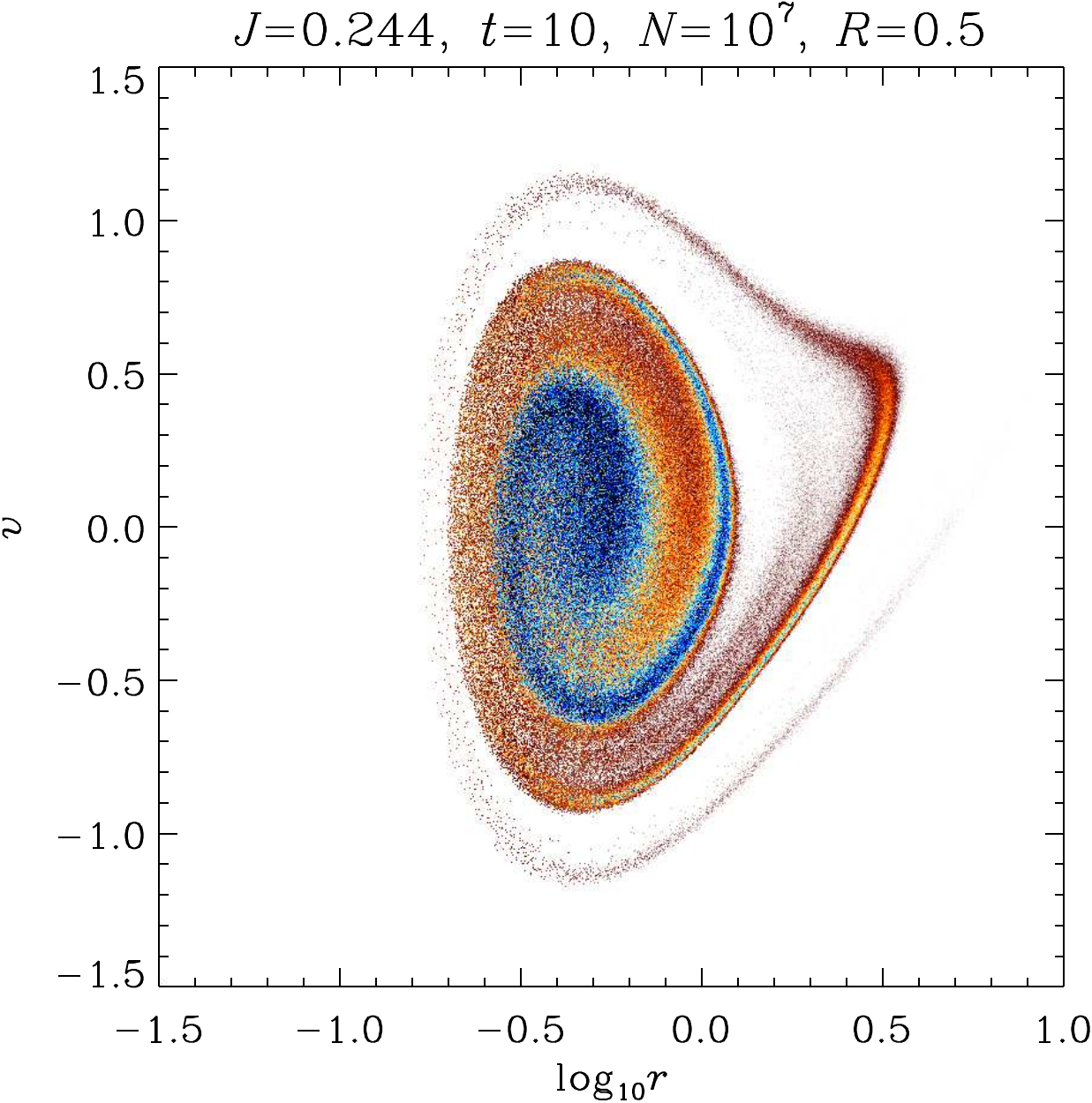}
\includegraphics[width=4.25cm]{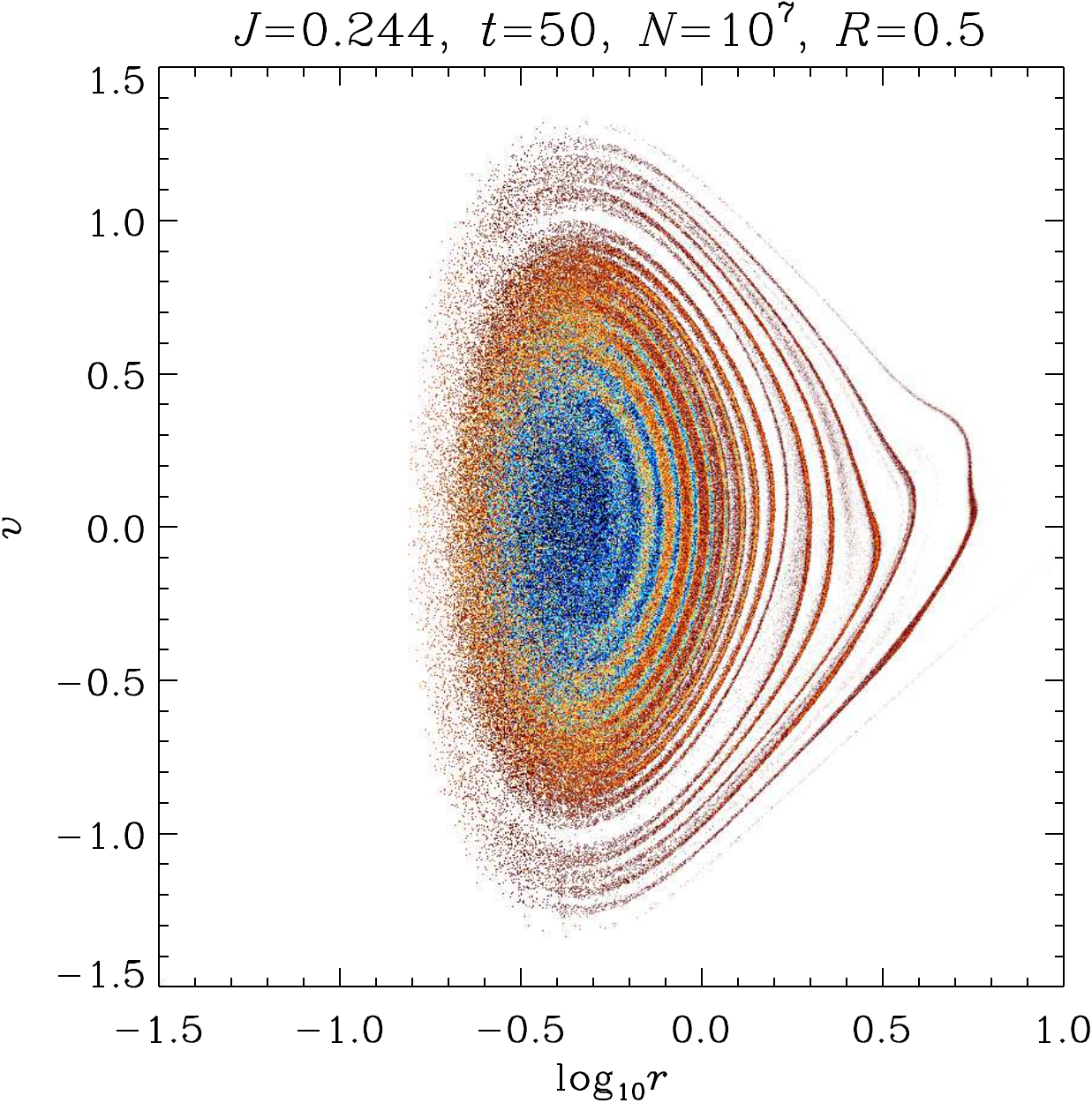}
\includegraphics[width=4.25cm]{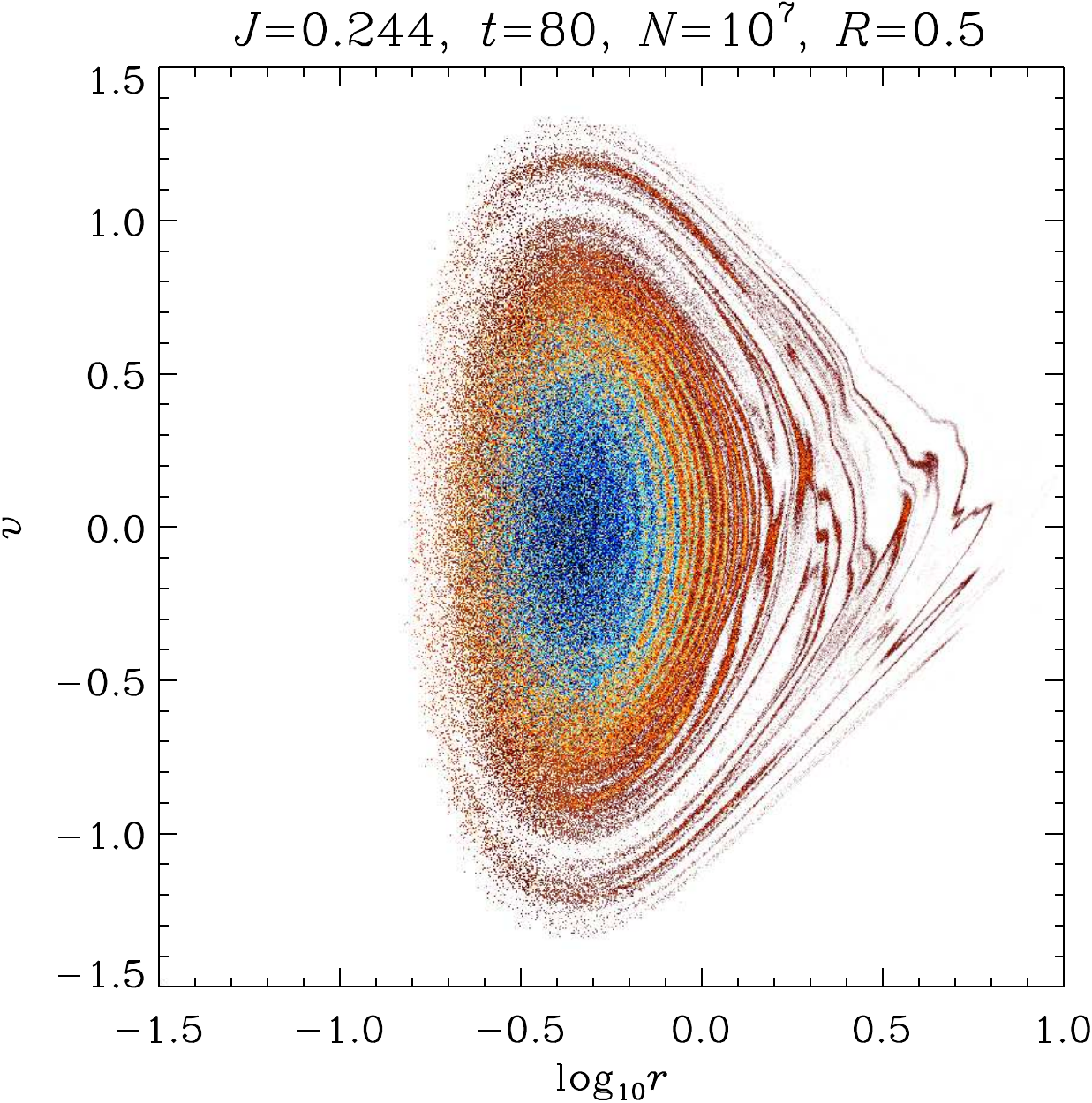}
\includegraphics[width=4.25cm]{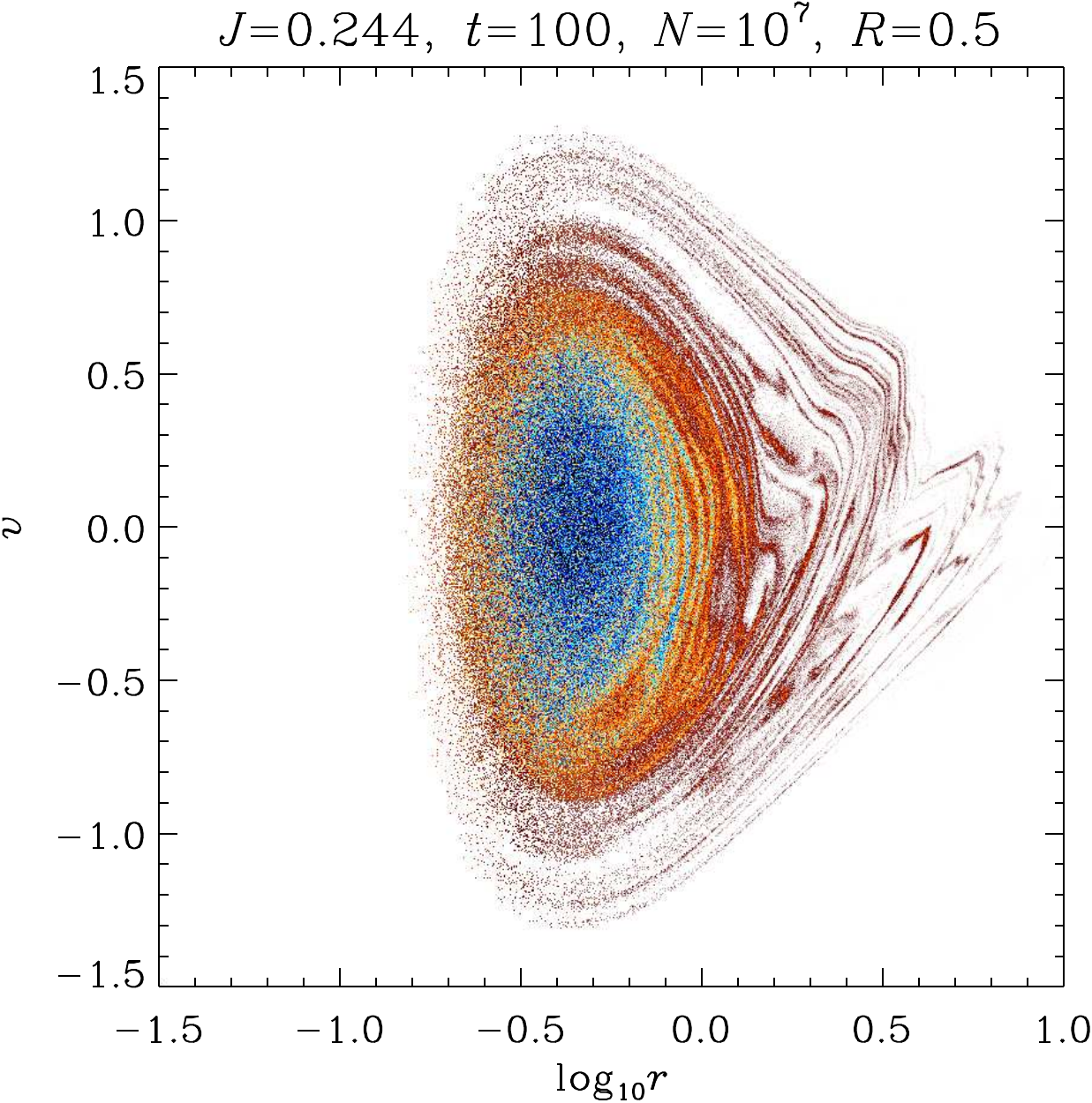}
}
\hbox{
\includegraphics[width=4.25cm]{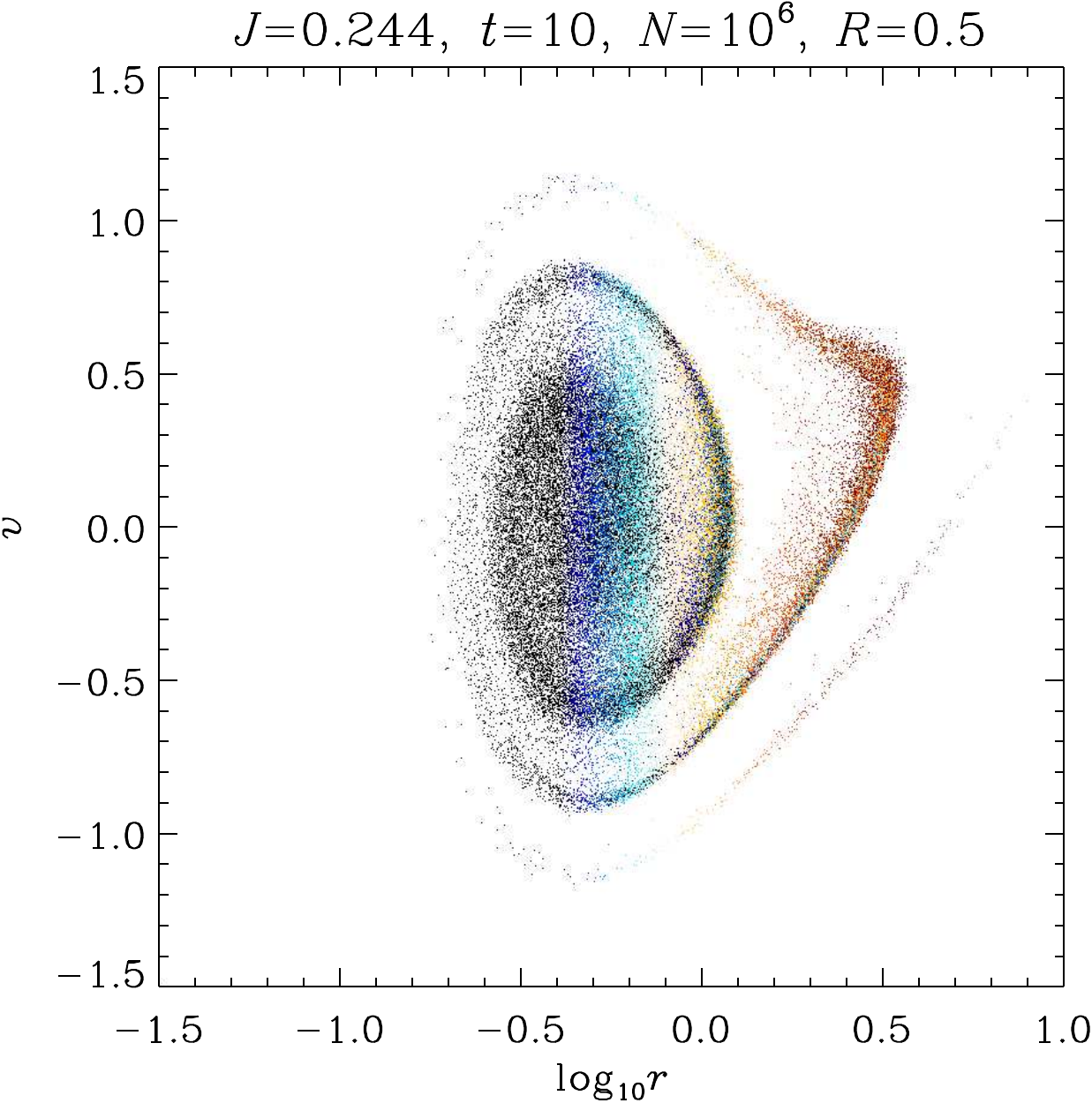}
\includegraphics[width=4.25cm]{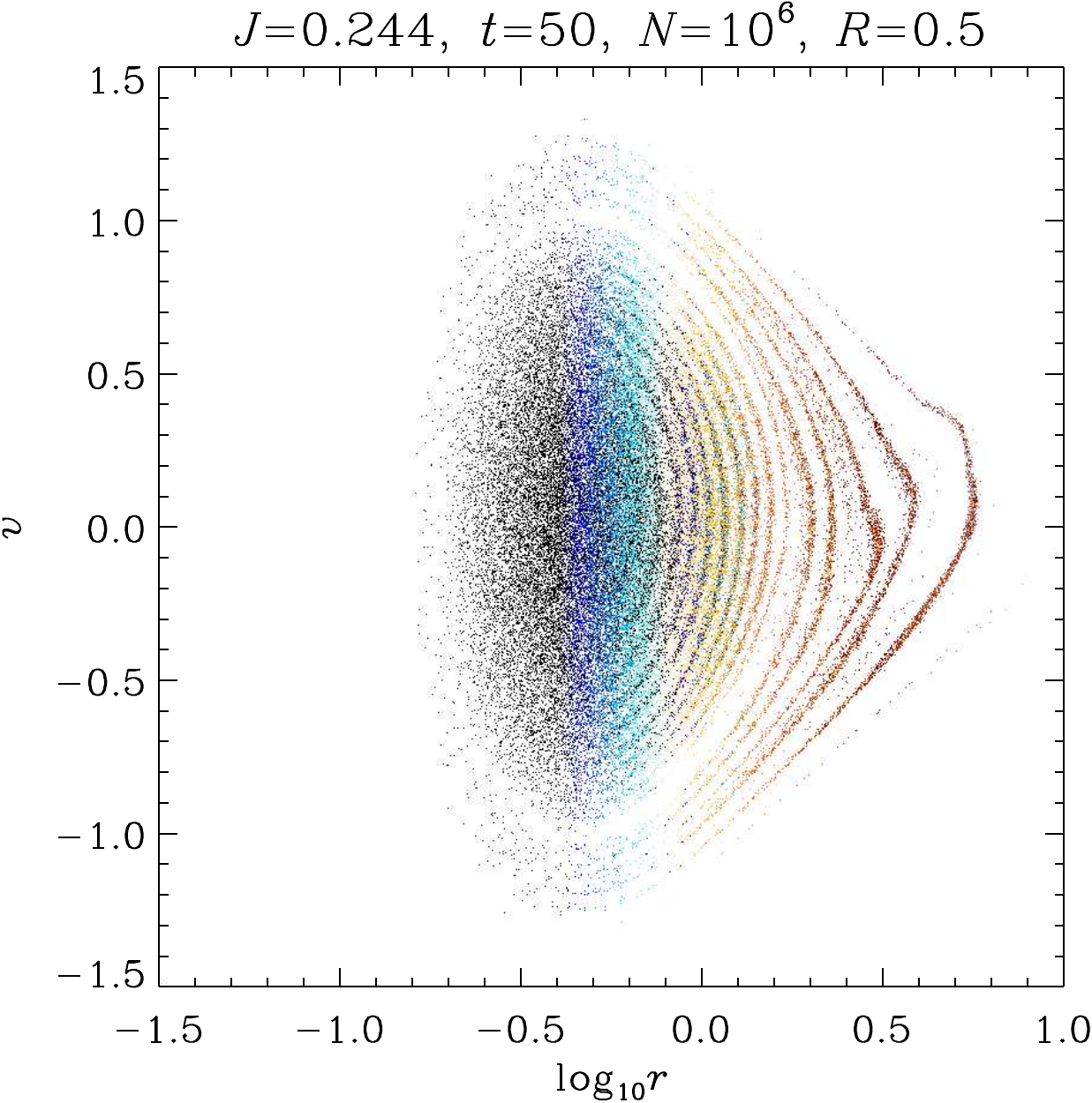}
\includegraphics[width=4.25cm]{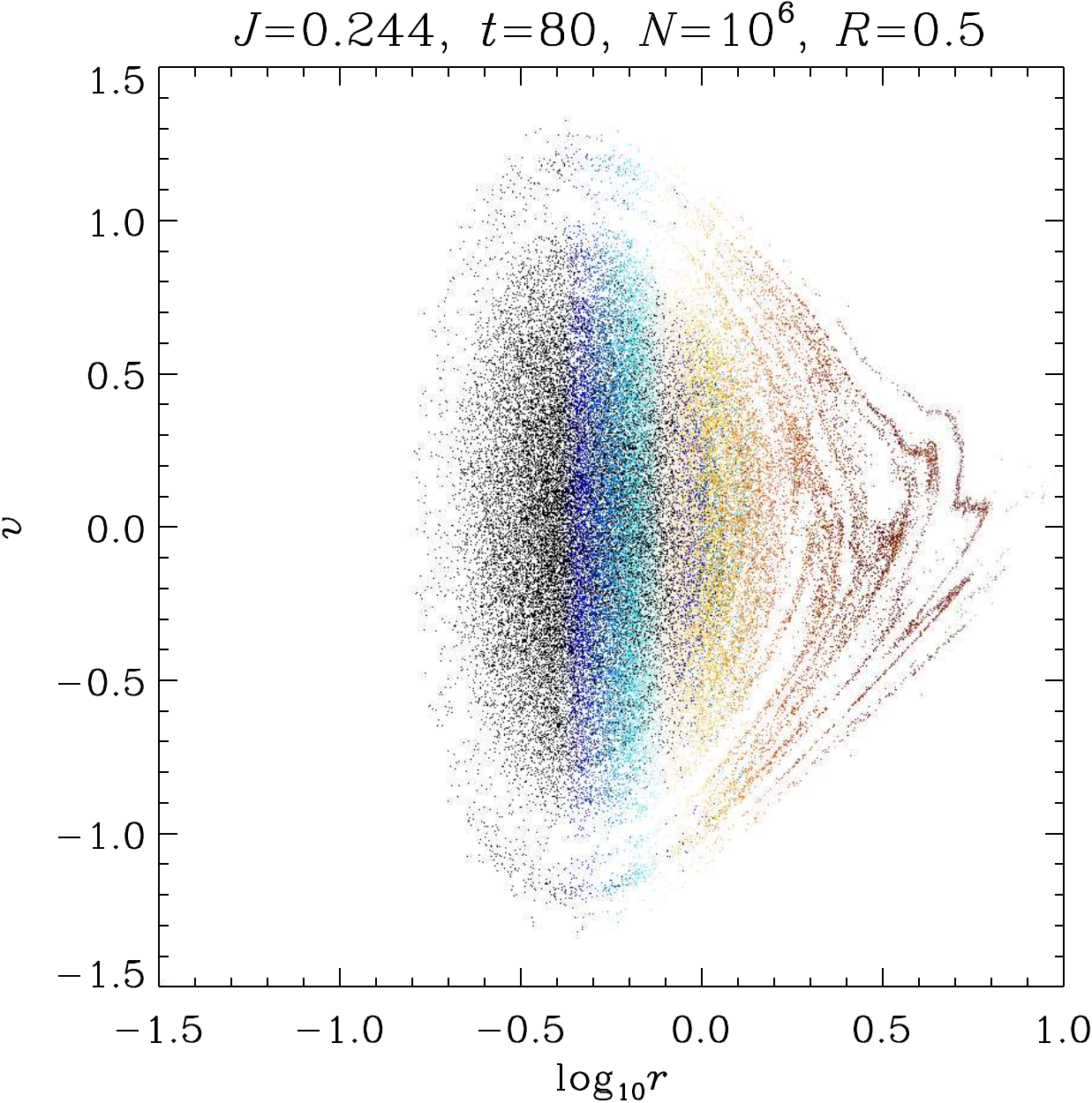}
\includegraphics[width=4.25cm]{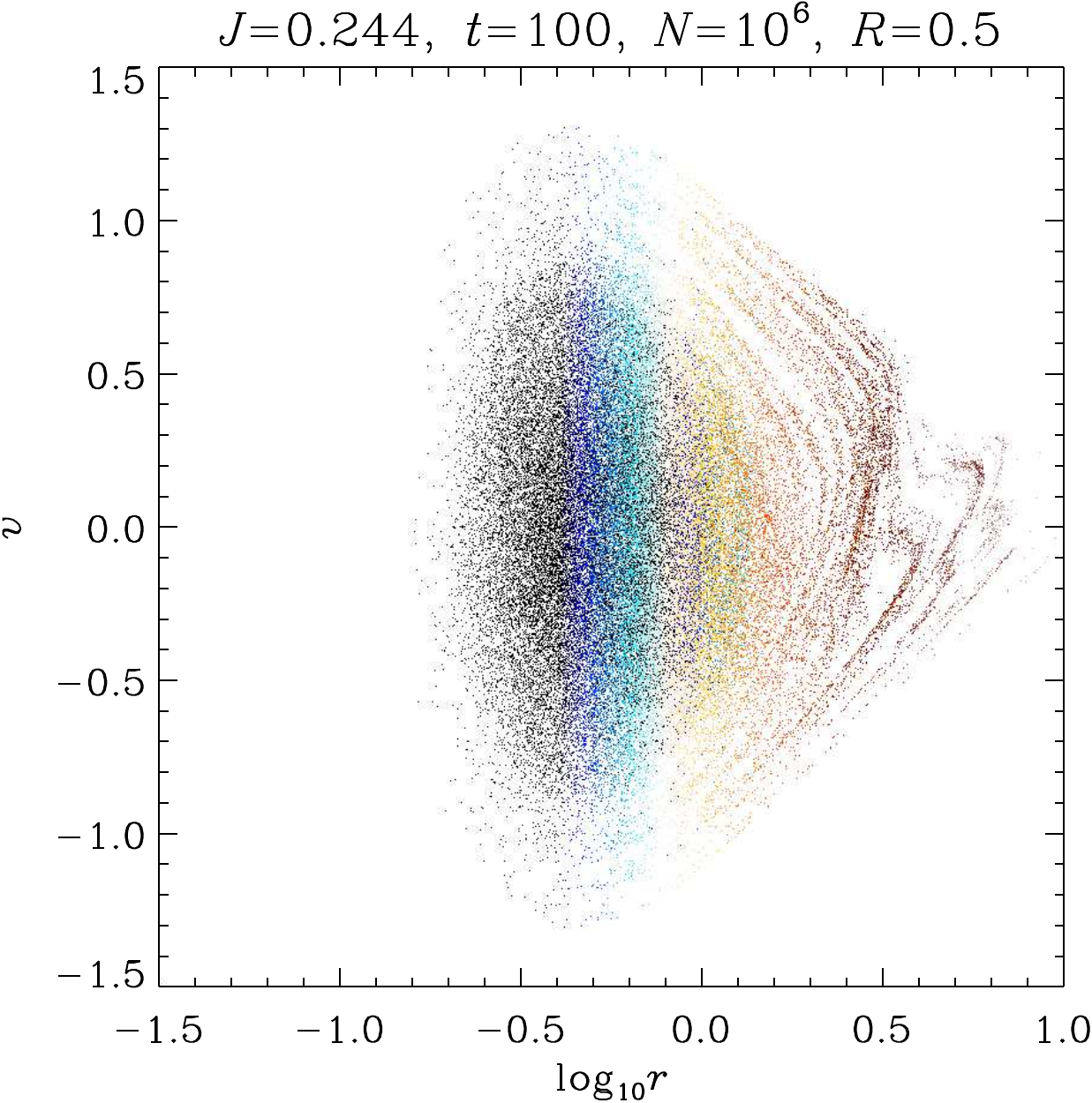}
}
\hbox{
\includegraphics[width=4.25cm]{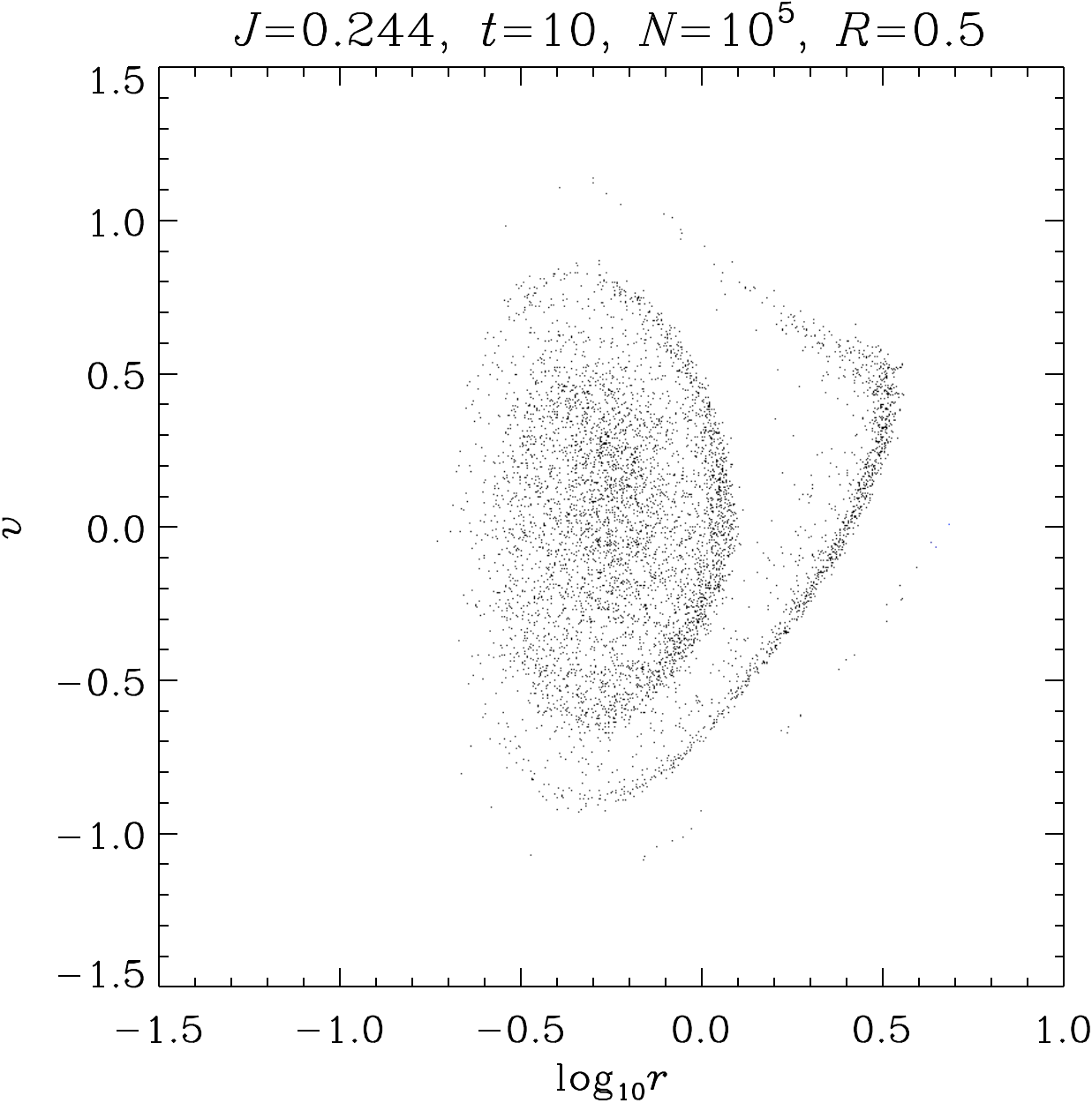}
\includegraphics[width=4.25cm]{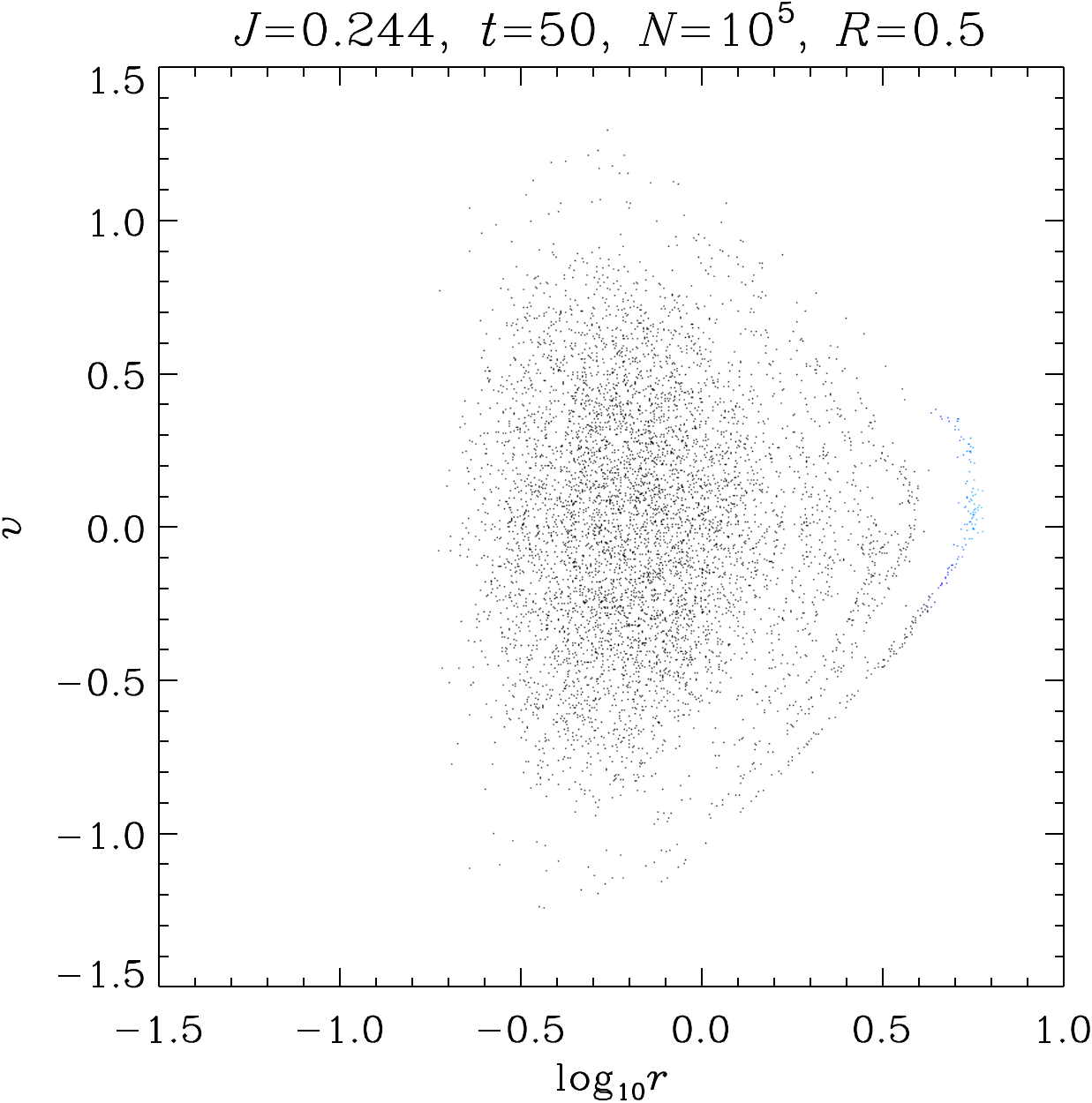}
\includegraphics[width=4.25cm]{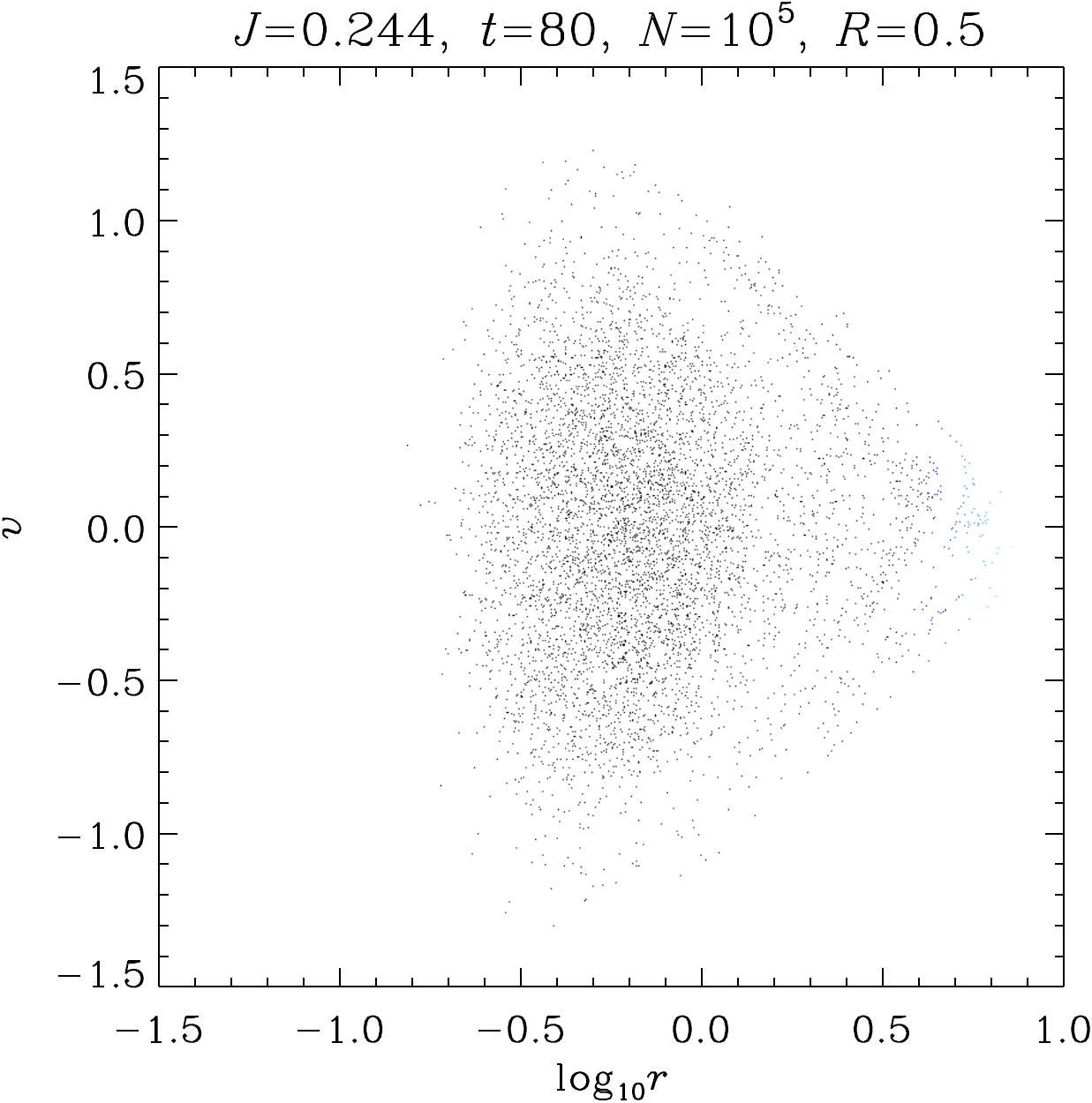}
\includegraphics[width=4.25cm]{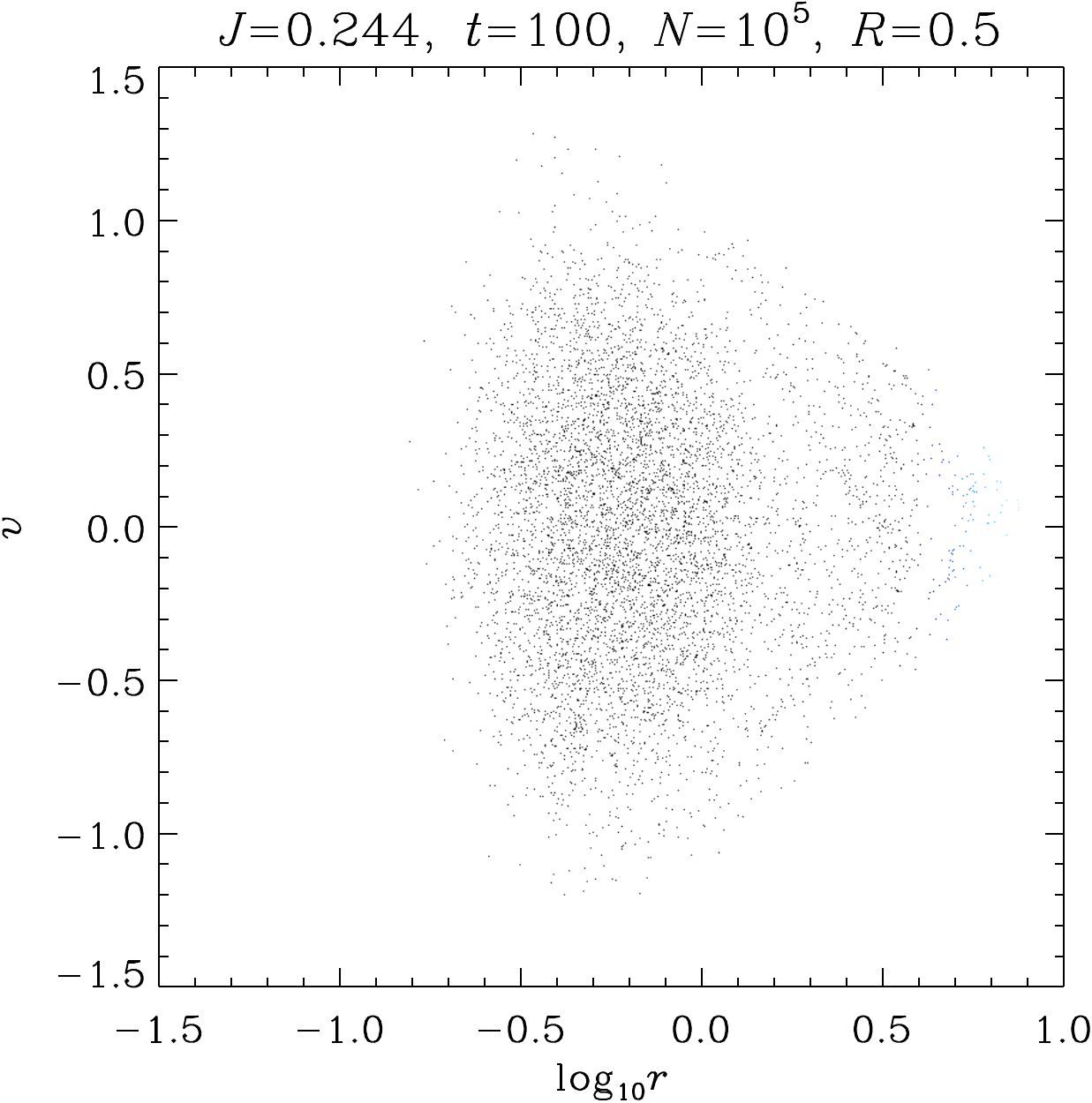}
}
\caption[]{ {\tt VlaSolve} versus {\tt Gadget} in phase-space:
  phase-space density for $R=0.5$ and averaged over $j \in
  I_J=[0.225,0.264]$. Each column of panels corresponds to a given
  value of time $t$, increasing from left to right. The first two
  lines of panels display $f(r,v,j)$ for {\tt VlaSolve} simulations
  with $(N_r,N_v,N_j)=(2048,2048,32)$ and $(1024,1024,512)$
  respectively, while the three bottom lines correspond to the
  $N$-body simulations, with various values of the number of particles
  $N$ as indicated on each panel. Note that the {\tt VlaSolve}
  simulation with $(N_r,N_v,N_j)=(2048,2048,32)$ has only one angular
  momentum slice, $J=0.244$, in the interval $I_J$, so there is no
  blurring of the filamentary details of $f(r,v,j)$ on the left side
  of the peak of the distribution function contrarily to the other
  cases. In the $N$-body case, $f(r,v,j)$ was computed on the same
  mesh as the $(1024,1024,32)$ {\tt VlaSolve} simulation using nearest
  grid point interpolation, which explains the artefacts on the color
  pattern in the last two lines of panels.}
\label{fig:0v5_12}
\end{figure*}

\begin{figure*}
\hbox{
\includegraphics[width=4.25cm]{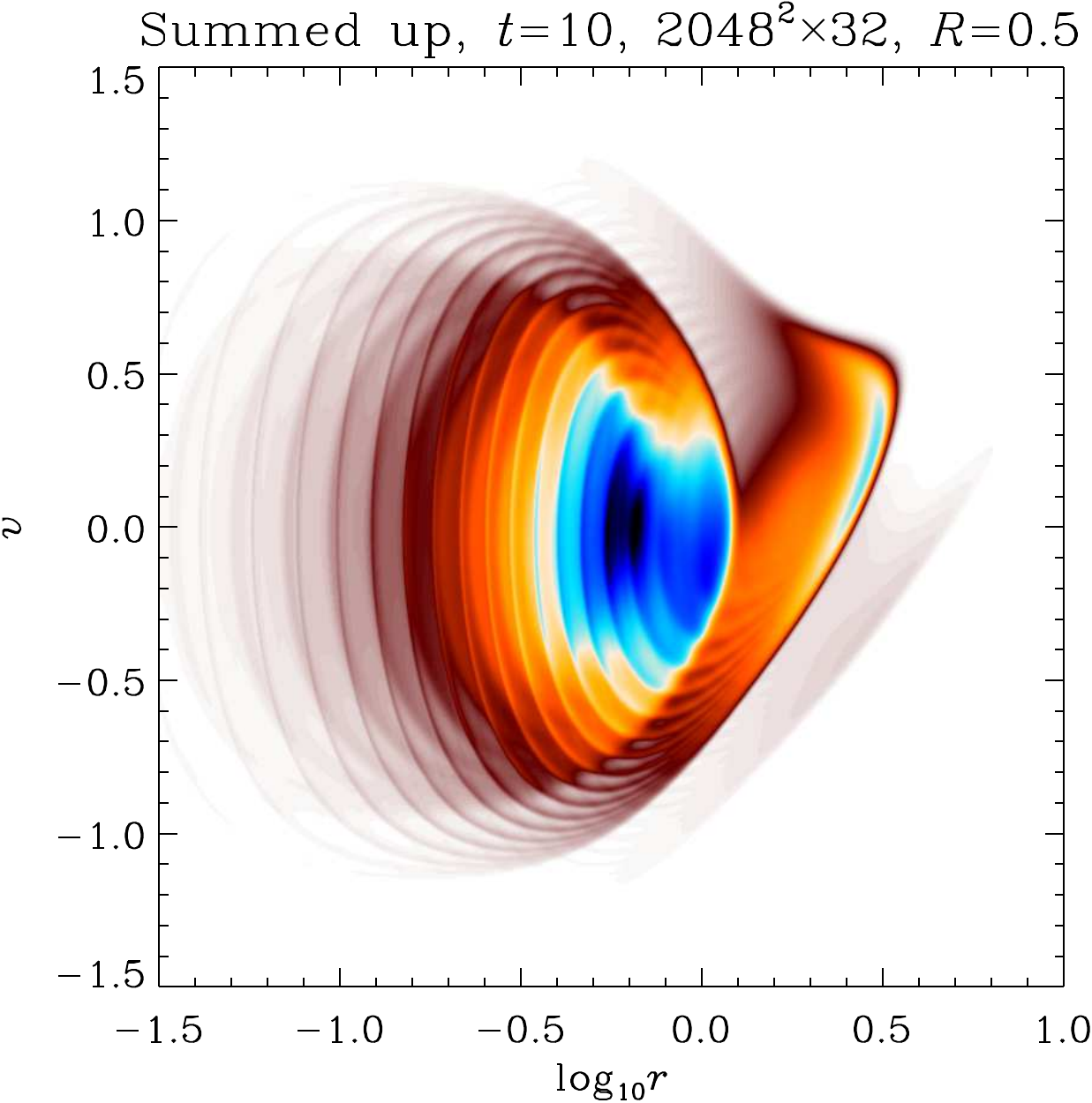}
\includegraphics[width=4.25cm]{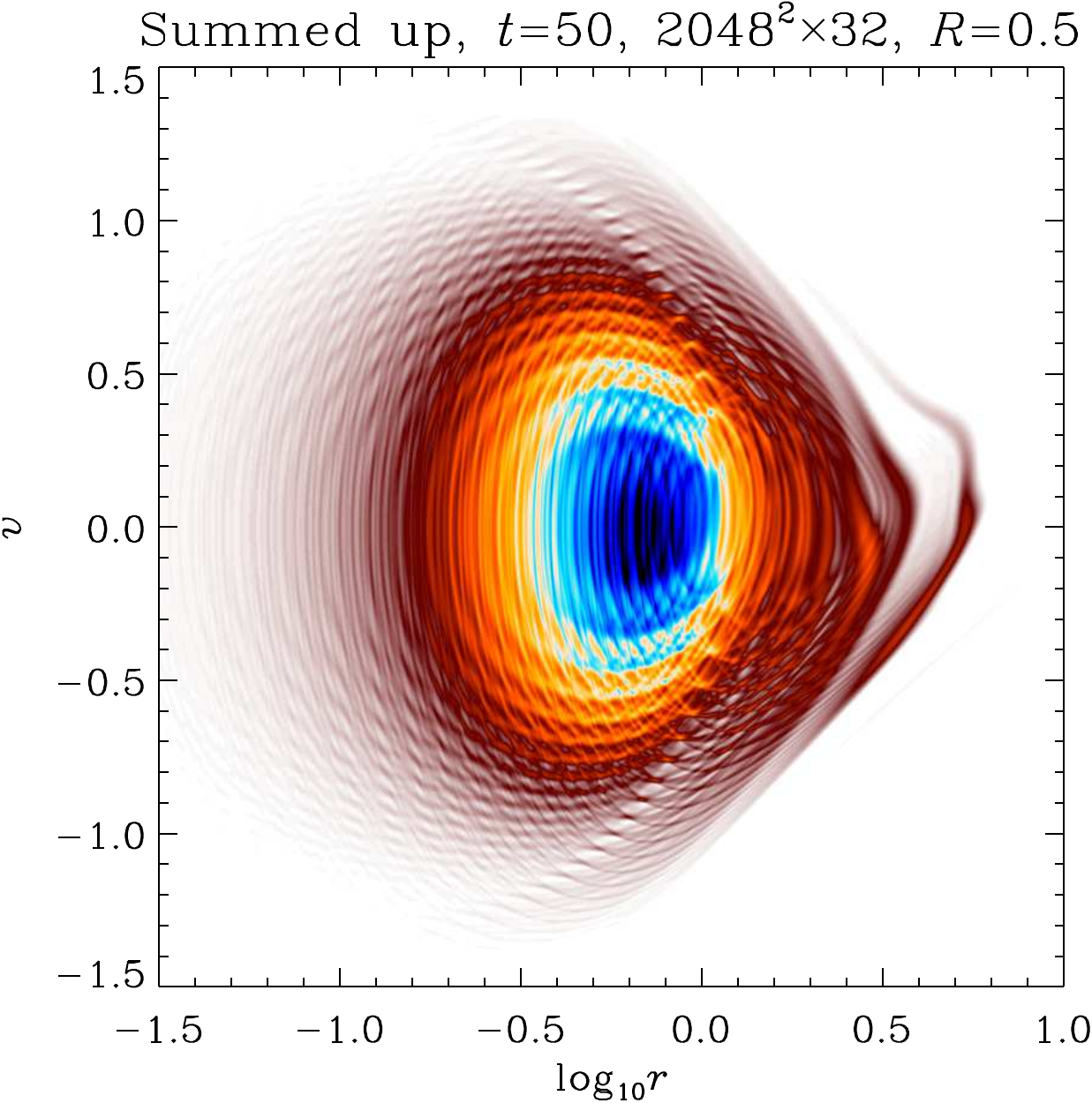}
\includegraphics[width=4.25cm]{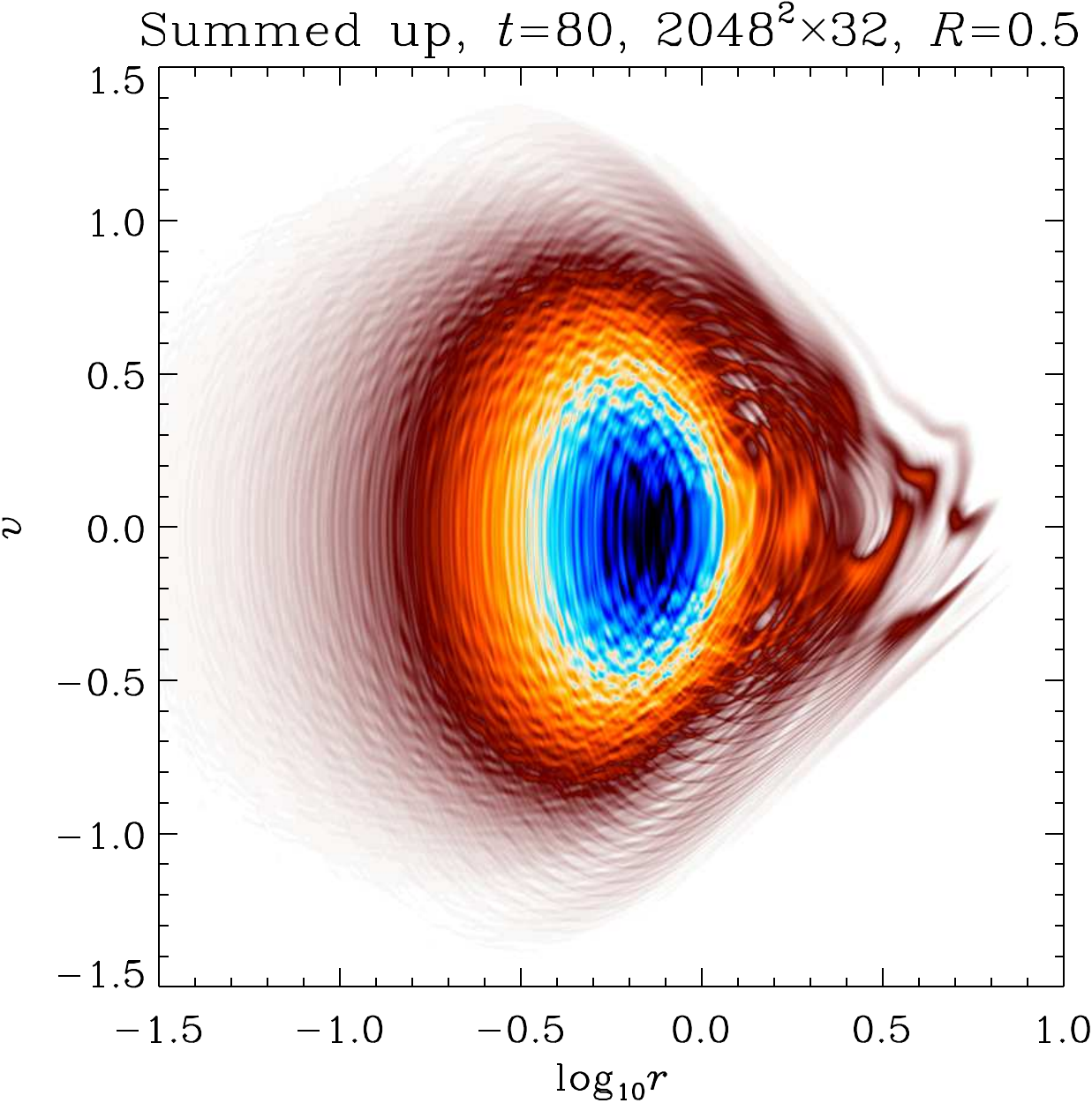}
\includegraphics[width=4.25cm]{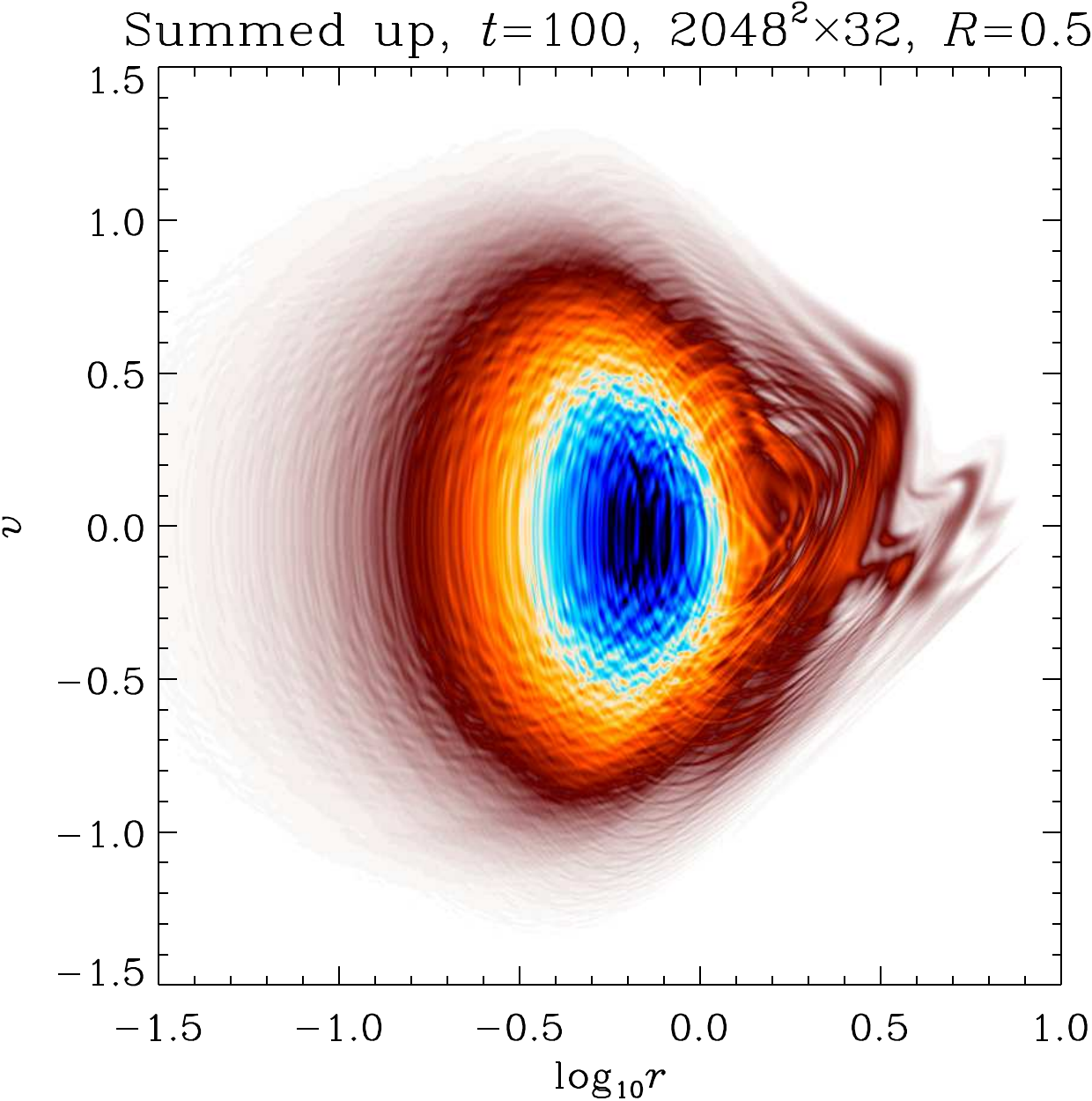}
}
\hbox{
\includegraphics[width=4.25cm]{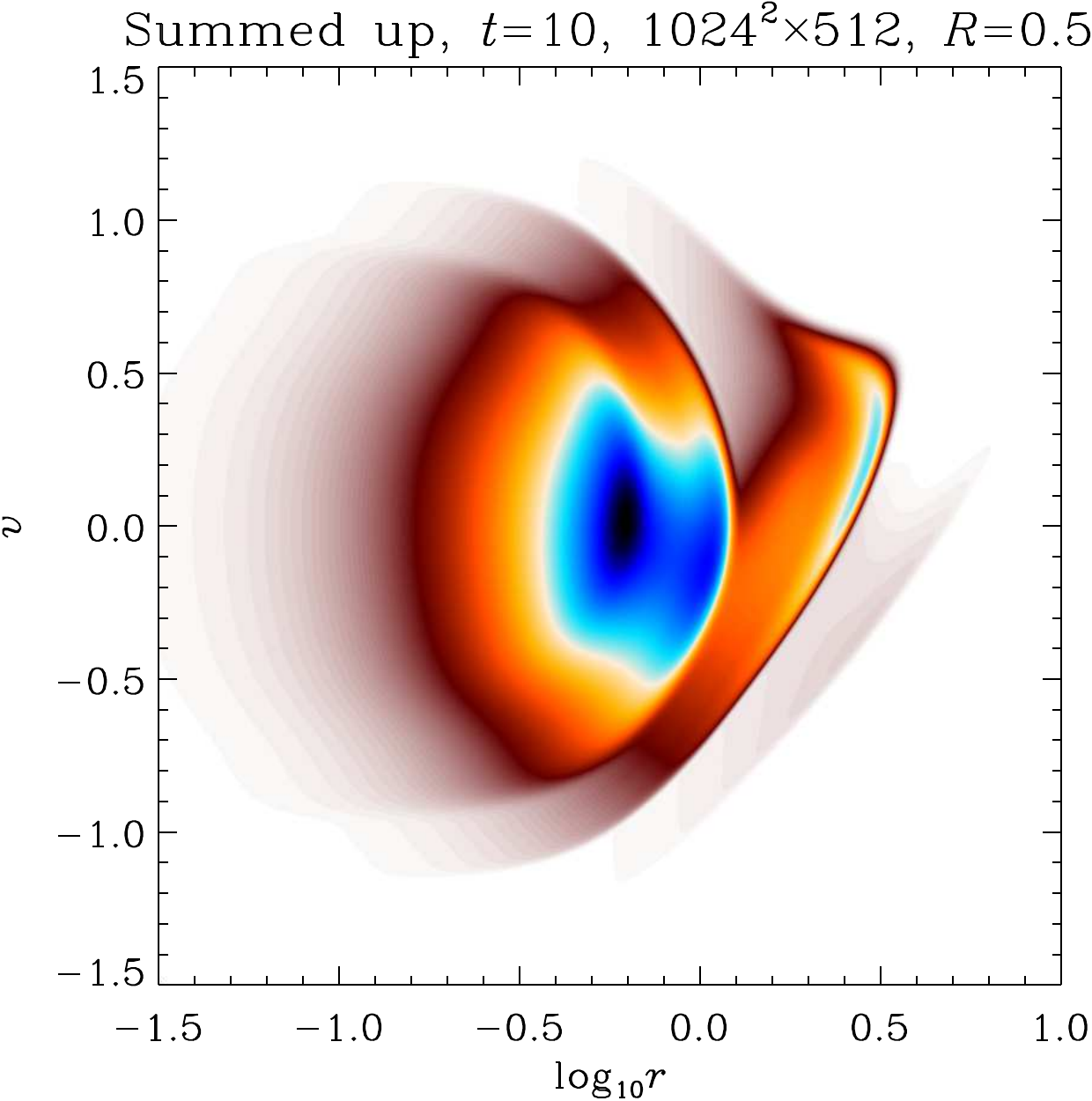}
\includegraphics[width=4.25cm]{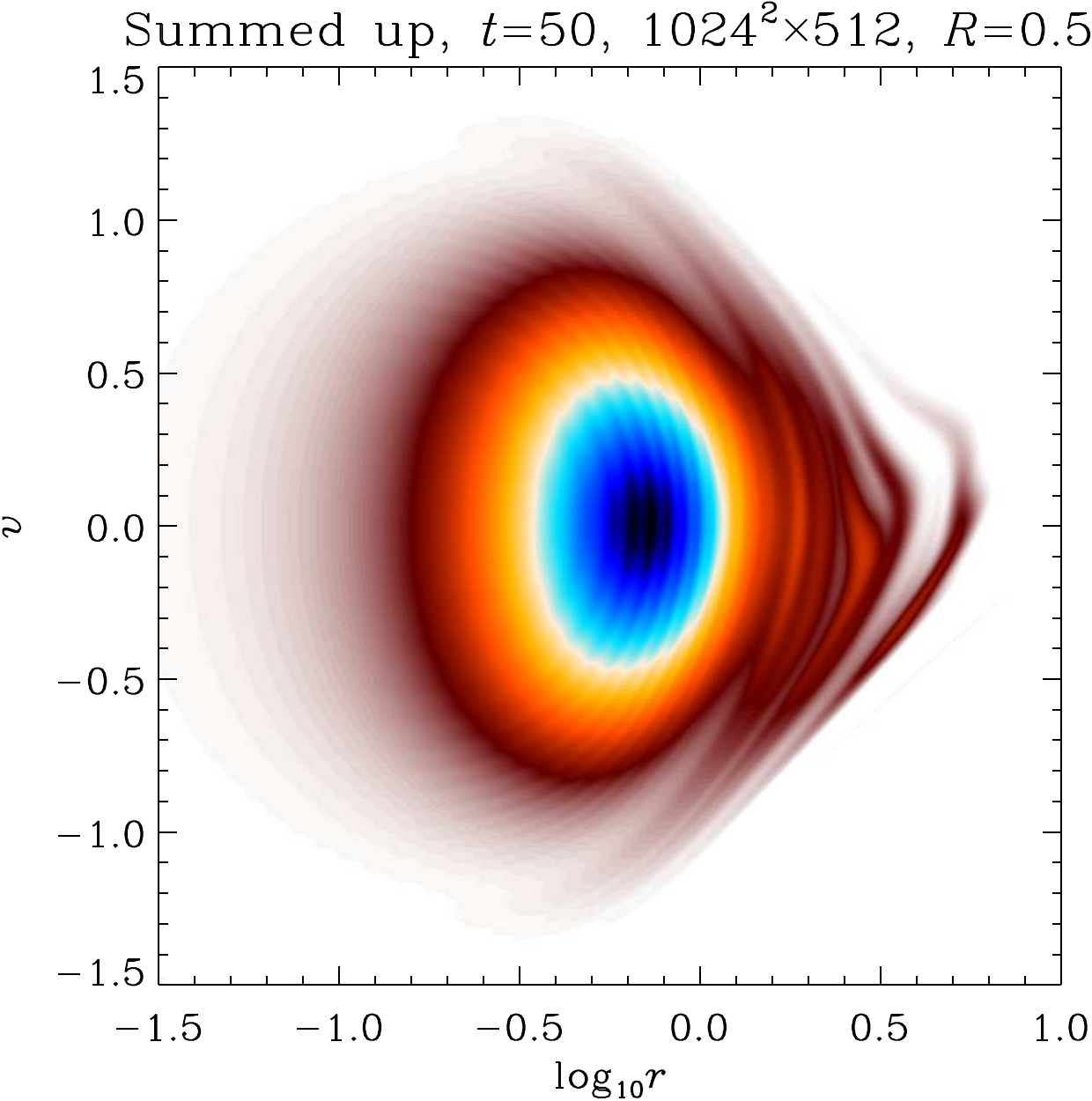}
\includegraphics[width=4.25cm]{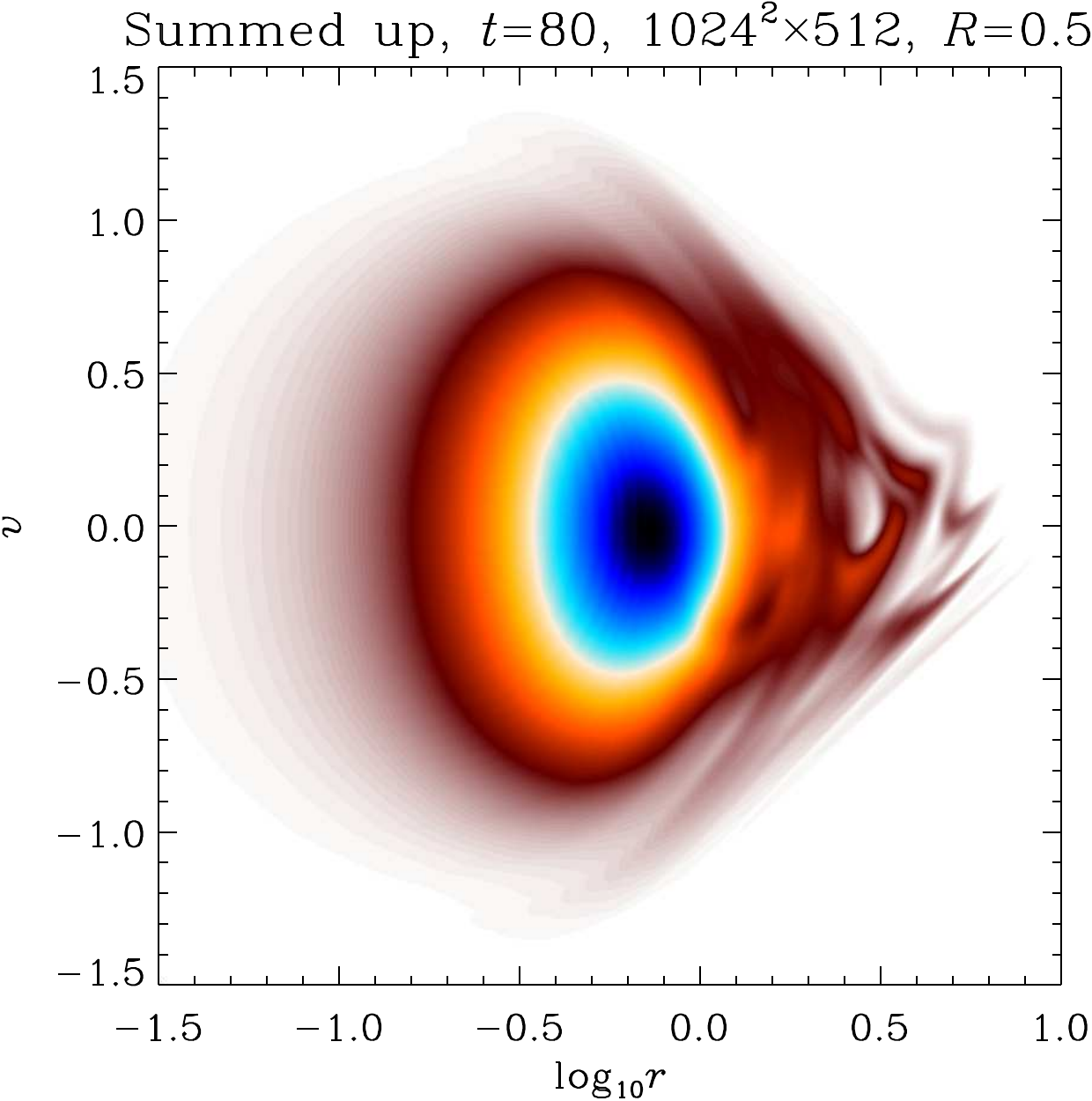}
\includegraphics[width=4.25cm]{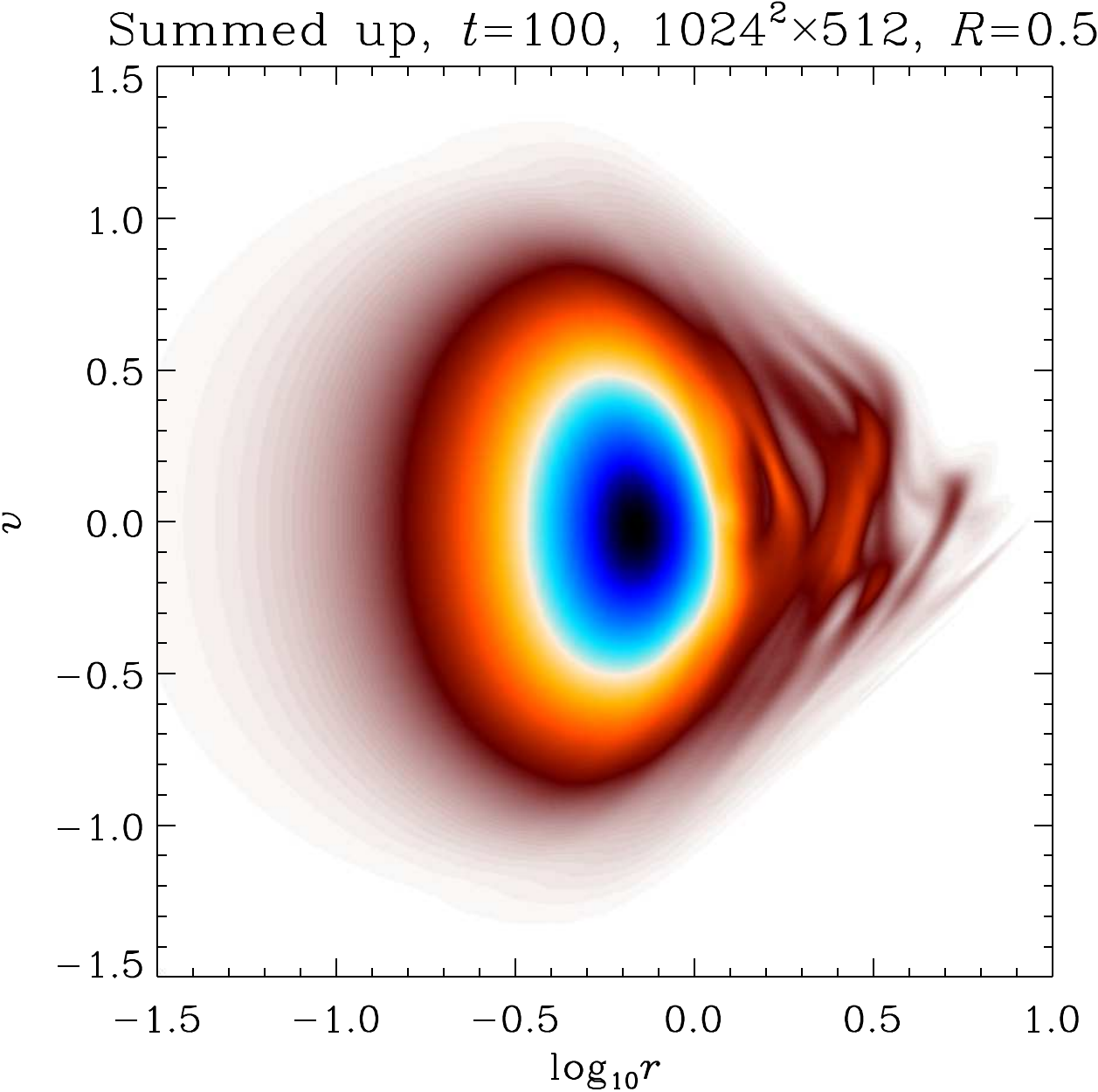}
}
\hbox{
\includegraphics[width=4.25cm]{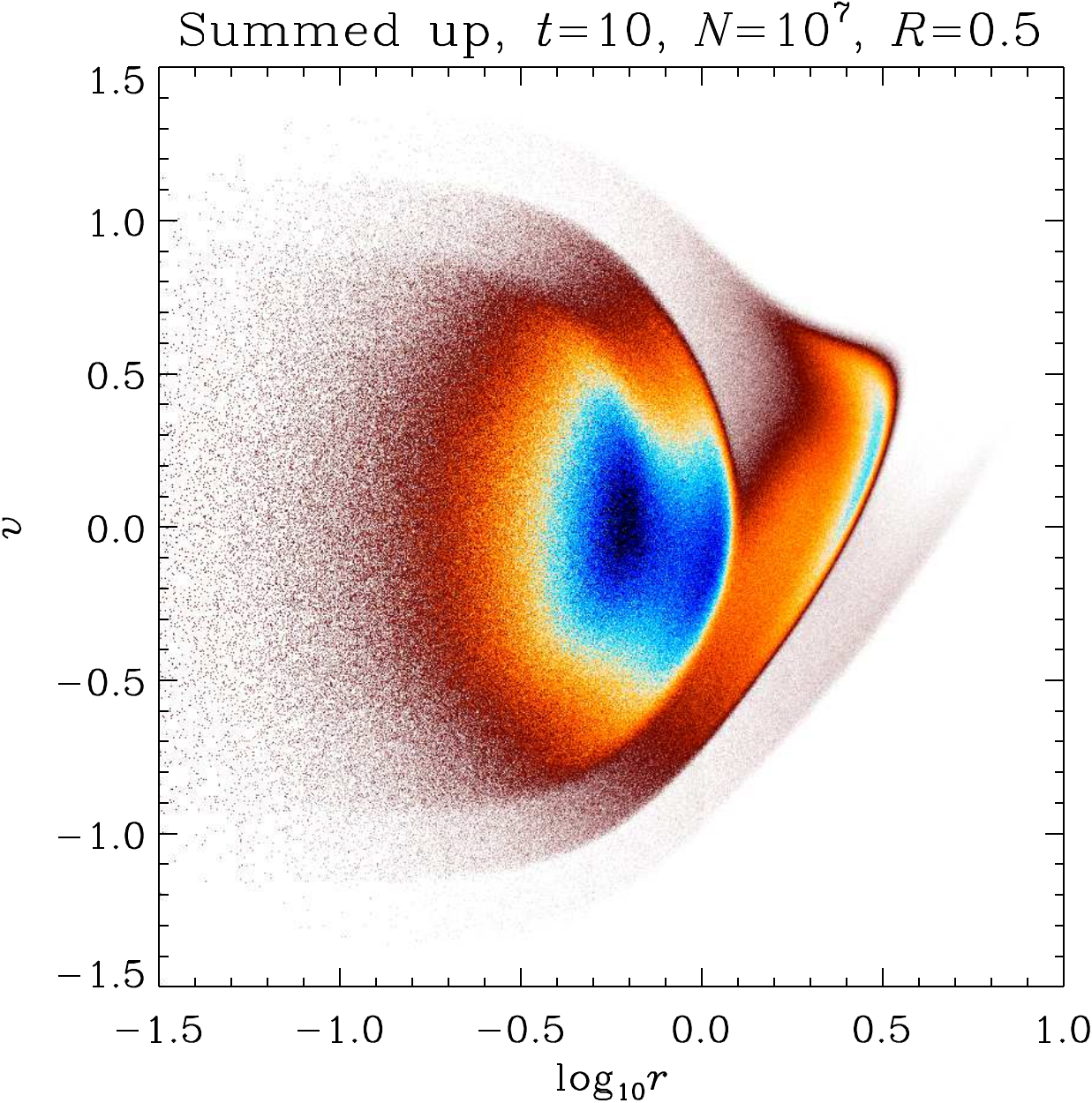}
\includegraphics[width=4.25cm]{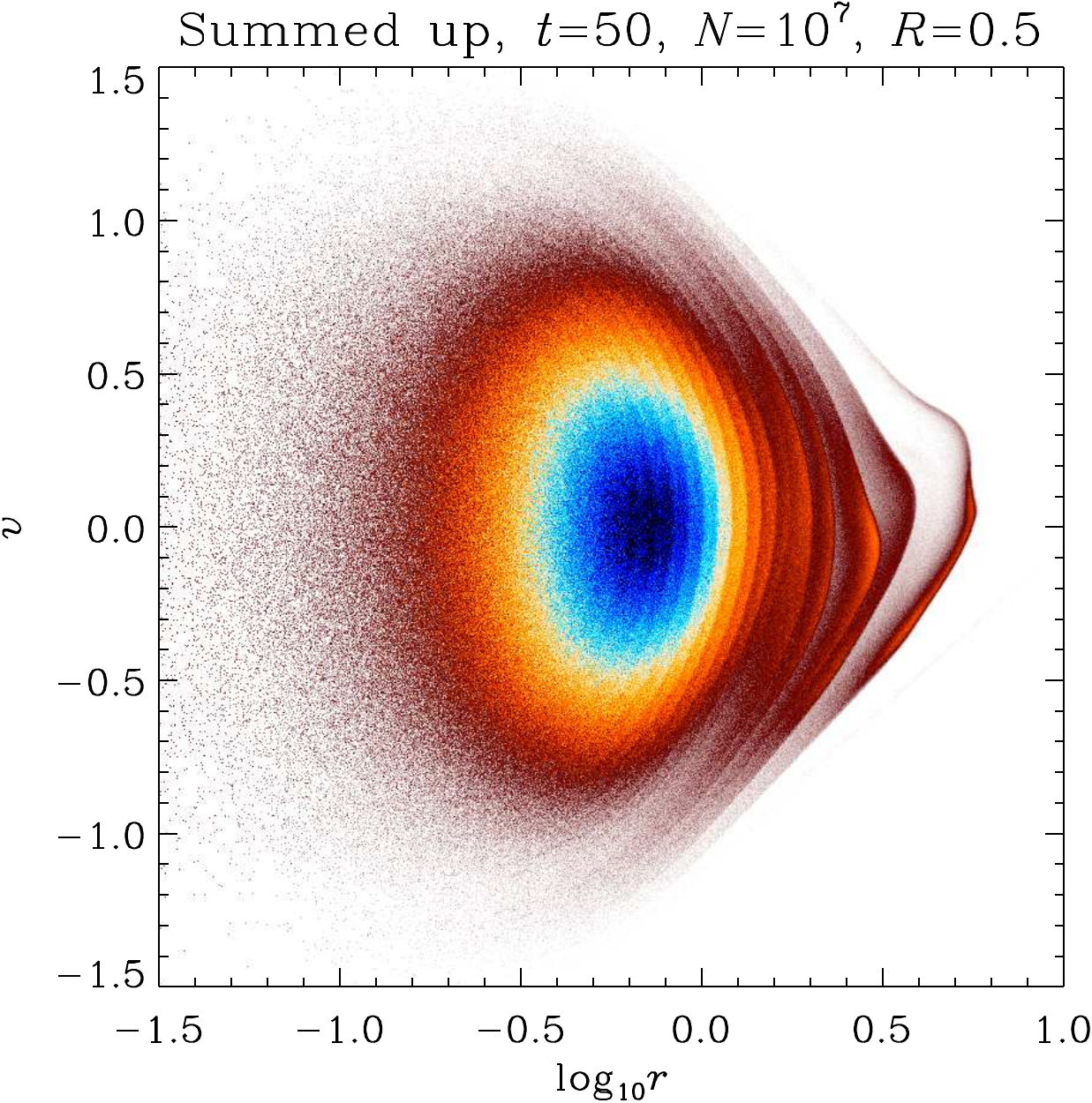}
\includegraphics[width=4.25cm]{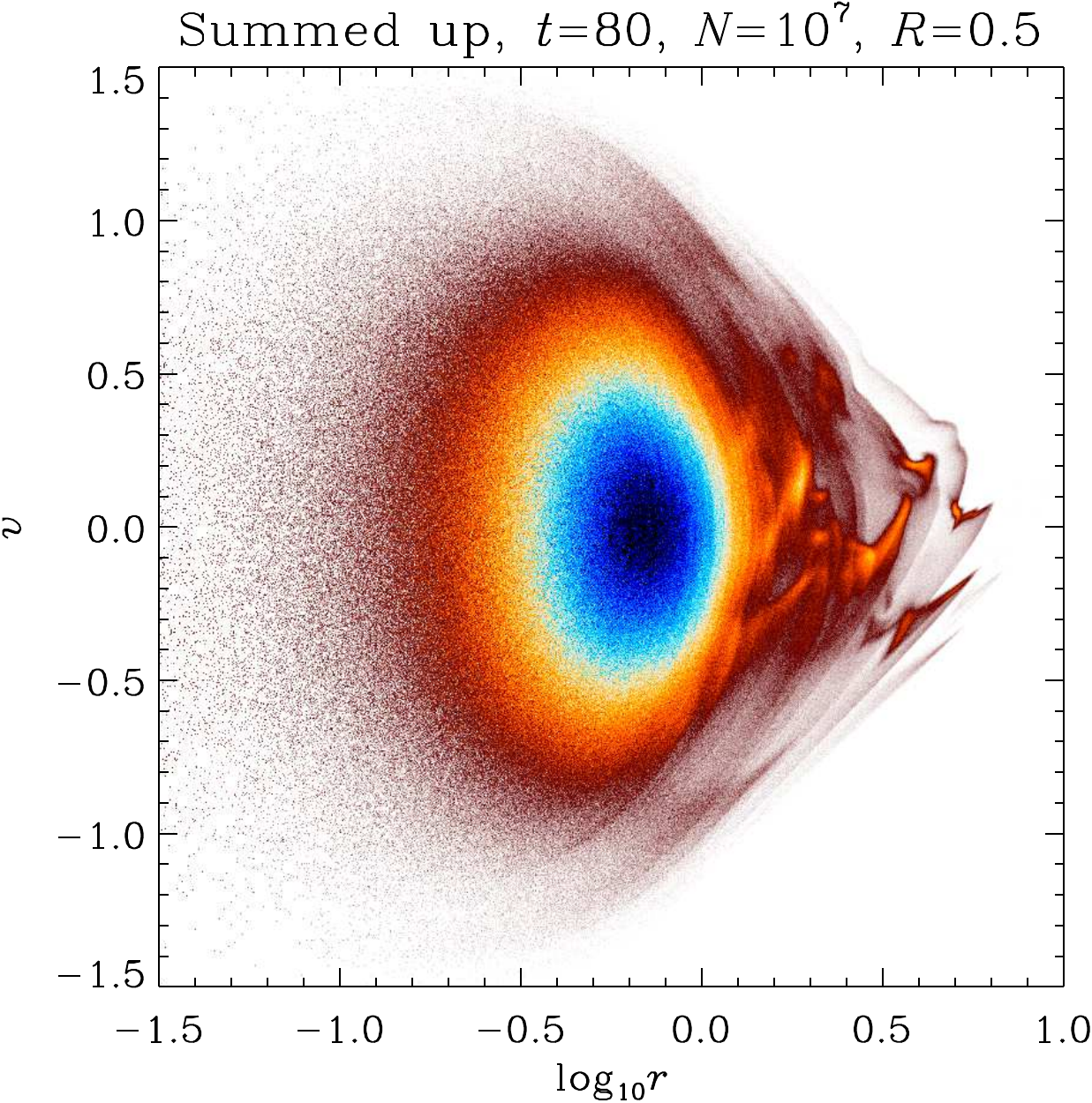}
\includegraphics[width=4.25cm]{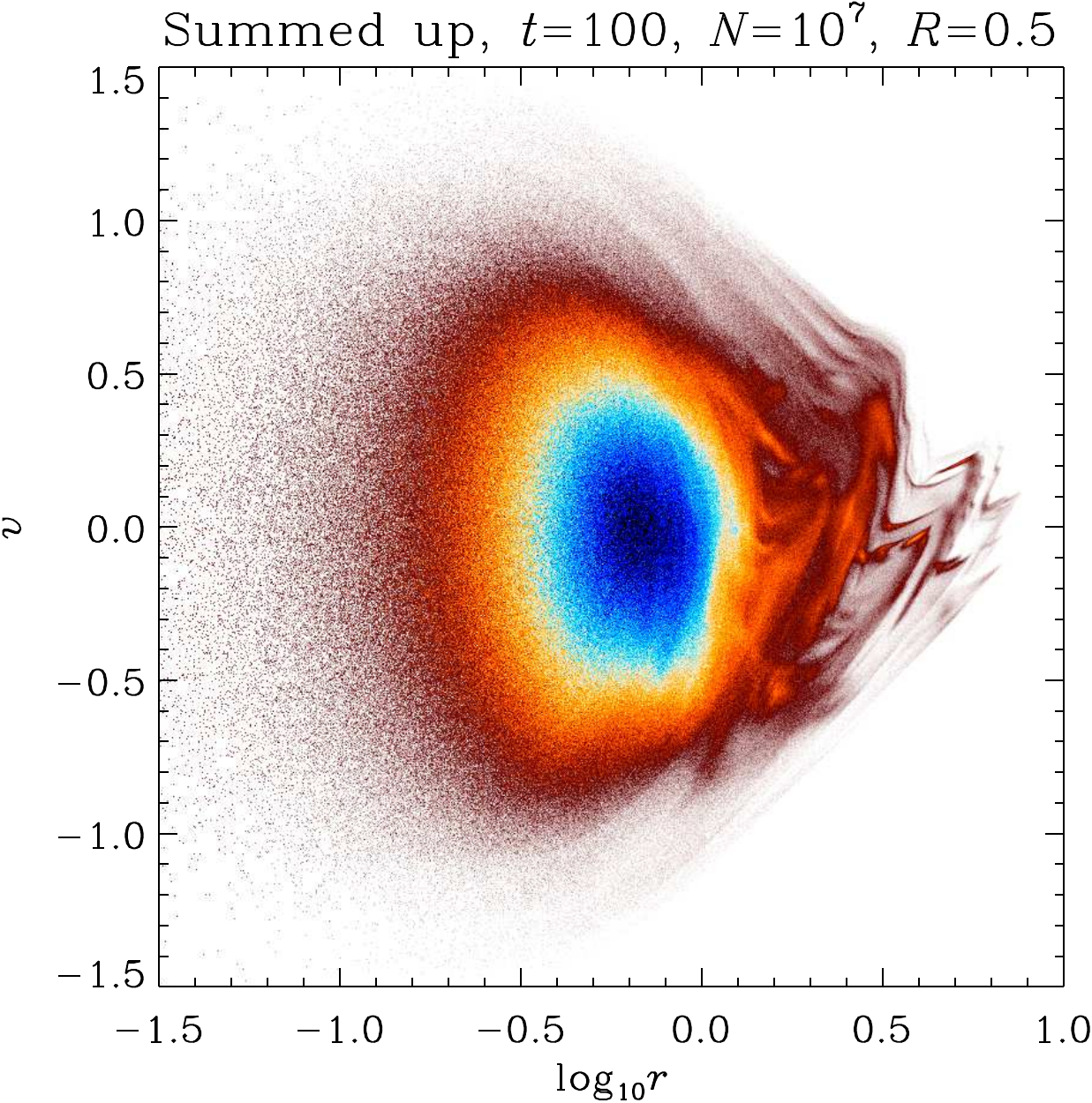}
}
\hbox{
\includegraphics[width=4.25cm]{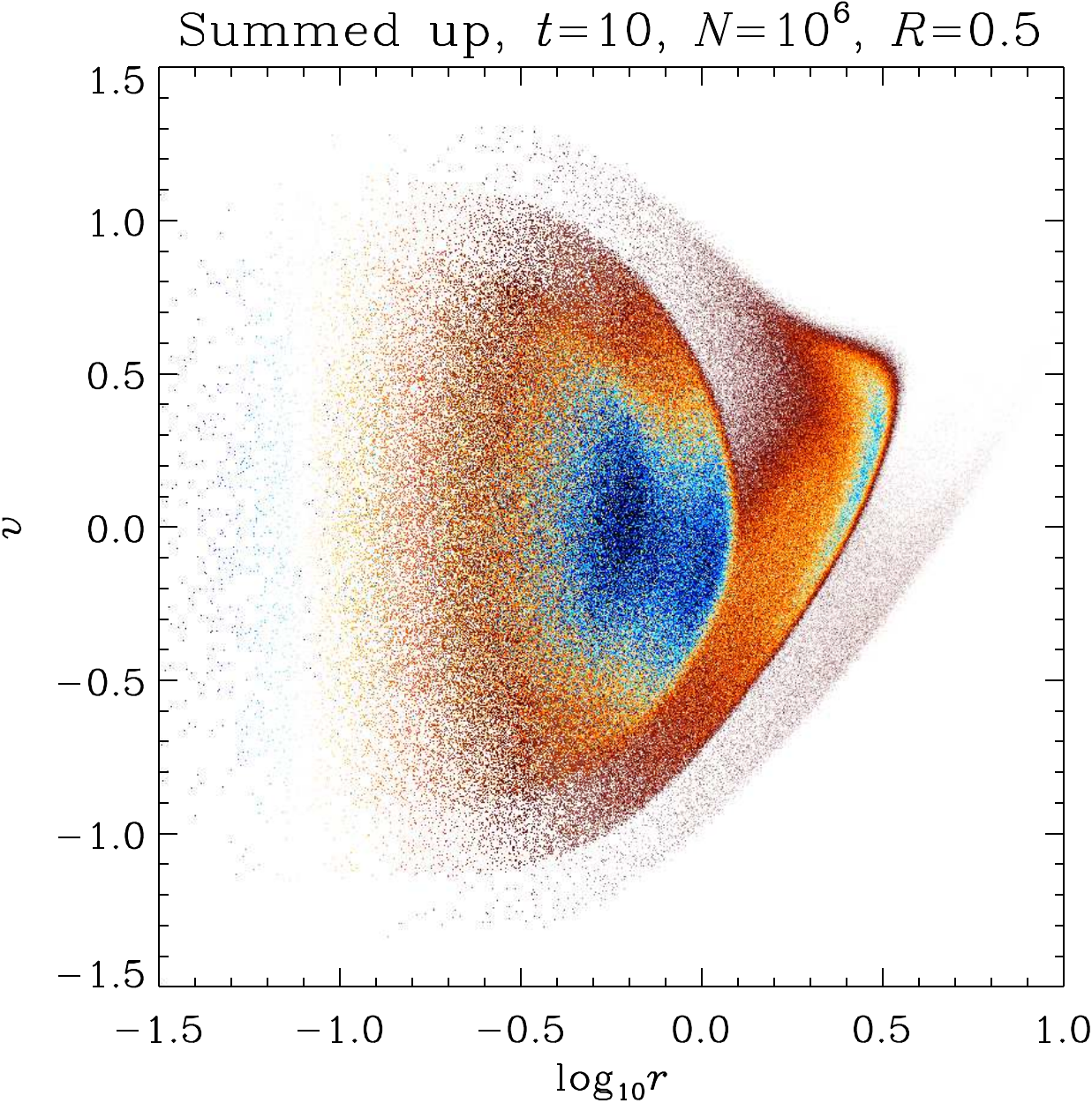}
\includegraphics[width=4.25cm]{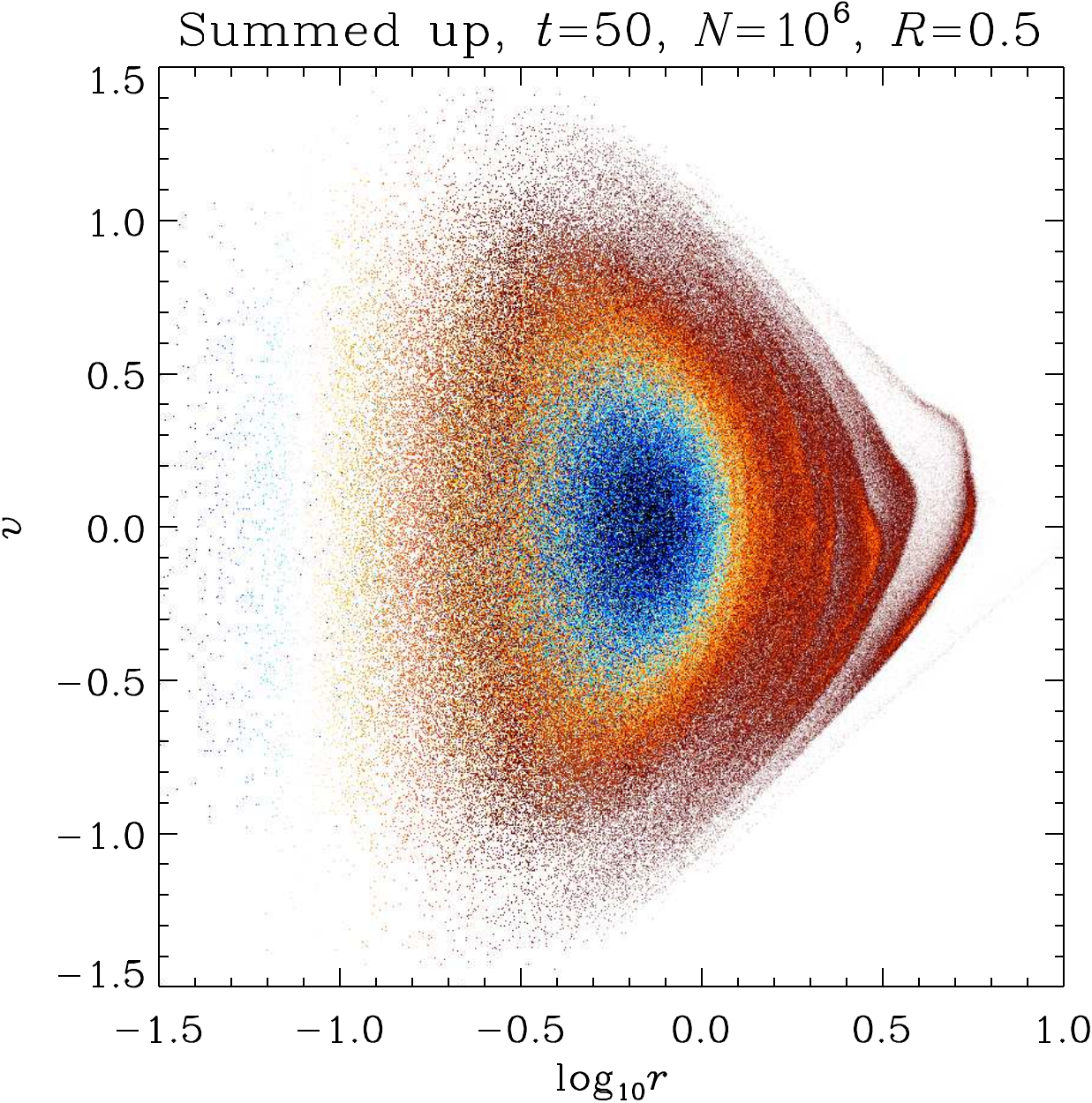}
\includegraphics[width=4.25cm]{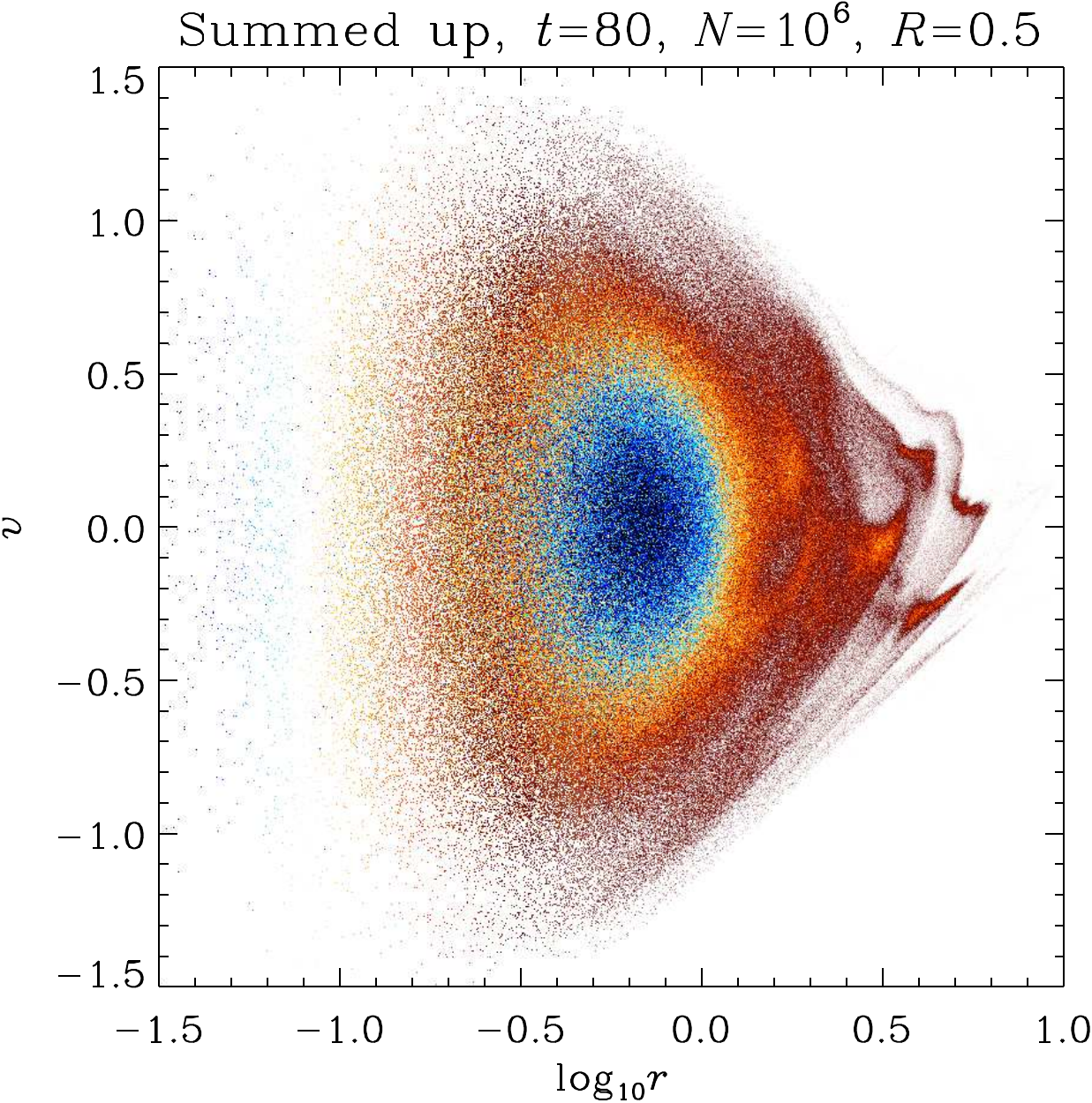}
\includegraphics[width=4.25cm]{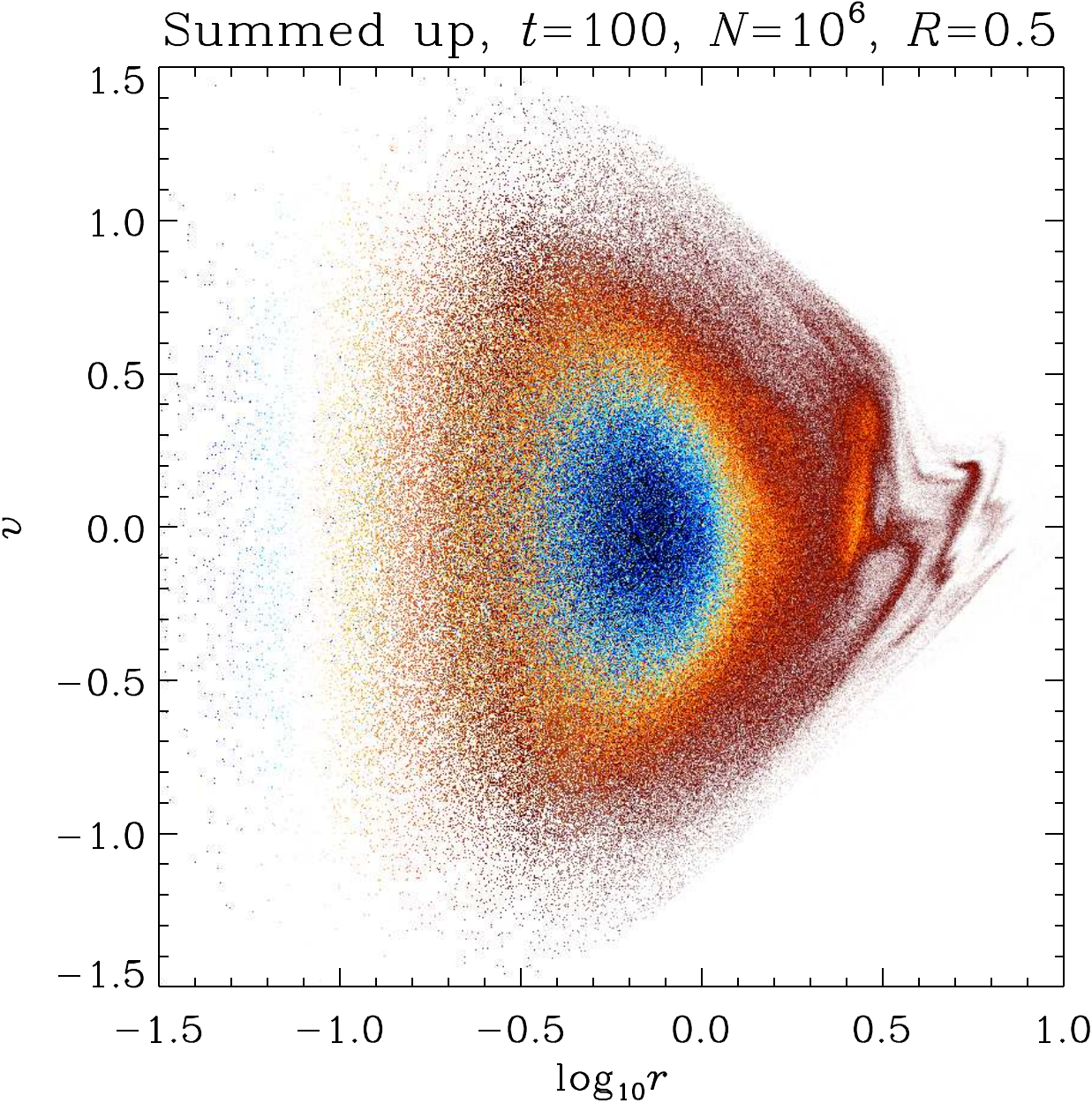}
}
\hbox{
\includegraphics[width=4.25cm]{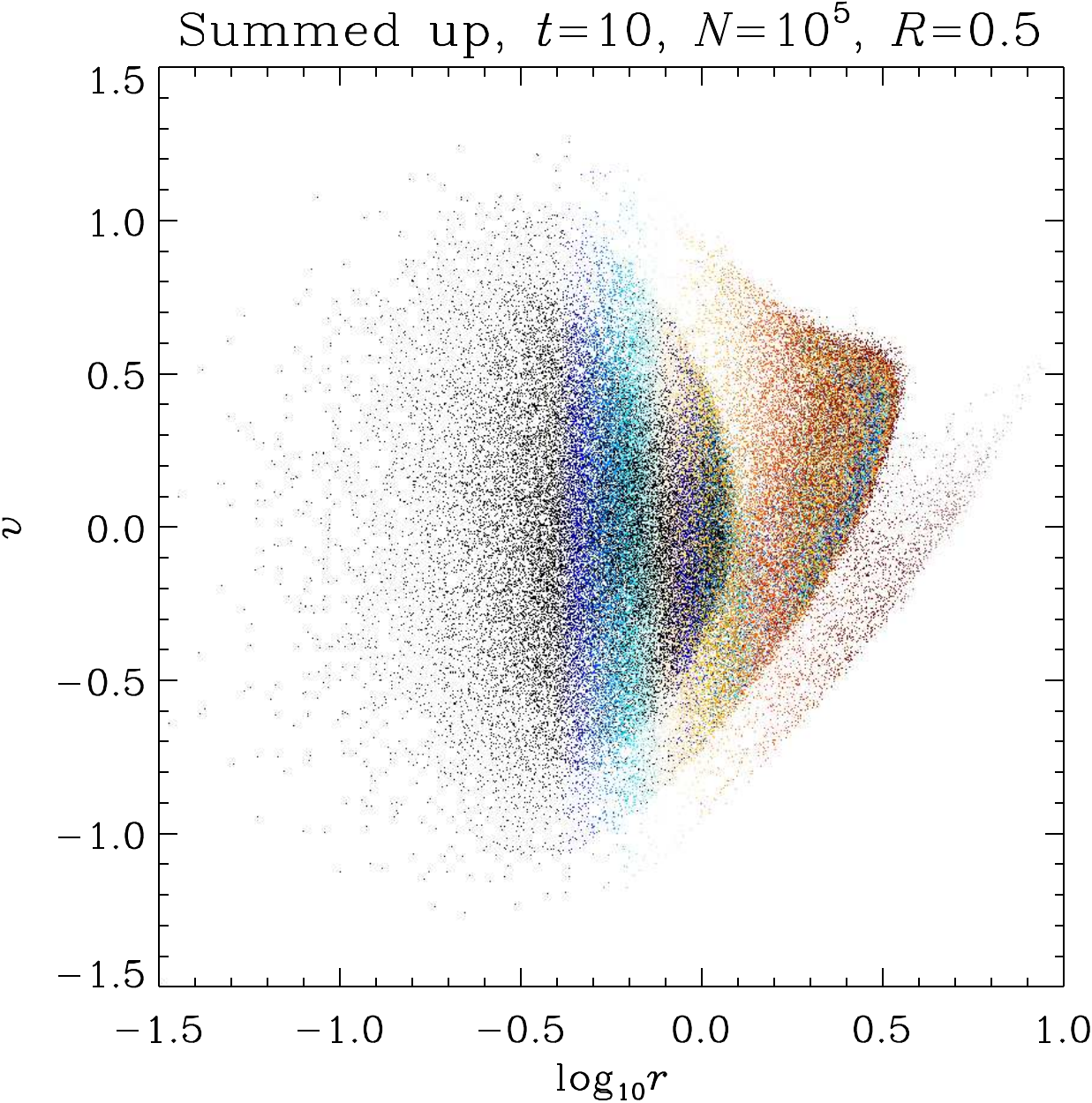}
\includegraphics[width=4.25cm]{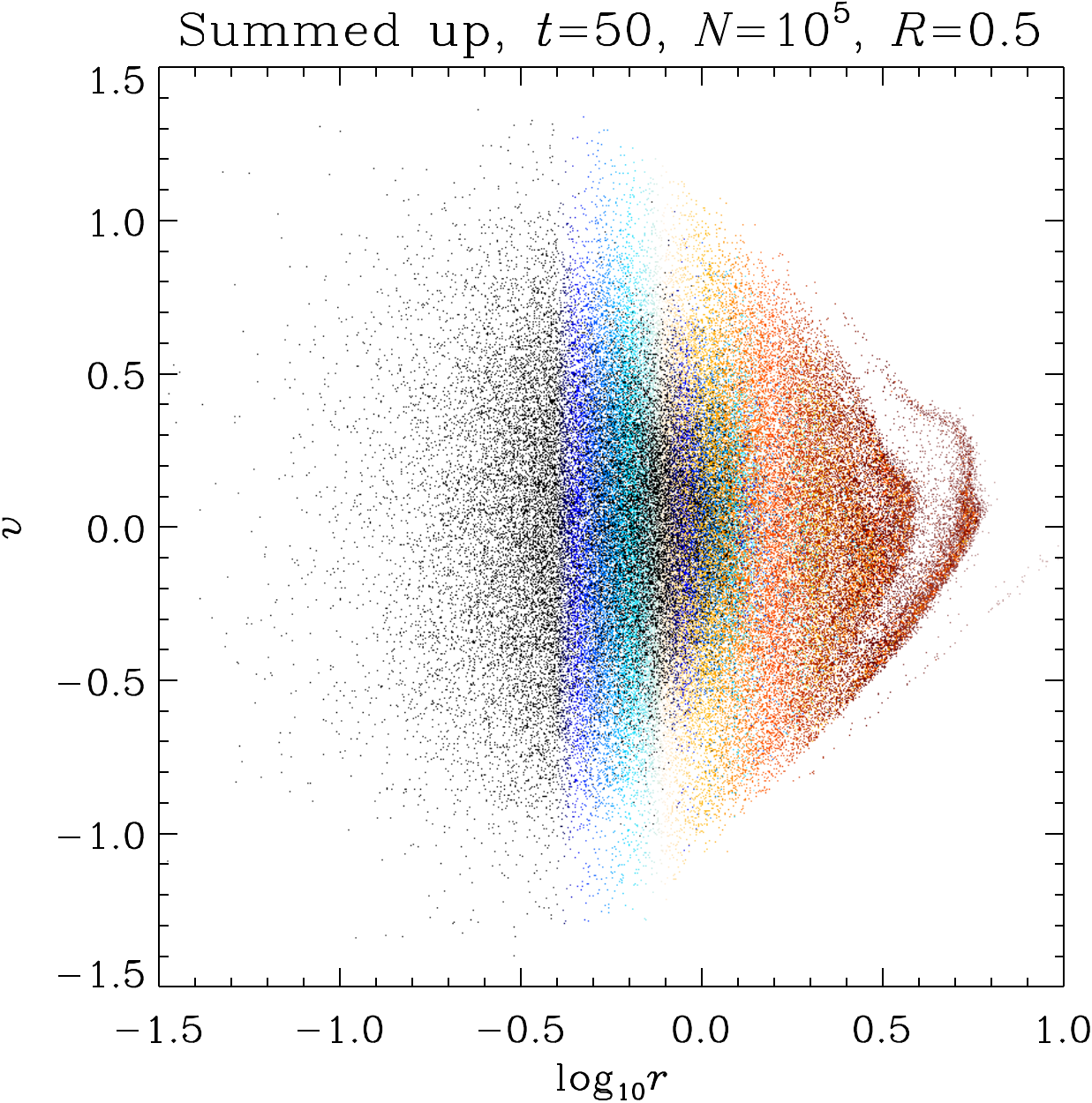}
\includegraphics[width=4.25cm]{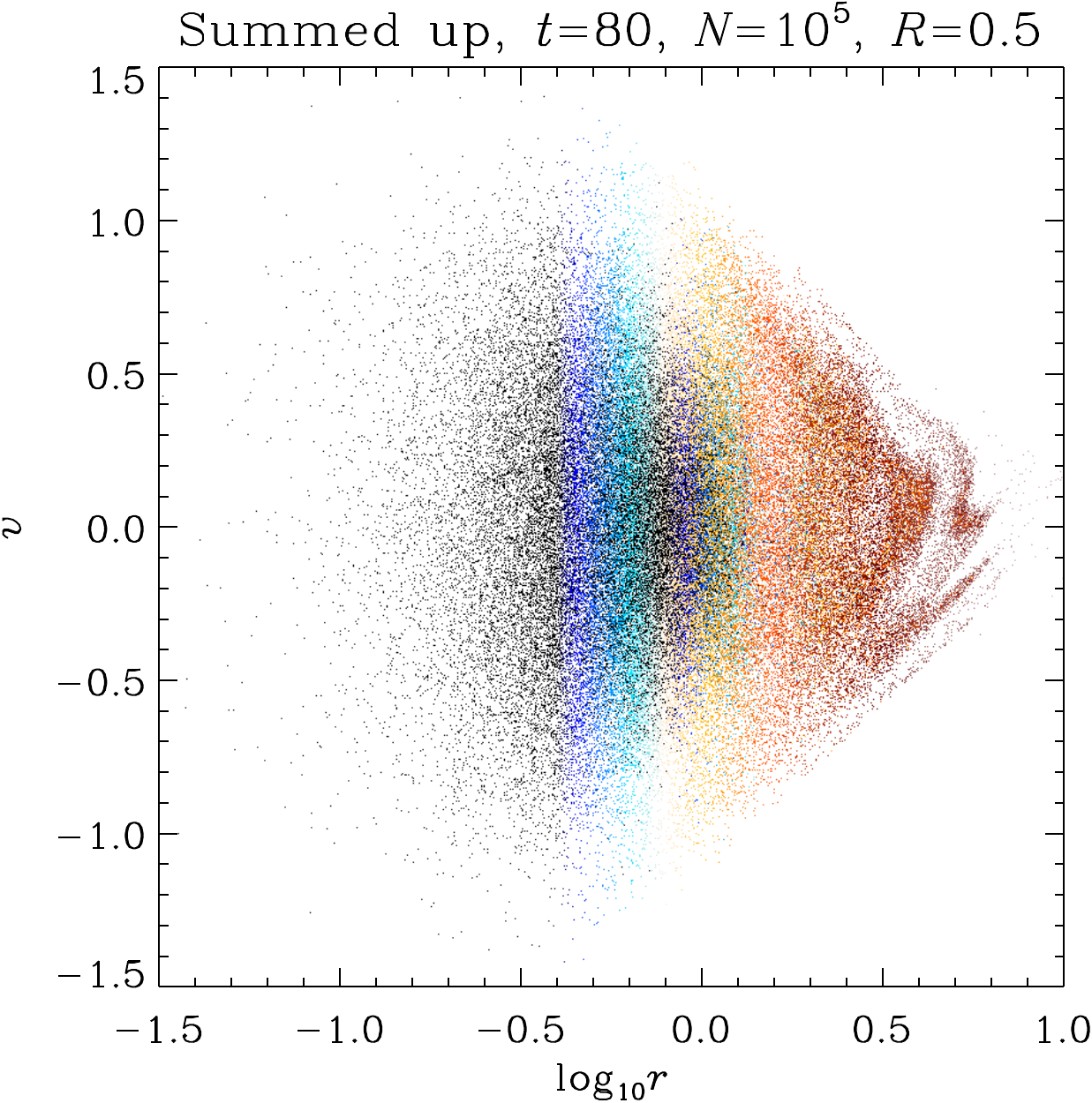}
\includegraphics[width=4.25cm]{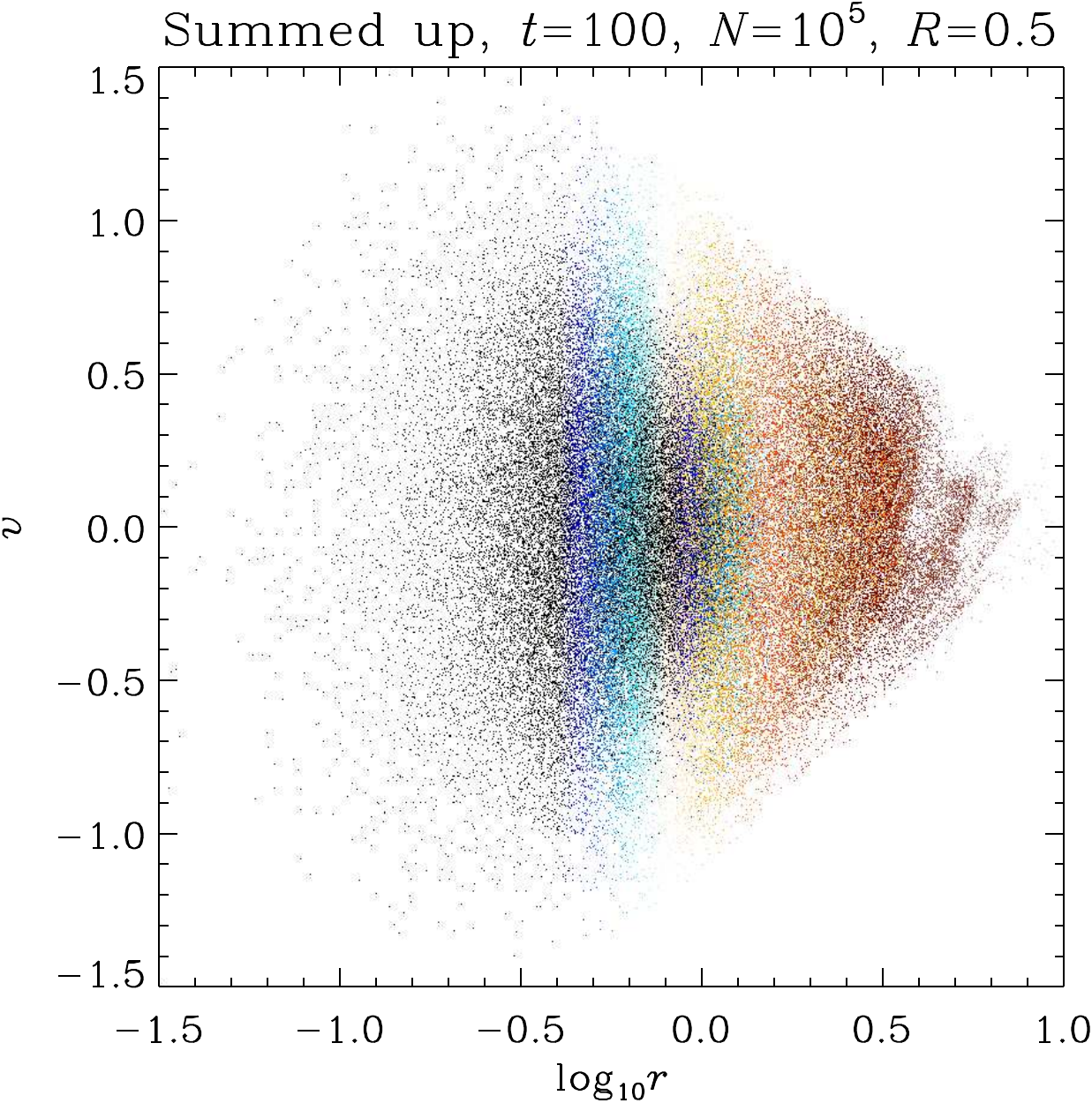}
}
\caption[]{Same as in Fig.~\ref{fig:0v5_12}, but the phase-space
  distribution function has now been summed up over the whole available
  range of values of $j \in [0,J_{\rm max}=1.6]$, where $J_{\rm max}$
  is the maximum sampled value of $j$ for the {\tt VlaSolve}
  simulations.}
\label{fig:0v5_ALLJ}
\end{figure*}

\begin{figure*}
\hbox{
\includegraphics[width=4.25cm]{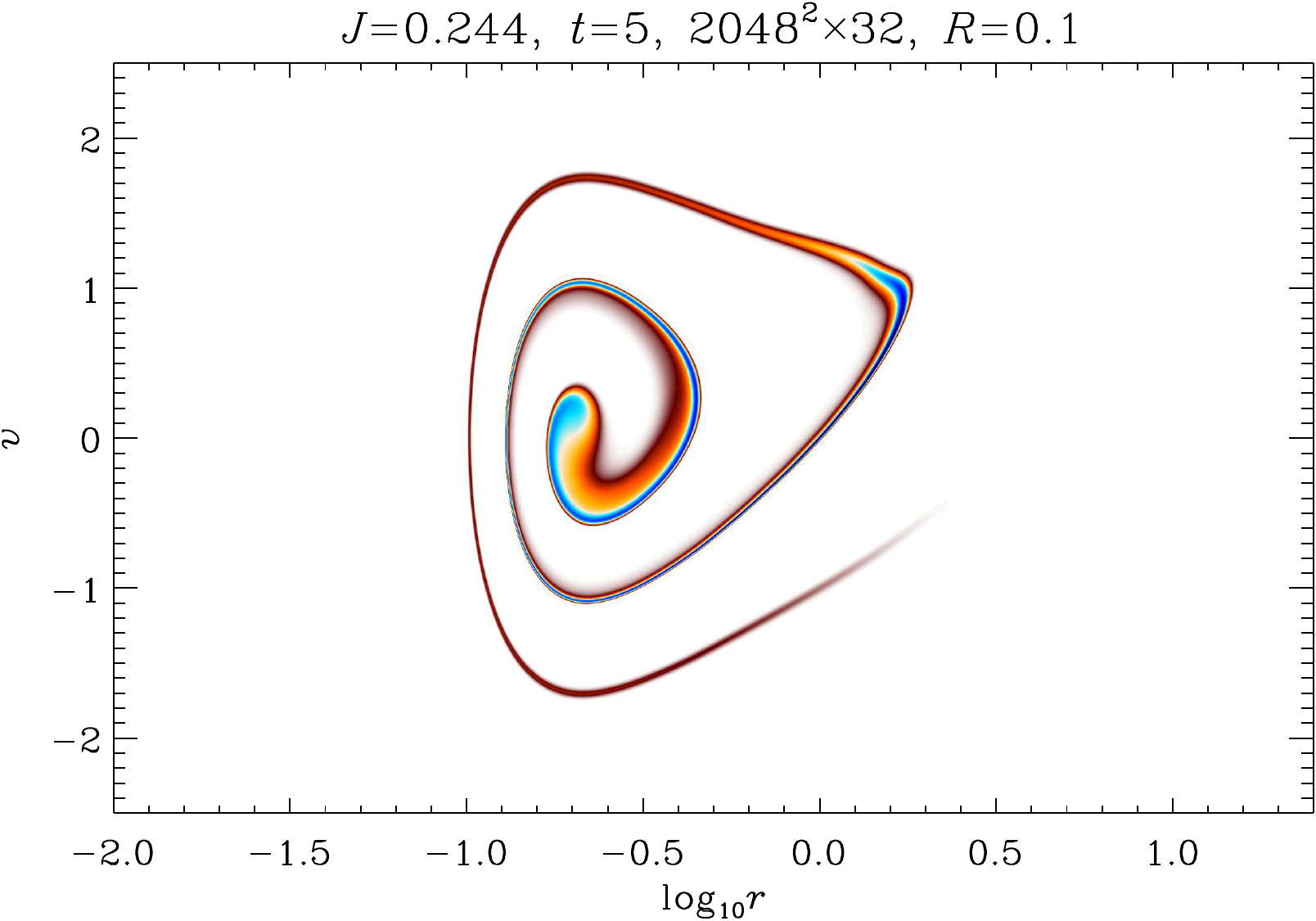}
\includegraphics[width=4.25cm]{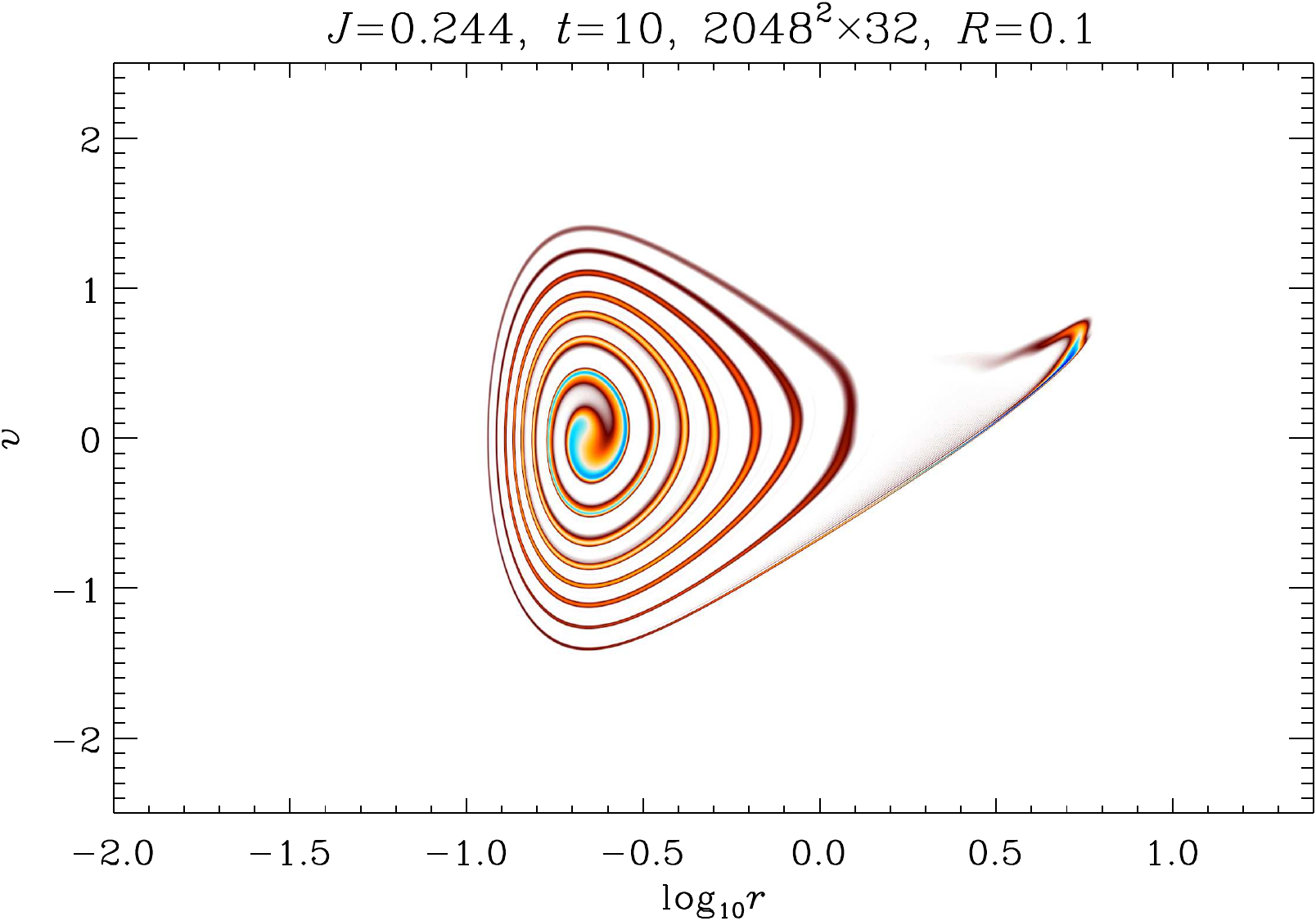}
\includegraphics[width=4.25cm]{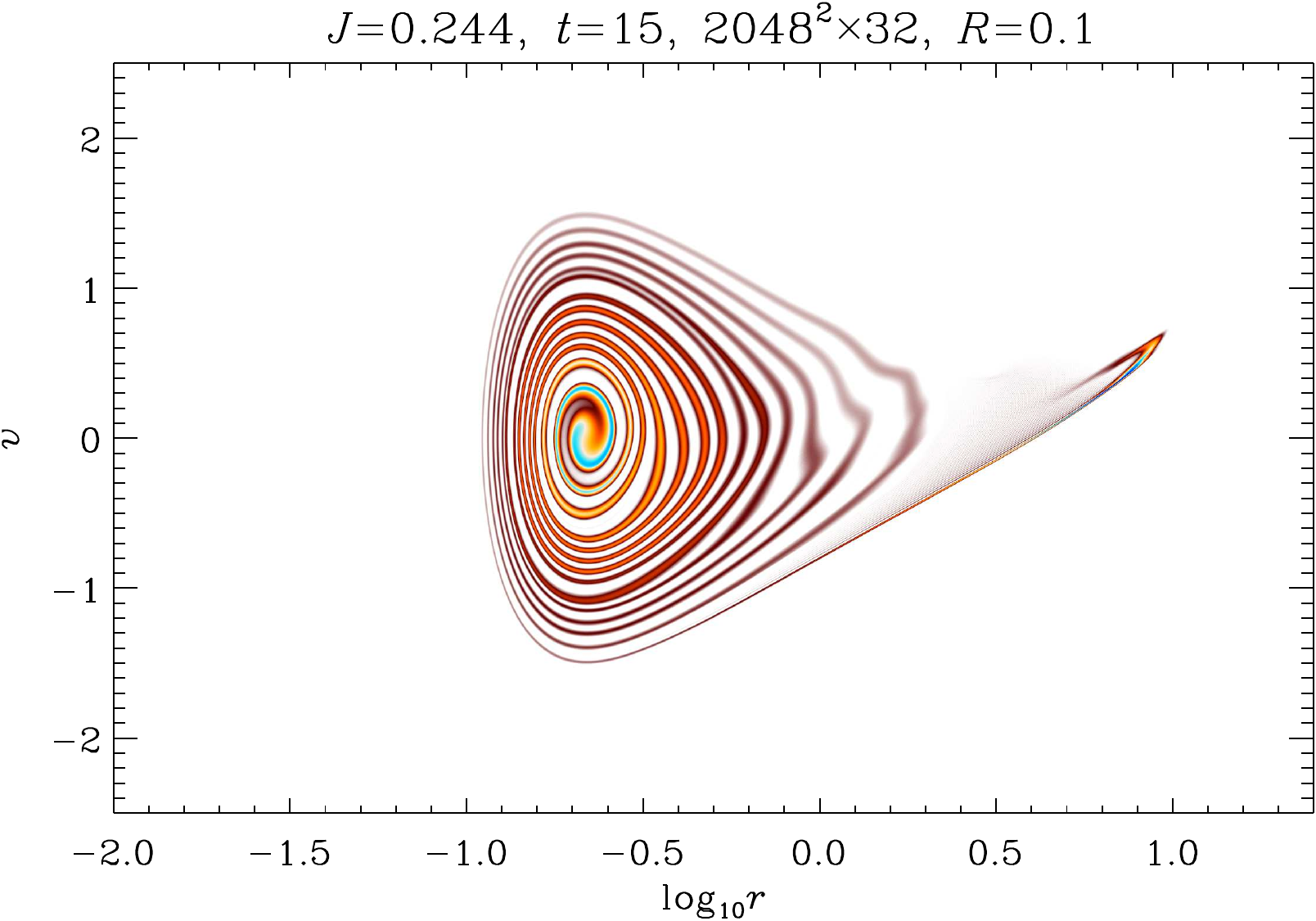}
\includegraphics[width=4.25cm]{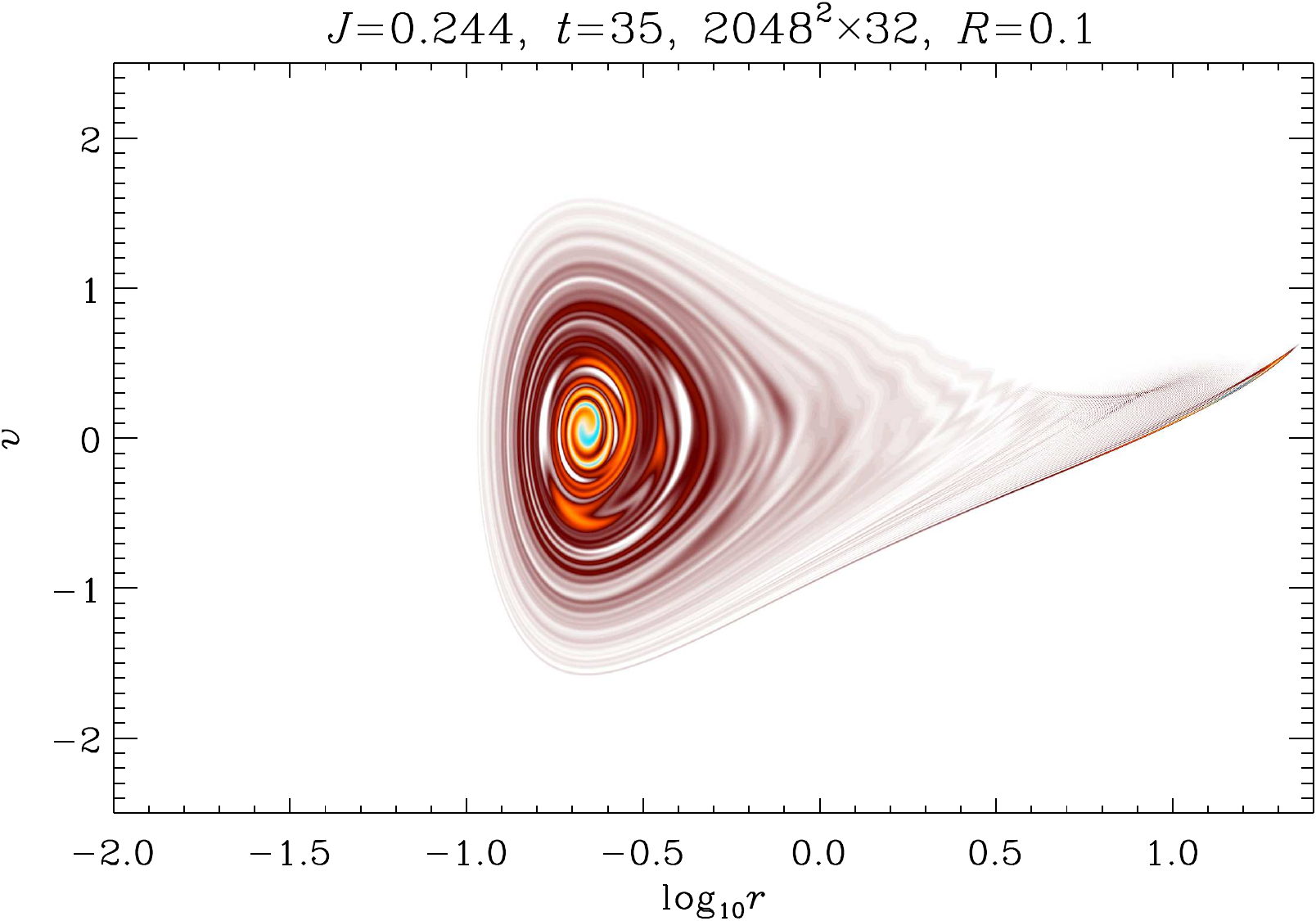}
}
\hbox{
\includegraphics[width=4.25cm]{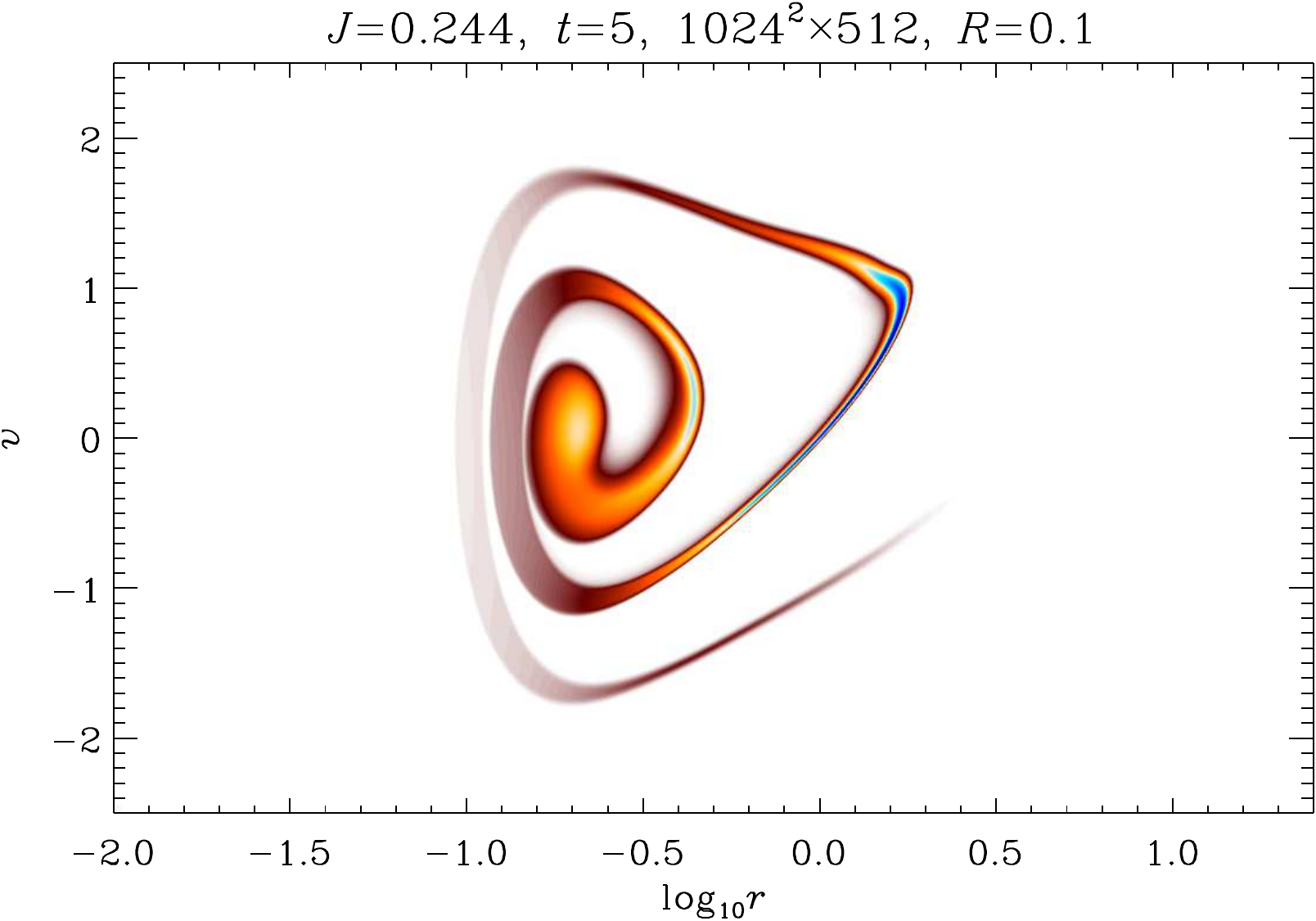}
\includegraphics[width=4.25cm]{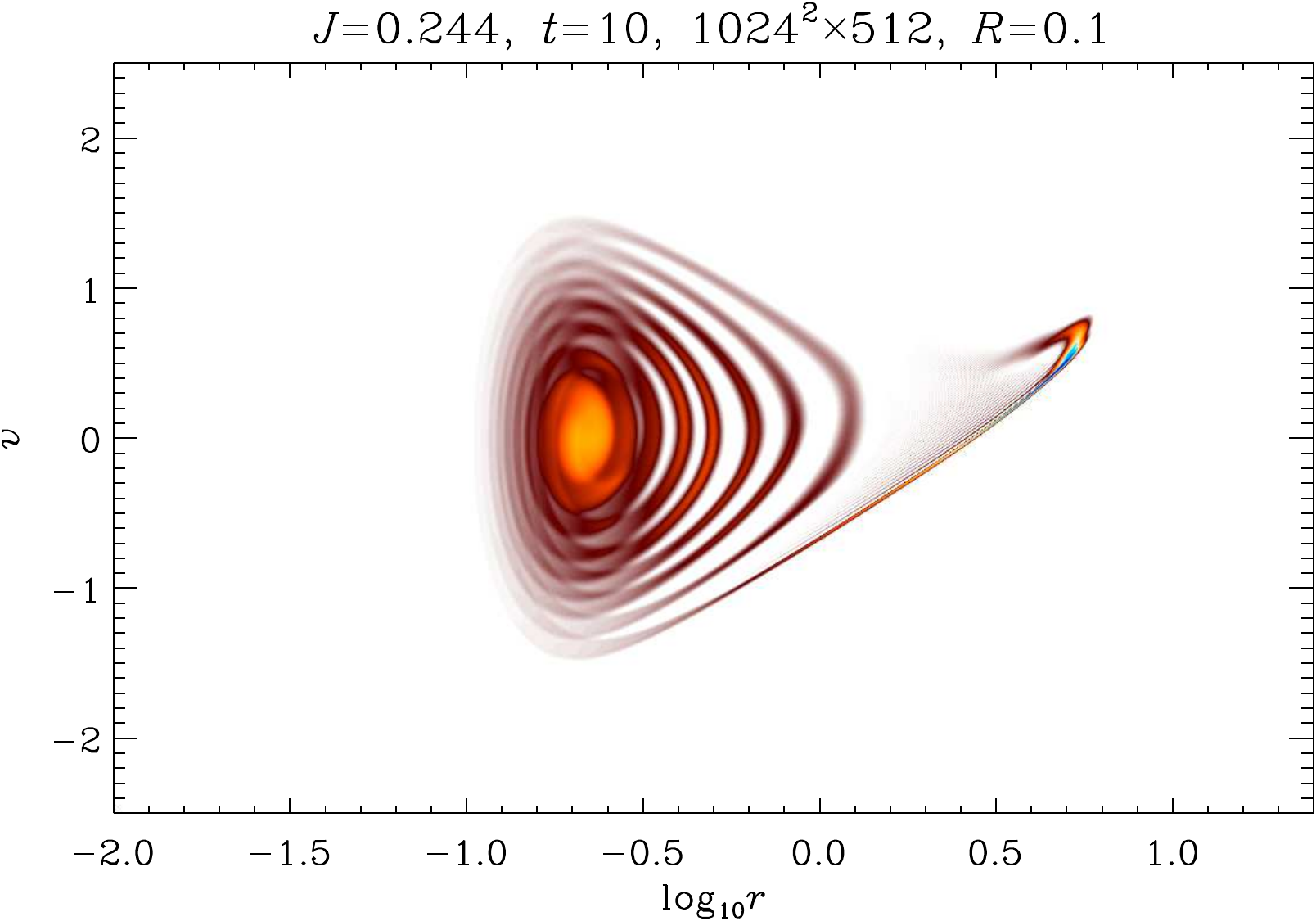}
\includegraphics[width=4.25cm]{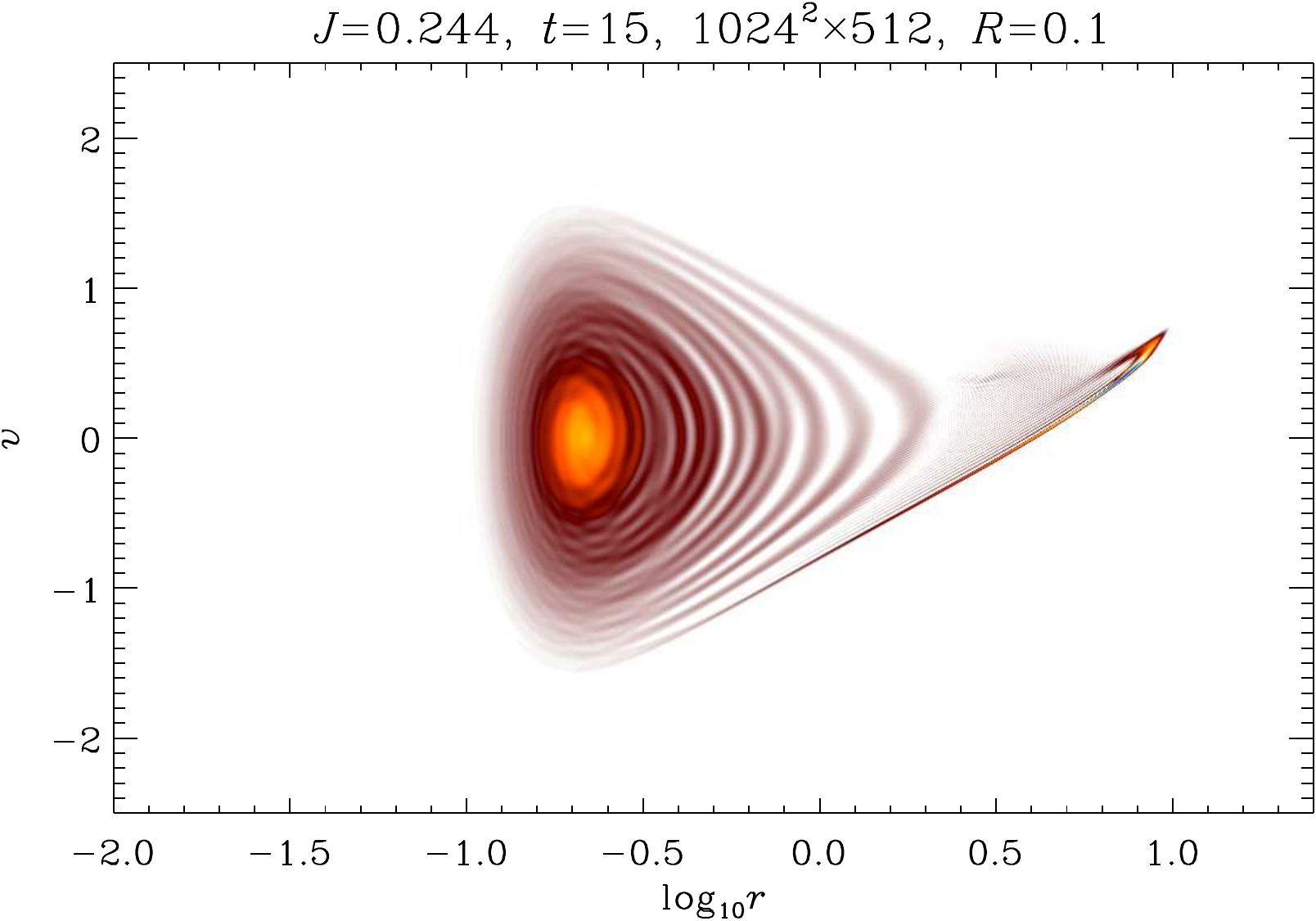}
\includegraphics[width=4.25cm]{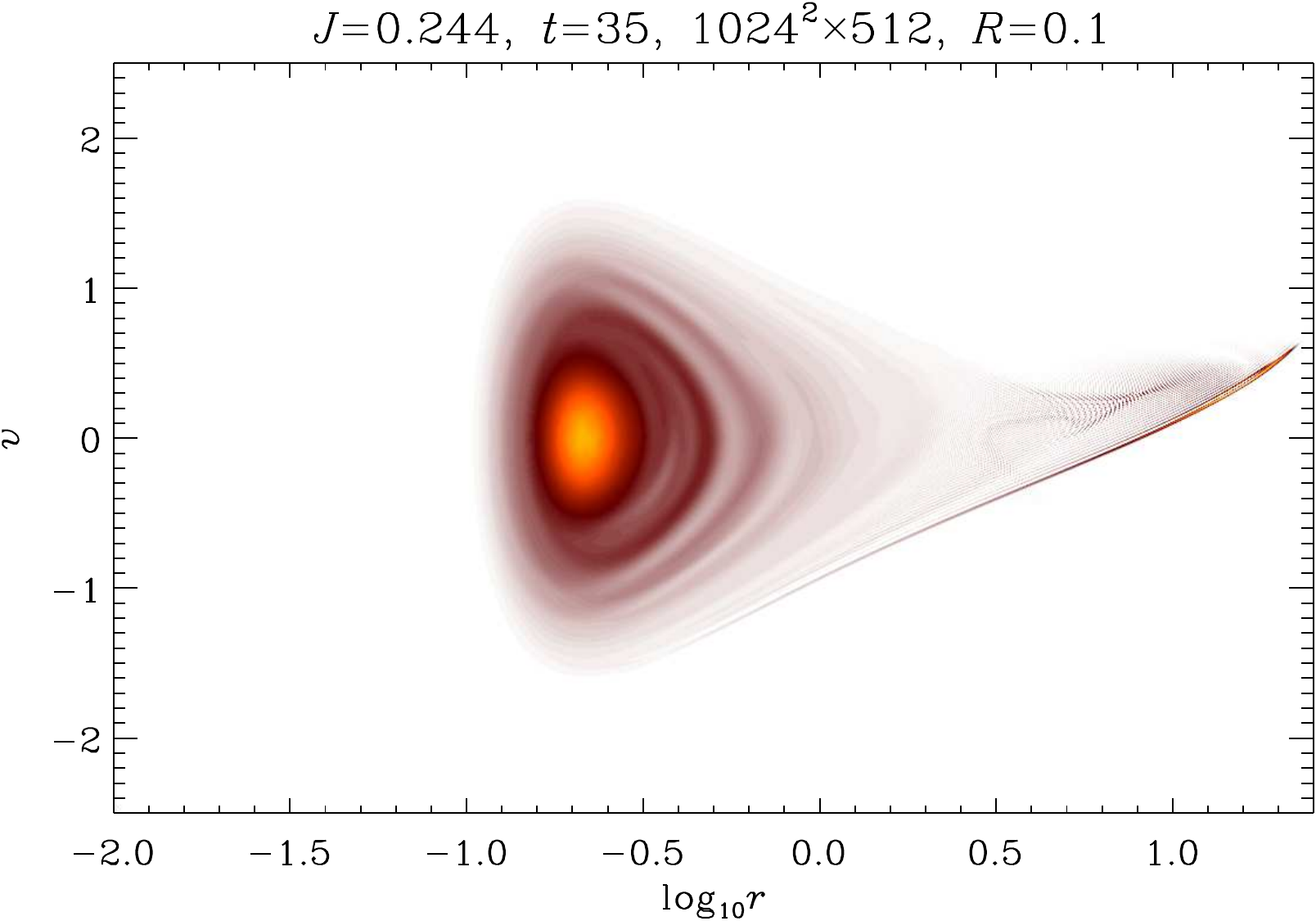}
}
\hbox{
\includegraphics[width=4.25cm]{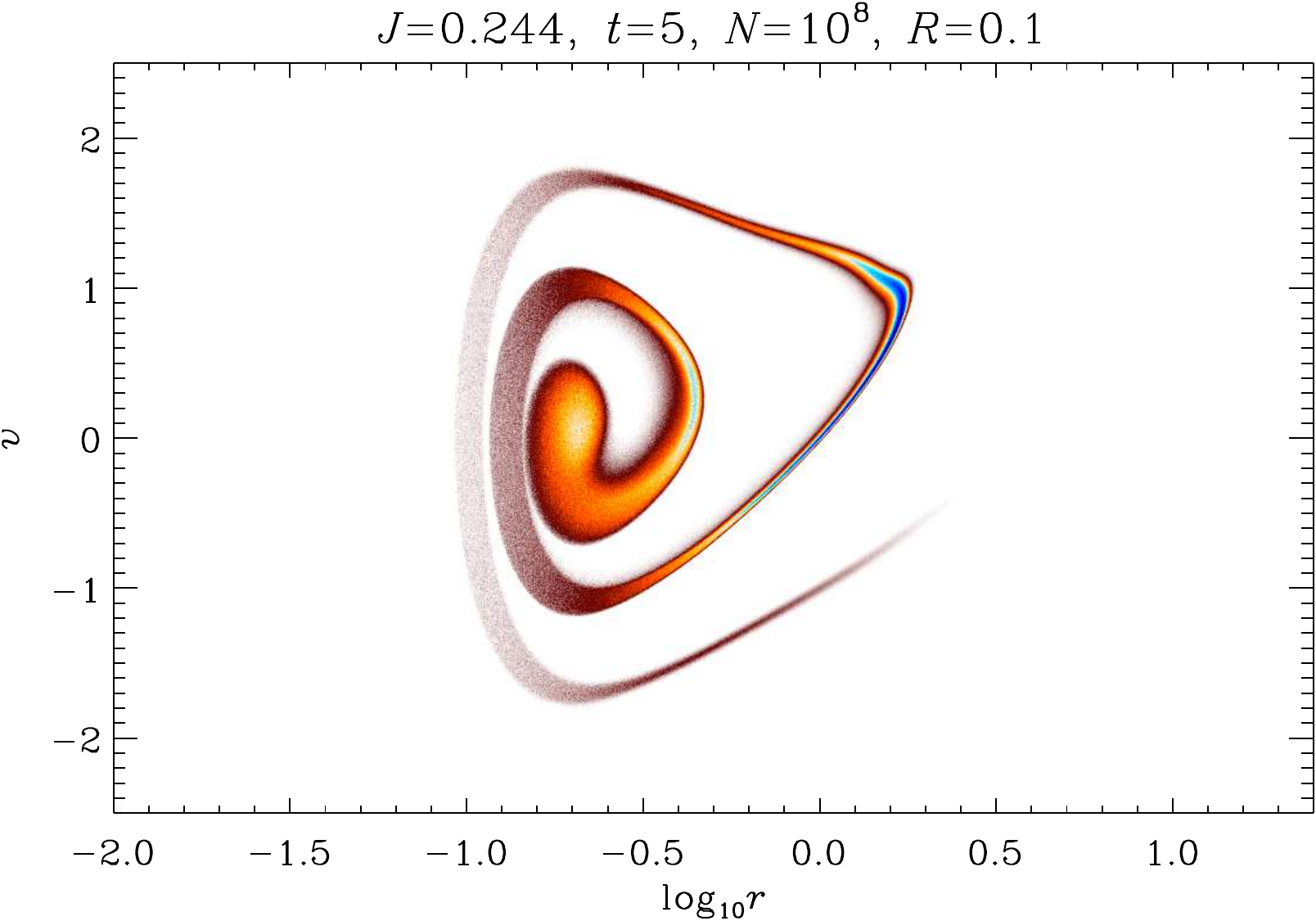}
\includegraphics[width=4.25cm]{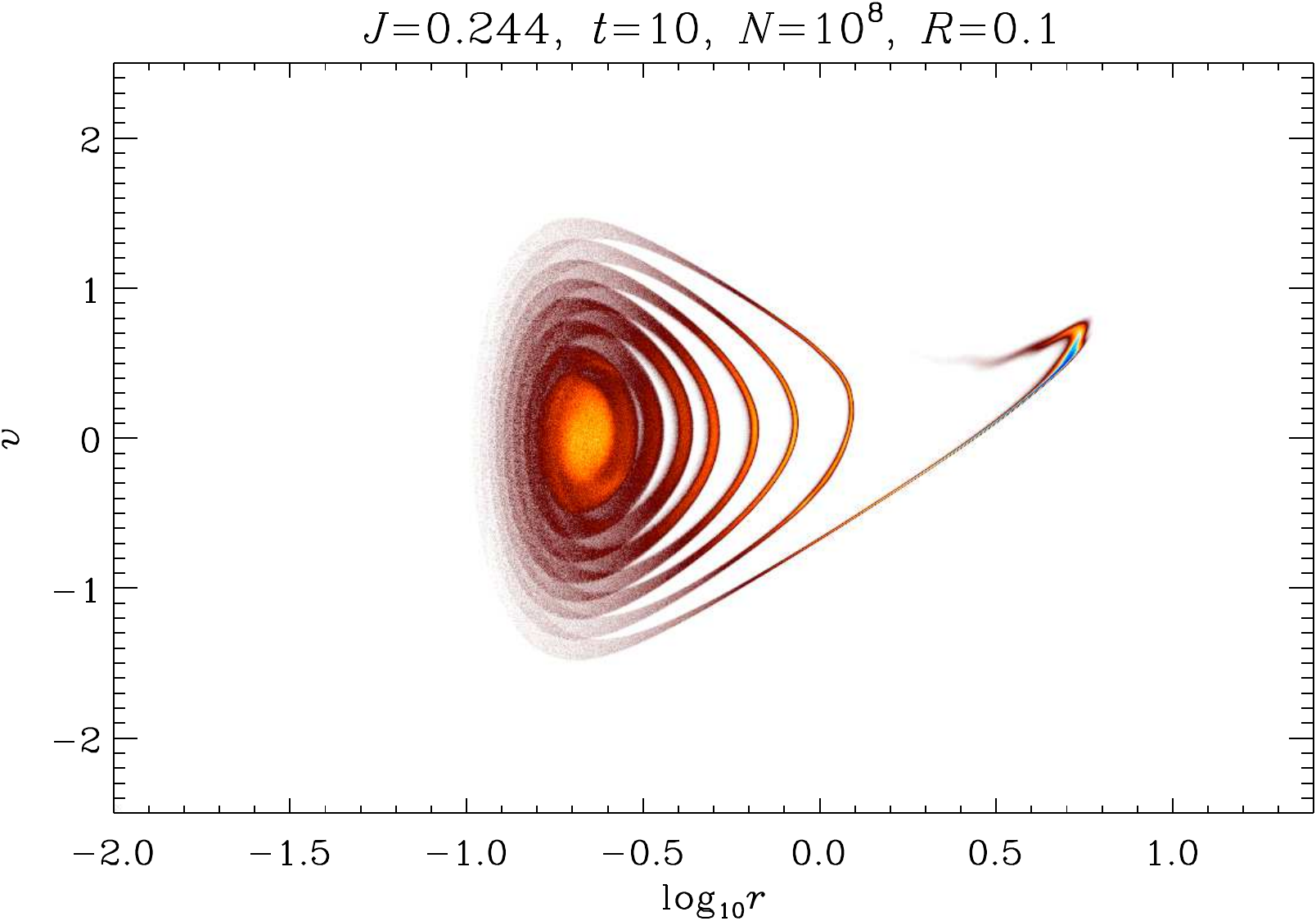}
\includegraphics[width=4.25cm]{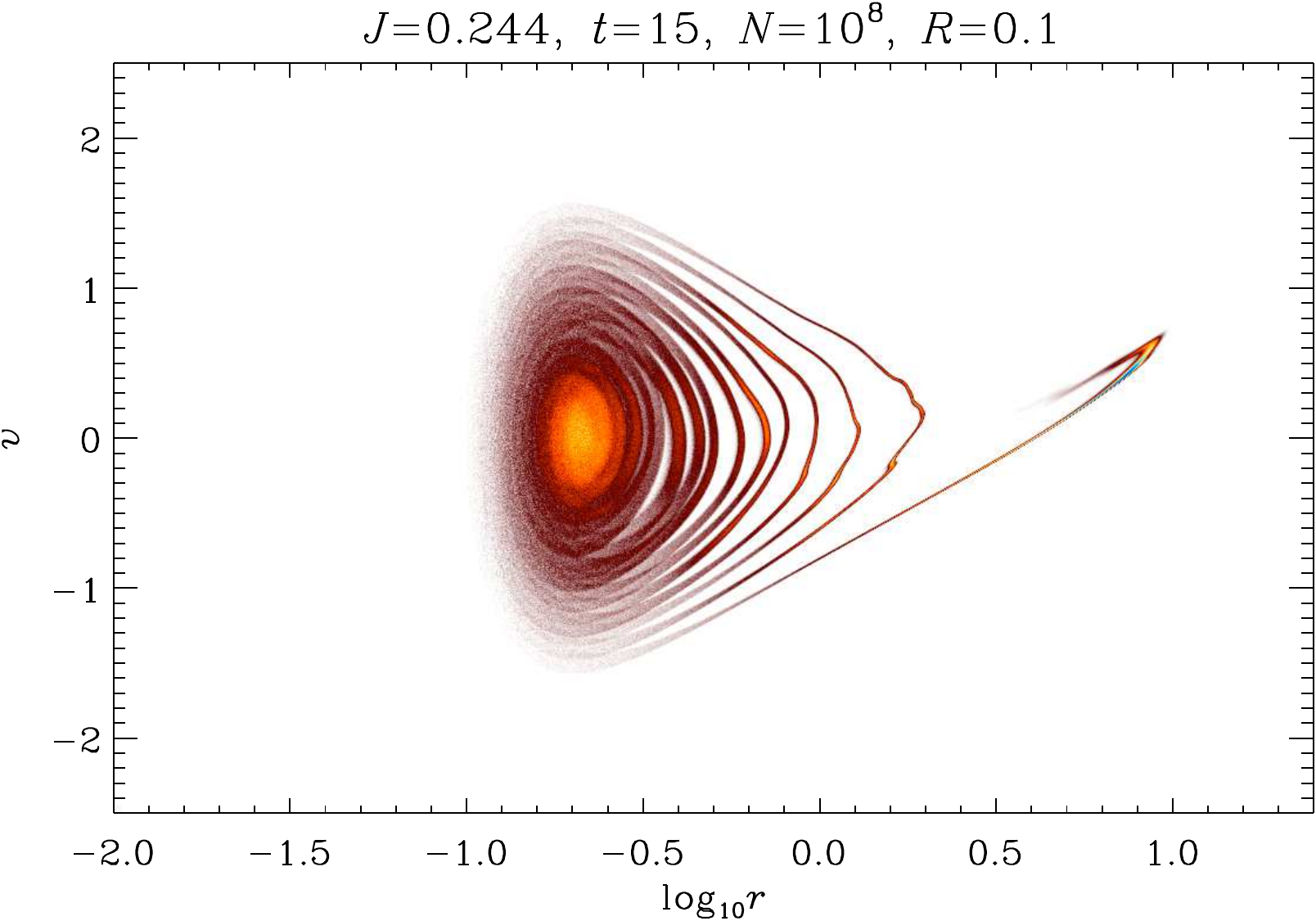}
\includegraphics[width=4.25cm]{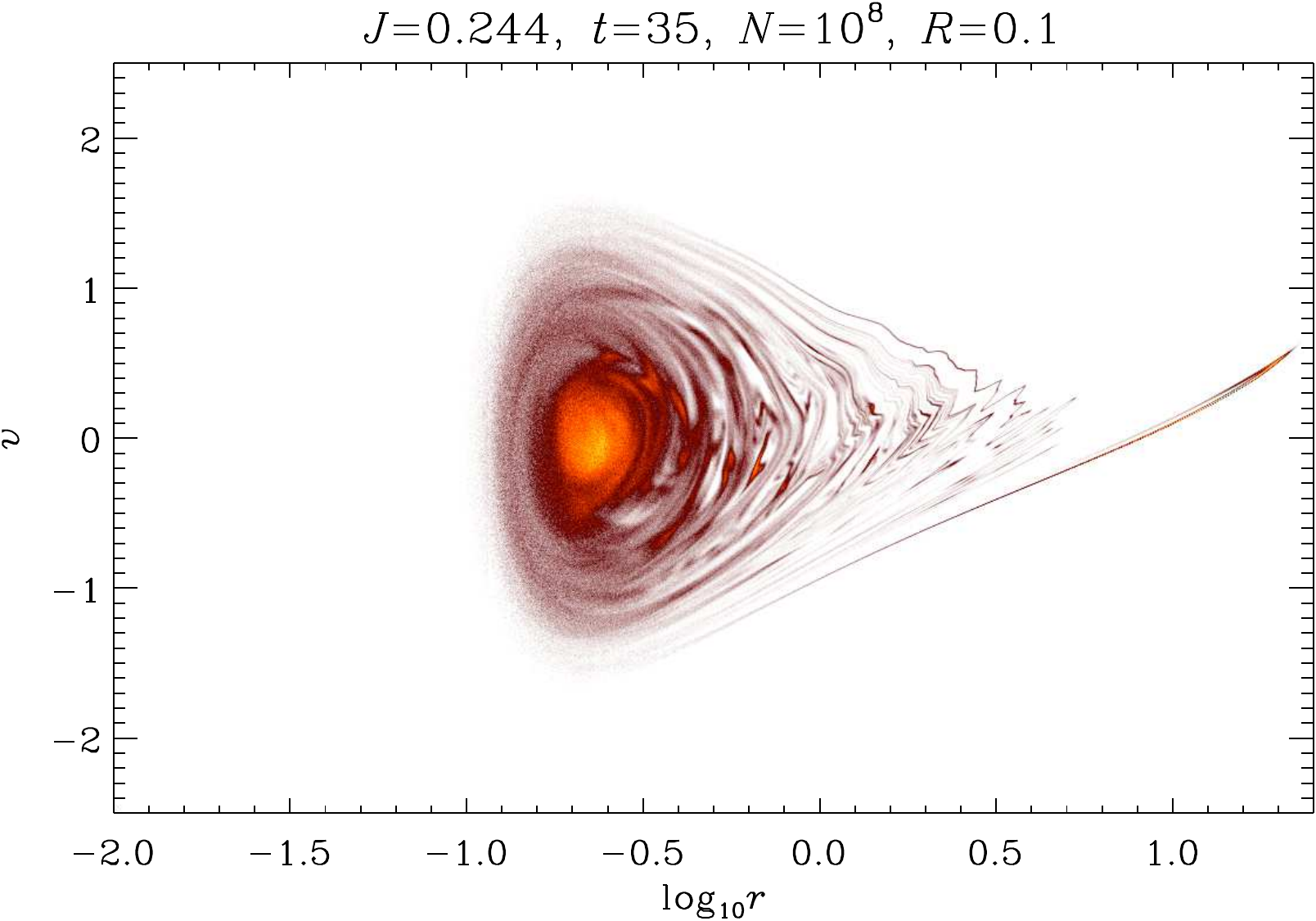}
}
\hbox{
\includegraphics[width=4.25cm]{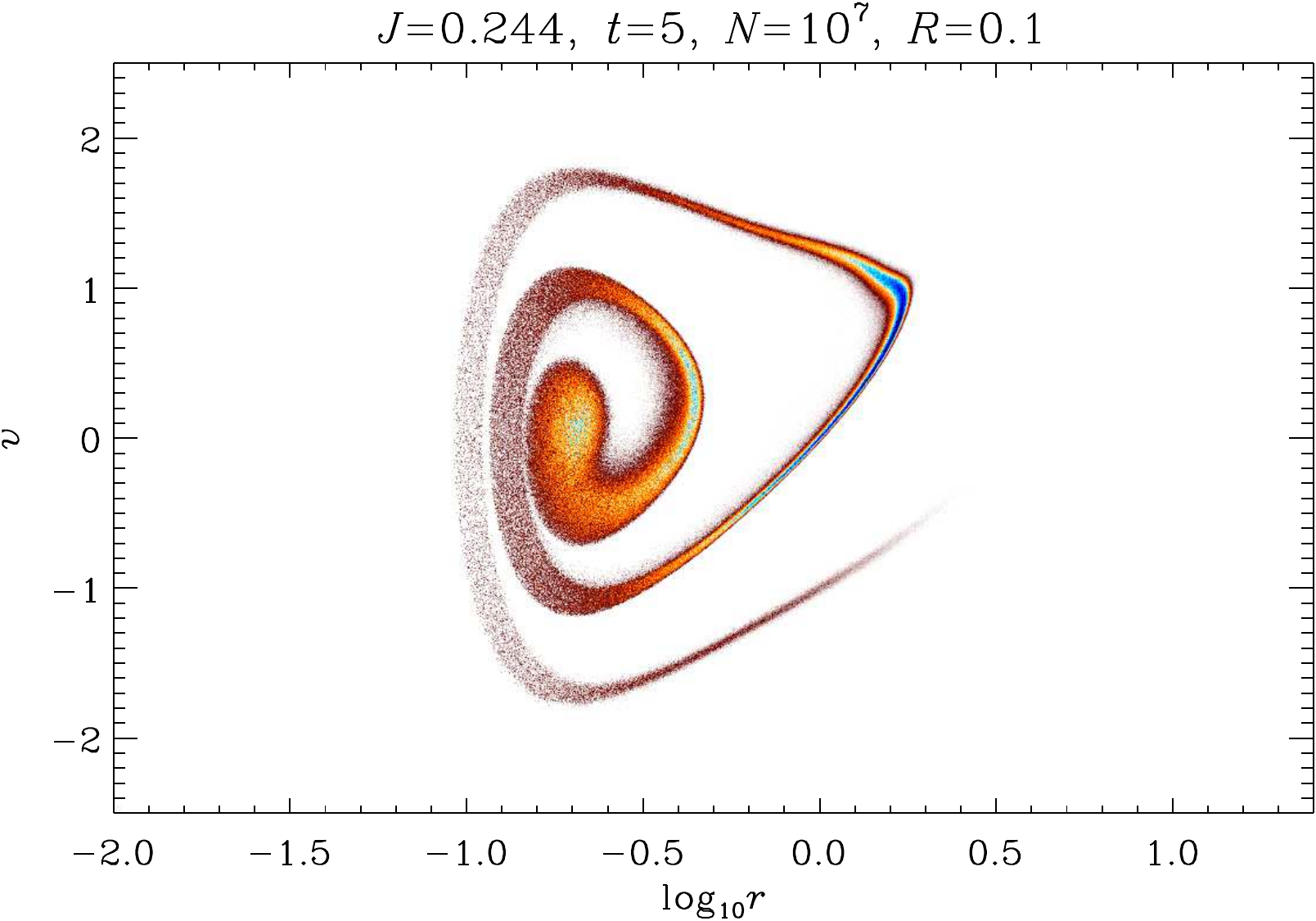}
\includegraphics[width=4.25cm]{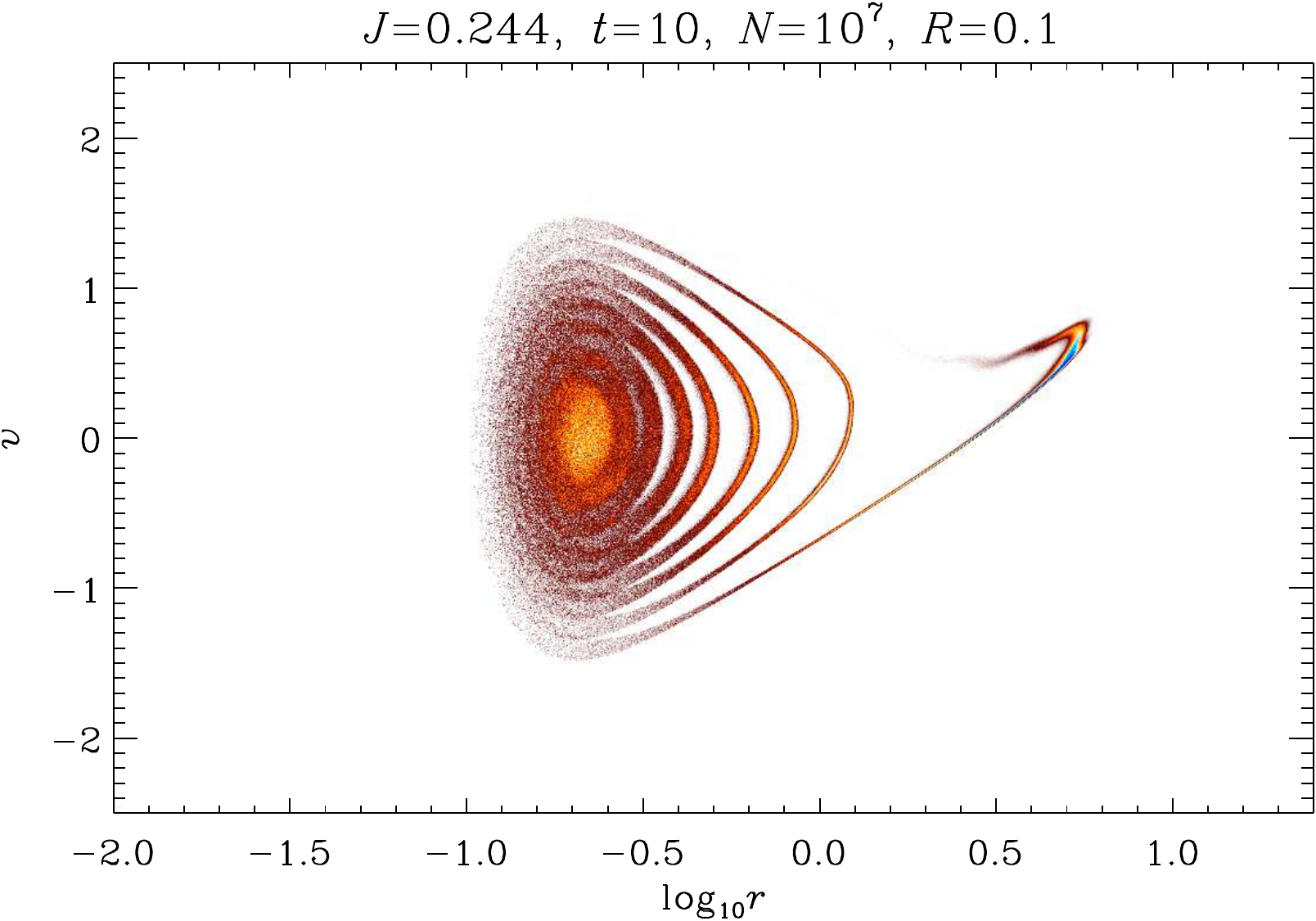}
\includegraphics[width=4.25cm]{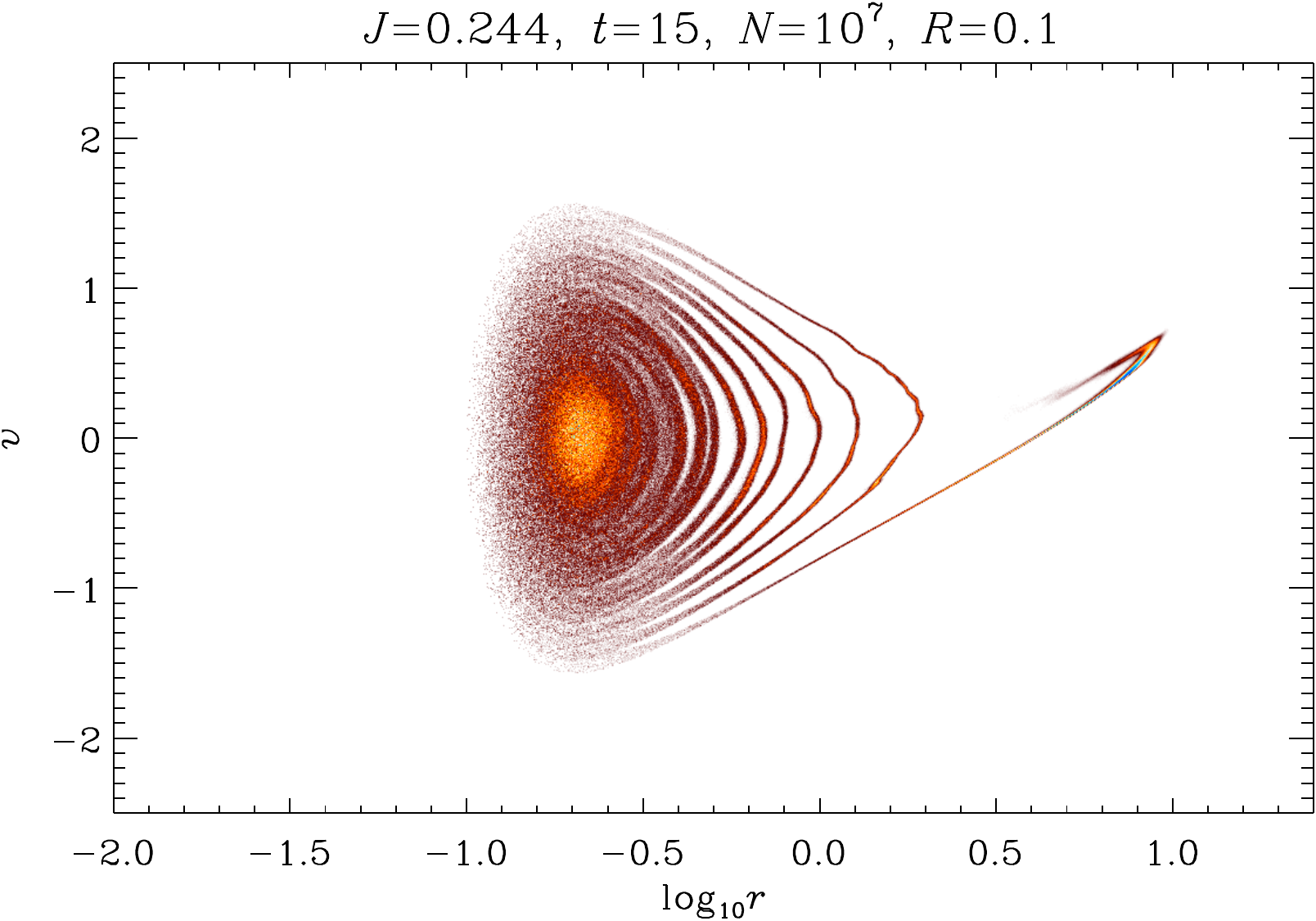}
\includegraphics[width=4.25cm]{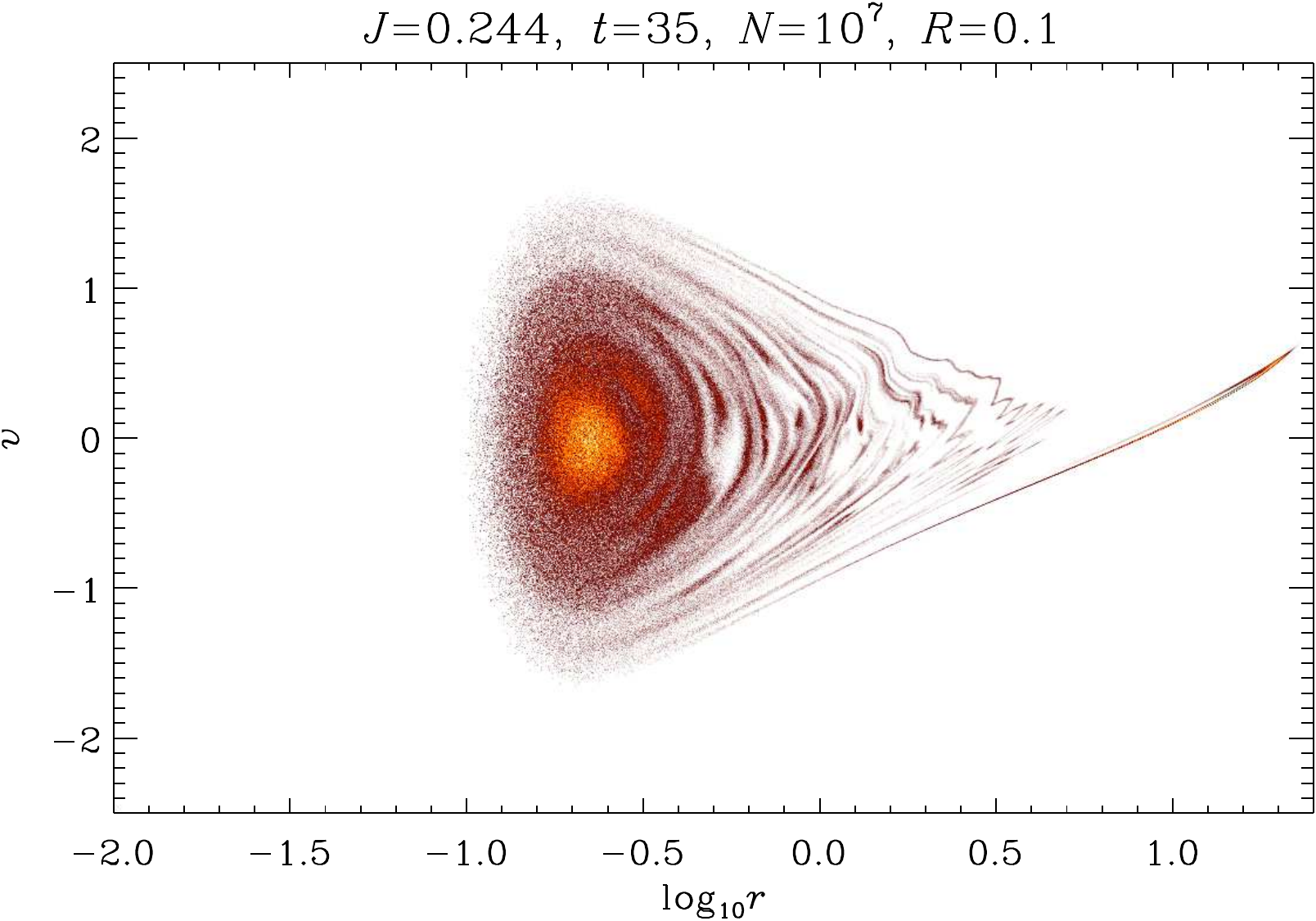}
}
\hbox{
\includegraphics[width=4.25cm]{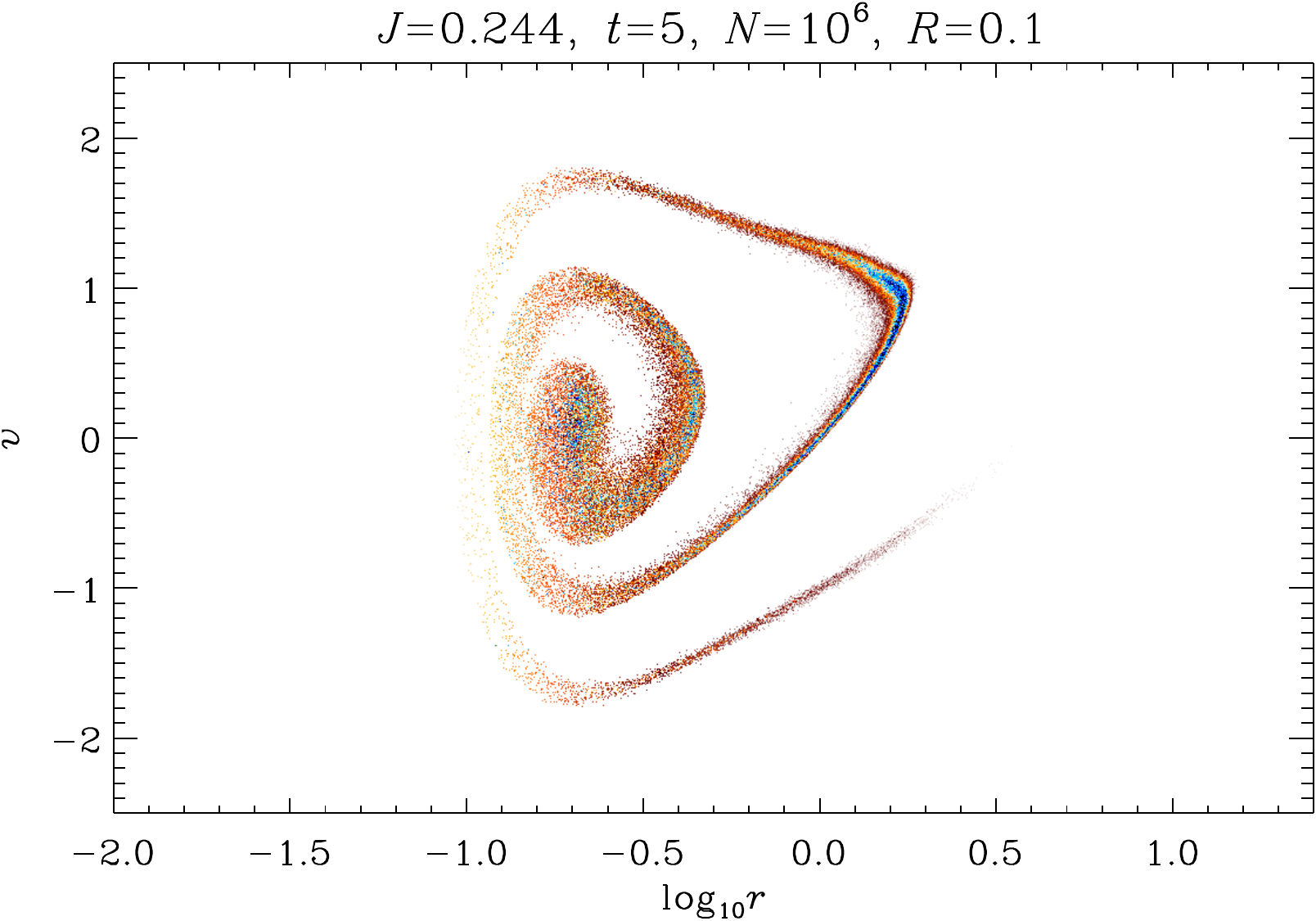}
\includegraphics[width=4.25cm]{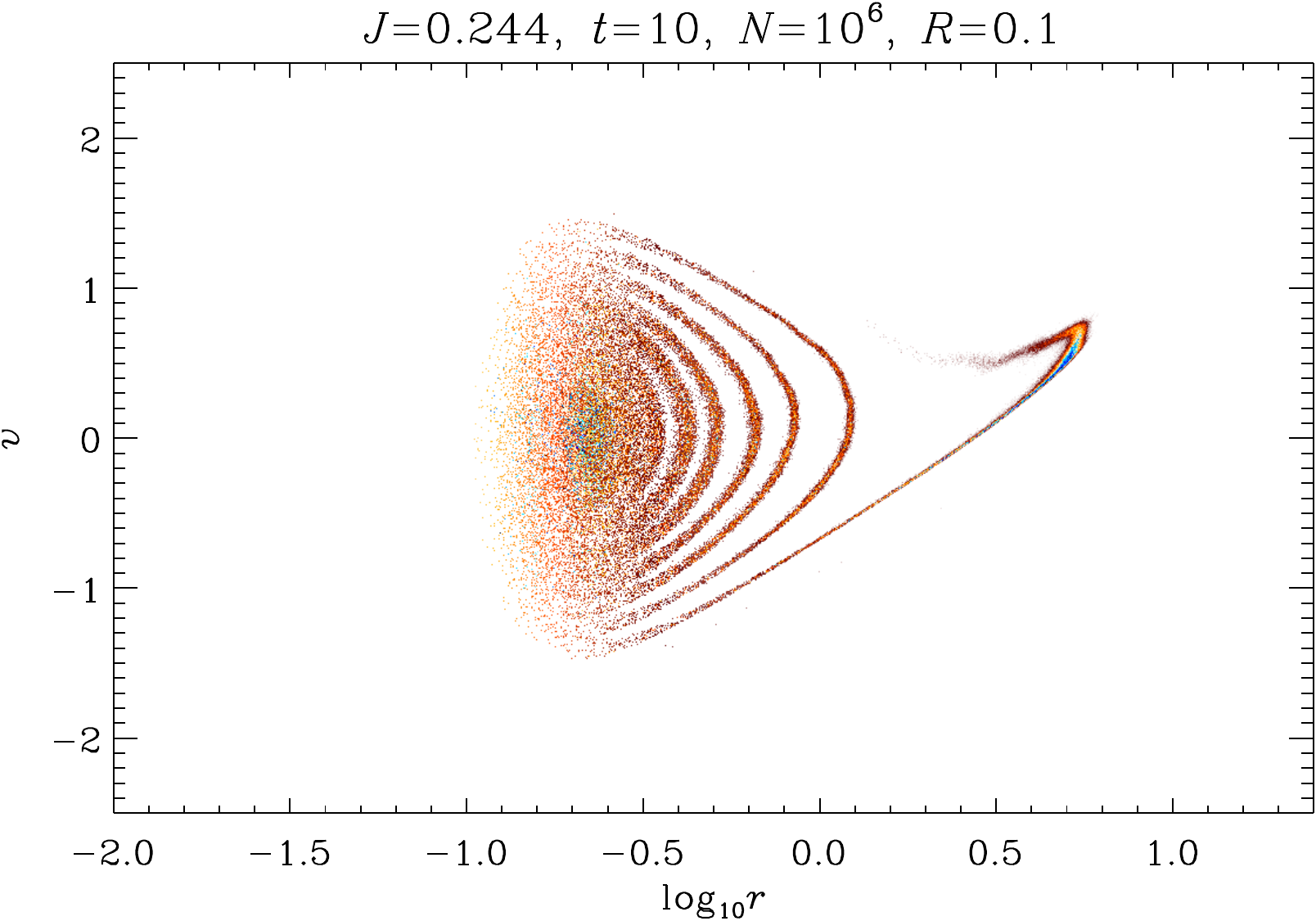}
\includegraphics[width=4.25cm]{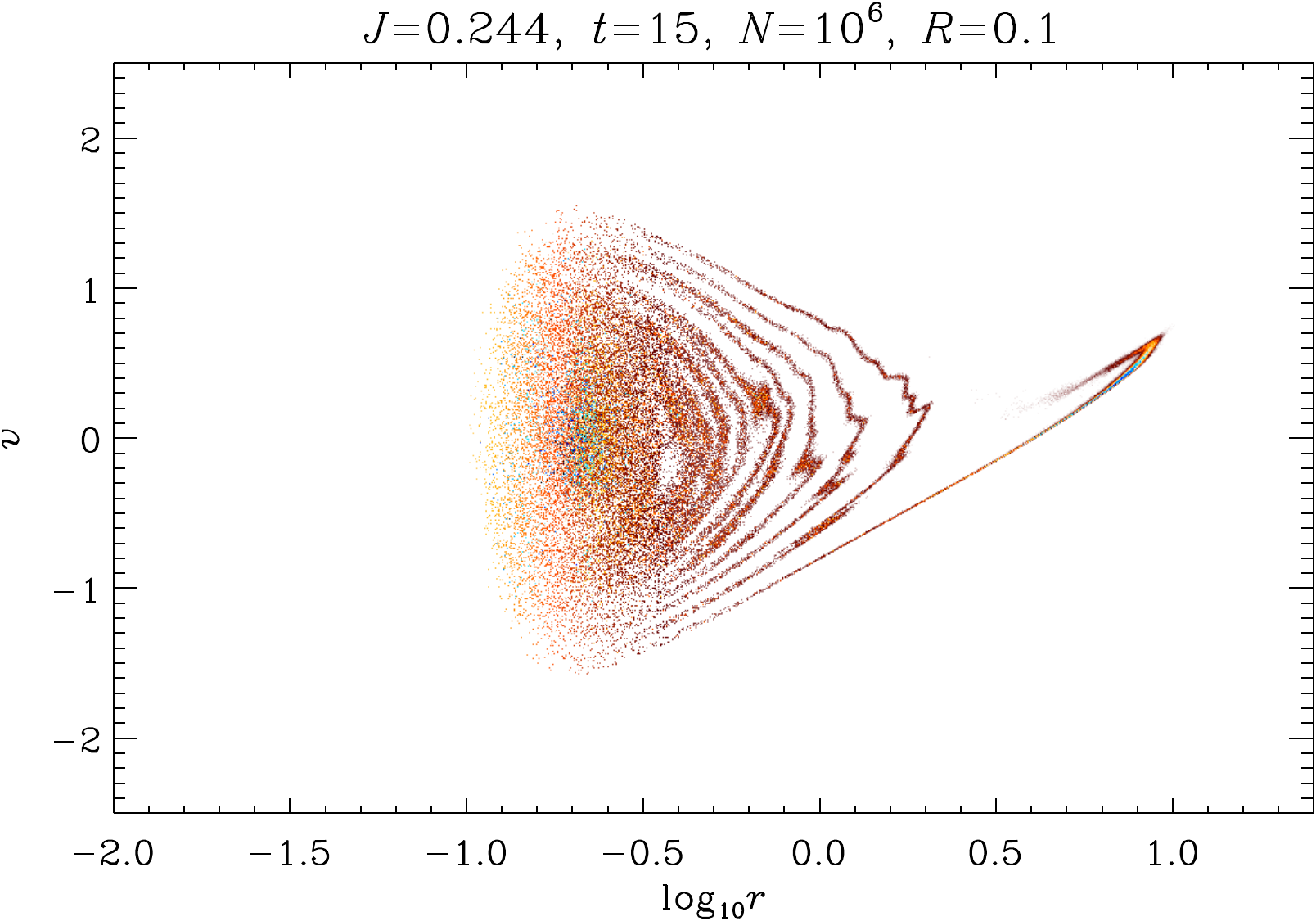}
\includegraphics[width=4.25cm]{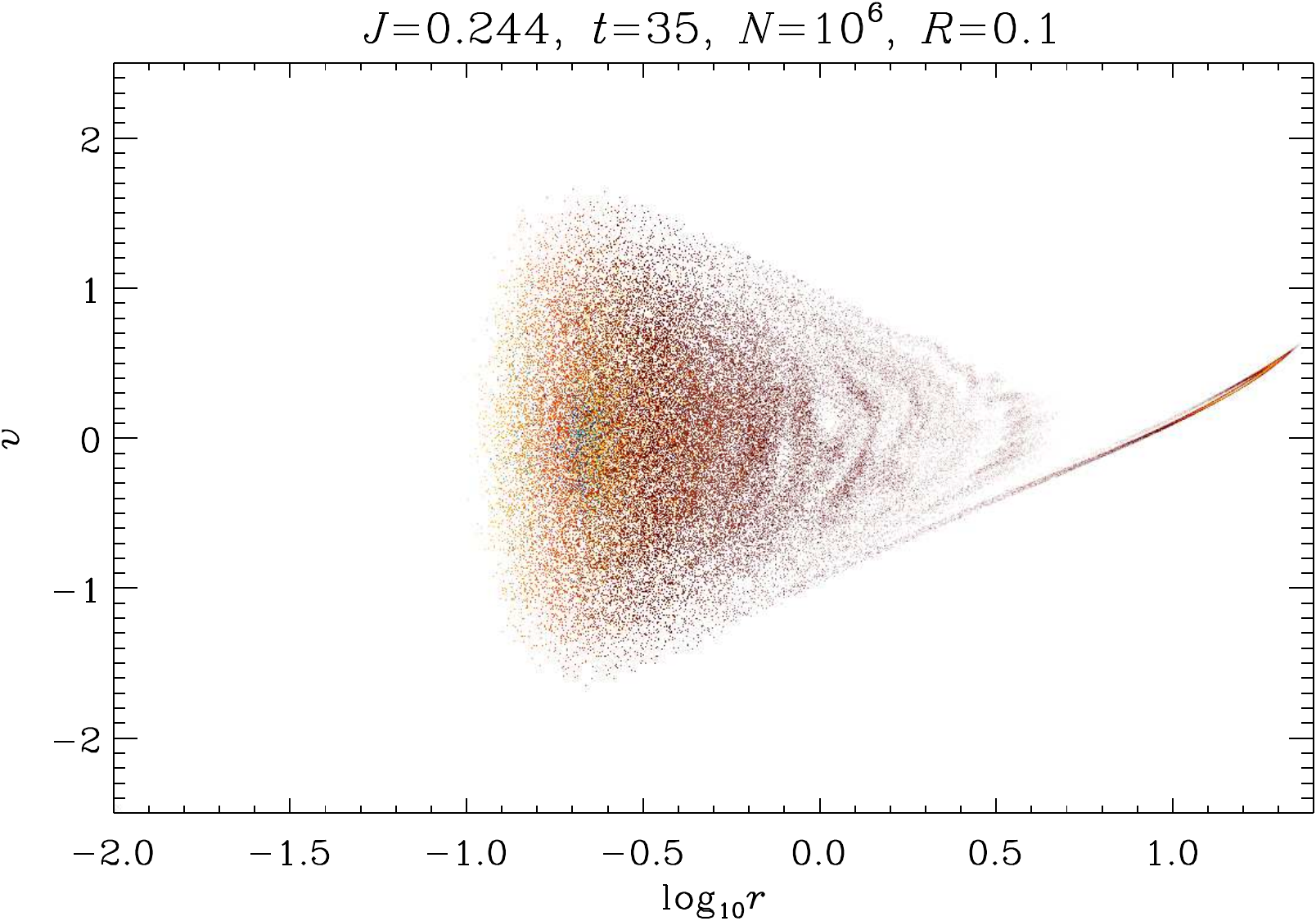}
}
\hbox{
\includegraphics[width=4.25cm]{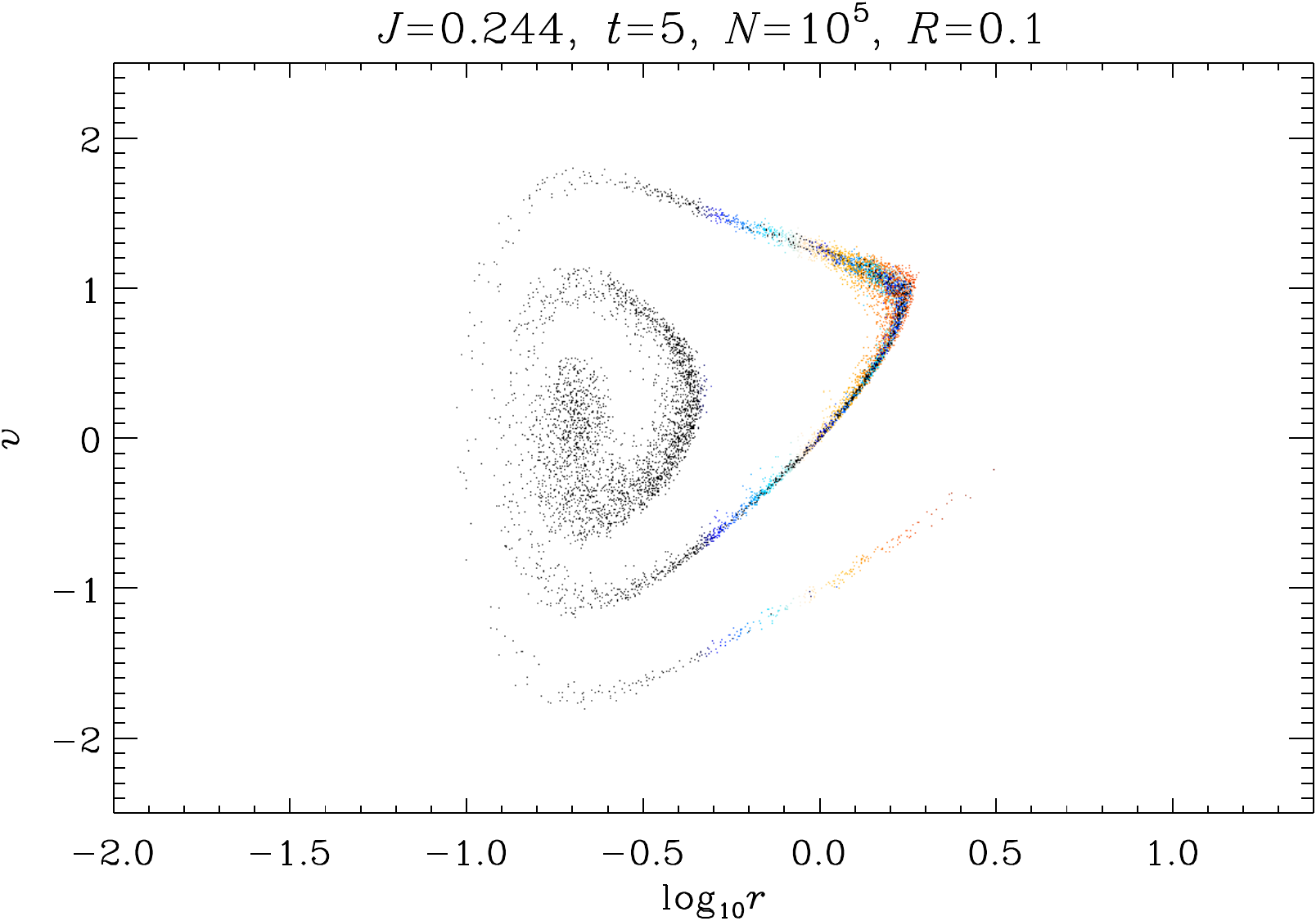}
\includegraphics[width=4.25cm]{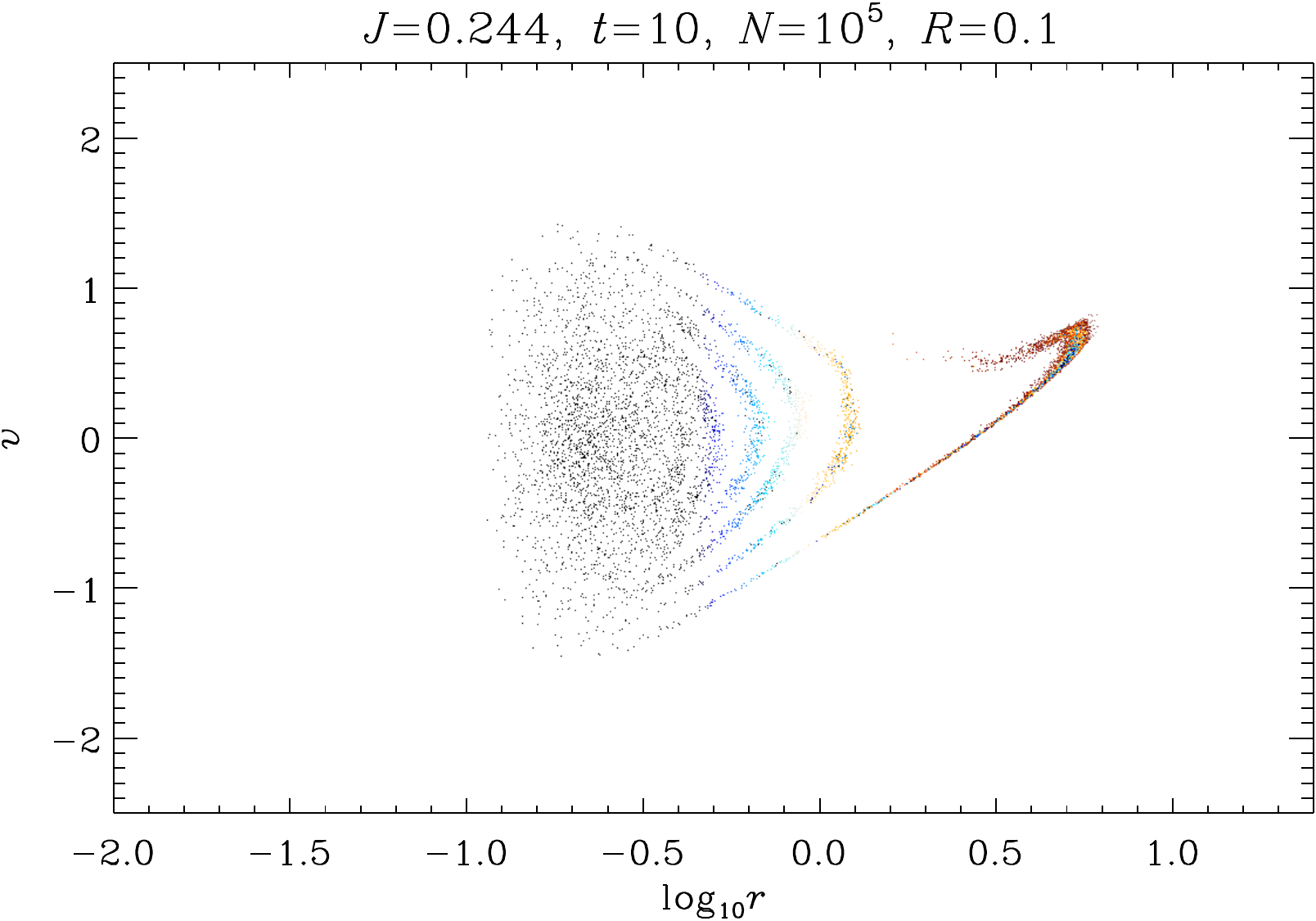}
\includegraphics[width=4.25cm]{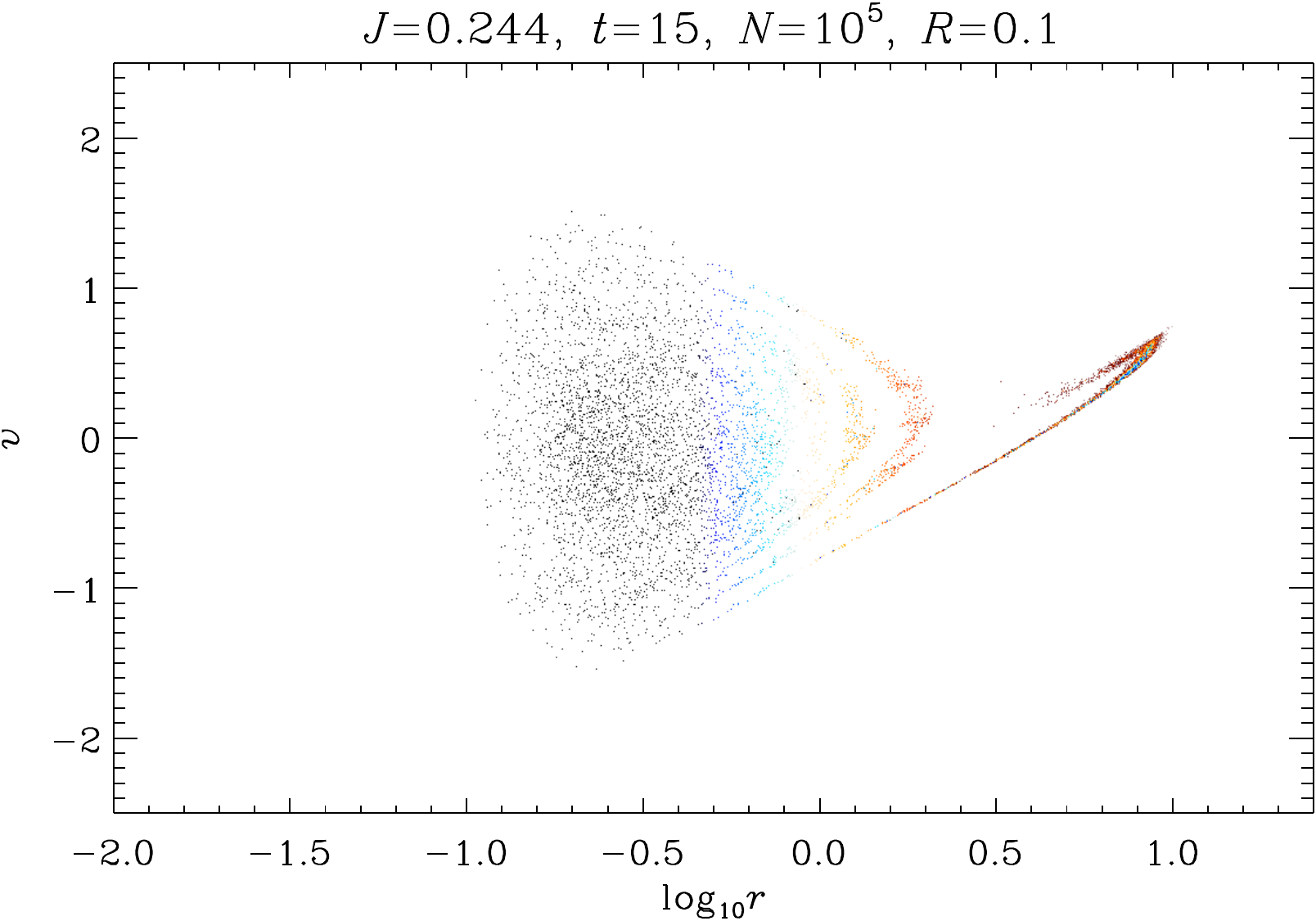}
\includegraphics[width=4.25cm]{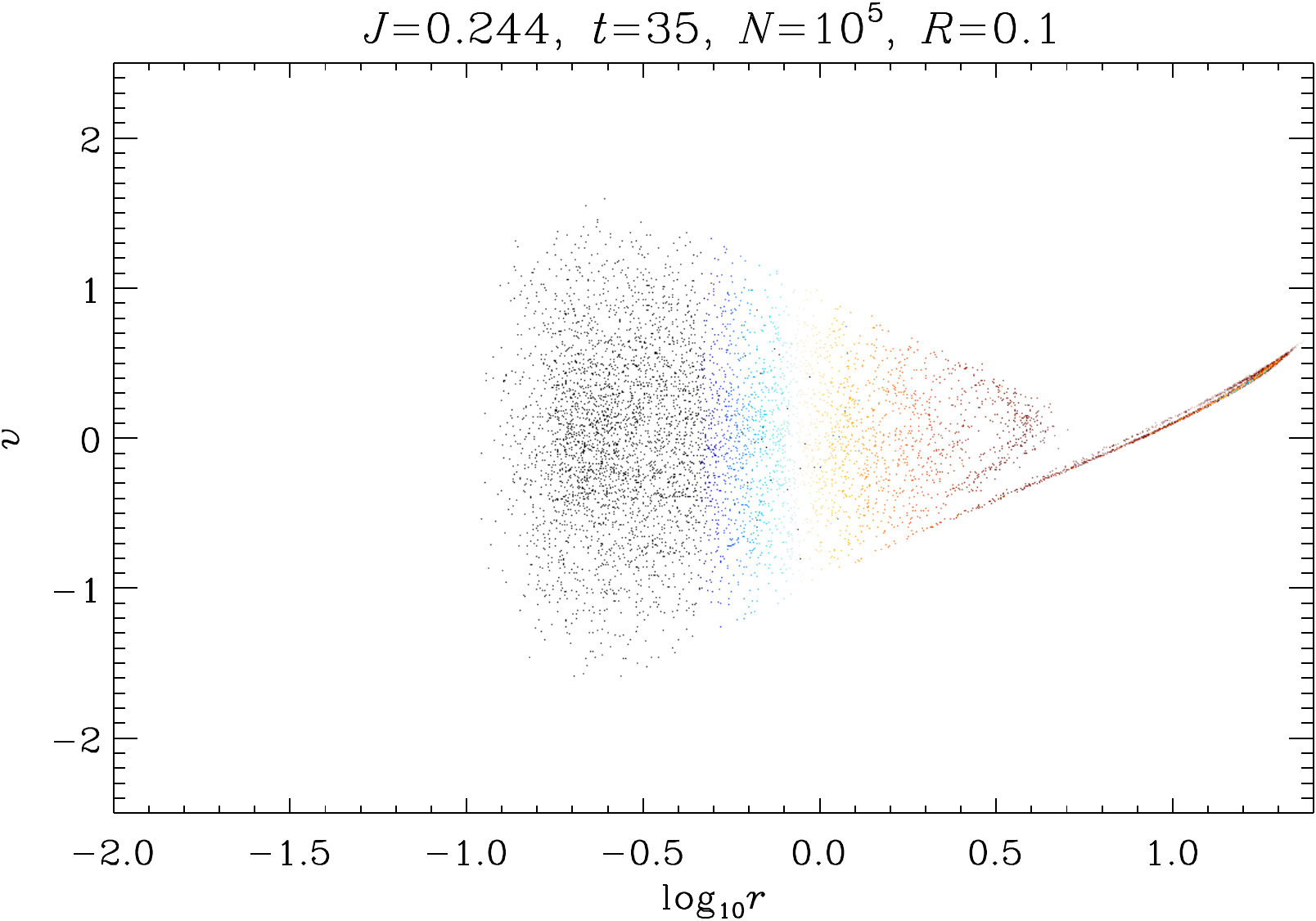}
}
\caption[]{Same as in Fig.~\ref{fig:0v5_12}, but for a colder initial
  configuration with virial ratio $R=0.1$. There is also an additional
  line of panels corresponding to the {\tt Gadget} simulation with
  $N=10^8$ particles. Note the large $R$ tail escaping from the
  system, corresponding to a fraction of the mass with positive energy
  \citep[see, e.g.,][]{VanAlbada1982,Joyce2009,SylosLabini2012}.}
\label{fig:0v1_12}
\end{figure*}

\begin{figure*}
\hbox{
\includegraphics[width=4.25cm]{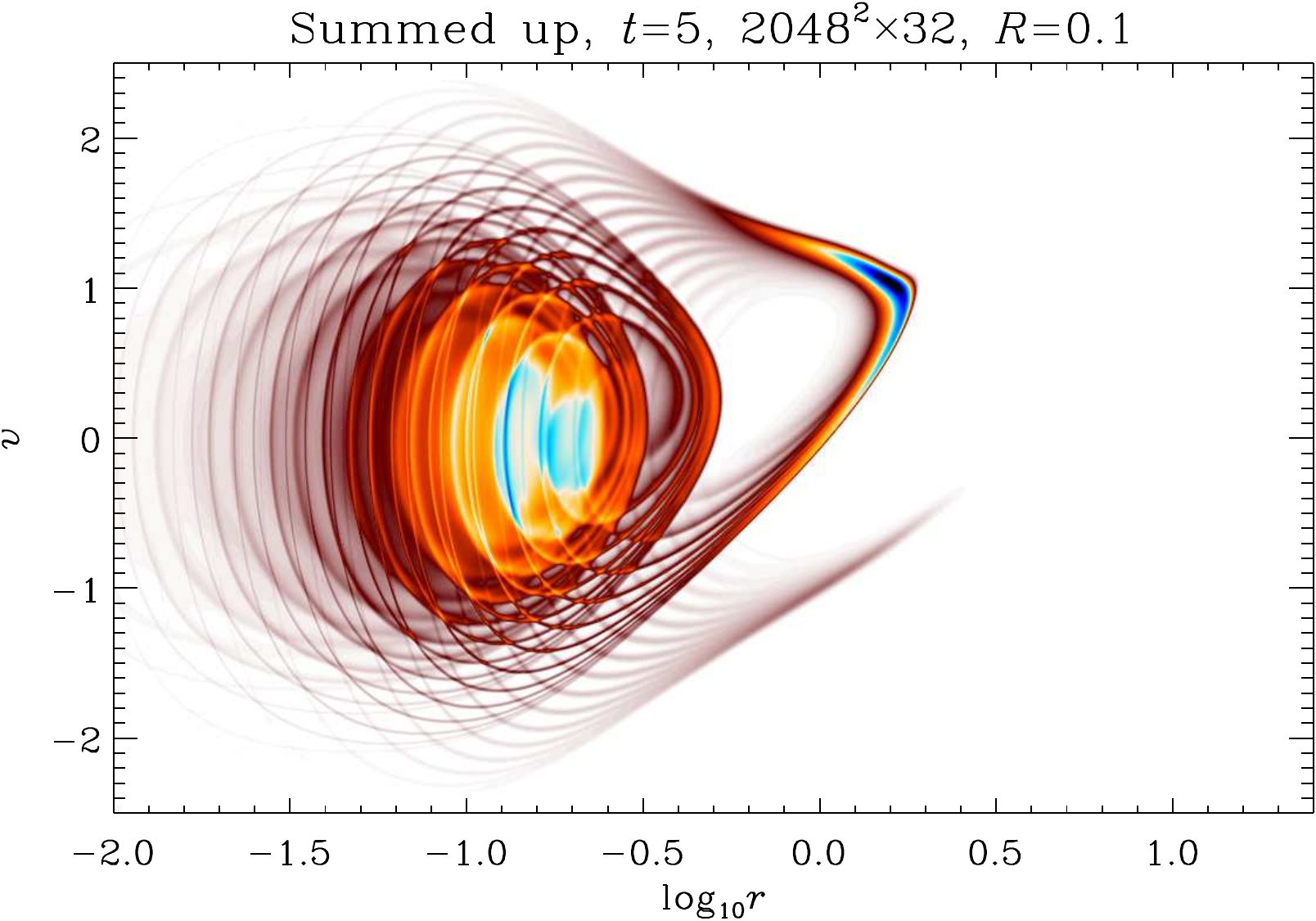}
\includegraphics[width=4.25cm]{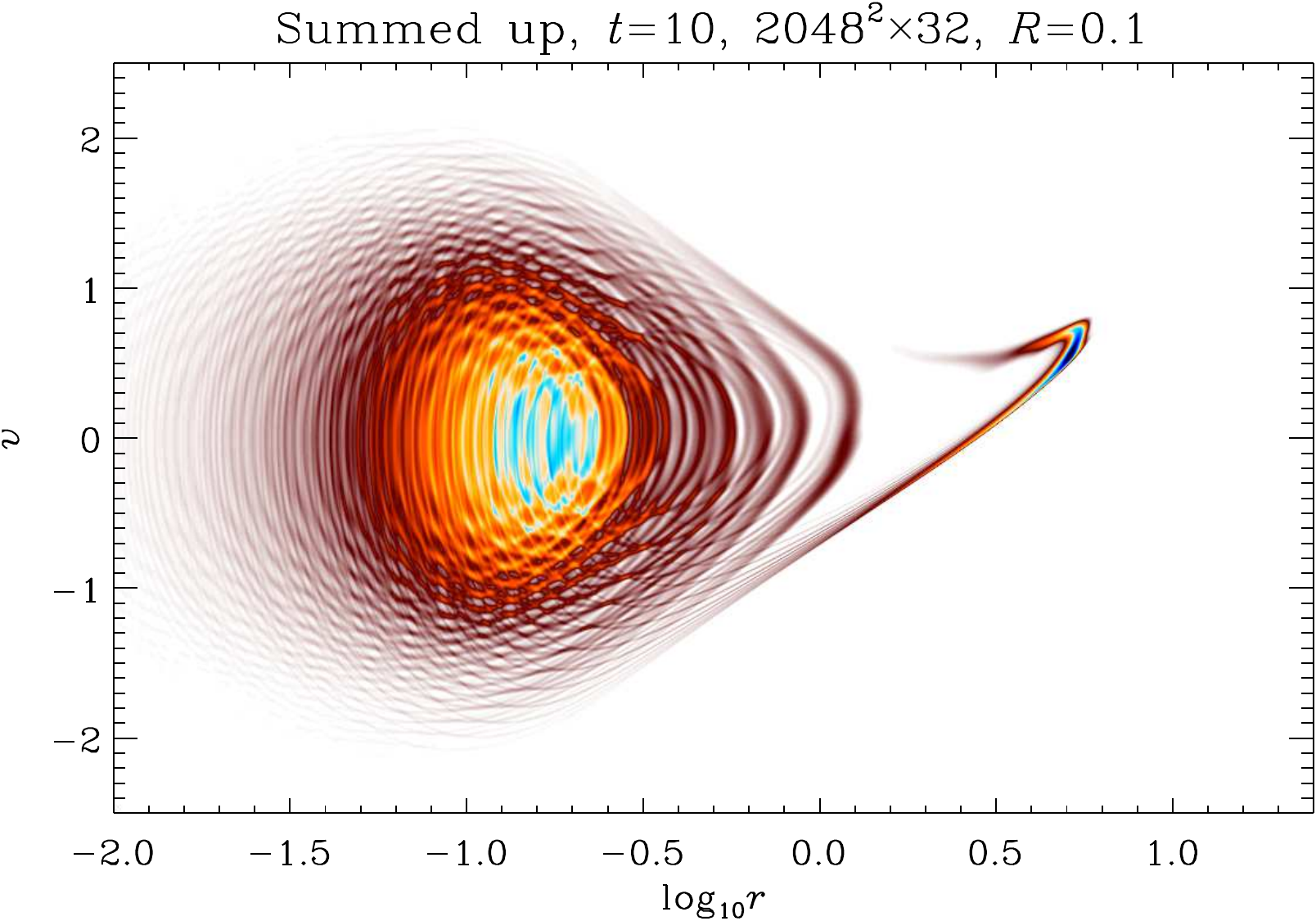}
\includegraphics[width=4.25cm]{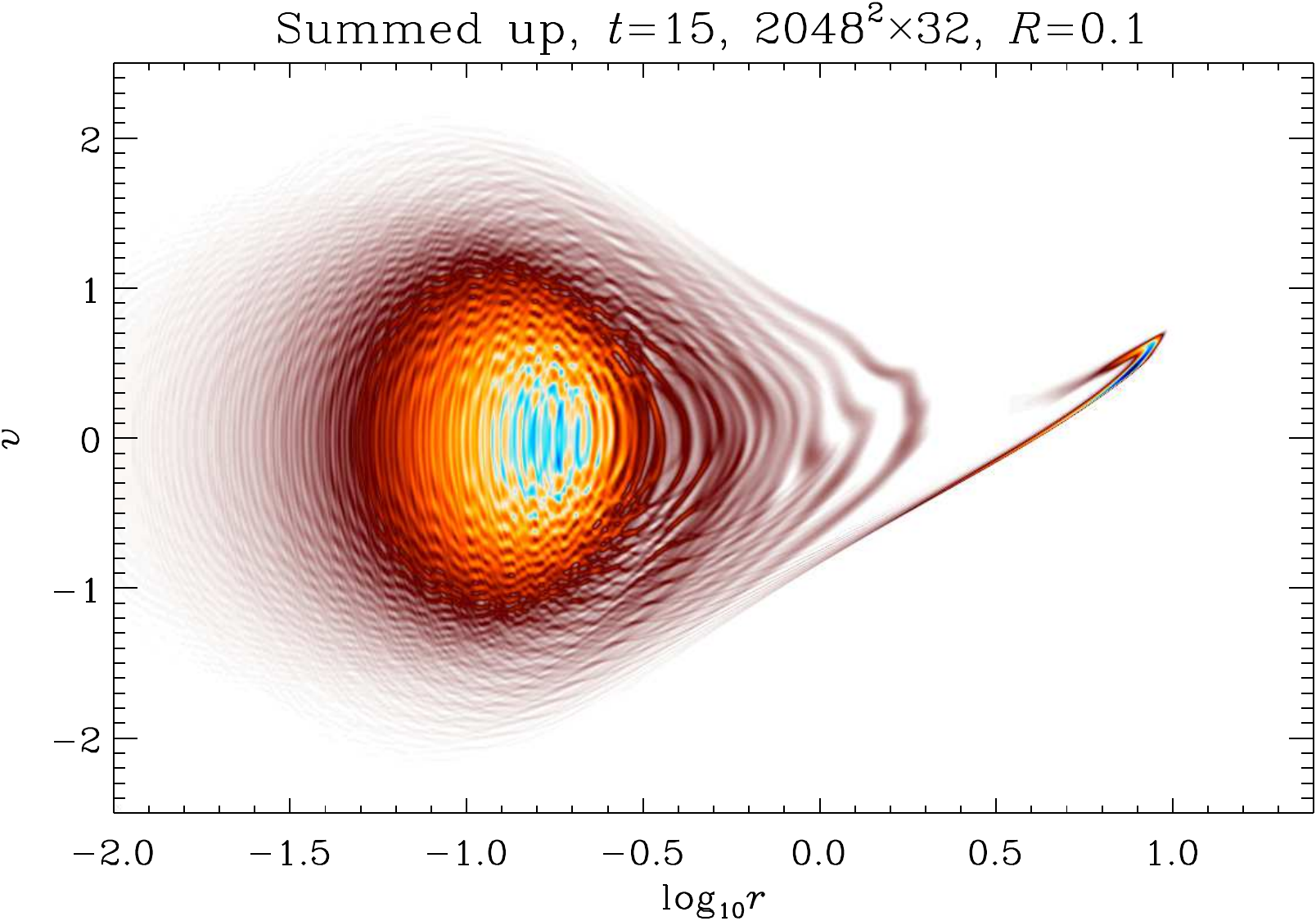}
\includegraphics[width=4.25cm]{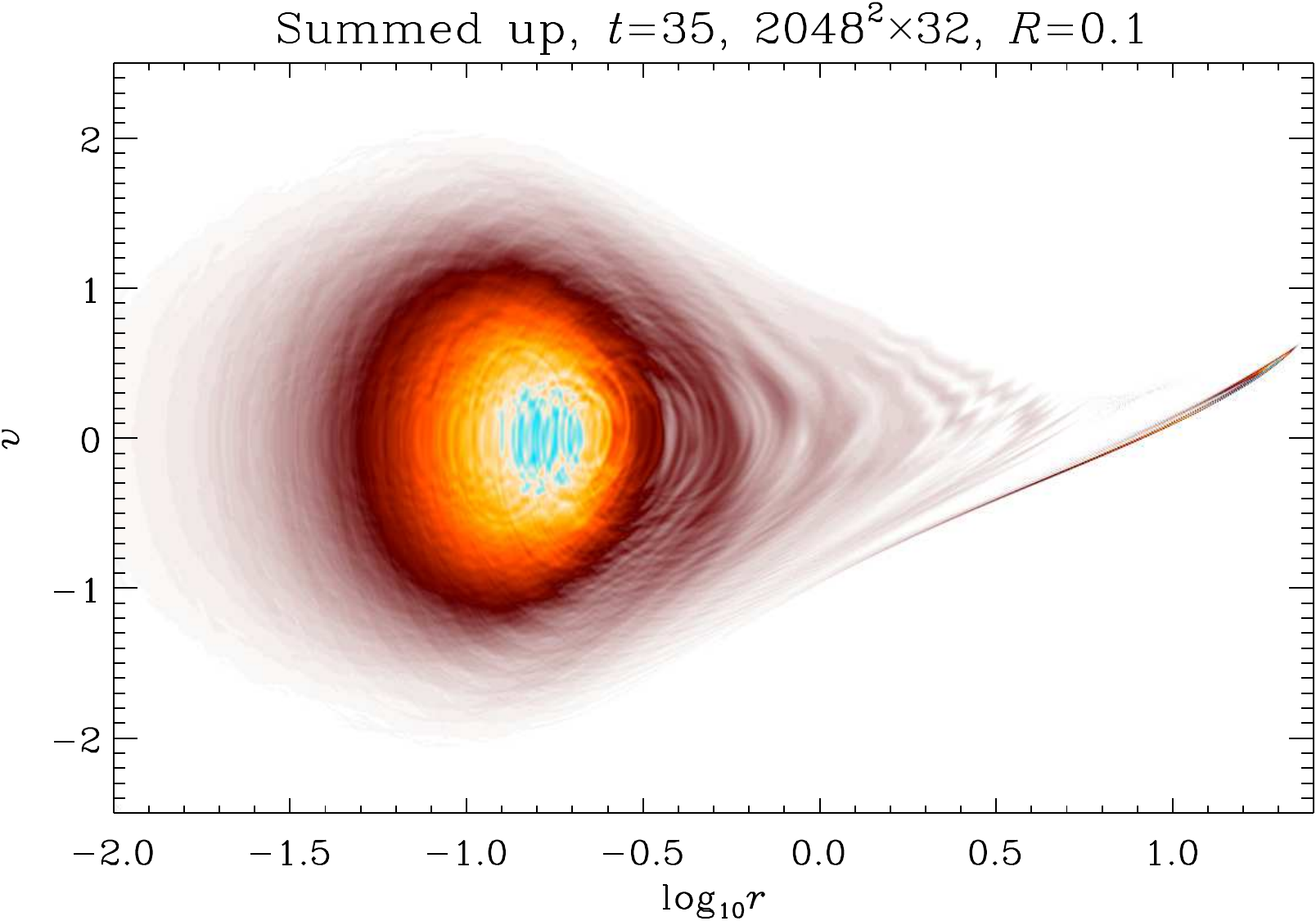}
}
\hbox{
\includegraphics[width=4.25cm]{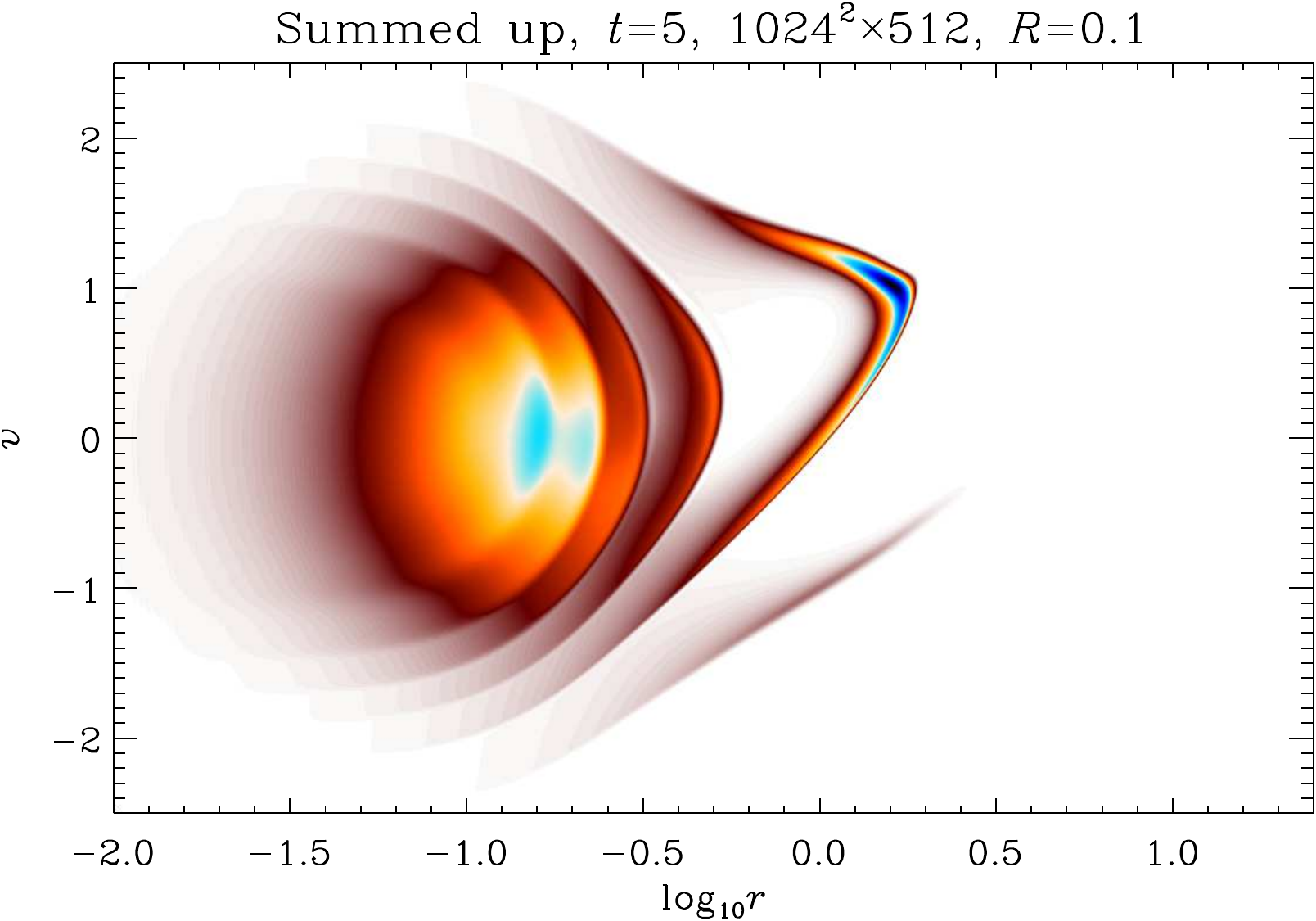}
\includegraphics[width=4.25cm]{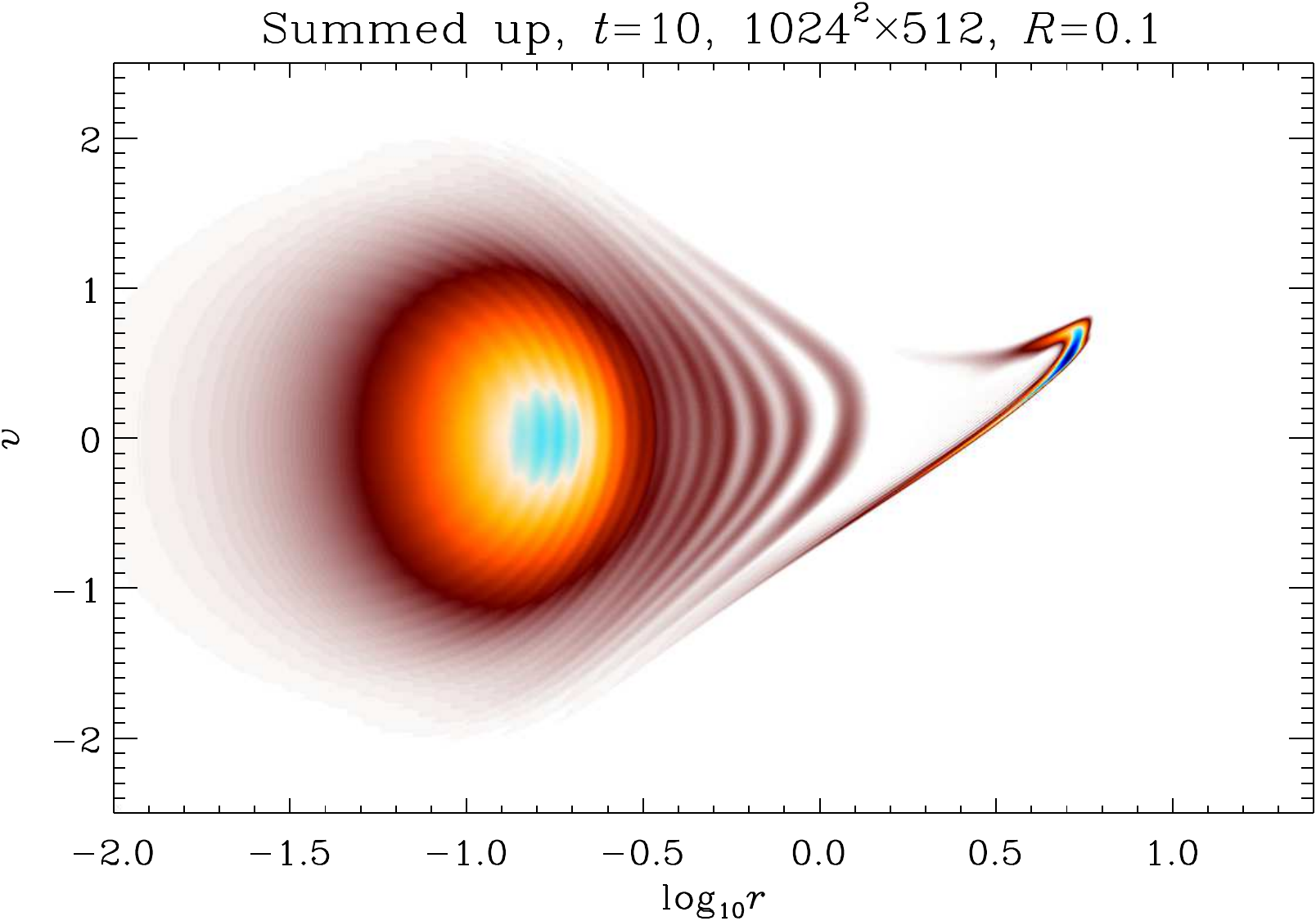}
\includegraphics[width=4.25cm]{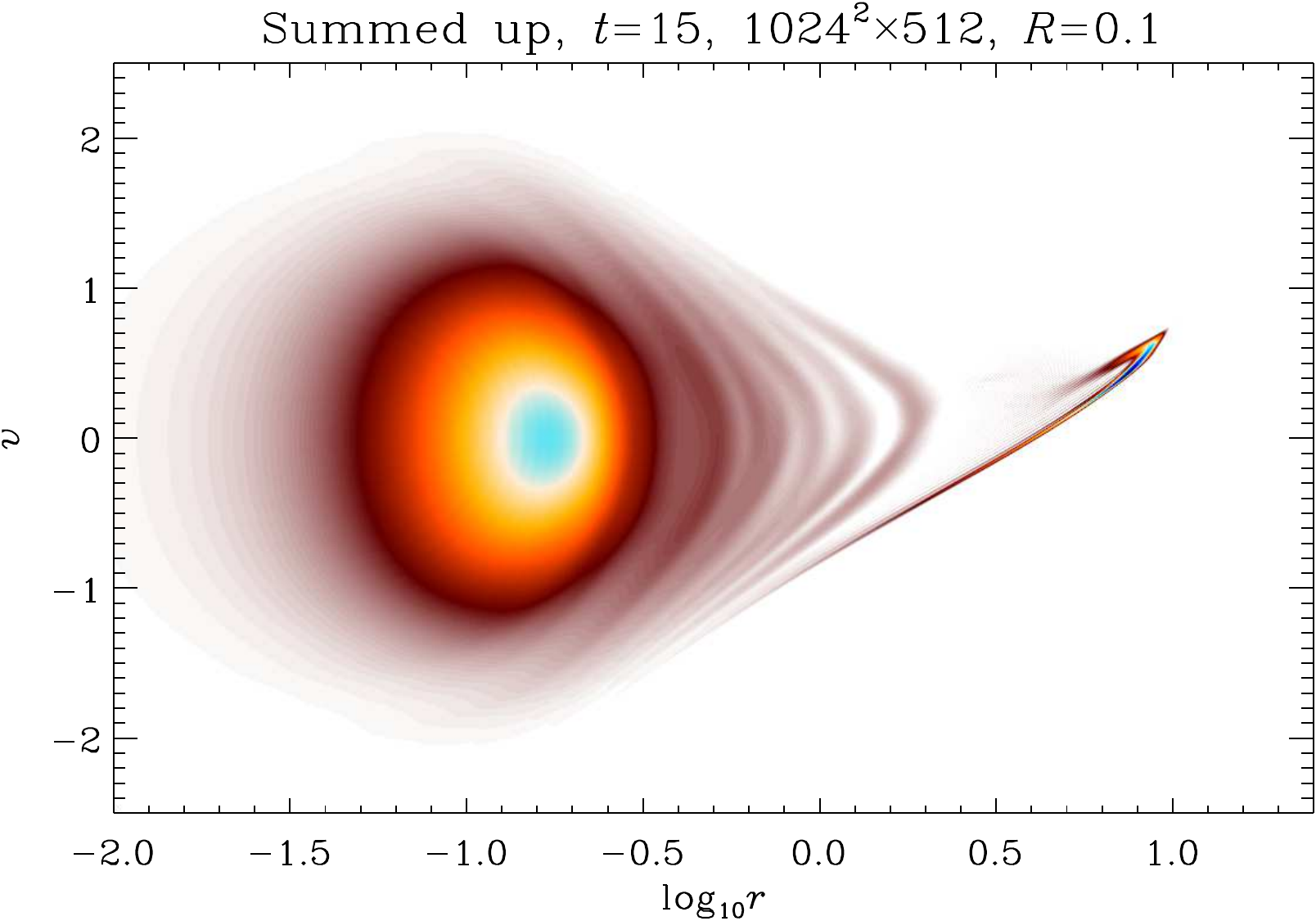}
\includegraphics[width=4.25cm]{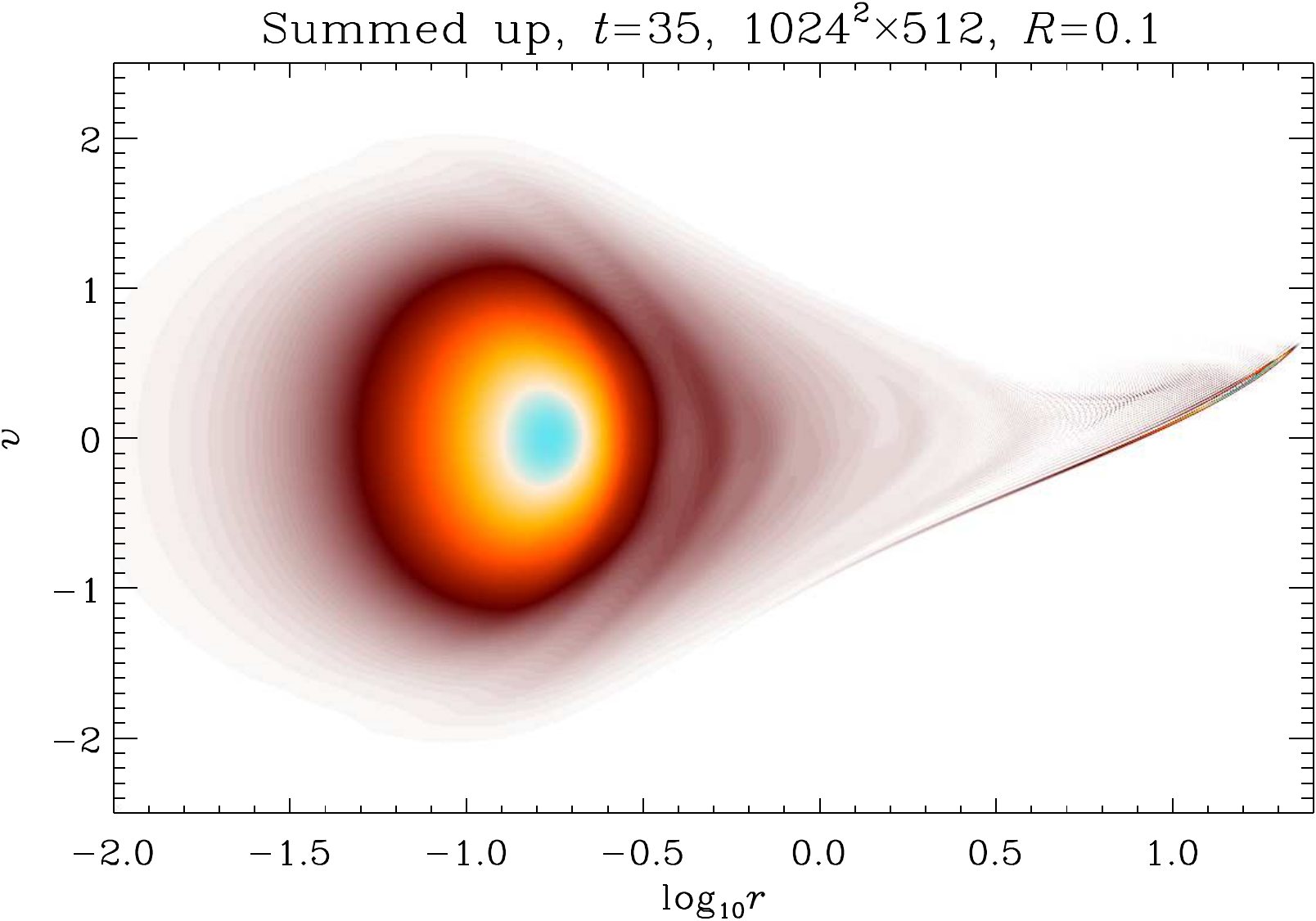}
}
\hbox{
\includegraphics[width=4.25cm]{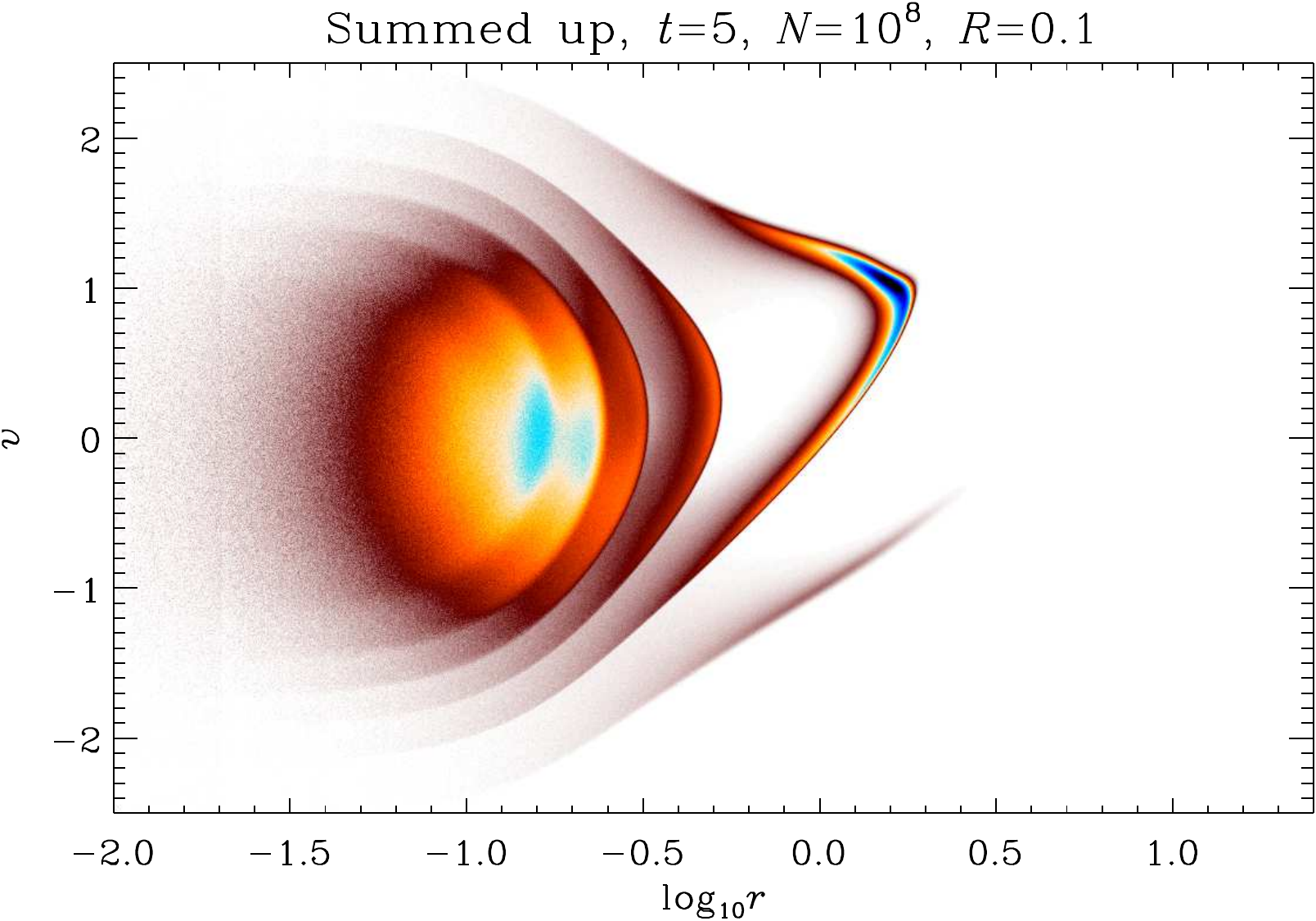}
\includegraphics[width=4.25cm]{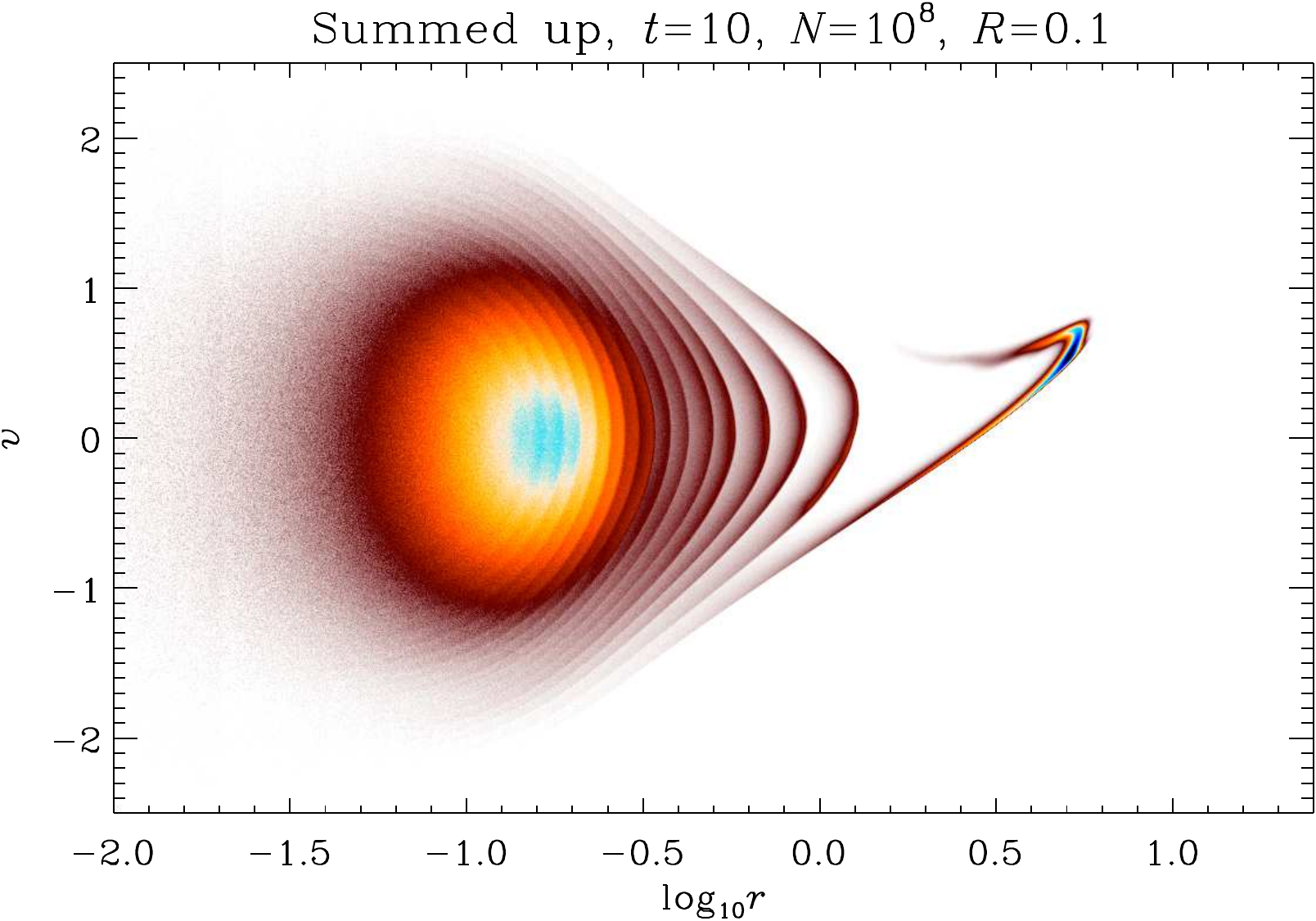}
\includegraphics[width=4.25cm]{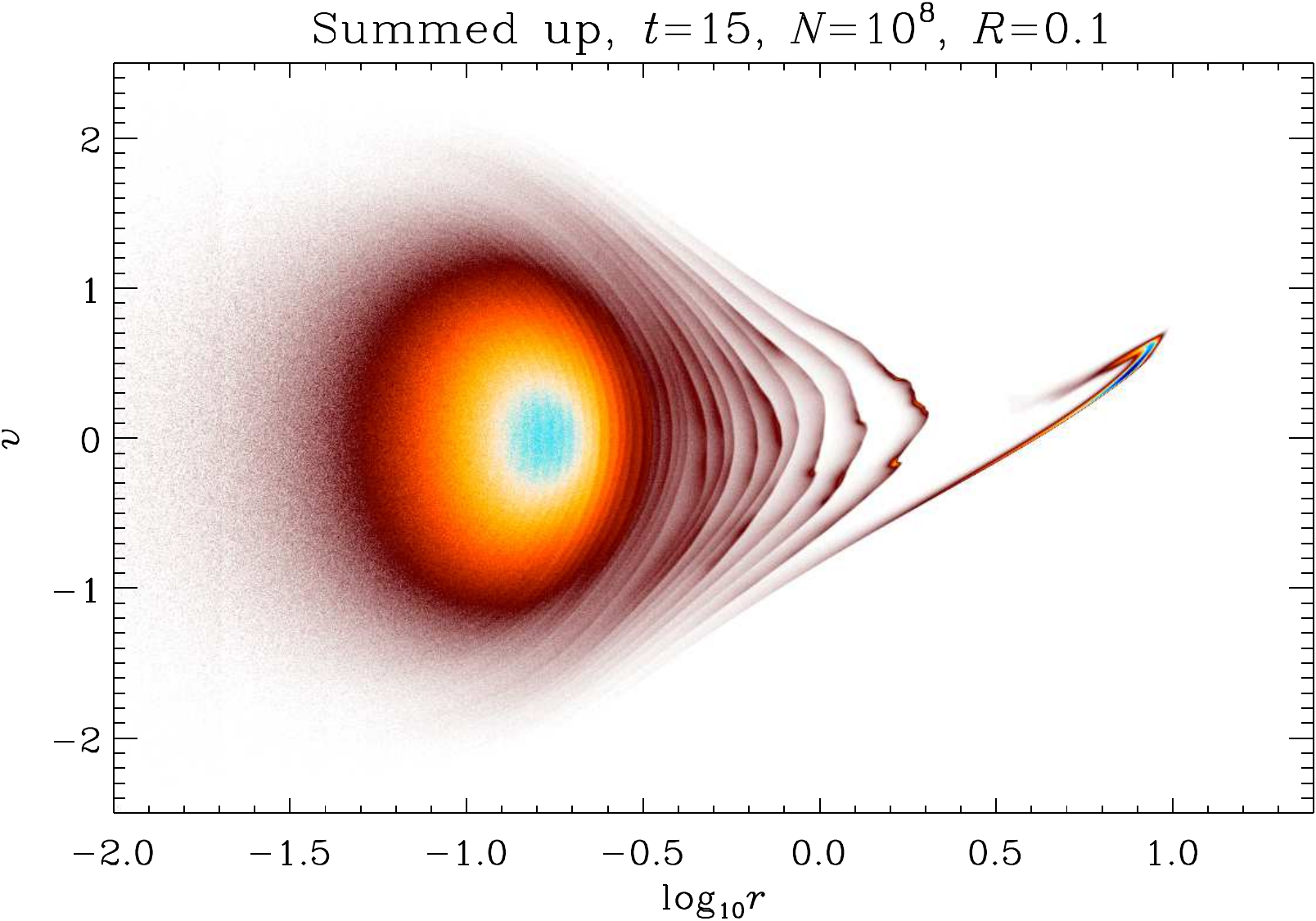}
\includegraphics[width=4.25cm]{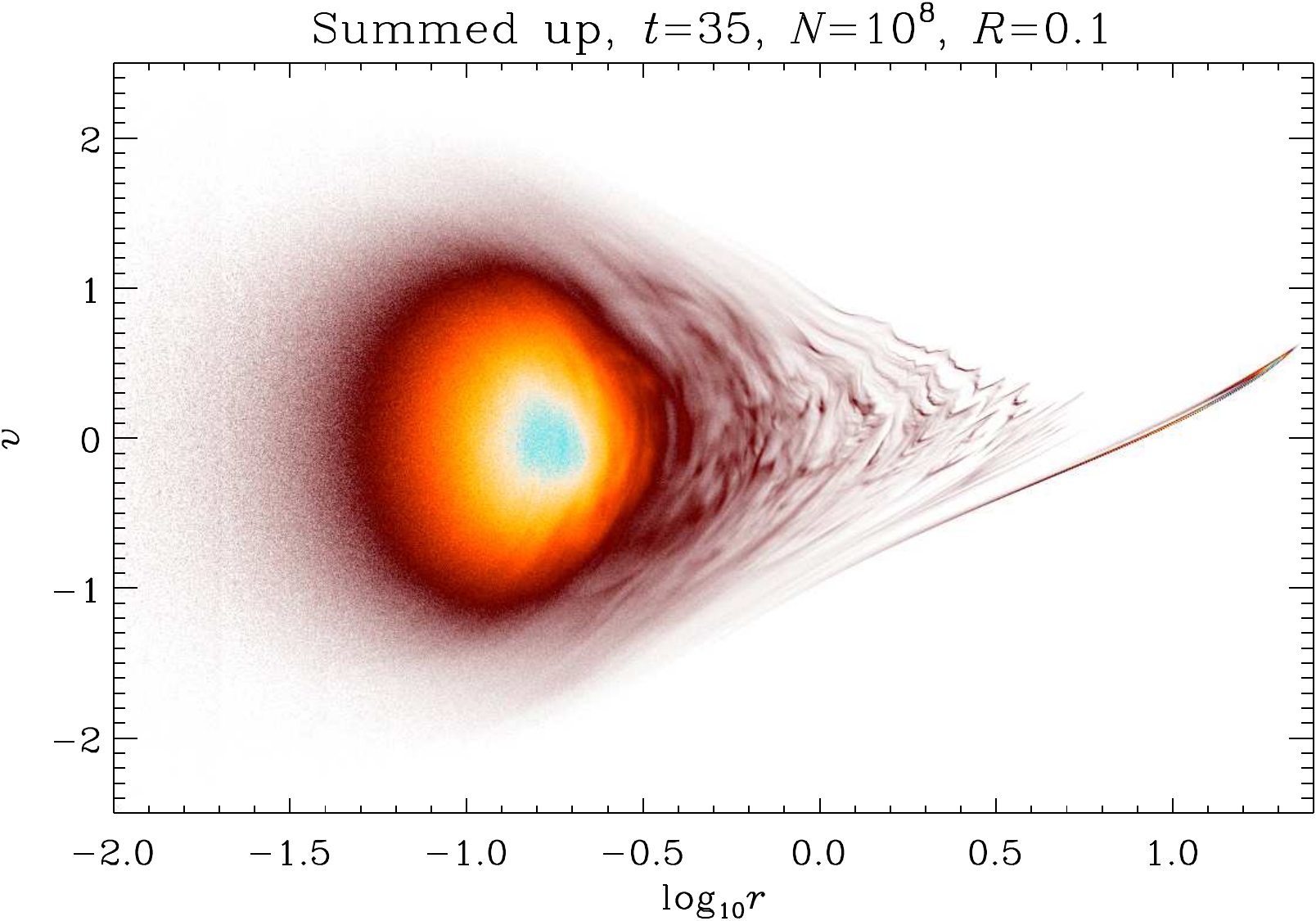}
}
\hbox{
\includegraphics[width=4.25cm]{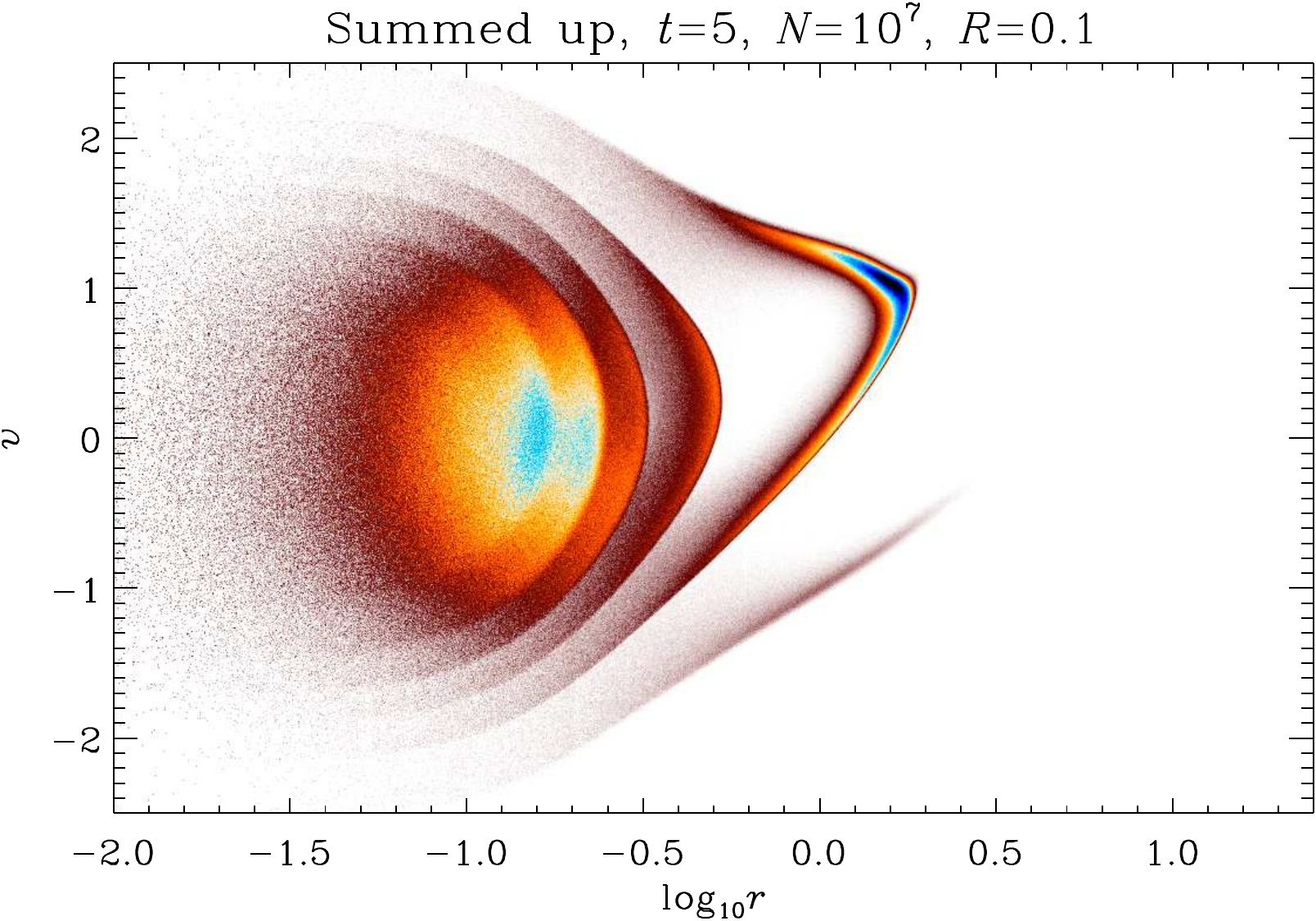}
\includegraphics[width=4.25cm]{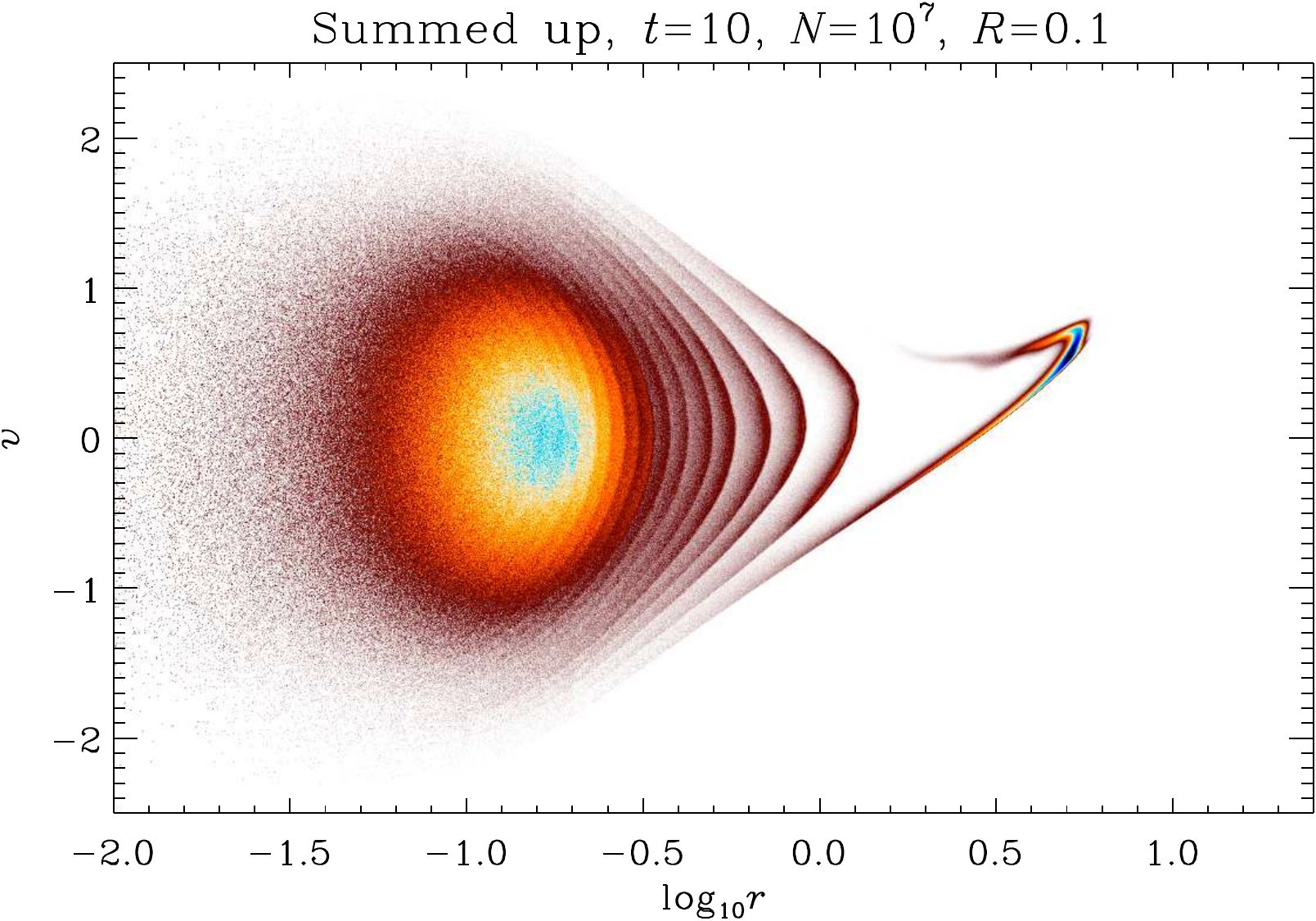}
\includegraphics[width=4.25cm]{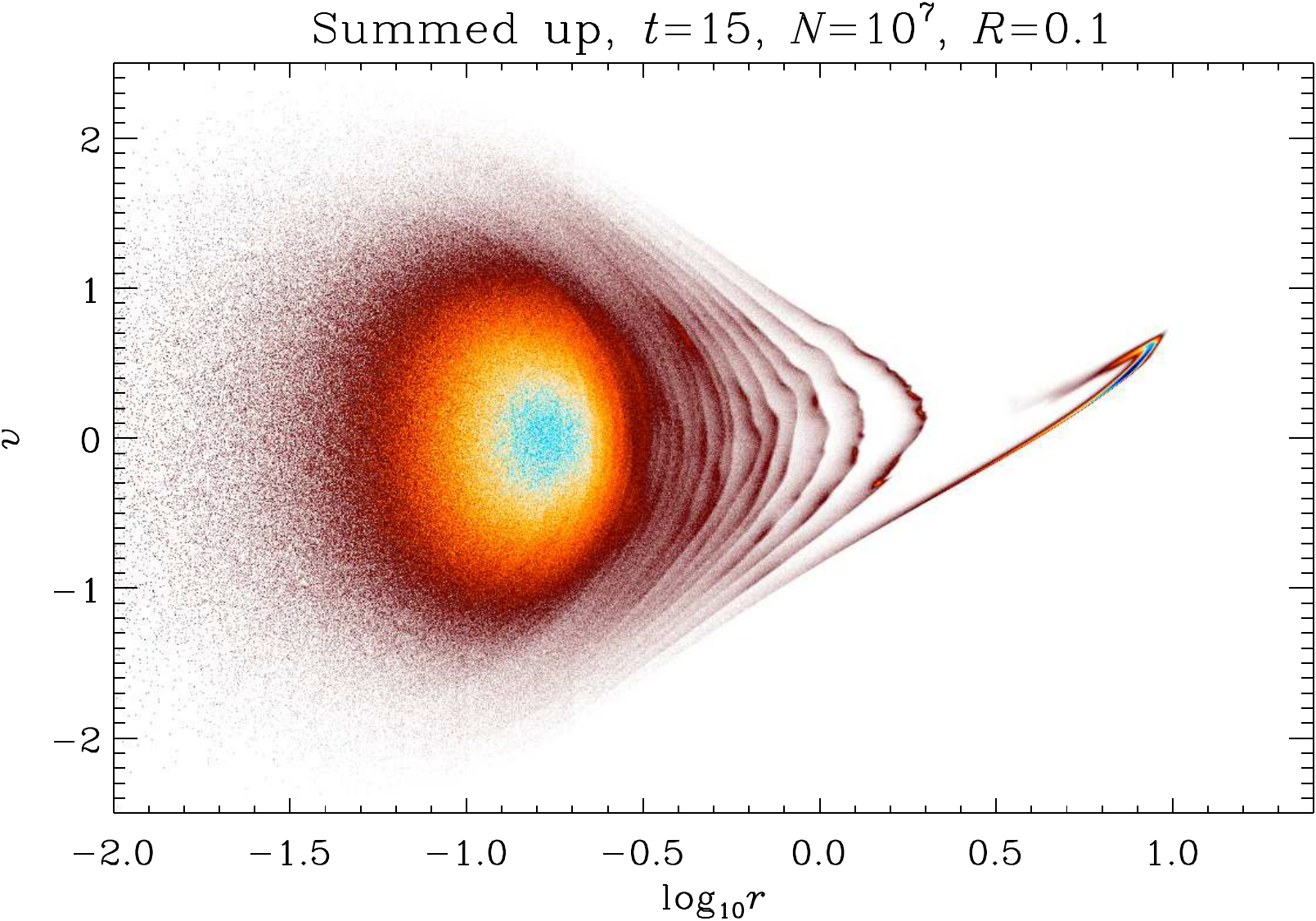}
\includegraphics[width=4.25cm]{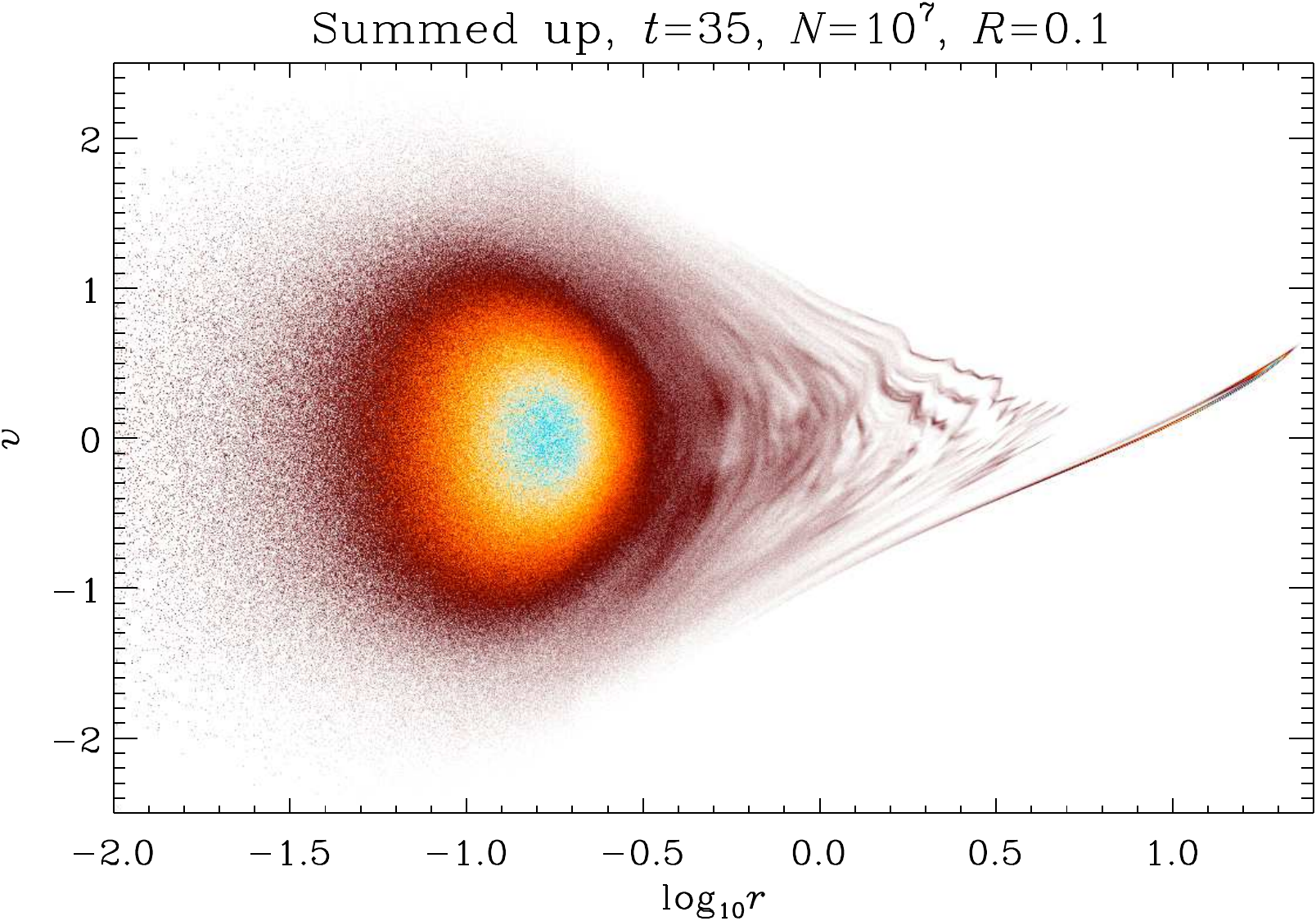}
}\hbox{
\includegraphics[width=4.25cm]{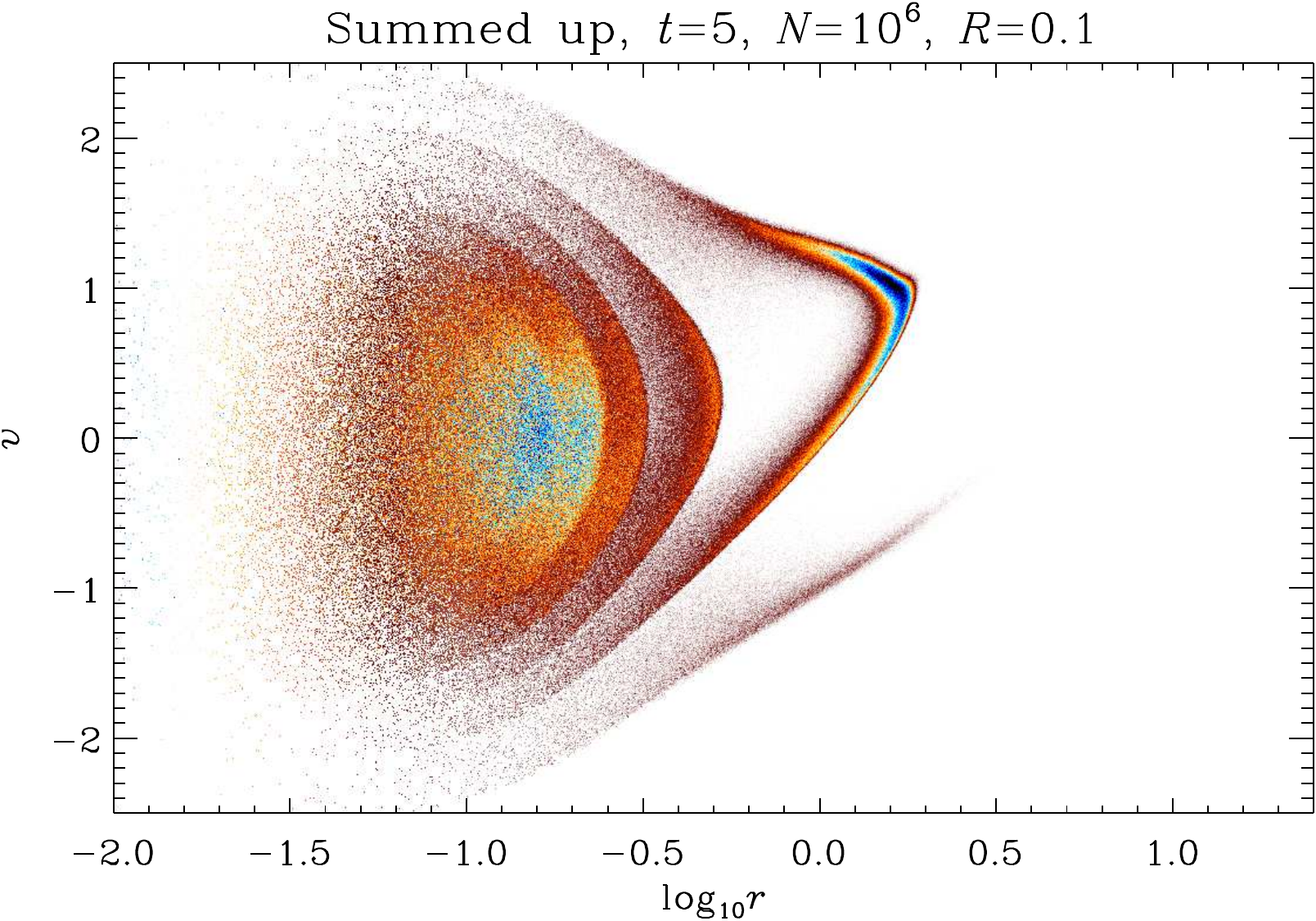}
\includegraphics[width=4.25cm]{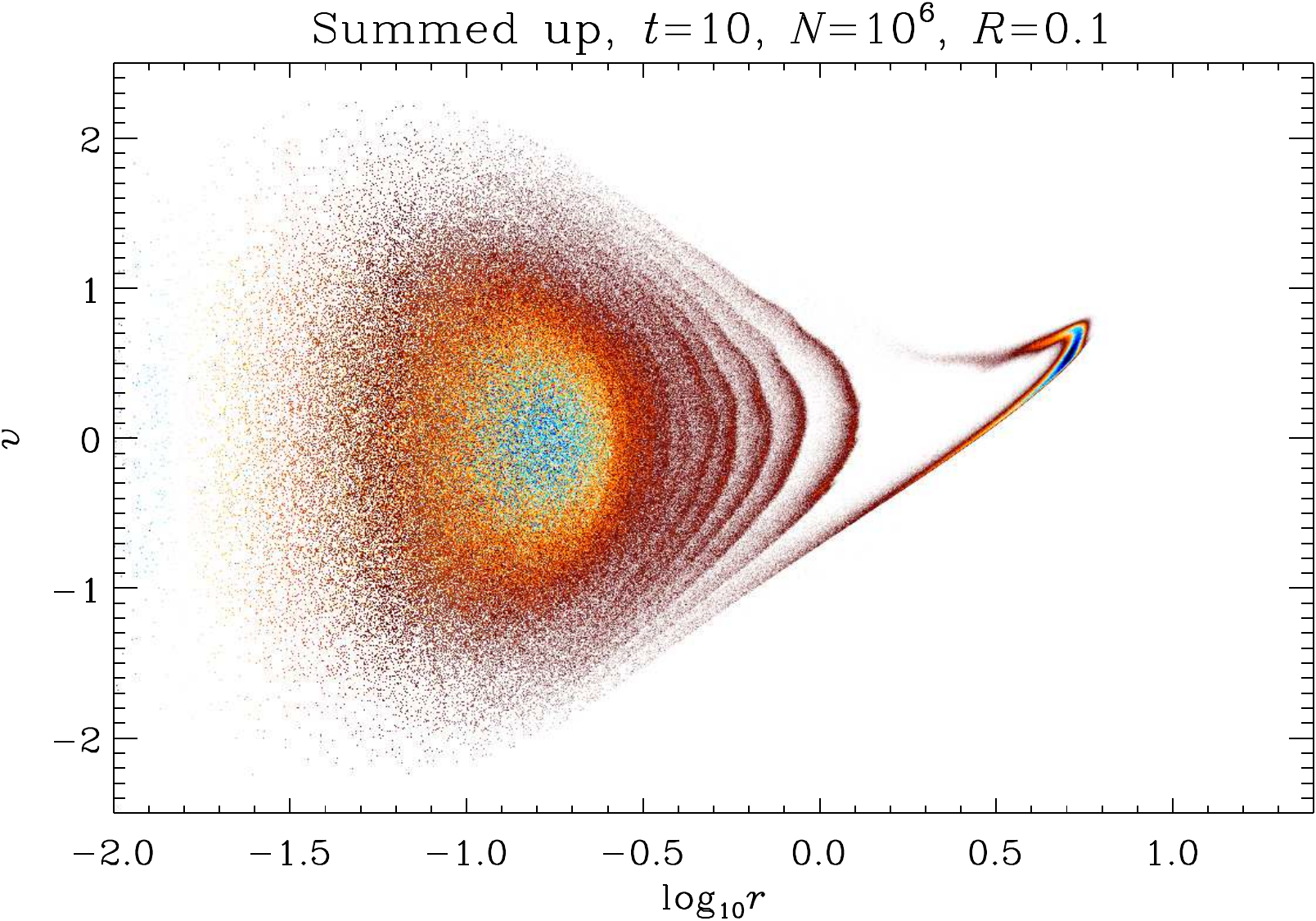}
\includegraphics[width=4.25cm]{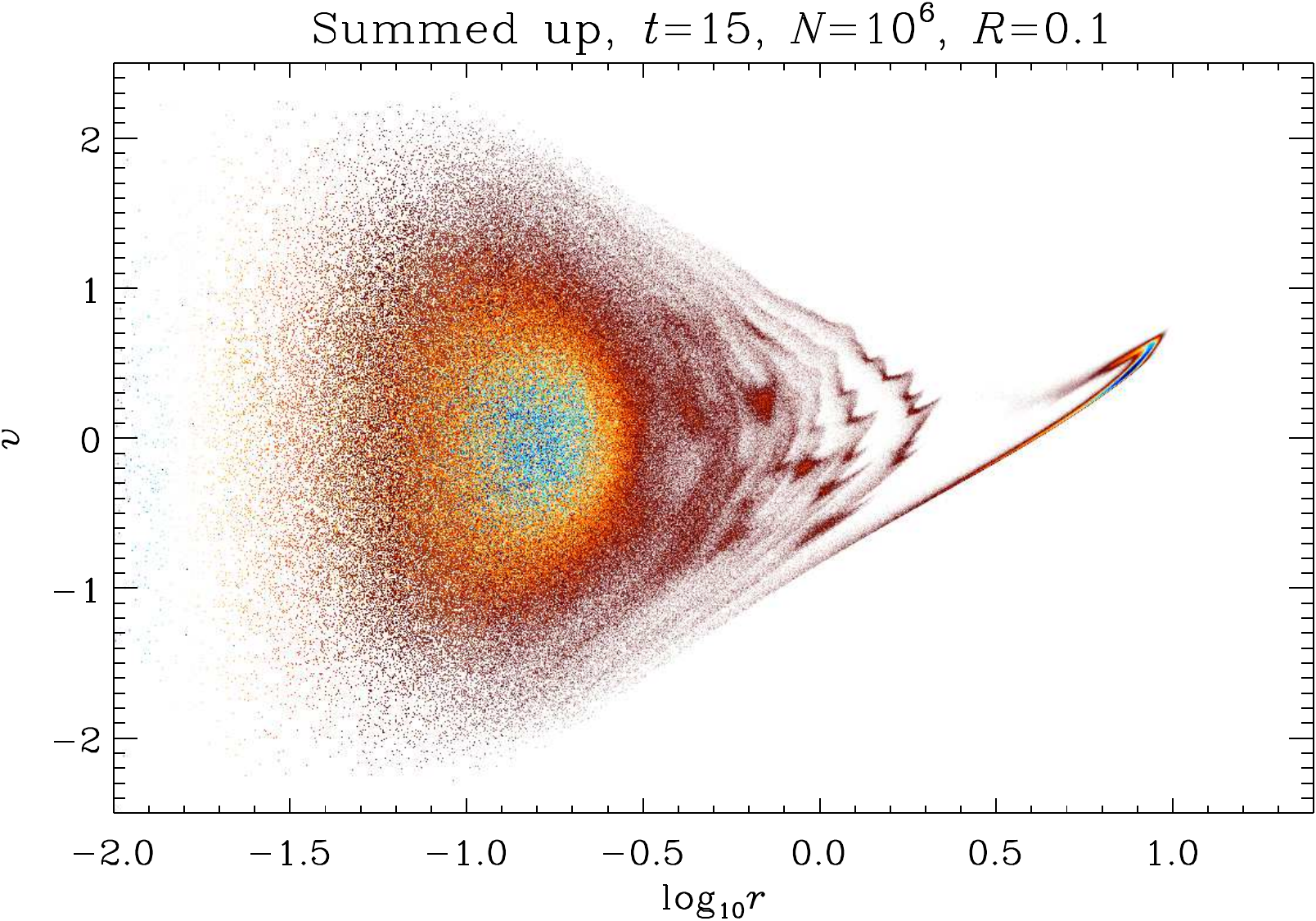}
\includegraphics[width=4.25cm]{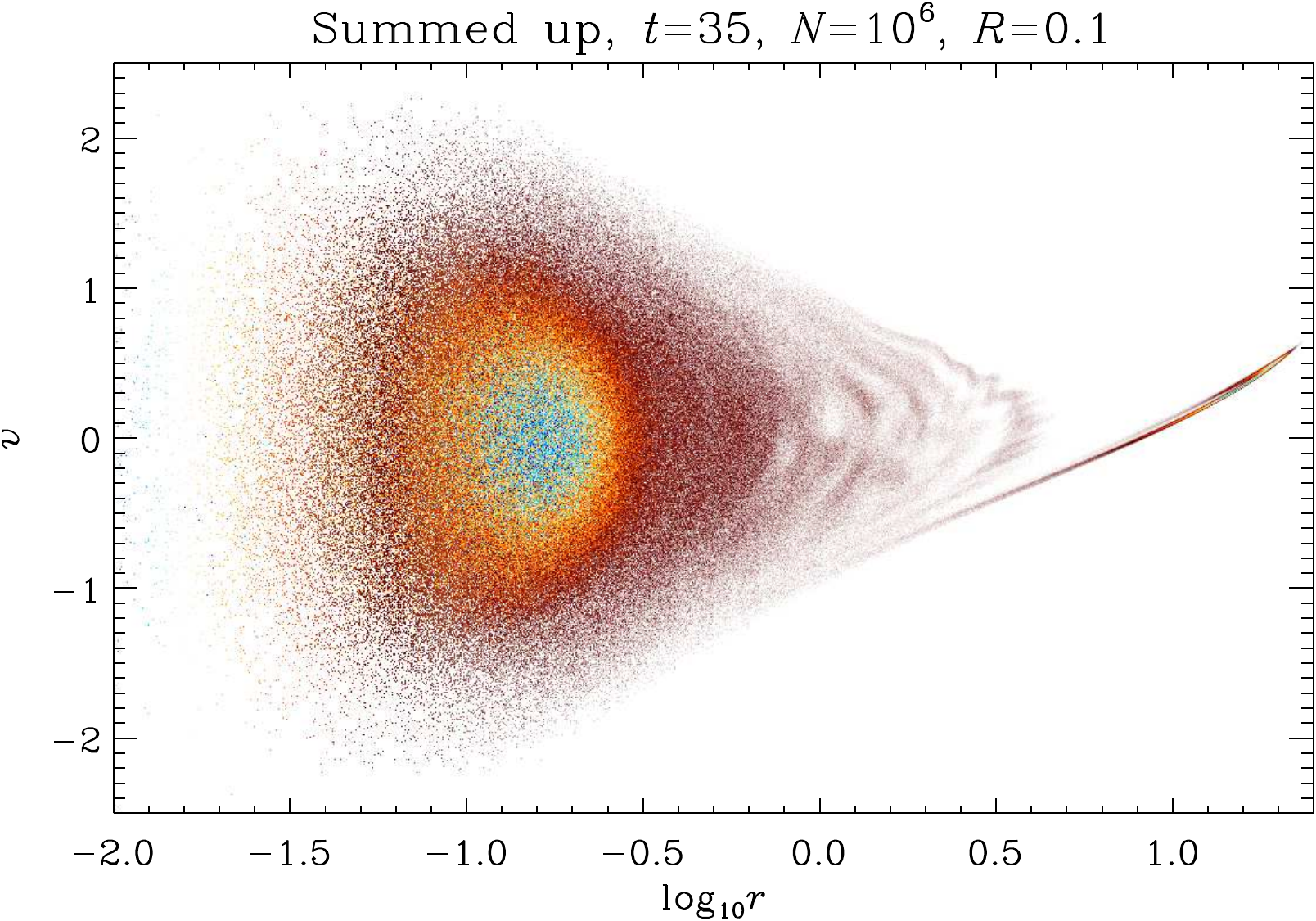}
}
\hbox{
\includegraphics[width=4.25cm]{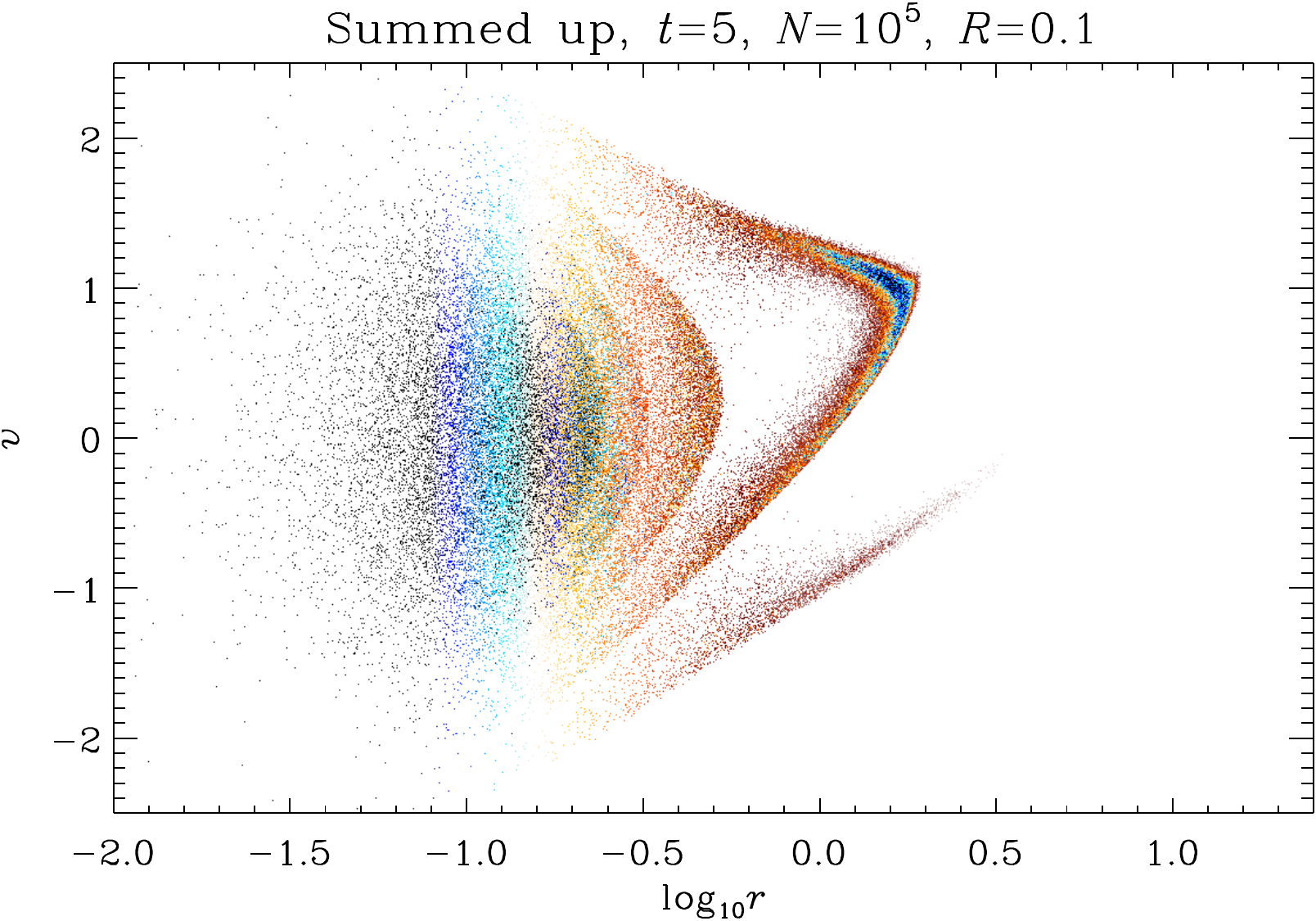}
\includegraphics[width=4.25cm]{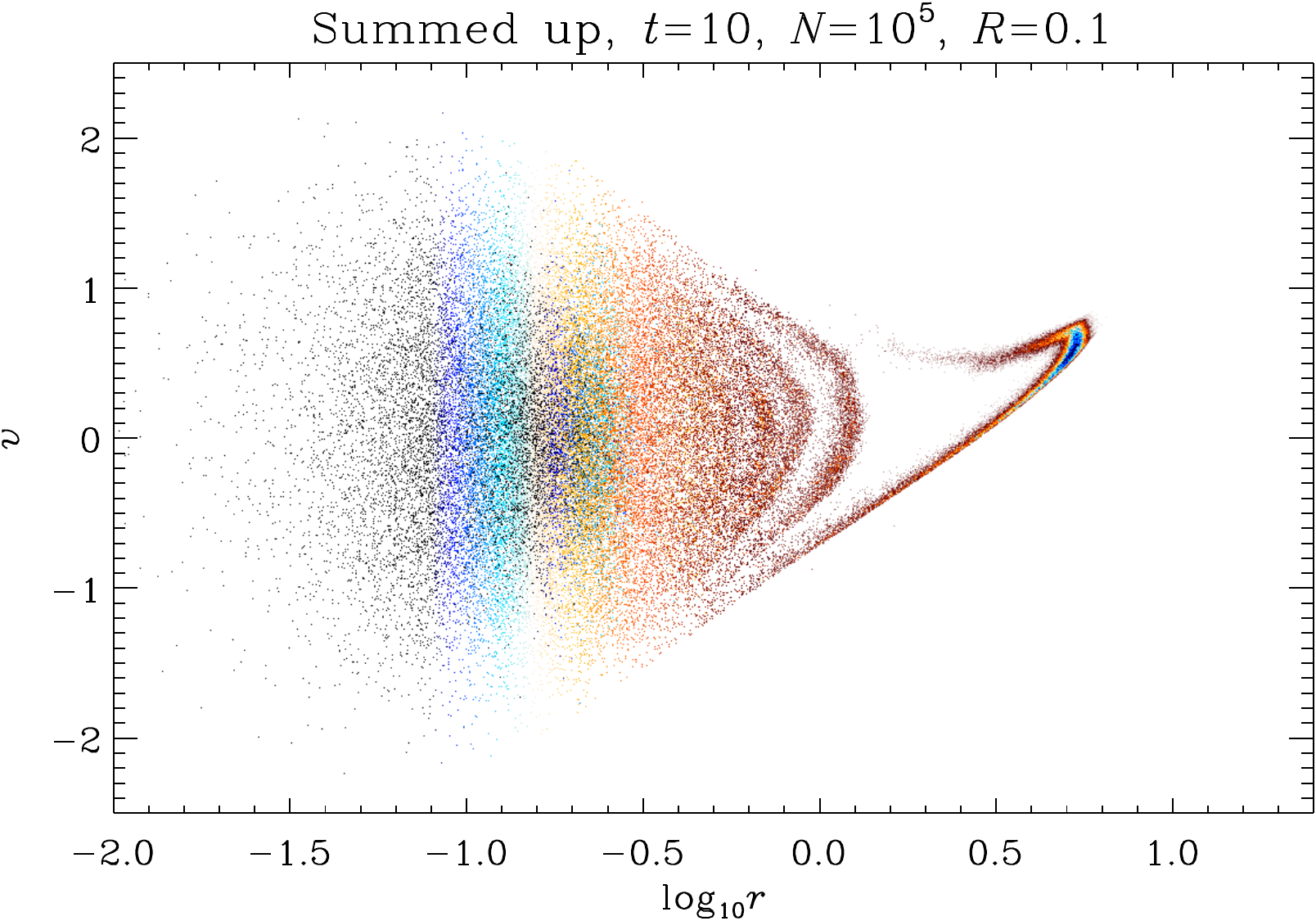}
\includegraphics[width=4.25cm]{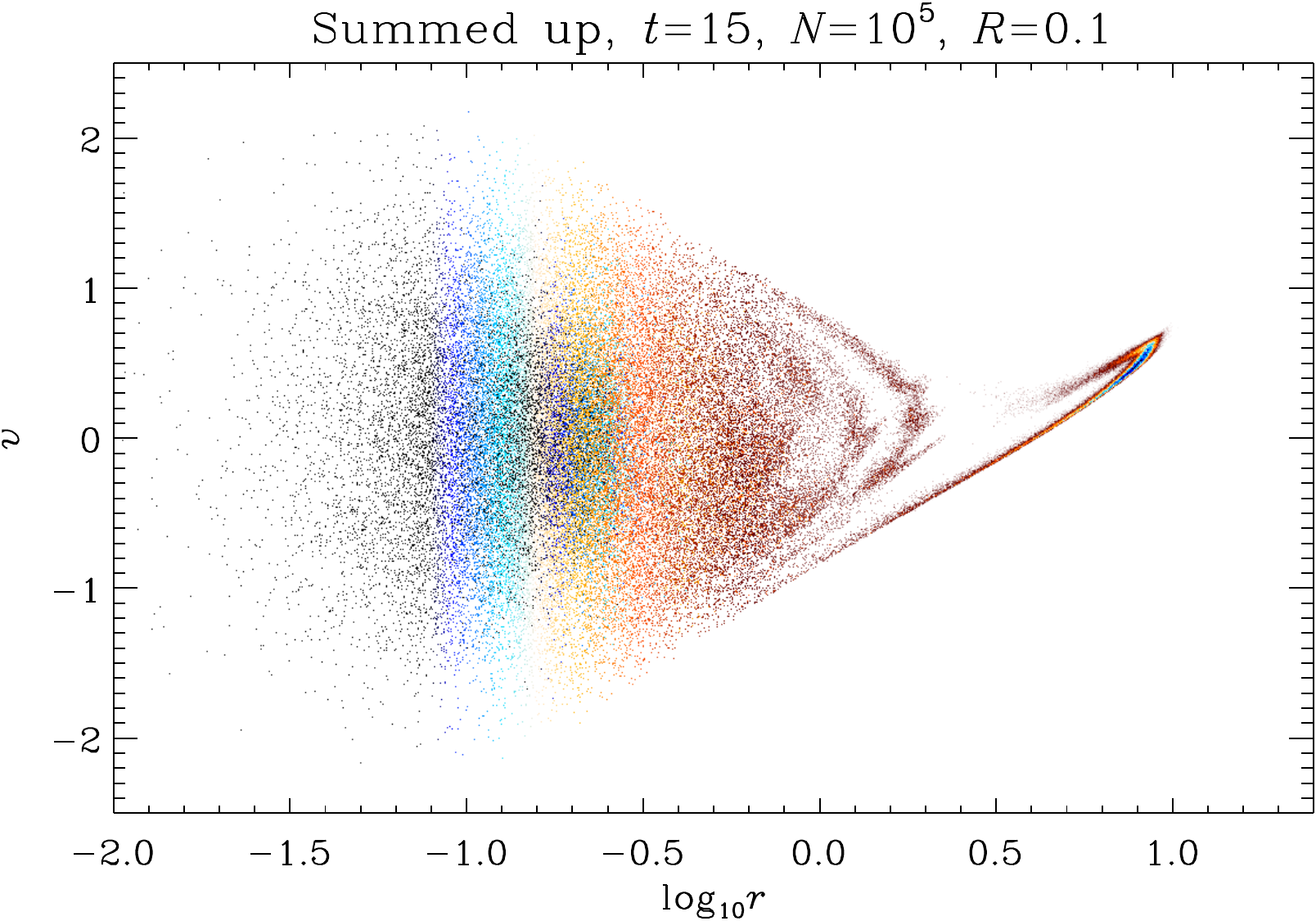}
\includegraphics[width=4.25cm]{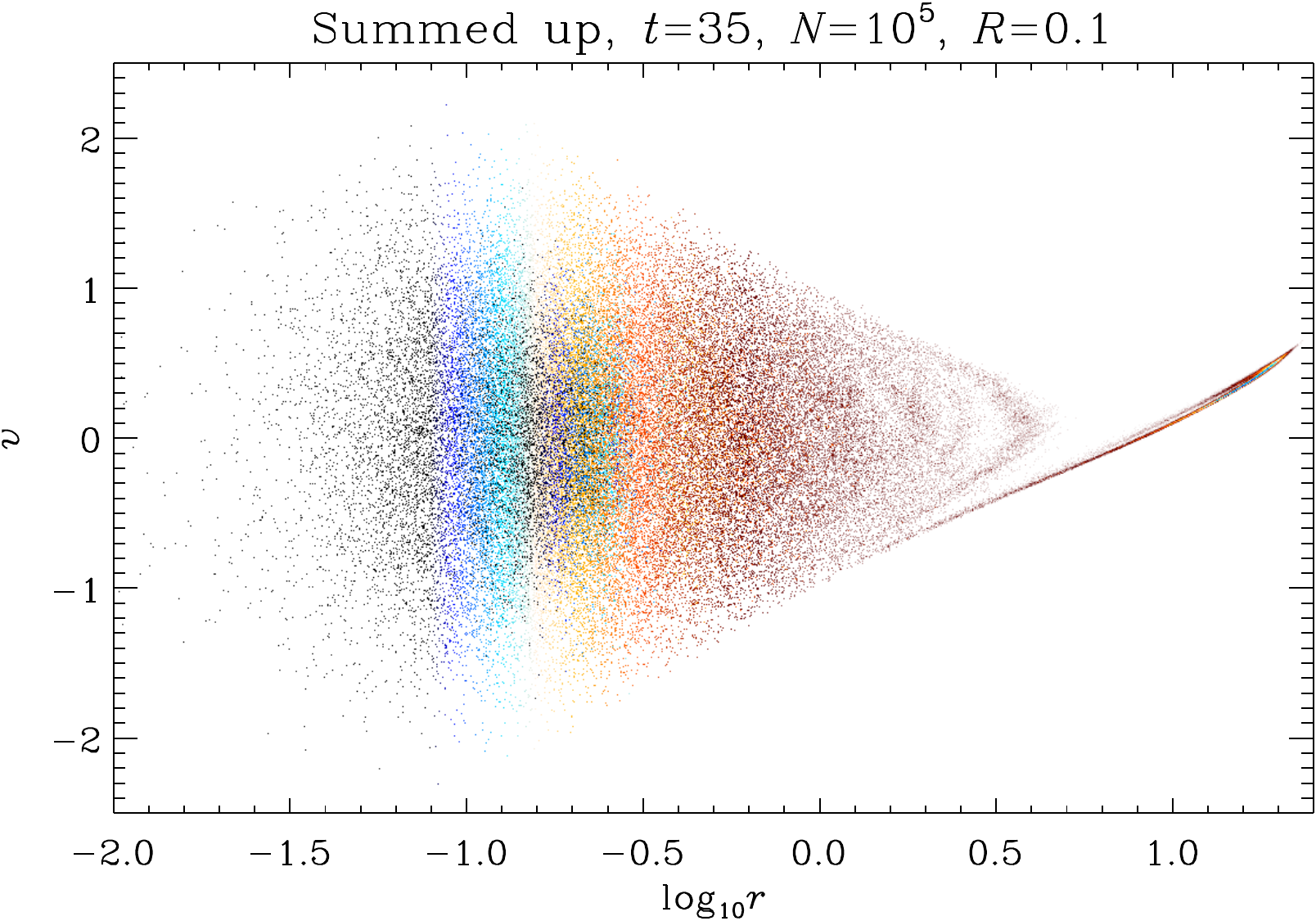}
}
\caption[]{Same as in Fig.~\ref{fig:0v5_ALLJ}, but for $R=0.1$ and with the additional $N$-body simulation involving $N=10^8$ particles.}
\label{fig:0v1_ALL}
\end{figure*}

The overall conclusion of the visual inspection of 
Figs.~\ref{fig:0v5_12} and \ref{fig:0v5_ALLJ} is that the Vlasov
solver and the $N$-body code exhibit very good agreement with each
other, probably even much more than
expected. In particular, both results present a
remarkably similar instability in the region $1 \la r \la 10^{0.8}$,
even in details, showing a surprising reliability of the conventional $N$-body
approach for these particular initial conditions.

However, before reaching this conclusion, one has 
to take into account several limiting factors. In particular, 
we should bear in mind the fact that the {\tt VlaSolve} simulations
are subject to significant diffusion, which smears out fine details of
the phase-space distribution function. This diffusion effect is
clearly visible at $t=50$, when comparing the outer filamentary
structures observed in the Vlasov simulations to the $N$-body
result. Putting aside this coarse-graining effect, the structures are
exactly similar in both the $N$-body and Vlasov simulations at $t \leq
50$, even including small gaps in the phase-space distribution
function related to nonlinear instabilities that start building
up. These instabilities grow further at later epochs.  They are
considerably smeared out in the $(1024,1024,512)$ {\tt VlaSolve}
simulation but unquestionably present. Adding resolution in $(r,v)$
space (at the cost of resolution in $j$) improves the agreement with
{\tt Gadget}, which
confirms that the instabilities observed in the {\tt Gadget}
simulations are physical and not of numerical nature.

Figure \ref{fig:0v5_12} indicates that lowering the number of
particles in the $N$-body simulations may be interpreted as a
coarse-graining: 
it makes finer details more fuzzy but still keeps global features of
the phase-space density correctly. We also note that using a small
number of slices in $j$ in the Vlasov solver does not seem to alter
the dynamical properties of the system despite the considerable level
of aliasing it introduces.

The situation is more complicated for the cold case, $R=0.1$
(Figs.~\ref{fig:0v1_12} and \ref{fig:0v1_ALL}).  Up to $t \simeq 10$,
the above conclusions for $R=0.5$ are still valid. However, some
instabilities emerge at $t \simeq 10$ in the {\tt Gadget} simulations
with $N \leq 10^6$ particles as well as the $(2048,2048,32)$ Vlasov
run. Until this epoch, the $N \geq 10^7$ and the
$(1024,1024,512)$ simulations agree perfectly with each other (modulo
the smearing effects already discussed above) and present a smooth
phase-space density without any sign of instability.  On the other
hand, the other simulations exhibit slightly irregular phase-space
density.  Such a trend is easily seen in Fig.~\ref{fig:0v1_12}, even
though not so obvious in the $(2048,2048,32)$ Vlasov simulation.
These irregularities appear as well in the $N \geq 10^7$ simulations
but at later epochs, and then develop in a dramatic way. A careful
inspection of successive snapshots of the simulations indeed suggests
that the onset of these irregular patterns comes later with
increasing $N$.

As discussed in Appendices~\ref{app:instab} and \ref{app:resvlaeff},
these instabilities result from the discrete nature of the system in
the $N$-body case, and from the aliasing effect due to sparse-sampling
of the angular momentum space in the Vlasov code.  Since the pattern
of the instabilities changes significantly from one simulation to
another unlike the $R=0.5$ case, they should be due to numerical, not
physical, origin.

As shown in Appendix~\ref{app:instab}, their presence is very
insensitive to the choice of softening, time step or parameters
controlling force accuracy in {\tt Gadget}.  They can therefore be
reduced only by increasing the number of particles and the resolution
in the {\tt Gadget} and {\tt VlaSolve} simulations, respectively.

It is important to notice that even the $N=10^8$ result might be
insufficient to describe properly the system at late epochs. In the
$(1024,1024,512)$ Vlasov simulation, the phase-space distribution
function seems to be rather smooth at all times and the system is free of
instability, contrarily to the $R=0.5$ case. However, it is difficult
at this point to know if actual physical instabilities build up at
late times in the $R=0.1$ case, because diffusion in the Vlasov
simulation might prevent the appearance of some unstable modes. 

While the irregular patterns observed in Figs.~\ref{fig:0v1_12} and
\ref{fig:0v1_ALL} are definitely of numerical nature, the fact that
they develop so easily may indicate that the system is prone to react
nonlinearly to small perturbations. Uneven gaps between the filaments
of the phase-space density can be observed at $t=15$ (third column of
Fig.~\ref{fig:0v1_12}), even in the $(1024,1024,512)$ Vlasov
simulation, and one might expect that they correspond to seeds of
actual physical instabilities.  In this respect, the system might
actually develop, at some point, physical unstable modes. These
results are quite suggestive of what was obtained previously with a
spherical shell code for cold and self-similar systems
\citep[]{Widrow97}.

Even with our $N=10^8$ particle simulation, it is not clear whether
these unstable modes dominate over collective effects due to
discreteness. A better understanding of the phenomenon would require
a convergence study using even higher-resolution simulations.

\section{Statistical analysis}
\label{sec:statana}

\subsection{Correlators and entropic estimators: definitions and concepts}
\label{sec:cordef}

To perform a more accurate analysis, one can try to quantify to which
extent the particle distribution in the $N$-body simulations can be
considered as a local Poisson process of the phase-space density
calculated in the semi-Lagrangian code. To do so, we use, in addition
to entropic measurements described further, the following correlators,
\begin{equation}
C_k \equiv  \frac{\mu_k}{\kappa_k}, \label{eq:cor1}
\end{equation}
with
\begin{eqnarray}
\mu_k & = & \frac{M}{N}\sum_{i=1}^N [f(\Omega_i)]^k, \label{eq:cor2} \\
\kappa_k & = & \int [f(\Omega)]^{k+1} {\rm d}\Omega. \label{eq:cor3}
\end{eqnarray}
In these equations, $k$ is a positive integer, $f$ the {\tt VlaSolve} phase-space density, $M$ the total mass, ${\rm d}\Omega \equiv 2\pi {\rm d}r\times{\rm d}v\times j {\rm d}j$ and $\Omega_i=(r_i,v_i,j_i)$, where $r_i$, $v_i$ and $j_i$ are respectively the radial position, radial velocity and angular momentum of each particle of the {\tt Gadget} simulation. 

For a point set randomly sampling a smooth density distribution $g$, the probability density $p(\Omega)$ of having  a given particle at phase-space position $\Omega$ is independent from the rest of the particle distribution and is simply proportional to $g(\Omega)$:
\begin{equation}
p(\Omega) {\rm d}\Omega = \frac{g(\Omega)}{M} {\rm d}\Omega.
\end{equation}
The density probability of having $N$ particles at respective positions $\Omega_1$, $\Omega_2$, $\ldots$, $\Omega_N$  is given by
\begin{equation}
{\cal P}(\Omega_1,\cdots,\Omega_N)=\prod_{i=1}^N p(\Omega_i).
\label{eq:lll}
\end{equation}
Ensemble averaging of $\mu_k$ under the law $g$ then reads
\begin{eqnarray}
\langle \mu_k \rangle_g & = & \frac{M}{N} \int \sum_{i=1}^N [f(\Omega_i)]^k\ {\cal P}(\Omega_1,\cdots,\Omega_N)\ {\rm d}\Omega_1 \cdots {\rm d}\Omega_N, \nonumber \\
 & & \\
 & = & \frac{M}{N} \sum_{i=1}^N \int_{\Omega} \frac{1}{M} [f(\Omega_i)]^{k} g(\Omega_i) {\rm d}\Omega_i, \nonumber \\
   & & \quad \quad \quad \quad \times \prod_{j \neq i} \int_{\Omega_j} \frac{g(\Omega_j)}{M} {\rm d}\Omega_j \\
 & = &  \int [f(\Omega)]^{k} g(\Omega) {\rm d}\Omega,
\end{eqnarray}
and
\begin{equation}
\langle C_k \rangle_g = \frac{\int [f(\Omega)]^{k} g(\Omega) {\rm d}\Omega}{\int [f(\Omega)]^{k+1} {\rm d}\Omega}. 
\label{eq:ensavck}
\end{equation}
Hence, if the distributions $g$ and $f$ coincide, i.e., in our case, if {\tt Gadget} actually Poisson samples the {\tt VlaSolve} phase-space density, one obtains $\langle C_k \rangle_{g=f} =1$ after ensemble averaging. 

When increasing $k$, more weight is given to regions in phase-space corresponding to larger values of $f$. For a point process totally anticorrelated with $f$, $C_k$ cancels, while its largest possible value is given by $C_k = (M \max f^k)/\kappa_k > 1$, when all the particles stay in the region where $f$ is maximal. 

An important issue is to compute properly the center of the system position in the {\tt Gadget} simulations. In order to do this, we find the coordinate origin {\em maximizing} $C_1$, even though the result of such a procedure can potentially lead to $C_1 > 1$,  to optimize the match between concentrations of particles and local extrema of $f$. 
 
The variance of $C_k$ can also be calculated in an analogous way to $\langle \mu_k \rangle_g$:
\begin{eqnarray}
\Delta C_k^2 & \equiv & \langle C_k^2 \rangle_g - \langle C_k \rangle^2_g \\
                   & = & \frac{1}{\kappa_k^2} \left[ \frac{M}{N} \langle \mu_{2k} \rangle_g - \frac{1}{N} \langle \mu_k \rangle^2_g \right],
\end{eqnarray}
which reduces to $\Delta C_k^2=(M/N) (\kappa_{2k}/\kappa_k^2) - 1/N$ when $f$ and $g$ coincide. 
In practice, we shall use the following estimator for this statistical error:
\begin{equation}
\Delta C_k^2 \simeq \frac{1}{\kappa_k^2} \left[ \frac{M}{N} \mu_{2k}  - \frac{1}{N}  \mu_k^2 \right],
\label{eq:varck}
\end{equation}
where $\mu_{2k}$ and $\mu_k$ are directly estimated from the $N$-body simulation. 

The log-likelihood that the {\tt Gadget} particle distribution locally Poisson samples the {\tt VlaSolve} phase-space density $f$ can be written, following the reasoning that leads to equation (\ref{eq:lll}), 
\begin{equation}
\ln {\cal L}=\sum_{i=1}^N \ln \left[ \frac{f(\Omega_i)}{M} \right].
\label{eq:simplogL}
\end{equation}
However, the region ${\cal D}$ where $f > 0$ being of finite extent, one expects $\ln {\cal L} =-\infty$ as soon as a particle escapes ${\cal D}$, which is very likely, due for instance to $N$-body relaxation. Furthermore, the Vlasov solver does not guaranty the positivity of $f$. To take into account in a fair way both the defects of the $N$-body and the Vlasov simulations, it is better to restrict to a region ${\cal D}_{\rm th}$ where $f$ is strictly positive:
\begin{equation}
D_{\rm th}\equiv \{ \Omega\ {\rm such}\ {\rm that}\  f(\Omega) \geq f_{\rm th},\ f_{\rm th} > 0\}.
\end{equation}
The log-likelihood of having $Q \leq N$ particles in the region ${\cal D}_{\rm th}$ and the rest outside it (leaving the freedom of the remaining particles to span all the space outside ${\cal D}_{\rm th}$) is given by a binomial law:
\begin{equation}
\ln {\cal L}_{\rm b}(Q,\nu)=\ln\left[ \frac{N!}{(N-Q)! Q!} \nu^Q (1-\nu)^{N-Q} \right],
\end{equation}
where $\nu$ is the fractional mass inside ${\cal D}_{\rm th}$ in the {\tt VlaSolve} simulation. Hence, equation (\ref{eq:simplogL}) simply becomes
\begin{equation}
\ln {\cal L}=\sum_{\Omega_i \in {\cal D}_{\rm th}} \ln \left[ \frac{f(\Omega_i)}{M_{\rm th}} \right] + \ln {\cal L}_{\rm b}(Q_{\rm th},\nu), 
\label{eq:logLcor}
\end{equation}
where $Q_{\rm th}$ is the number of particles of the ${\tt Gadget}$ simulation inside ${\cal D}_{\rm th}$ and $M_{\rm th}=\int_{D_{\rm th}} {\rm d}\Omega\ f(\Omega)$.

Note that the distribution of particles which maximizes the first term in equation (\ref{eq:logLcor})  corresponds again to the case where all the particles of  ${\cal D}_{\rm th}$ stay in the region where $f$ is maximal, similarly to the case when the correlator $C_k$ is equal to its maximum possible value. Clearly, this situation is not typical, but it is in fact the most likely to consider when it can take place: this is why we maximize $C_1$ to estimate the center of the $N$-body system, even though it might turn to be larger than unity. 

The expectation value of $\ln {\cal L}$ under the law $f$ can be obtained by ensemble averaging:
\begin{eqnarray}
S(f_{\rm th}) &\equiv & -\frac{1}{\nu N}\langle \ln {\cal L} \rangle_f  =  S_{f}(f_{\rm th})+S_{\rm b}(f_{\rm th}), \label{eq:lik} \\
S_{f}(f_{\rm th}) & \equiv & -\int_{D_{\rm th}} \frac{f(\Omega)}{M_{\rm th}} \ln\left[ \frac{f(\Omega)}{M_{\rm th}} \right] {\rm d}\Omega, \label{eq:likf}\\
S_{\rm b}(f_{\rm th}) & \equiv & -\frac{1}{\nu N}  \sum_{Q=0}^{N} {\cal L}_{\rm b}(Q,\nu) \ln {\cal L}_{\rm b}(Q,\nu).
\end{eqnarray} 
In the limit $f_{\rm th} \rightarrow 0$, the quantity $S_f(f_{\rm th})$ reduces to the Gibbs entropy of the system, which explains the choice of notations. Moreover, if $N \gg 1$ and if the fractional mass $\nu$ inside the domain of interest $D_{\rm th}$ is of order of unity, which is the case for our analyses, the term $S_{\rm b}(f_{\rm th})$ is in practice negligible compared to $S_f(f_{\rm th})$, so $S(f_{\rm th})$ depends only weakly on the total number of particles, as expected. 

The variance of $\ln {\cal L}$ can be calculated likewise
\begin{eqnarray}
\sigma_L^2 & \equiv & \frac{1}{(\nu N)^2} \left[ \langle \ln{\cal L}^2\rangle_f - \langle \ln {\cal L} \rangle_f^2\right] \\
                 & \simeq & \frac{1}{\nu N} \left\{ \int_{D_{\rm th}}\frac{f(\Omega)}{M_{\rm th}} \ln^2\left[ \frac{f(\Omega)}{M_{\rm th}} \right] {\rm d}\Omega - \nu [ S(f_{\rm th})]^2 \right\},\nonumber \\
& & 
\label{eq:errstat}
\end{eqnarray}
where we have neglected, following the arguments developed earlier, the contributions of $S_{\rm b}$ to the error. 

To understand better the interest of using the statistics given by equation (\ref{eq:logLcor}), one can introduce the difference between the measured value of the log-likelihood and its expectation under the law $f$:
\begin{equation}
\delta S=\frac{1}{\nu N} \left[ \langle \ln {\cal L} \rangle_f - \ln{\cal L} \right],
\end{equation}
where ${\cal L}$ is given by expression (\ref{eq:logLcor}) calculated for $\Omega_i$ extracted from a {\tt Gadget} simulation.  The
quantity $\delta S$ estimates the magnitude of the difference between the underlying smooth phase-space density $g$ sampled by {\tt Gadget} and the {\tt VlaSolve} phase space density, $f$: its ensemble average other many {\tt Gadget} realizations indeed reads, when neglecting the binomial term in equation (\ref{eq:logLcor}),
\begin{equation}
\langle \delta S \rangle_g \simeq \int_{f \geq f_{\rm th}}\frac{1}{M_{\rm th}} [g(\Omega)-f(\Omega)] \ln \left[ \frac{f(\Omega)}{M_{\rm th}} \right] {\rm d} \Omega.
\end{equation}
Under the assumption that the $N$-body simulation Poisson samples the distribution $f$, the magnitude of $\delta S$ should be of the same order of $\sigma_L$. 
\subsection{Correlators and entropic estimators: measurements}
\begin{figure*}
\centerline{\hbox{
\includegraphics[height=6cm]{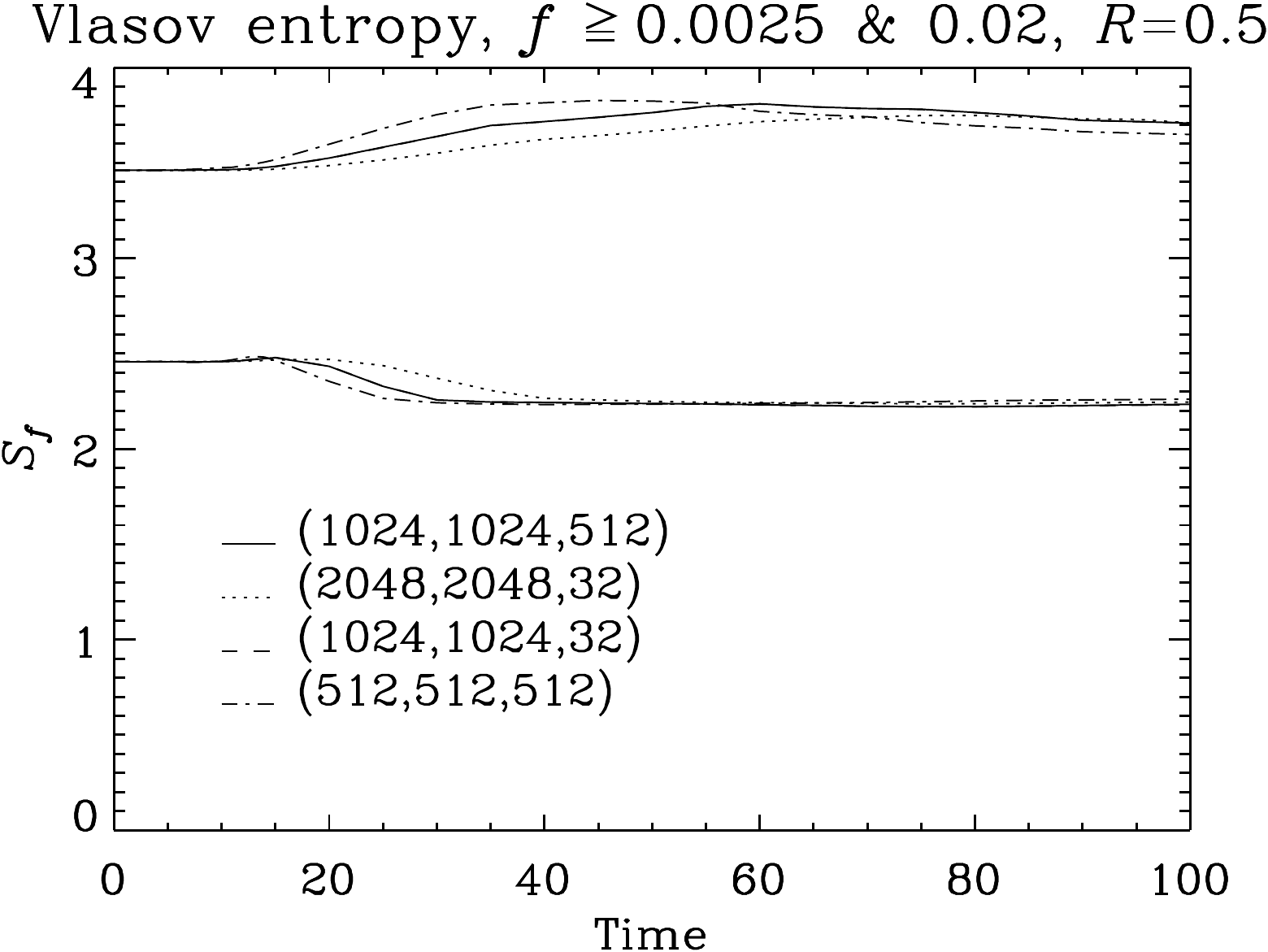}
\includegraphics[height=6cm]{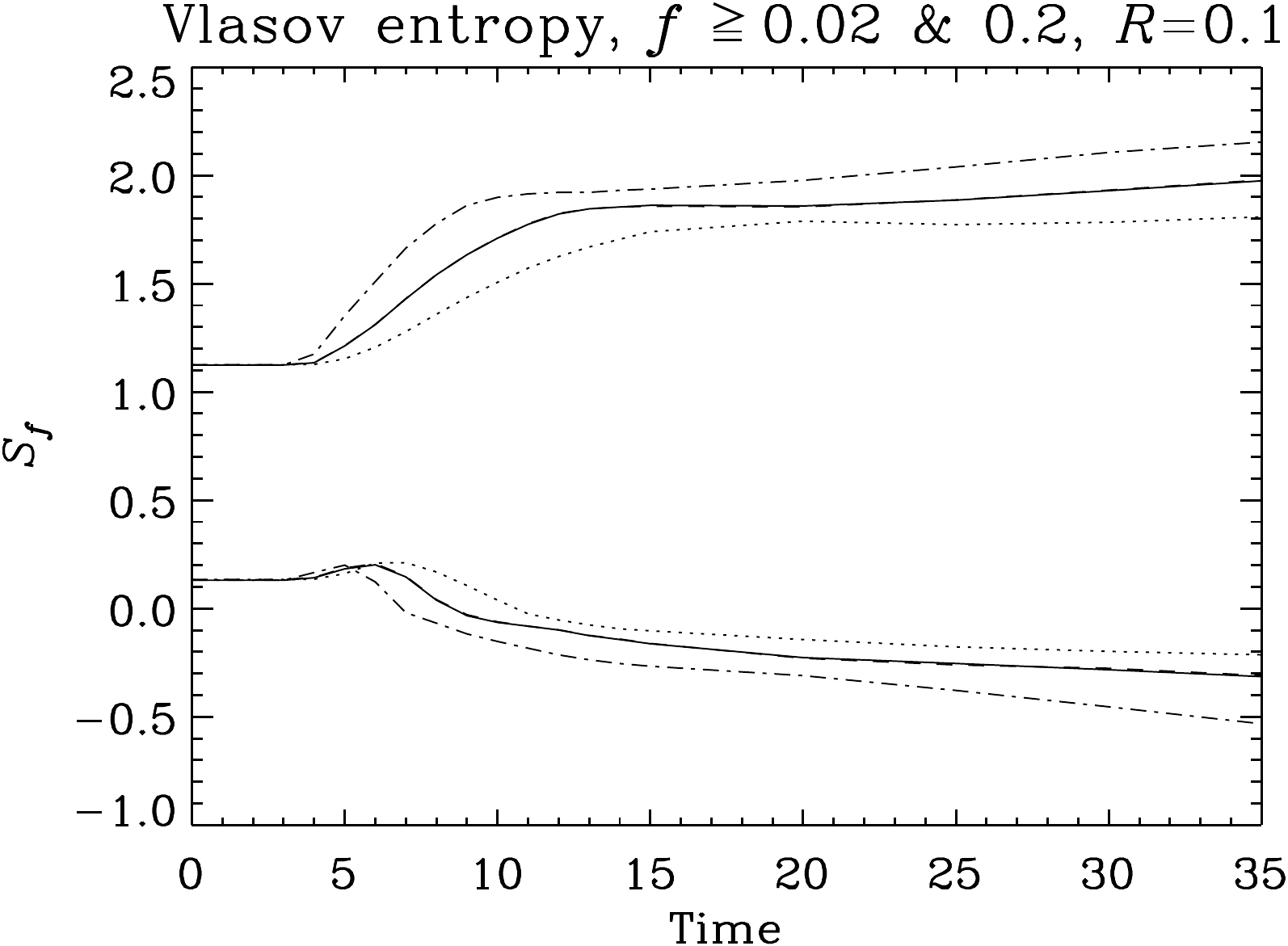}
}}
\centerline{\hbox{
\null \hskip 0.18cm\includegraphics[height=6cm]{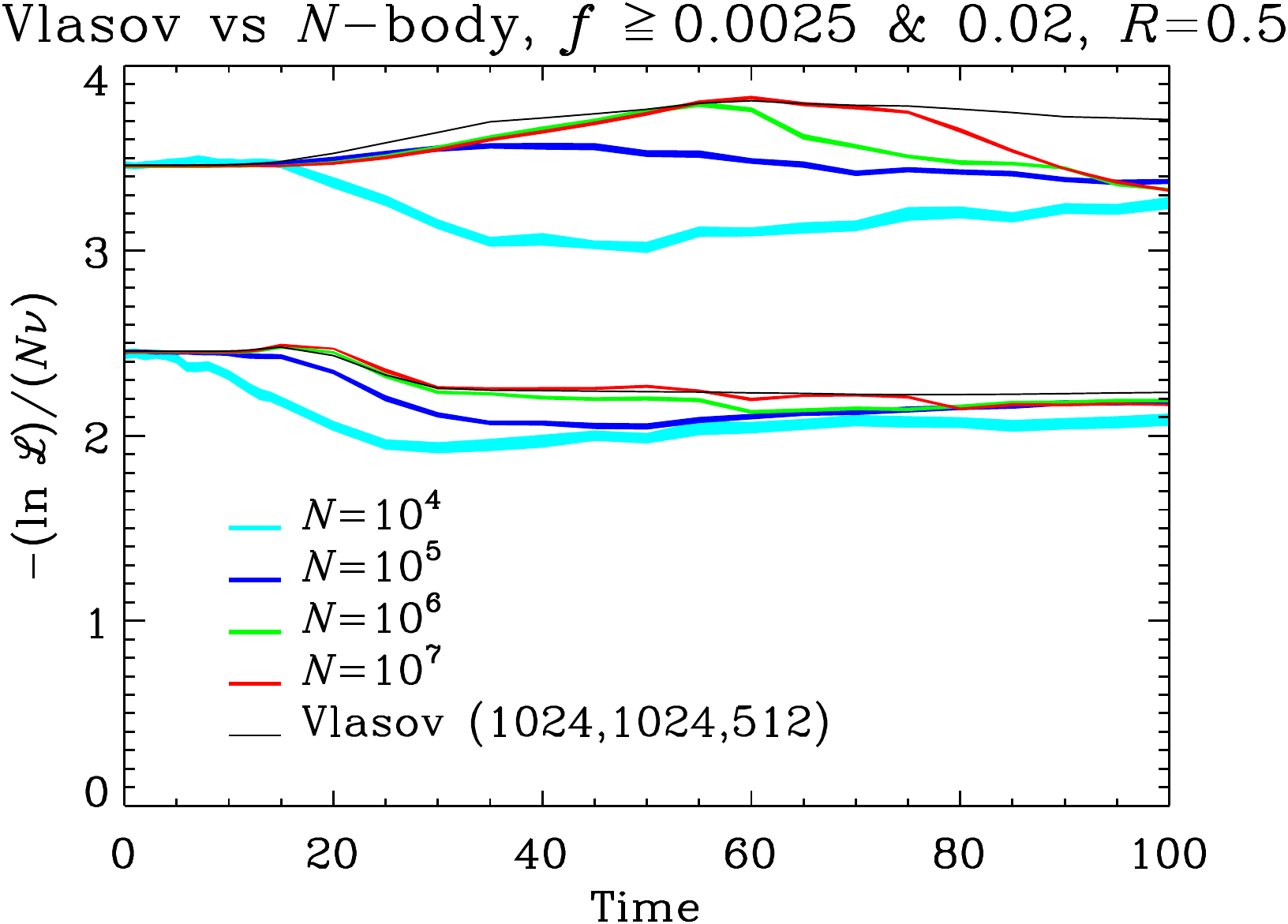}
\hskip -0.28cm\includegraphics[height=6cm]{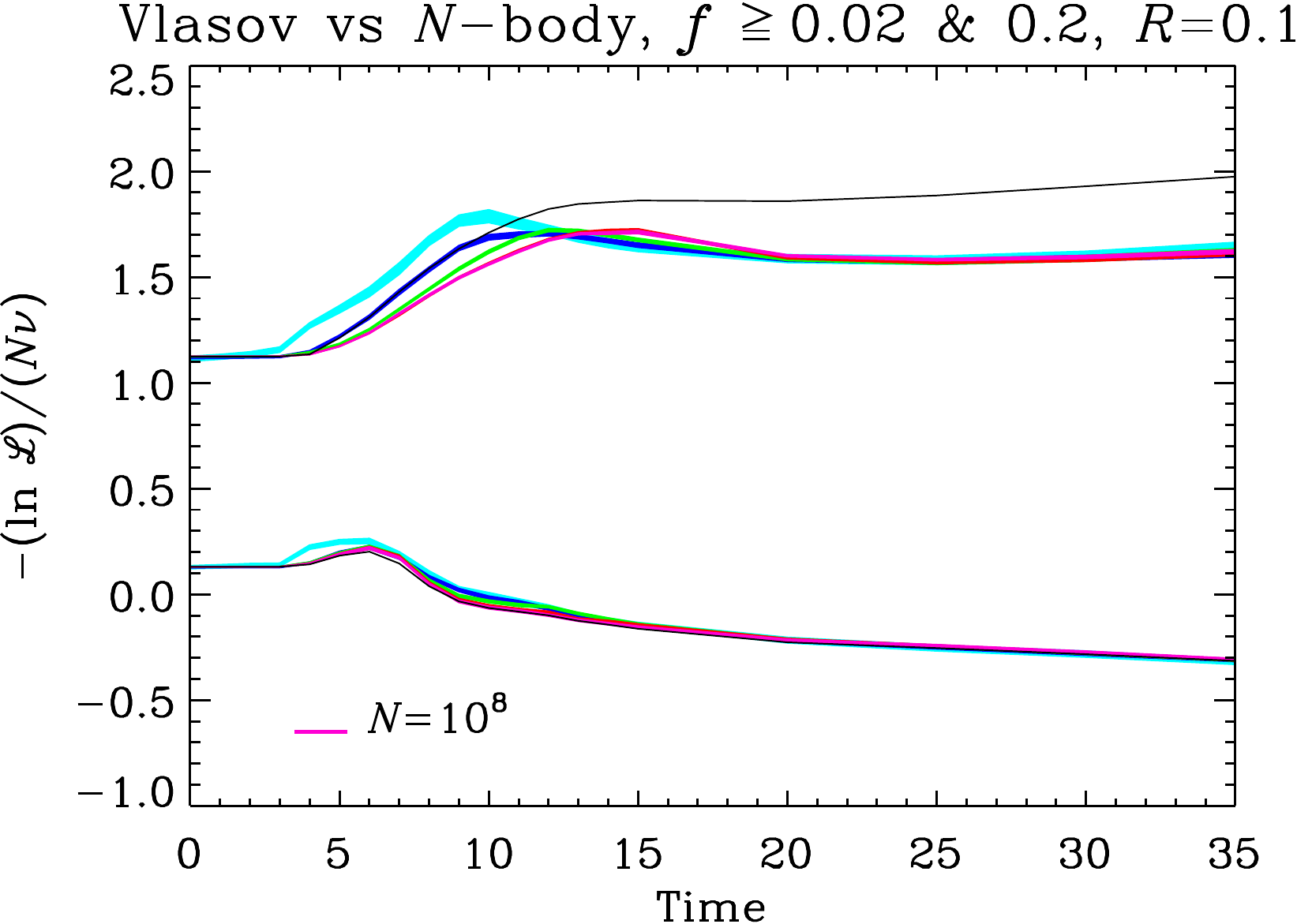}
}}
\centerline{\hbox{
\includegraphics[height=6cm]{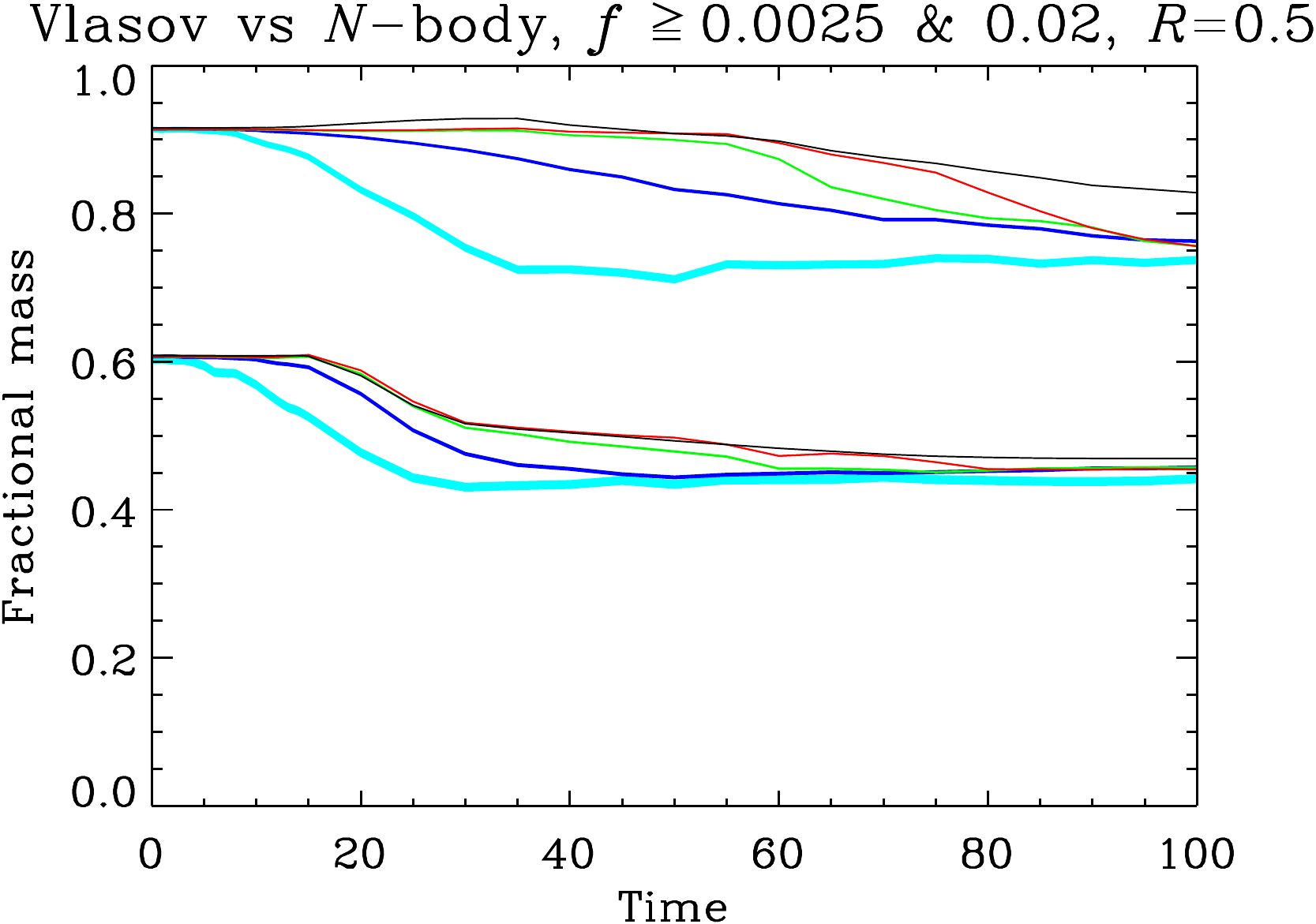}
\includegraphics[height=6cm]{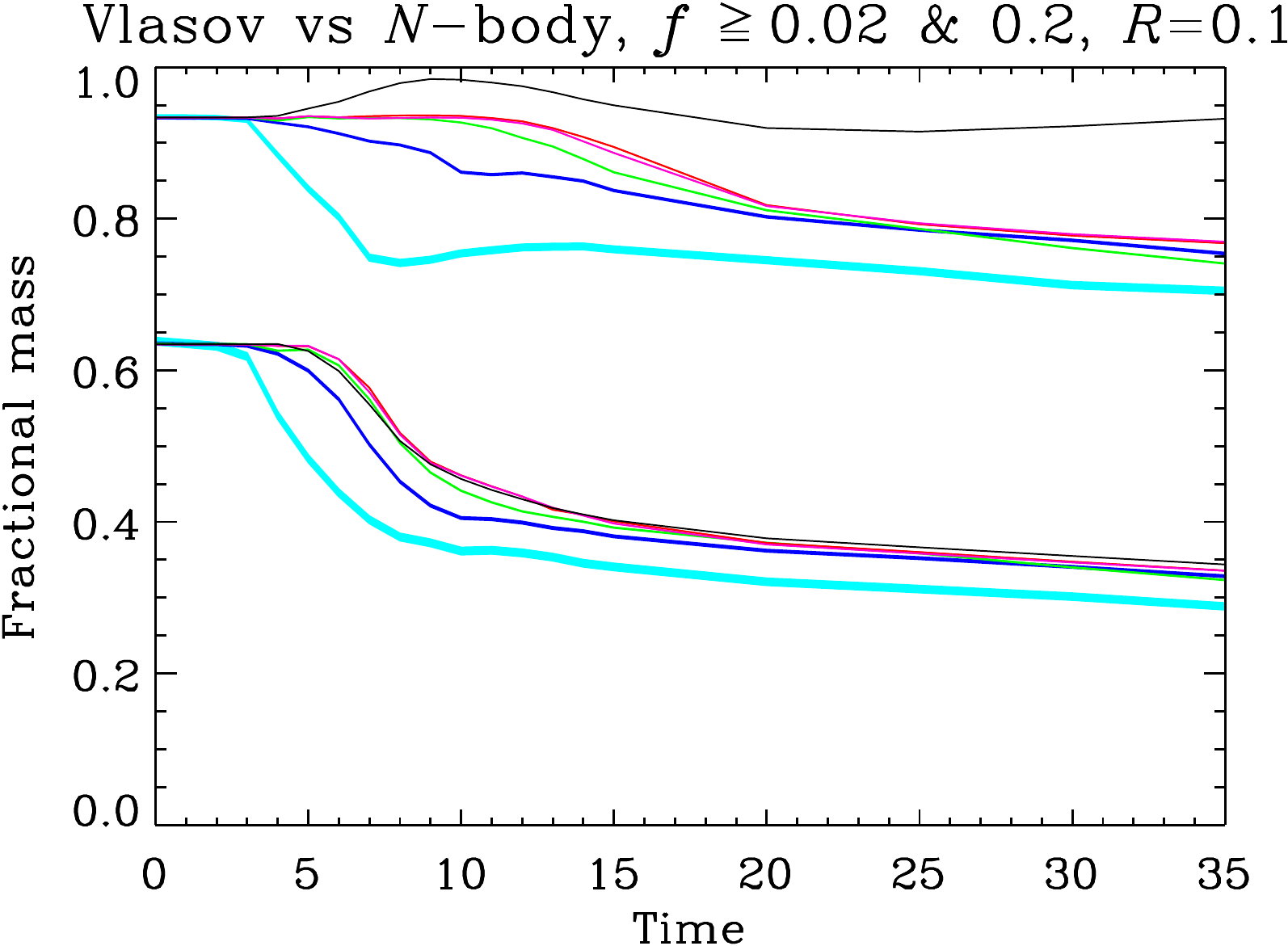}
}}
\caption[]{Entropic measurements: effects of {\tt VlaSolve} resolution (top two panels) and {\tt Gadget} number of particles (four bottom panels). The left and right panels correspond respectively to $R=0.5$ and $0.1$. On the {\em top panels}, the quantity $S_f(f_{\rm th})$ given by equation (\ref{eq:likf}) is plotted as a function of time for the Vlasov simulations and for two values of $f_{\rm th}$ indicated on each panel corresponding to approximately initially keeping 90 and 60 percent of the mass inside the excursion. Each curve corresponds to a given resolution as indicated on each panel (the dashes are nearly superposed to the solid line). The top/bottom group of four curves correspond to a smaller/larger value of $f_{\rm th}$. On the {\em middle panels}, the solid line is the same as on the top panels, while the colored curves display, for each value of the particle number $N$ in the {\tt Gadget} simulations, the quantity $-\ln {\cal L}/(N \nu)$ as a function of time, where $\ln {\cal L}$ is given by equation (\ref{eq:logLcor}). If the $N$-body simulations would Poisson sample the {\tt VlaSolve} phase-space density, the ensemble average of this quantity over many {\tt Gadget} realizations should match the solid line (except for a negligible correction due to the $S_{\rm b}$ term in equation \ref{eq:likf}). Finally, the {\em bottom panels} show the fractional mass as a function of time for the two values of $f_{\rm th}$ considered. On the two bottom right panels, there is an additional purple curve nearly indistinguishable from the red one, corresponding to the additional simulation with 100 millions particles we performed for $R=0.1$. In the four bottom panels, the thickness of each colored curve takes into account statistical errors (equation \ref{eq:errstat} for $\ln {\cal L}$). In addition, for the middle panels, systematic errors due to the interpolation of the phase-space distribution function in the {\tt VlaSolve} simulations also contribute to the estimated errors. In the latter case, we compute $f(\Omega_i)$ both using nearest grid point and linear interpolation from the values of $f$ on the computational mesh. The difference between the two interpolating methods adds to the thickness of the curves. Note that we use the $(1024,1024,512)$ {\tt VlaSolve} simulation to perform the comparison to $N$-body results, to minimize the effects of interpolation. }
\label{fig:entro}
\end{figure*}

Top panels of Fig.~\ref{fig:entro} show the quantity $S_f(f_{\rm th})$ as a function of time for the various {\tt VlaSolve} simulations we performed and two values of $f_{\rm th}$ chosen such that approximately 90 percent and 60 percent of the total mass is initially inside the excursion $D_{\rm th}$, respectively.  The quantity $S_f(f_{\rm th})$ is a Casimir invariant --that is an integral over a function of $f$-- and should thus be conserved during runtime if the code was perfect. This not the case because of diffusion and aliasing effects in $(r,v)$ space: deviation from conservation of $S_f$ happens shortly after collapse time. Then there is a strong mixing phase during which $S_f$ increases, then possibly decreases, according to the value of $f_{\rm th}$, and finally reaches an approximate plateau. Deviation from conservation of $S_f$ naturally happens sooner when resolution in $(r,v)$ space is smaller. Resolution in $j$ space does not have much influence on $S_{f}$ because angular momentum is an invariant of the dynamics. However, as clearly shown in \S~\ref{sec:visu} and in Appendix~\ref{app:resvlaeff} for $R=0.1$, we already know that sparse sampling in $j$ space is not recommended since it can introduce some instabilities in the dynamics, even though this effect does not affect much our likelihood measurements. 

Middle panels of Fig.~\ref{fig:entro} show the quantity $-\ln {\cal
  L}/(N \nu)$ measured in {\tt Gadget} from the particles belonging to
the excursion $D_{\rm th}$ as a function of time, where $\ln {\cal L}$
is given by equation (\ref{eq:logLcor}). For a given value of the
threshold $f_{\rm th}$, if the {\tt Gadget} simulations would actually
behave like Poisson realizations of the {\tt VlaSolve} ones, all the
colored curves should be close to the solid line, which corresponds to
$S_f$. This is clearly not the case for small $f_{\rm th}$ (upper
group of curves), except a early times. Increasing the number of
particles in the $N$-body simulation improves the agreement with the
Vlasov code for $R=0.5$ but does not seem to have a convincing impact
in the $R=0.1$ case: for $f_{\rm th}=0.02$, all the $N$-body
simulations converge to the same plateau somewhat below the Vlasov
code result. On the contrary, for $f_{\rm th}=0.2$ and $R=0.1$, the
agreement between {\tt Gadget} and {\tt VlaSolve} is striking at all
times, except may be for the $N=10^4$ simulation during the strong
mixing phase. Note also, that at late times, all the $N$-body
simulations converge which each other, independently of $f_{\rm th}$
and $R$, except again for the $N=10^4$ simulation with $R=0.1$, but we know that this latter presents significant deviations from spherical symmetry and should be probably discarded for the analyses performed here. 

To complete the analyses and understand better the results obtained for the log-likelihood, the fractional mass inside the excursions $f \geq f_{\rm th}$ is shown in bottom panels of Fig.~\ref{fig:entro}. Again, this quantity is a Casimir, so it should not change with time in the idealistic case. In practice, while it is difficult to predict the effects of aliasing on the ${\tt VlaSolve}$ mass inside $D_{\rm th}$, diffusion effects are more likely to decrease it, especially by dilution of filamentary structures that build up during the course of dynamics. In the $R=0.5$ case, most of the disagreement between {\tt Gadget} likelihood and its expectation given by {\tt VlaSolve} can be understood in terms of fractional mass: effects related to the discrete nature of  the $N$-body simulations seem to spread particles away from $D_{\rm th}$. However this process is subtle and seems to remain local as suggested by visual inspection of Figs.~\ref{fig:0v5_12} and \ref{fig:0v5_ALLJ}. We also checked that it does not affect dramatically the projected density, $\rho(r)$.

In the $R=0.1$ case, the interpretation of the results is slightly
more complicated. For $f_{\rm th}=0.2$, the {\tt Gadget} fractional
mass inside the excursion $D_{\rm th}$ behaves similarly as in the
$R=0.5$ case as a function of particle number. On the other hand, when
examining the quantity $-\ln {\cal L}/(N \nu)$, the $N$-body
measurements converge with each other and with {\tt VlaSolve} much
better, especially after relaxation. This means that particles left in
$D_{\rm th}$ are redistributed in a non trivial way, such that the
effects of the excursion mass loss are compensated. For $f_{\rm
  th}=0.02$, even the $N=10^8$ {\tt Gadget} sample disagrees with the
{\tt VlaSolve} simulation. Clearly, the Vlasov simulation becomes
quickly  defective in regions where $f$ is small. On the other hand,
convergence of the {\tt Gadget} simulations at late times might be
misleading. Indeed, we noticed from visual inspection of
Figs.~\ref{fig:0v1_12} and \ref{fig:0v1_ALL} that some instabilities
appeared in all of them as soon as $t \ga 15$, although later when $N$
is larger. Interestingly, the measurements in the $N=10^7$ and
$N=10^8$ simulations are nearly indistinguishable from each other,
which is a sign that we are nevertheless close to numerical convergence. 

\begin{figure*}
\centerline{\hbox{
\includegraphics[height=6cm]{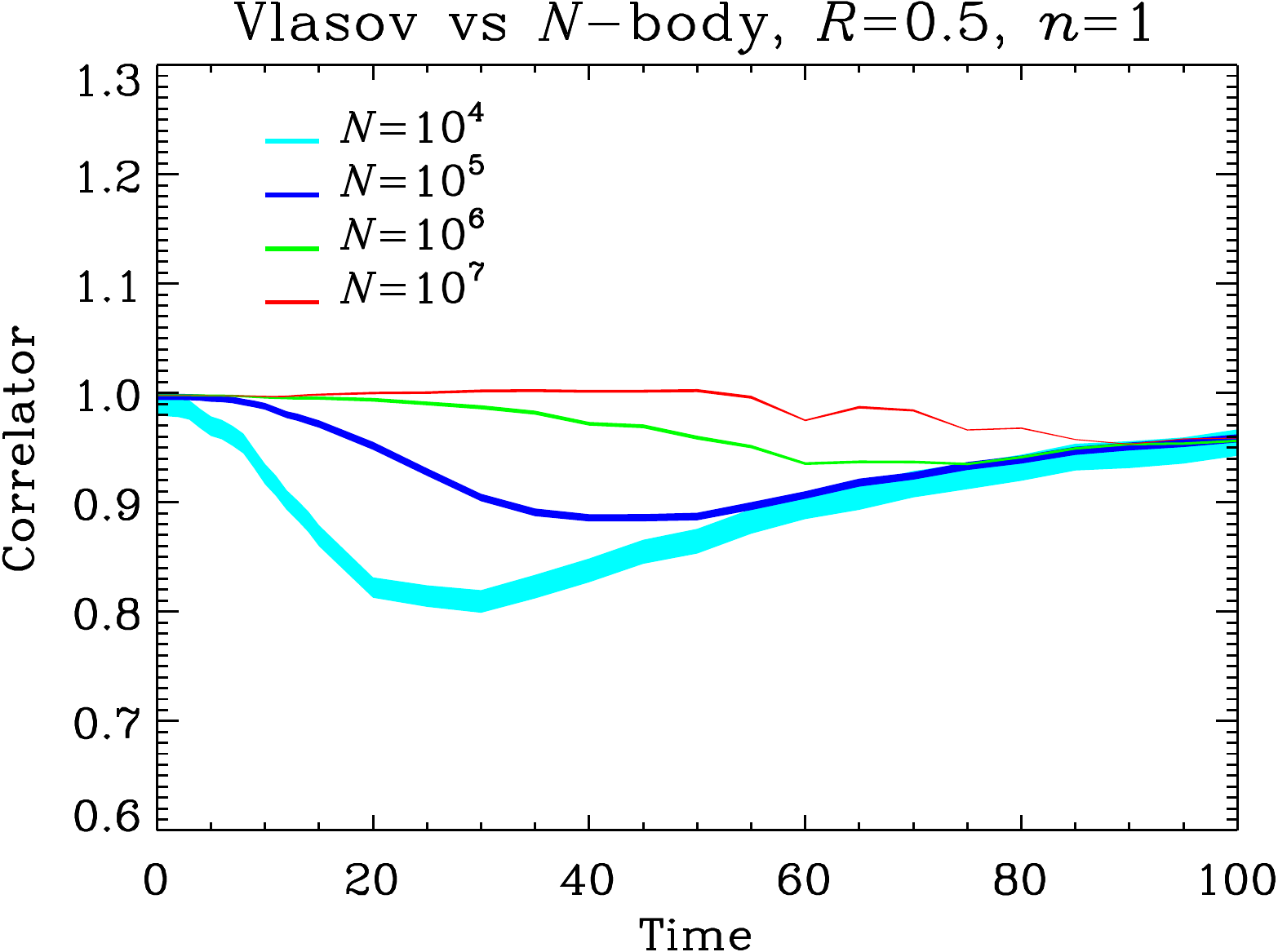}
\includegraphics[height=6cm]{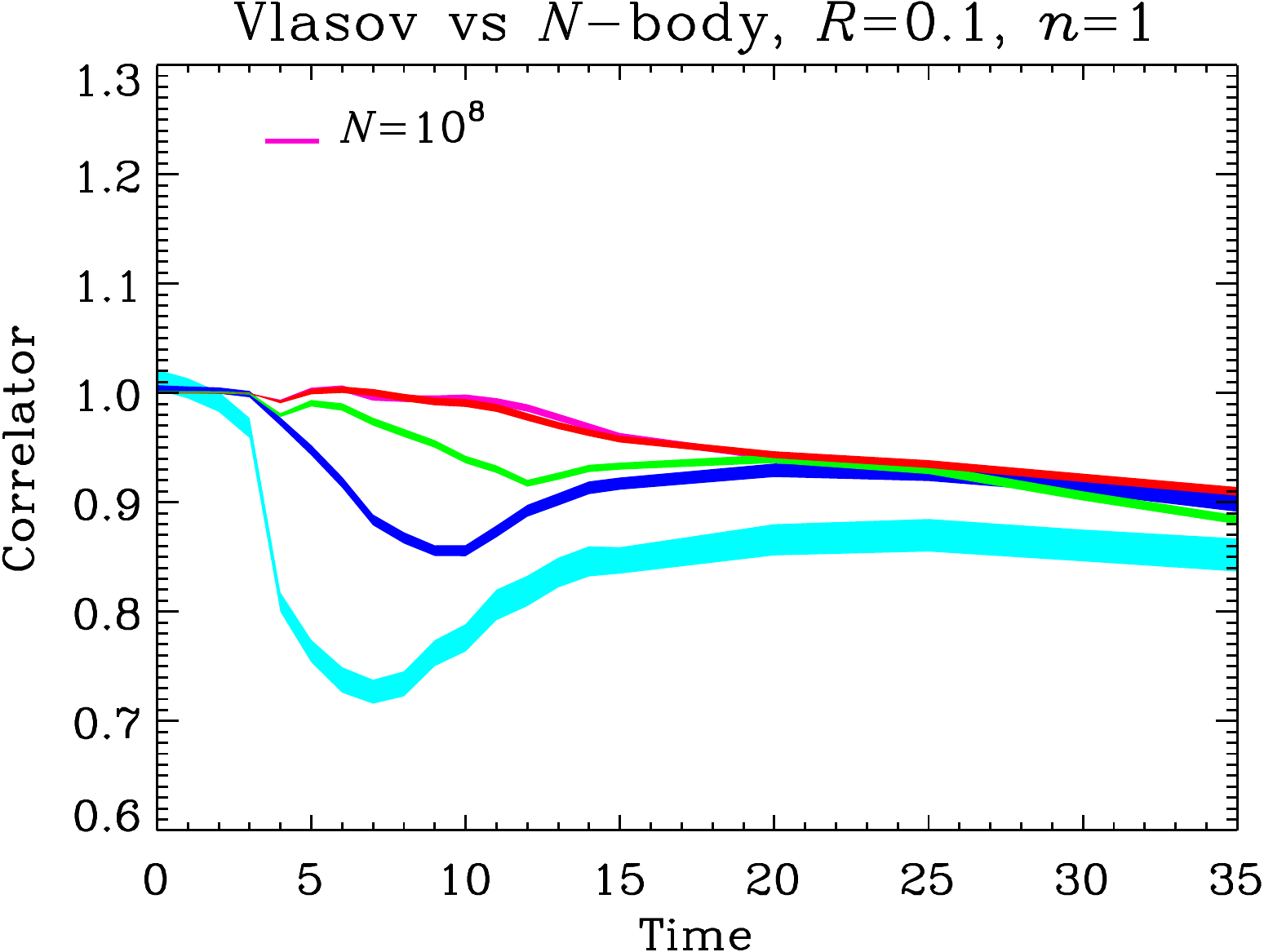}
}}
\centerline{\hbox{
\includegraphics[height=6cm]{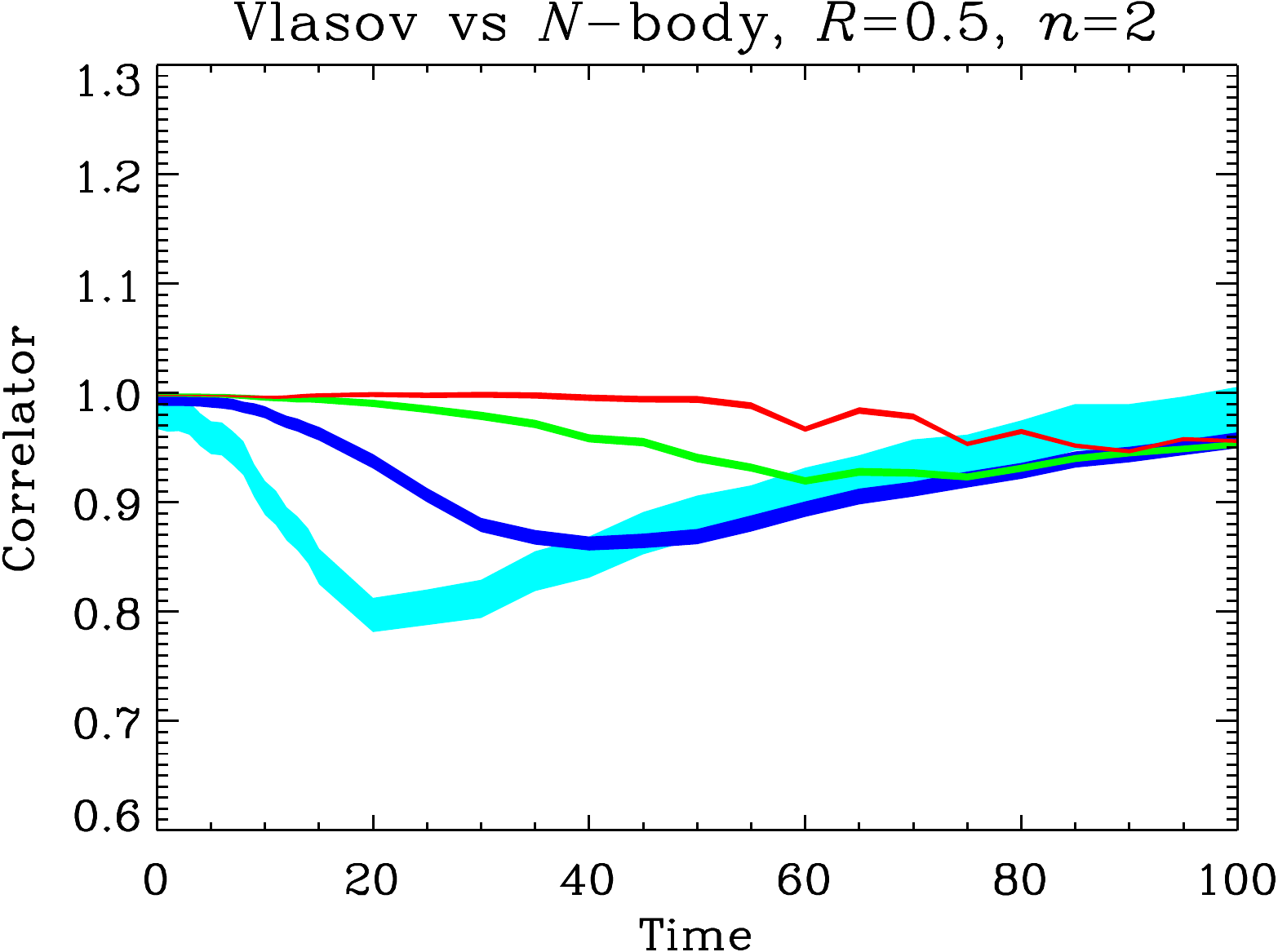}
\includegraphics[height=6cm]{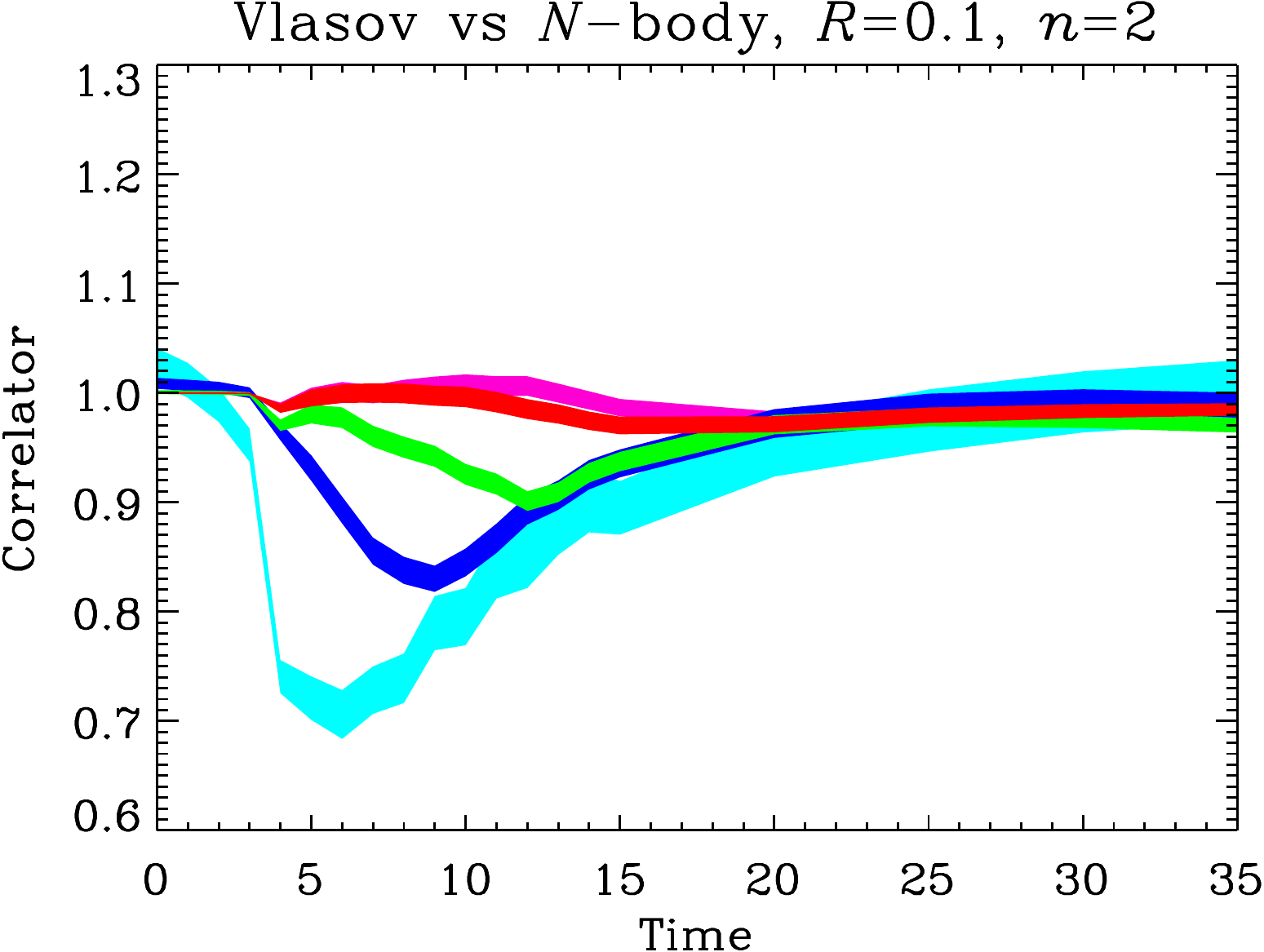}
}}
\centerline{\hbox{
\includegraphics[height=6cm]{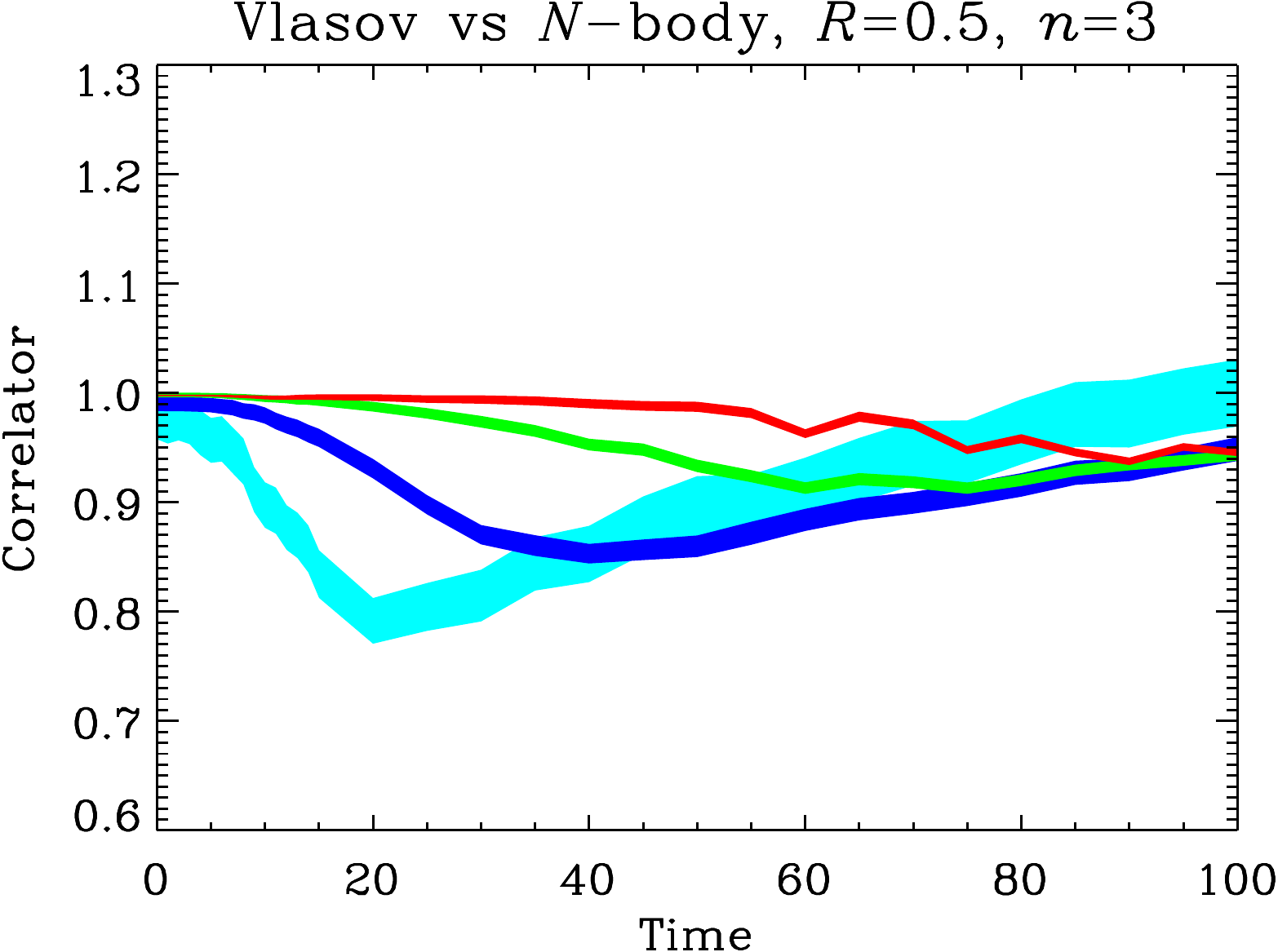}
\includegraphics[height=6cm]{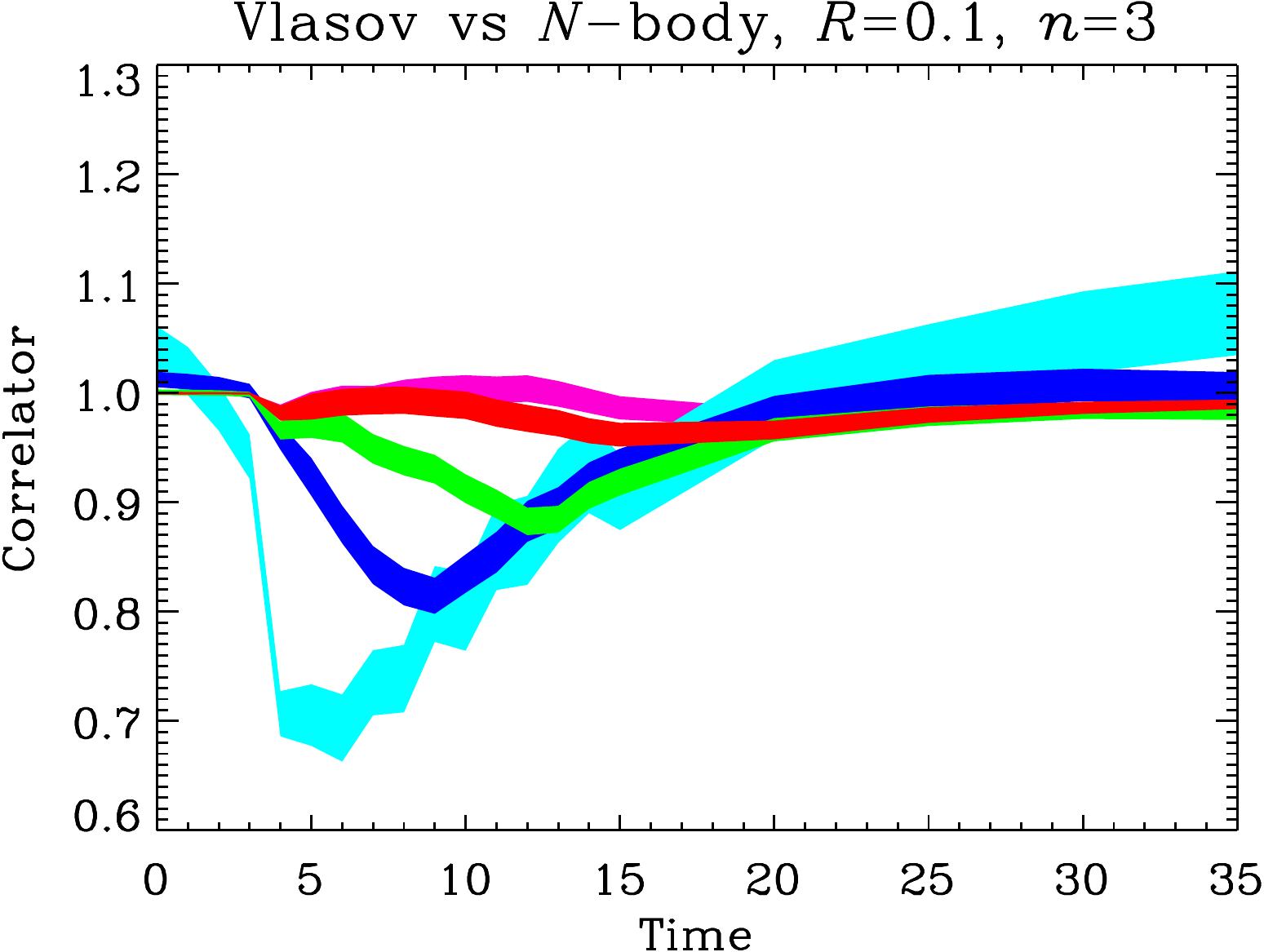}
}}
\caption[]{Correlators between {\tt VlaSolve} and {\tt Gadget} as functions of time. These quantities, defined in equations (\ref{eq:cor1}), (\ref{eq:cor2}) and (\ref{eq:cor3}), are plotted for $k = 1,\ 2,\ 3$ increasing from top to bottom, while left and right panels correspond to $R=0.5$ and $0.1$, respectively. The thickness of the curves, analogously to Fig.~\ref{fig:entro}, takes into account statistical errors according to equation~(\ref{eq:varck}) using the measured value of $\nu_{2k}$ and $\nu_k$ and systematic errors due to the interpolation of the phase-space density in the {\tt VlaSolve} samples. Note that there is an additional purple curve on each panel of the right column corresponding to the 100 millions particles simulation.}
\label{fig:correlators}
\end{figure*}

Entropic measurements of Fig.~\ref{fig:entro} are confirmed, at least partly, by Fig.~\ref{fig:correlators}. 
In particular, a depression of which the depth depends on the number of particles in the $N$-body simulation appears on all the curves. When increasing $N$, the amplitude of the depression decreases and the occurrence of its maximum amplitude is delayed, independently of the actual dynamical state of the system. Again, it can certainly be attributed to collective effects due to Poisson noise. Overall agreement between $N$-body and Vlasov codes improves when increasing the number of particles in the $N$-body simulation. For $R=0.5$, this is rather independent of $k$ in equation (\ref{eq:cor1}), i.e. of the fact of putting more or less weight to overdense regions in phase-space. In the $R=0.1$ case, putting aside the depression of which the depth depends on the number of particles, the correlator $C_1$ starts to decrease with time at $t \sim 10$. This can be mainly attributed to defects in the Vlasov simulation in underdense regions as discussed earlier. For $k \geq 2$, which gives more weight to higher values of the phase-space density, the correlator indeed stays steady as a function of time (again putting aside the $N$-dependent depression). However, one notices for $k=3$ a net increase with time of the correlator for the simulation with $N=10^4$ particles, but let us remind that this simulation presents significant deviations from spherical symmetry.

\section{Conclusion}
\label{sec:discussion}

In this paper we have compared the phase-space distribution function
traced by the particle distribution in {\tt Gadget} simulations to the
results obtained with our new Vlasov code {\tt VlaSolve} for spherical
systems, an improved version of the splitting algorithm of
\cite{Fuji83}.  For the specific comparison, we have chosen (apodized)
H\'enon spheres, which are known to be insensitive to radial orbit
instability and in particular to preserve the spherical nature of the
system. The latter property is confirmed from simulations run with
three-dimensional $N$-body codes. We considered two values of the
initial virial ratio of the spheres, $R=0.5$ and $R=0.1$, corresponding
to ``warm'' and  ``cold'' configurations, respectively.

We have plotted detailed structures of the phase-space distribution
functions varying the spatial/mass resolution of the numerical code in a
systematic fashion. we have conducted further a quantitative
analysis by introducing two new statistical tools. The first one is of
entropic nature and corresponds to the log-likelihood quantifying to
which extent the $N$-body results represent a local Poisson sampling of
the Vlasov phase-space density. The second tool is a correlator of order
$k$, proportional to the integral over phase-space of the product
between the Vlasov phase-space density raised to the power $k$ and the
particle distribution function.

The overall conclusion is that both the Vlasov and
$N$-body methods agree remarkably well with each other, both from the
visual and statistical points of view, if sufficient
resolution is employed. Given the completely different numerical
approaches to collisionless dynamics, this is not trivial at
all, and the degree of agreement that we have shown for the first time
is perhaps even better than what had been expected before. This is
reassuring for numerous previous results that have been almost
exclusively obtained from the $N$-body method.

Nevertheless there are still unsolved subtle issues in details:
\begin{itemize}
\item When performing a visual inspection of the phase-space
  distribution function in the cold case, $R=0.1$, although still good at the coarse
level, we find that the level of agreement between the $N$-body and the Vlasov codes
worsens at small scales after a few dynamical times. This is mainly due
to collective effects induced by the shot noise of the particles in the
$N$-body simulations (and not to
close particle encounters). Even with $N=10^8$ particles, we are not
able to prove numerical convergence of the $N$-body results. 
The comparison at this level, however, is made difficult by the fact that the Vlasov code is
significantly diffusive, which might prevent the development of a
variety of physical unstable modes.
\item While the statistical tools do not provide as rich and intuitive
information as visual inspection, they identify some subtle effects. In
particular, when taking into account general trends due to diffusion in
the Vlasov code, significant for $R=0.1$, we notice that the match
between {\tt Gadget} and {\tt VlaSolve} worsens with time, then
improves.  The degree of the mismatch increases, and it shows up
earlier, when reducing the number of particles in the
$N$-body simulation. Again, this may be ascribed to 
collective effects due to the shot noise of the particles.
Nevertheless, the very
good match between the {\tt Gadget} simulations with $N=10^7$ and
$N=10^8$ particles may  suggest that convergence is nearly reached in terms
of number of particles and information theory, even if it is not fully
proved.
\end{itemize}

It is worth mentioning again that the collective effect mentioned above
is not related to $N$-body relaxation, but rather results from random
Poisson fluctuations. This can be formulated as follows \citep[see][for
similar arguments]{Aarseth1988,Widrow97,Boily2002,Joyce2009}: a given
particle at some distance $r$ from the center of the system feels a
force proportional to the number $N_{\rm in}$ of particles inside the
sphere of radius $r$. Poisson fluctuations imply thus that there is a
relative error of order of $1/\sqrt{N_{\rm in}}$ on this
force. Importantly, the inner number of particles $N_{\rm in}$ changes
with time with random fluctuations around the mean behavior: these
fluctuations can be considered as a correlated random walk. Indeed,
because of the finite velocity dispersion, particles cross both inwards
and outwards the frontier of the sphere of radius $r$. A larger velocity
dispersion weakens the amount of correlation, thus makes the errors on
the force more random, which should have a fuzzy effect on the
phase-space density, similarly as collisional relaxation: this is what
we can expect for $R=0.5$ and as observed on Fig.~\ref{fig:0v5_ALLJ}. On
the contrary, a smaller velocity dispersion makes the error on the force
more {\em systematic} which should induce coherent distortions of the
phase-space density: this is what we can expect for $R=0.1$ and
confirmed by visual inspection of Fig.~\ref{fig:0v1_ALL}. This effect
has non-trivial consequences on the energy spectrum of the particles,
particularly in cold configurations \citep[]{Joyce2009}. It certainly
explains as well the deviations between {\tt VlaSolve} and {\tt Gadget}
observed when measuring the statistical estimators defined in this
paper. According to \citet[]{Aarseth1988}, this {\em collective} effect
is dominant over $N$-body relaxation, and, as confirmed by our detailed
numerical tests in Appendix~\ref{app:instab}, is not significantly
influenced by softening.

Note as well that shot noise creates anisotropies in the system,
i.e. deviations from spherical symmetry that may be eventually
amplified. \citet[]{Aarseth1988} argue that this effect is subdominant
compared to the radial component of the noise-induced
perturbation when considering the collapse of an homogeneous sphere.  
Although their calculation is performed only prior to
collapse and in the cold case, we believe that the conclusion still
remains valid for the kind of initial conditions studied in this paper,
as suggested by our numerical experiments that seem
to preserve well spherical symmetry. 

Clearly, the collective effect due to particle shot noise is a real problem for simulations of
close to cold spherical systems when it comes to examine fine structures
of the phase-space density. We were not able to prove convergence of the
phase-space density in the $R=0.1$ case even for an $N=10^8$ particle
simulation. Notably, this may have non-trivial consequences on the fine
structure of simulated dark matter halos, where numerical convergence in
terms of number of particles might not have been reached yet despite the
numerous intensive studies. Indeed, convergence toward the continuous
limit might be much slower than expected, hence giving the false
impression that it is achieved.

\section*{Acknowledgements}

We thank Christophe Alard, Ana\"elle Hall\'e, J\'er\^ome Perez and
Simon Prunet for useful
discussions. This work has been funded in part by ANR grant
ANR-13-MONU-0003 and was granted access to the HPC resources of The
Institute for scientific Computing and Simulation financed by Region Ile
de France and the project Equip@Meso (ANR-10-EQPX-29-01) overseen by the
French National Research Agency (ANR) as part of the ``Investissements
d'Avenir'' program.  Y.S. gratefully acknowledges the support from
Grant-in Aid for Scientific Research by JSPS (Japan Society for
Promotion of Science) No. 24340035.


\appendix
\section{Vlasov solver: details on the algorithm}
\subsection{Reflecting boundaries with time delay}
\label{sec:monR0}
\begin{figure*}
\centerline{\hbox{
\includegraphics[width=7cm]{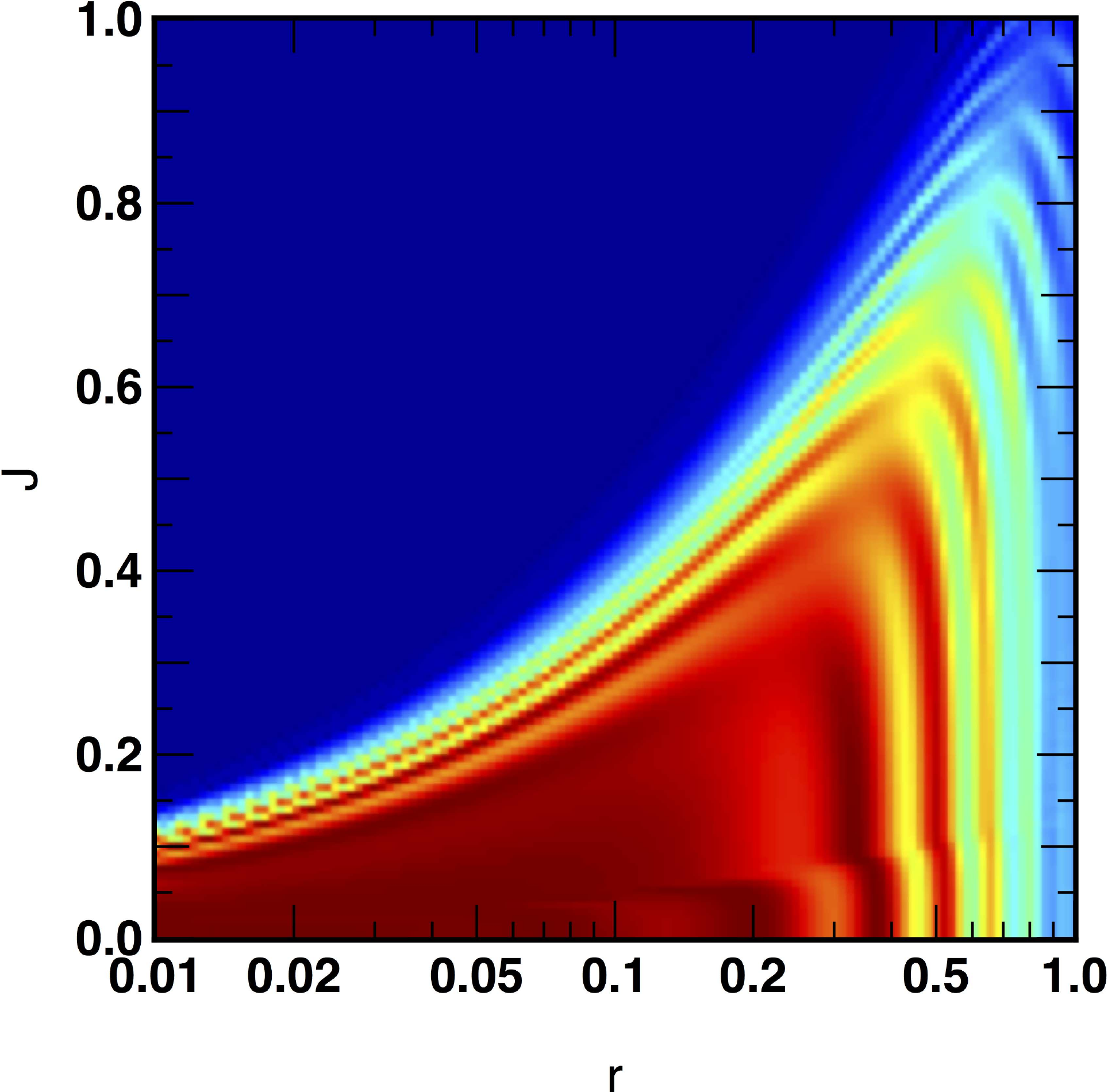}
\hskip 1cm \includegraphics[width=7cm]{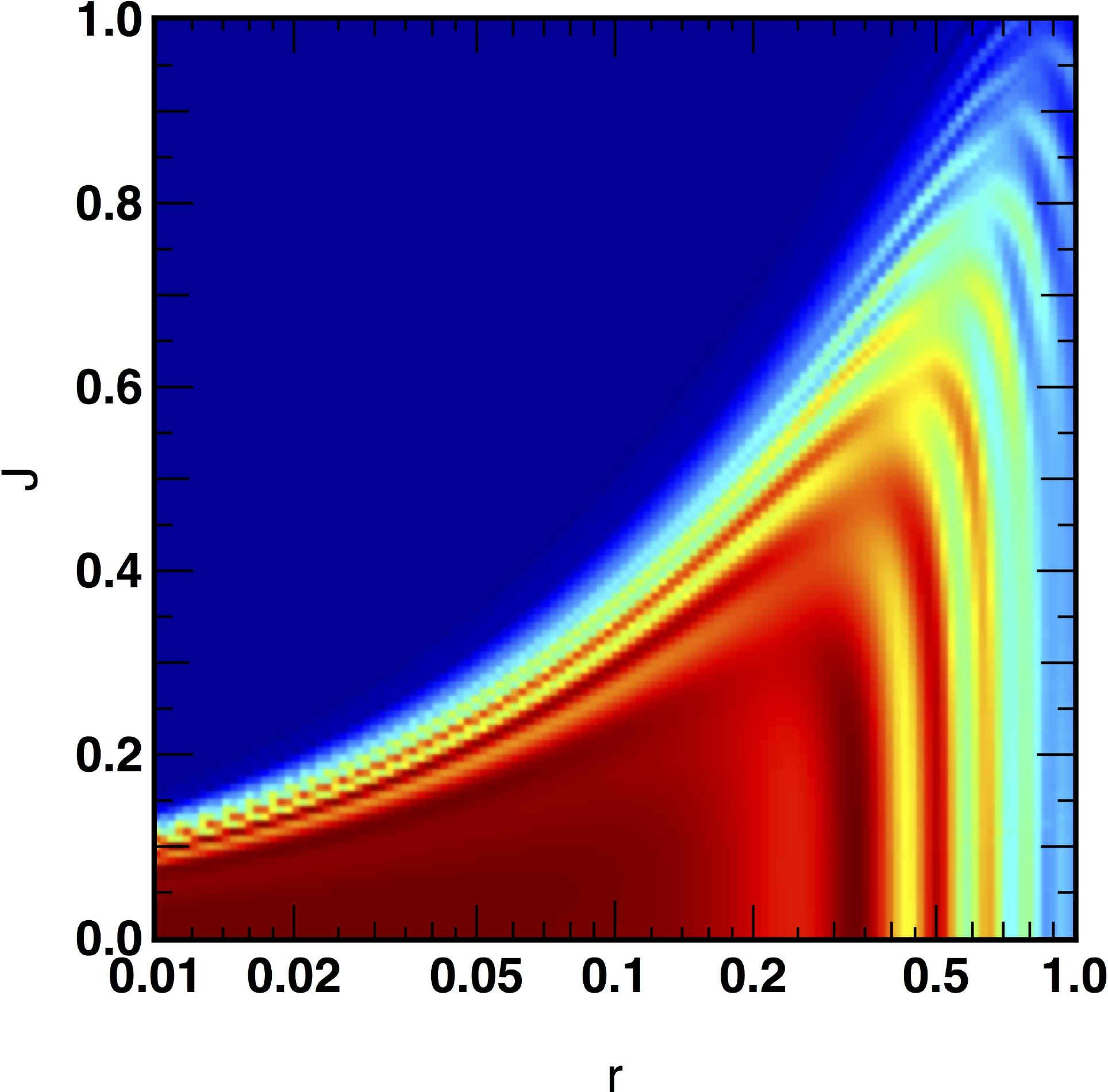}
}}
\caption{Comparison between the reflecting central sphere method (left panel) and our improved delayed central sphere implementation (right panel). A simulation of a H\'enon sphere with $(N_r,N_v,N_j)=(200,200,200)$ and a virial ratio $R=0.5$ is shown at $t=30$ in the $\left(r,u=0,j\right)$ plane. The systematic artificial speed increase undergone by orbits that penetrate the central region compared to their higher angular momentum counterparts can clearly be observed at low $j$ on the left panel where a reflective sphere is used, while the distribution function does not exhibit such spurious features when a delayed kernel is used (right panel).}
\label{fig:delayed}
\end{figure*}
\begin{figure*}
\hbox{
\includegraphics[width=4.25cm]{myfig/image_rvir0.1_2047x2047x32_12_5.0000.ND12-crop.pdf}
\includegraphics[width=4.25cm]{myfig/image_rvir0.1_2047x2047x32_12_10.0000.ND12-crop.pdf}
\includegraphics[width=4.25cm]{myfig/image_rvir0.1_2047x2047x32_12_15.0000.ND12-crop.pdf}
\includegraphics[width=4.25cm]{myfig/image_rvir0.1_2047x2047x32_12_35.0000.ND12-crop.pdf}
}
\hbox{
\includegraphics[width=4.25cm]{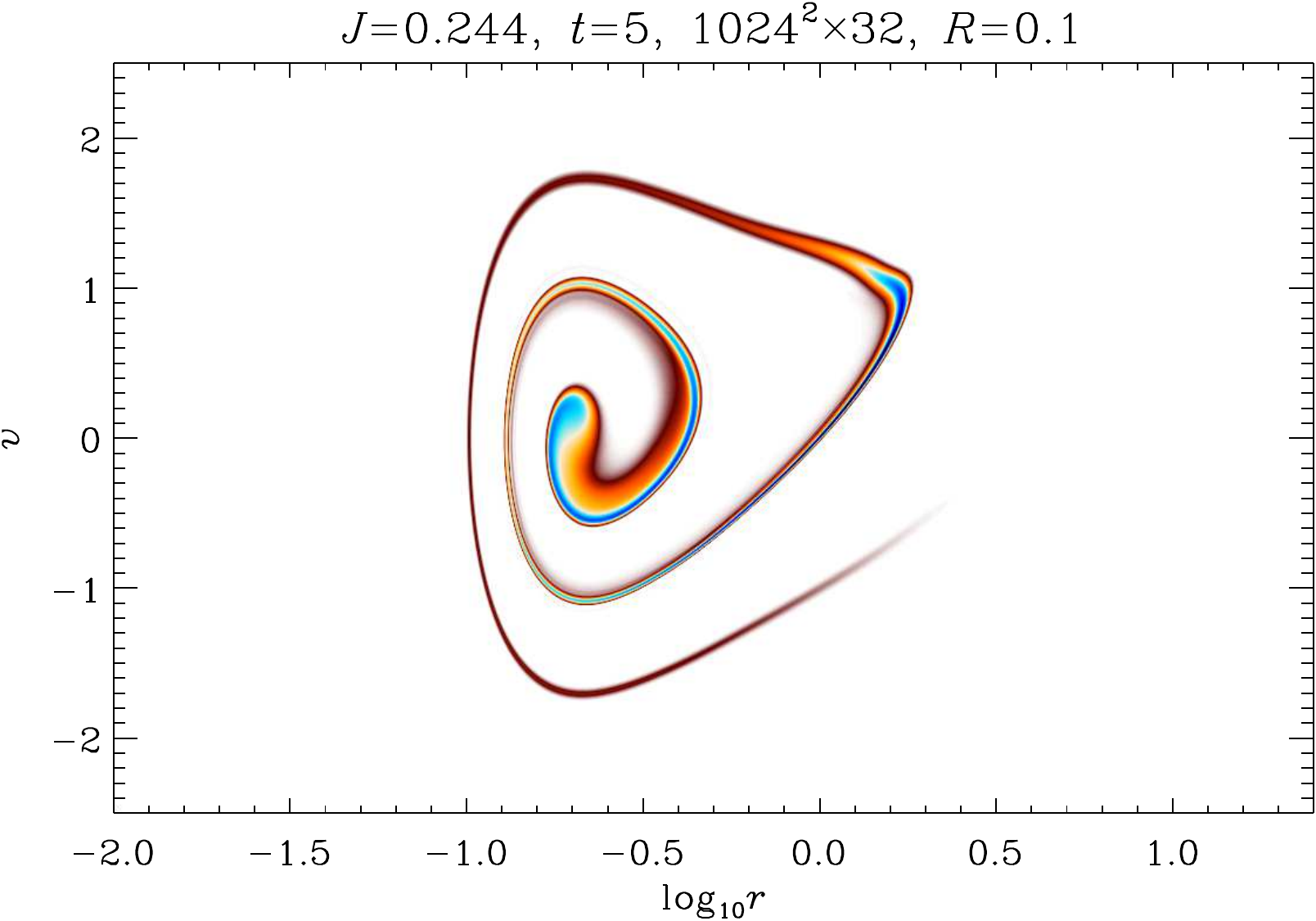}
\includegraphics[width=4.25cm]{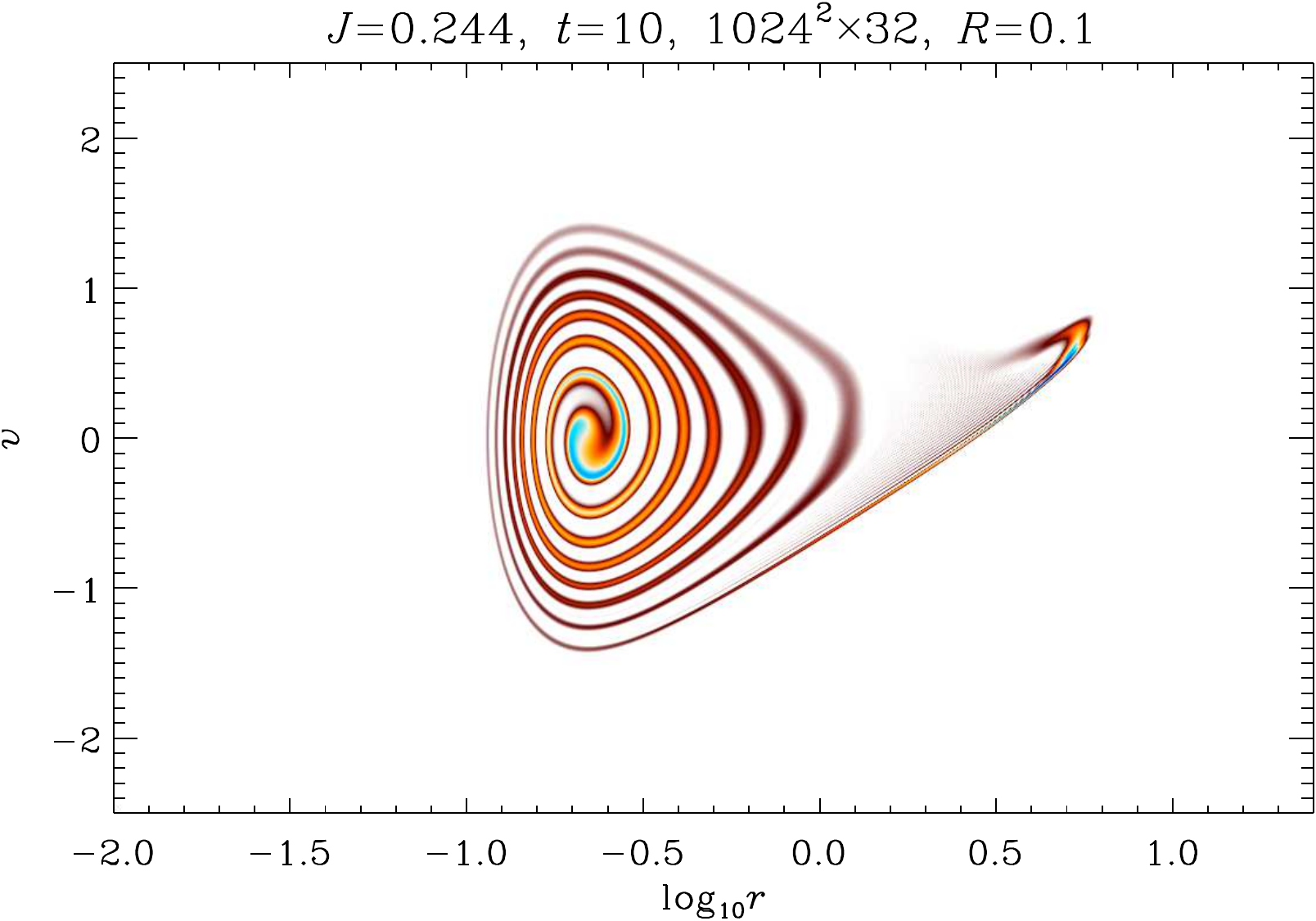}
\includegraphics[width=4.25cm]{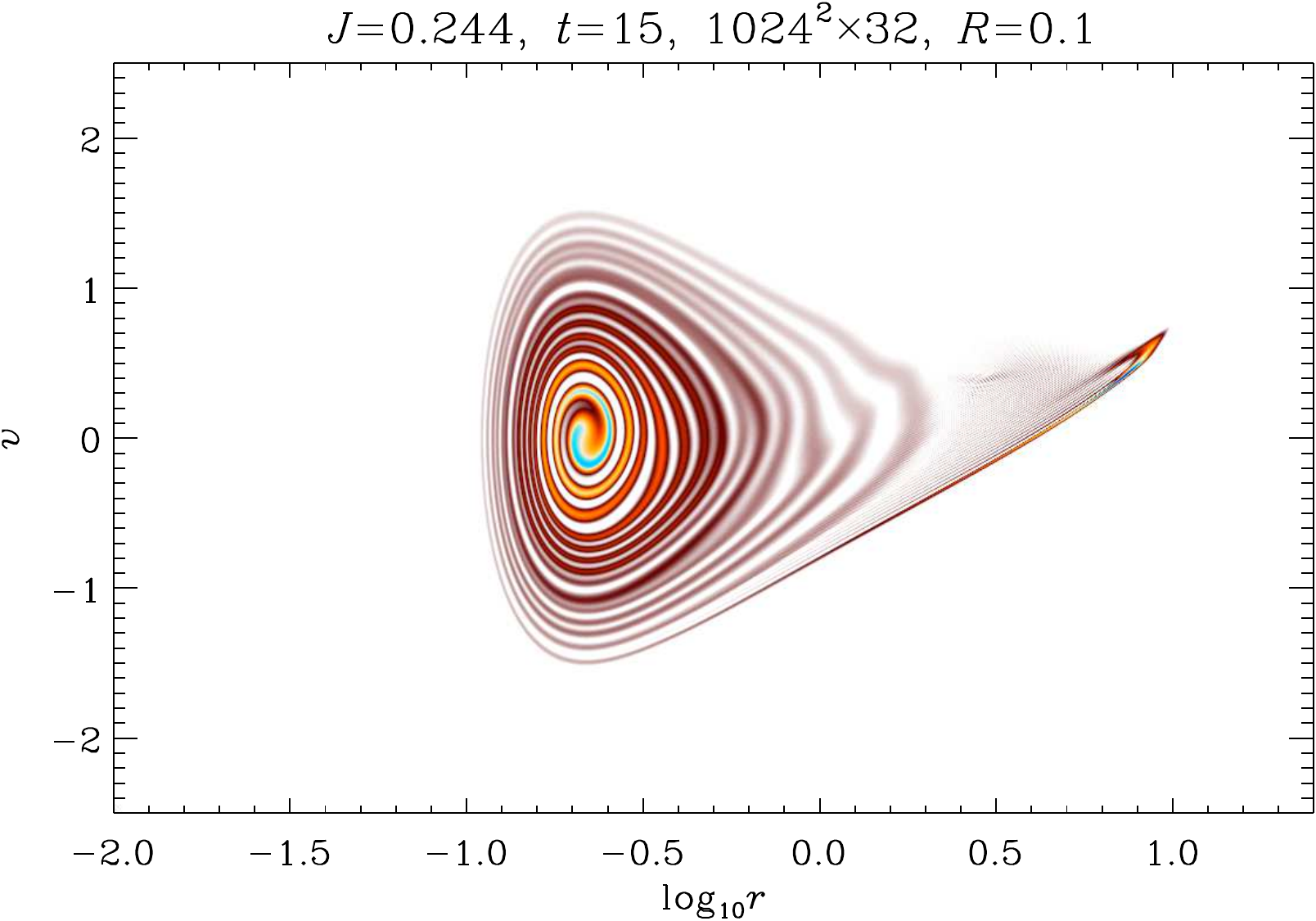}
\includegraphics[width=4.25cm]{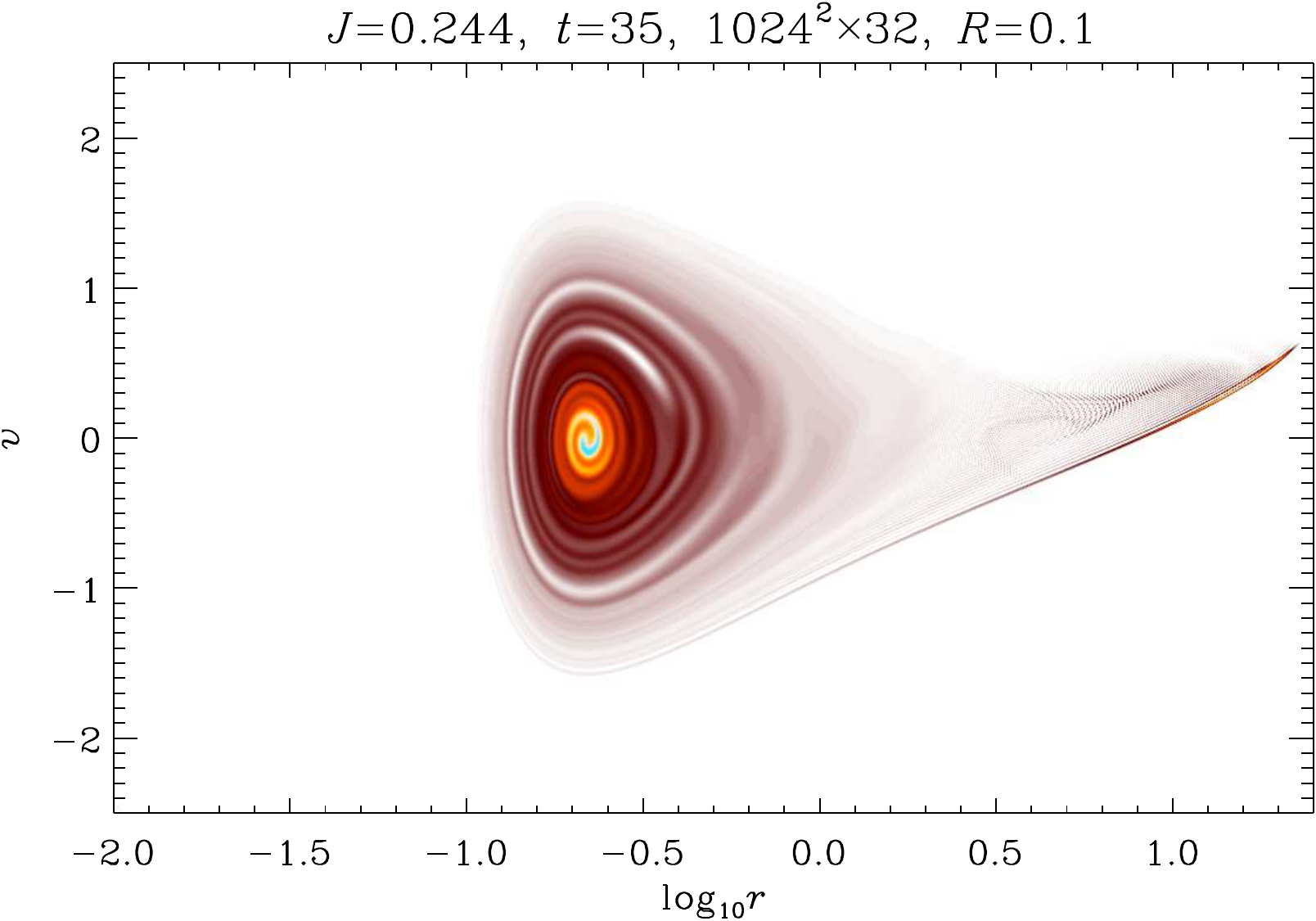}
}
\hbox{
\includegraphics[width=4.25cm]{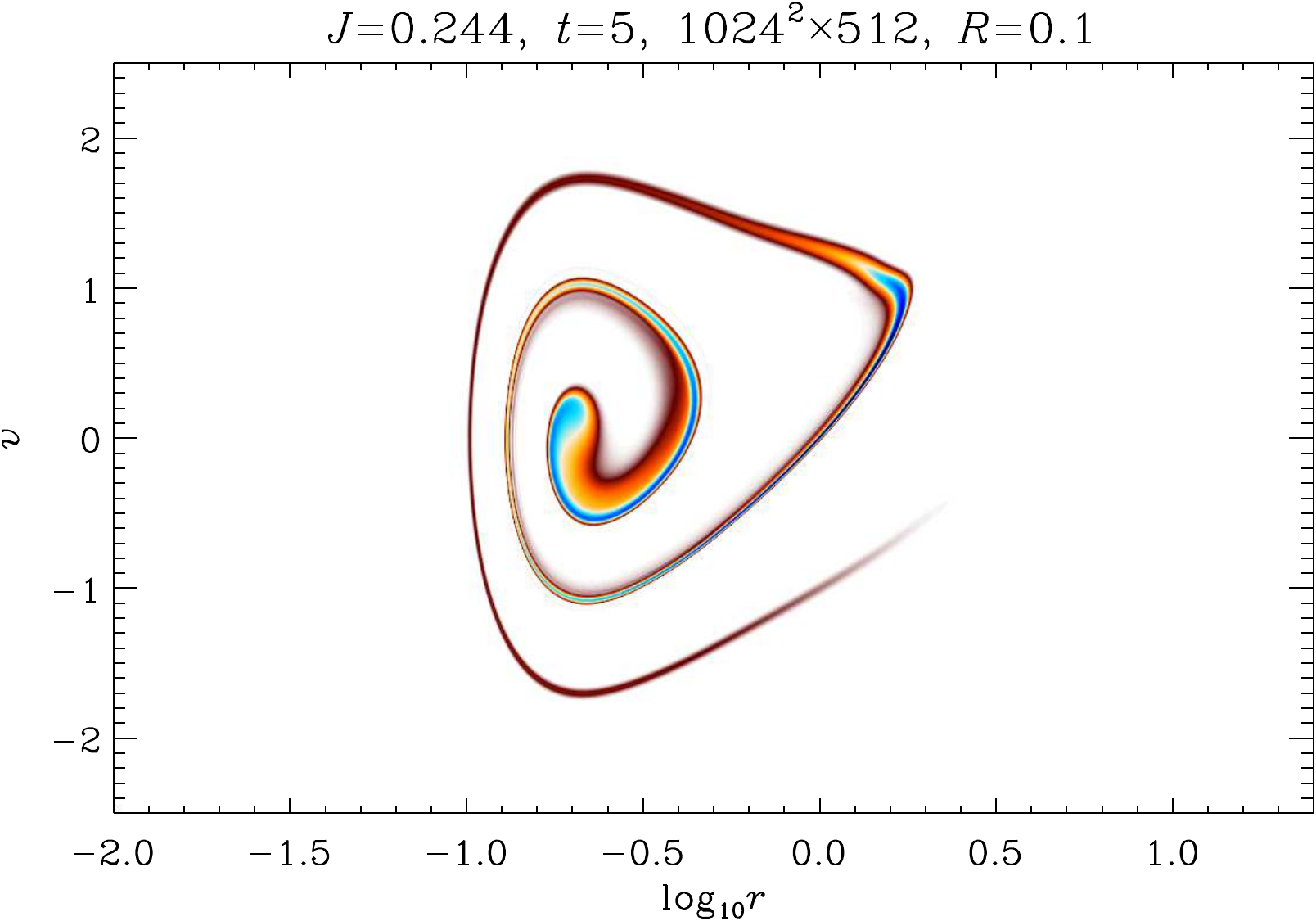}
\includegraphics[width=4.25cm]{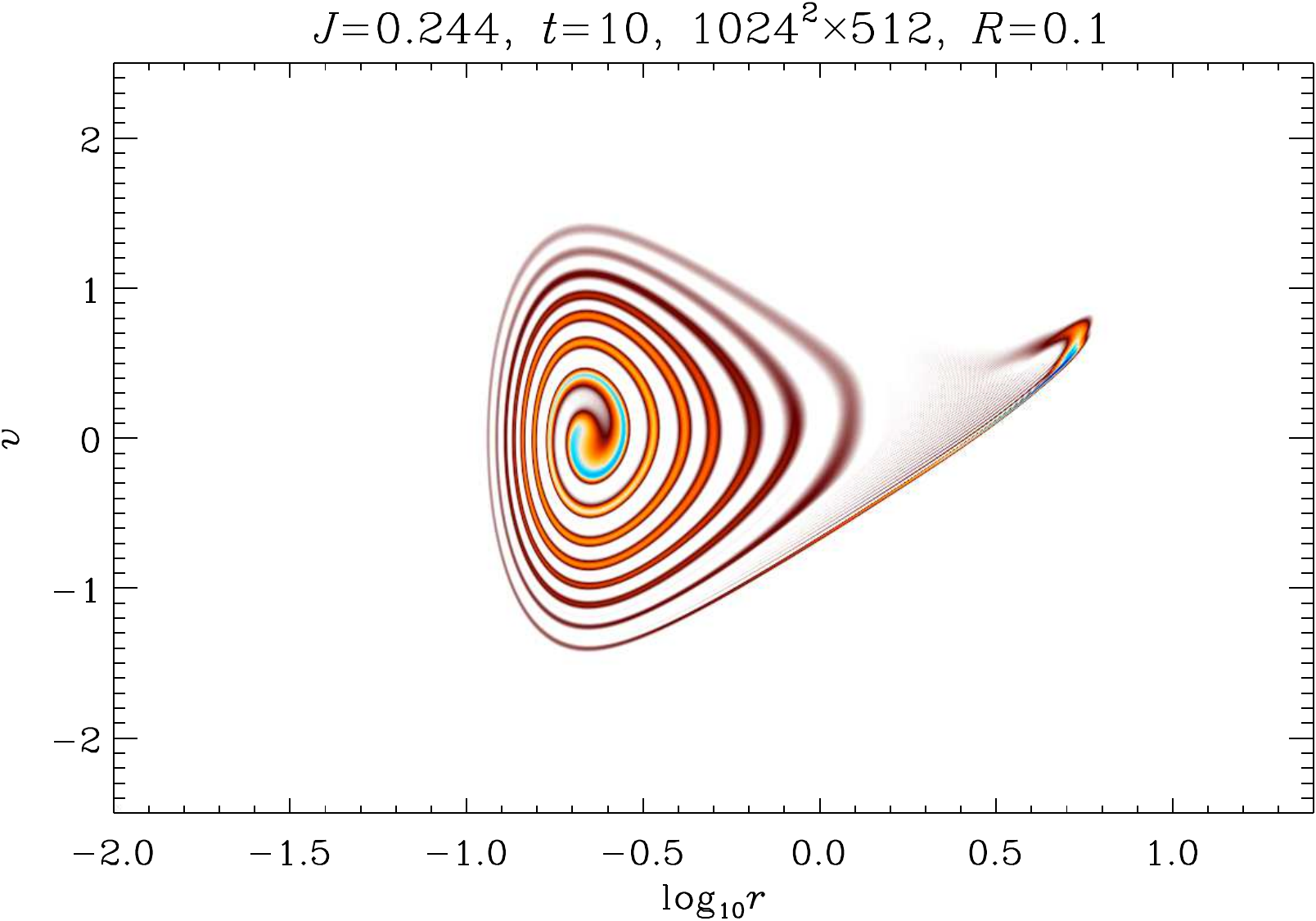}
\includegraphics[width=4.25cm]{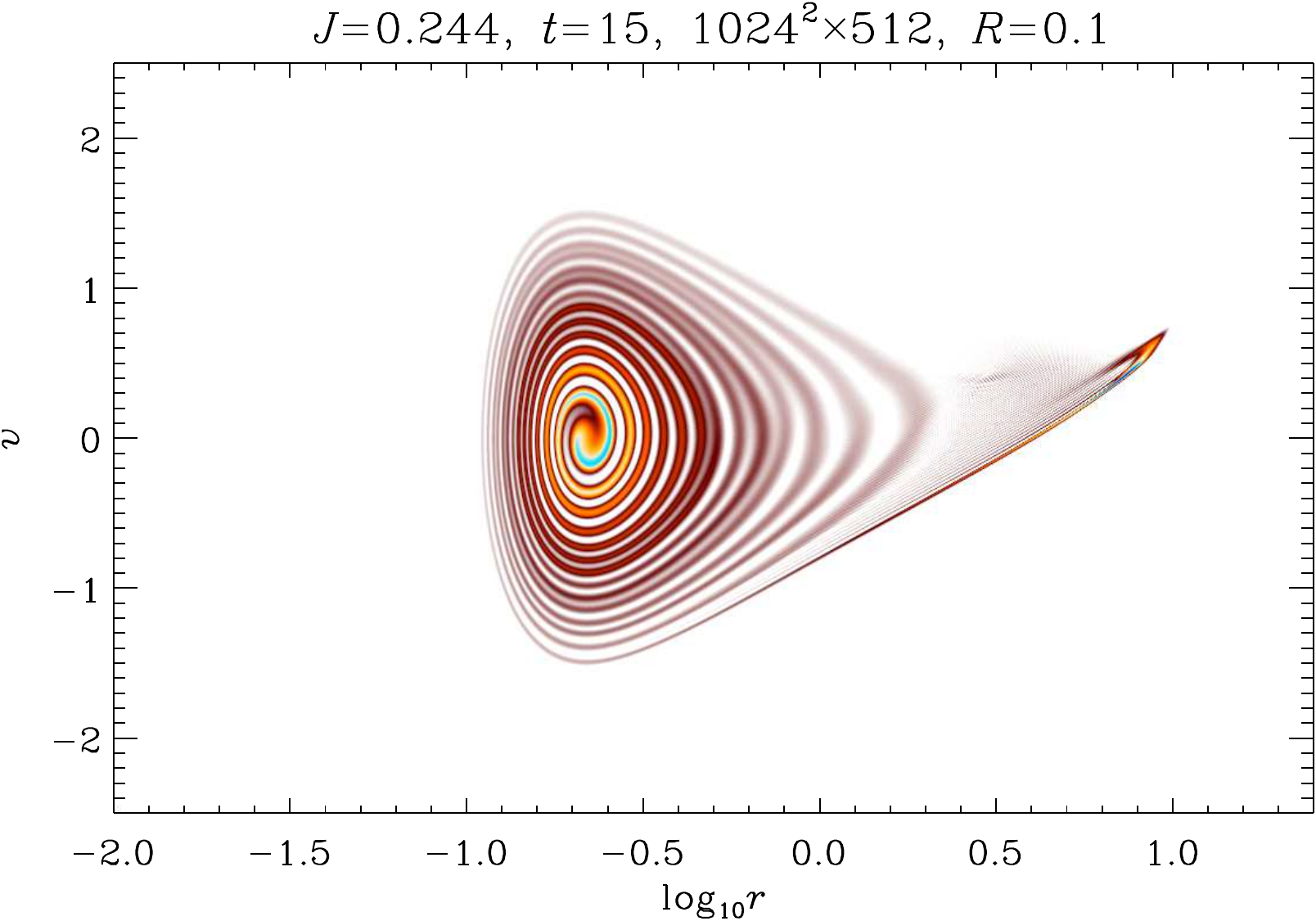}
\includegraphics[width=4.25cm]{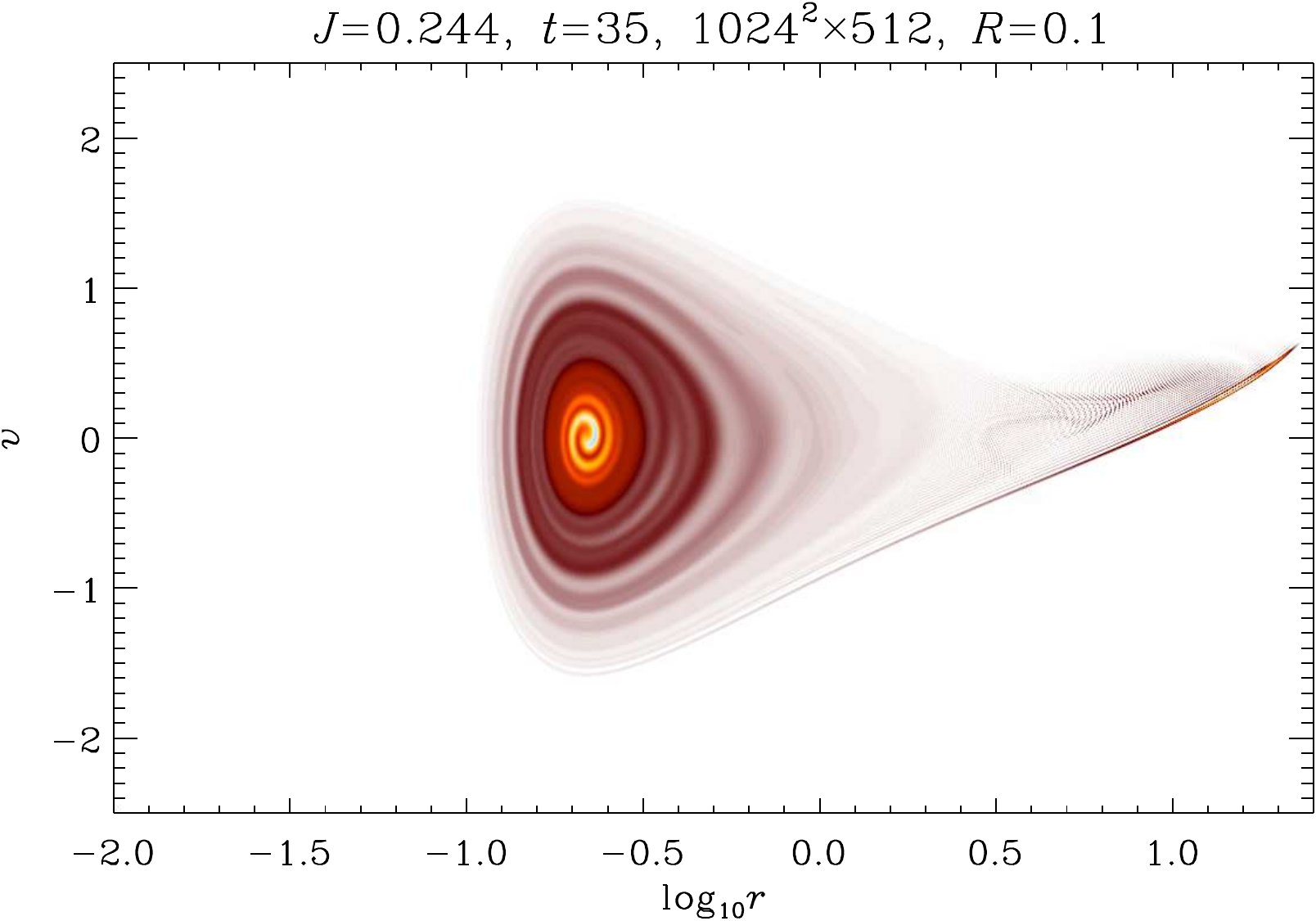}
}
\hbox{
\includegraphics[width=4.25cm]{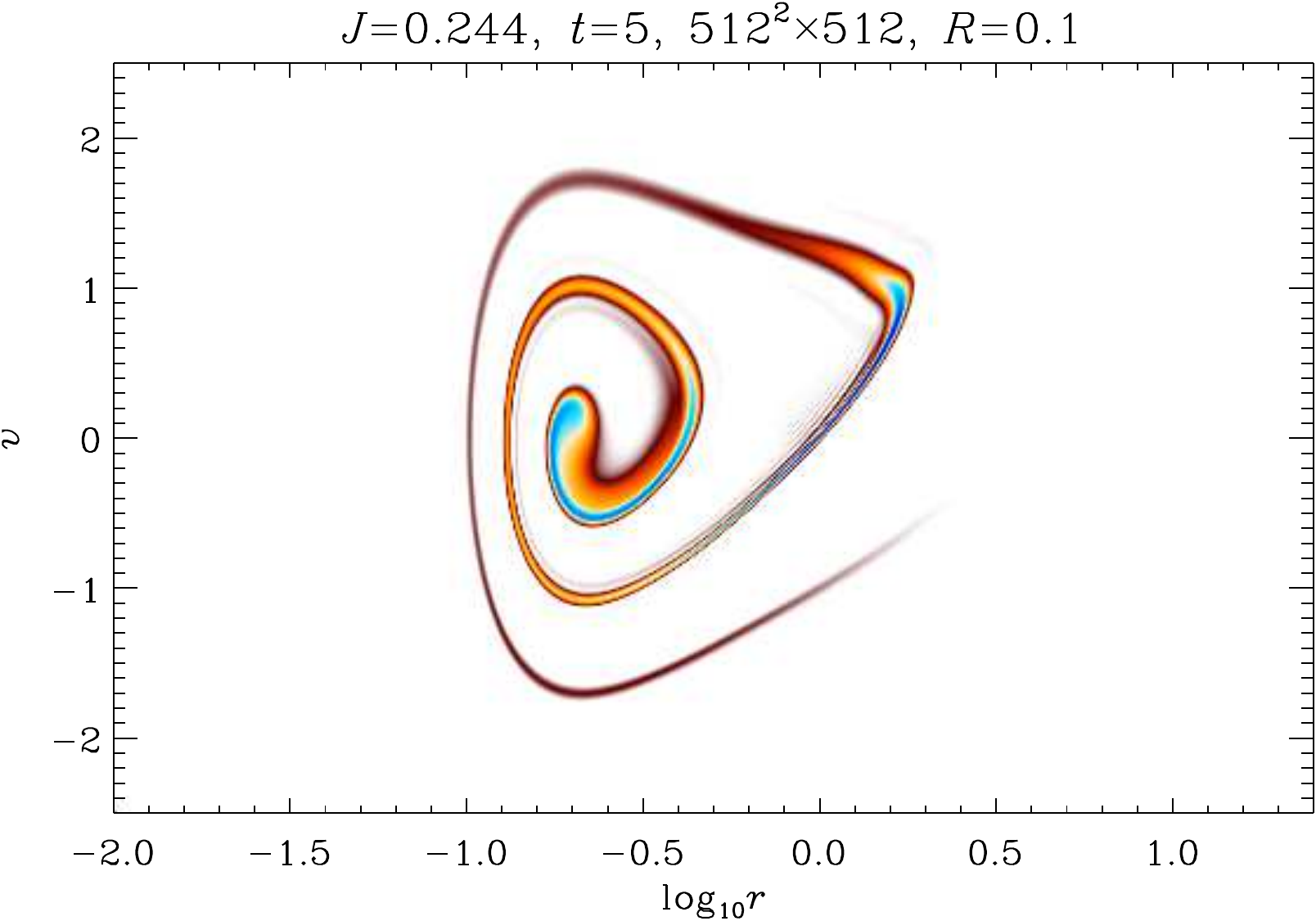}
\includegraphics[width=4.25cm]{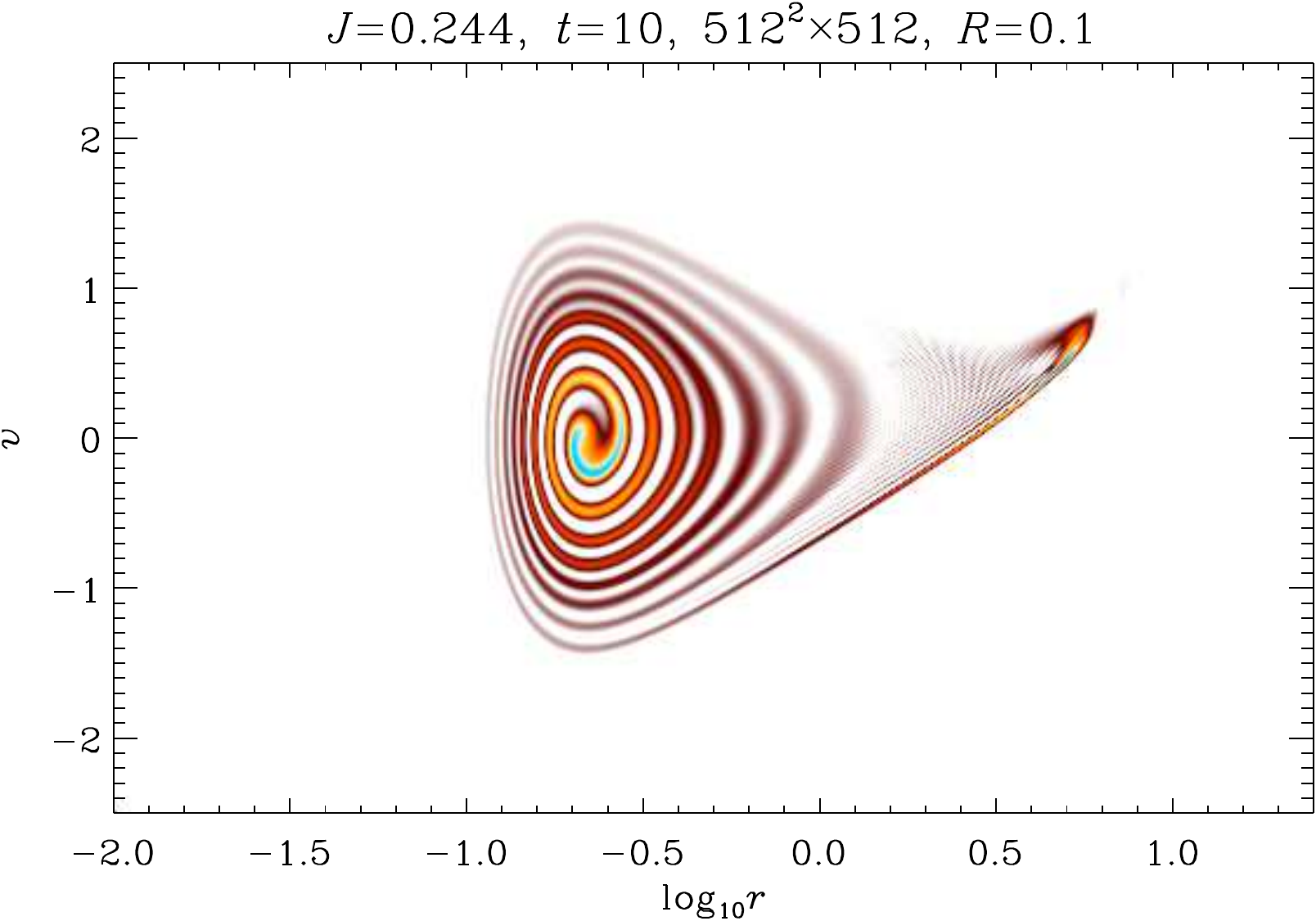}
\includegraphics[width=4.25cm]{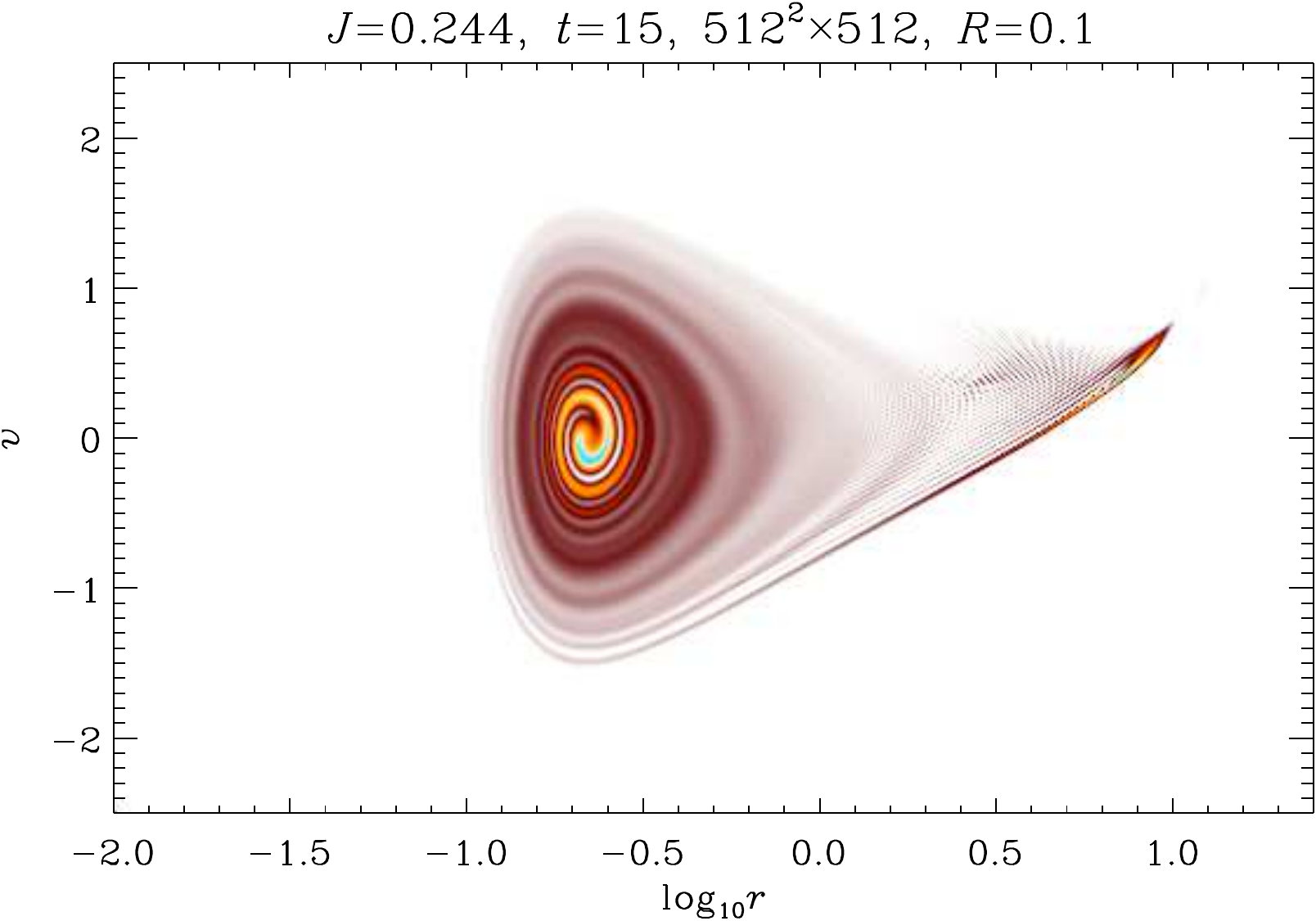}
\includegraphics[width=4.25cm]{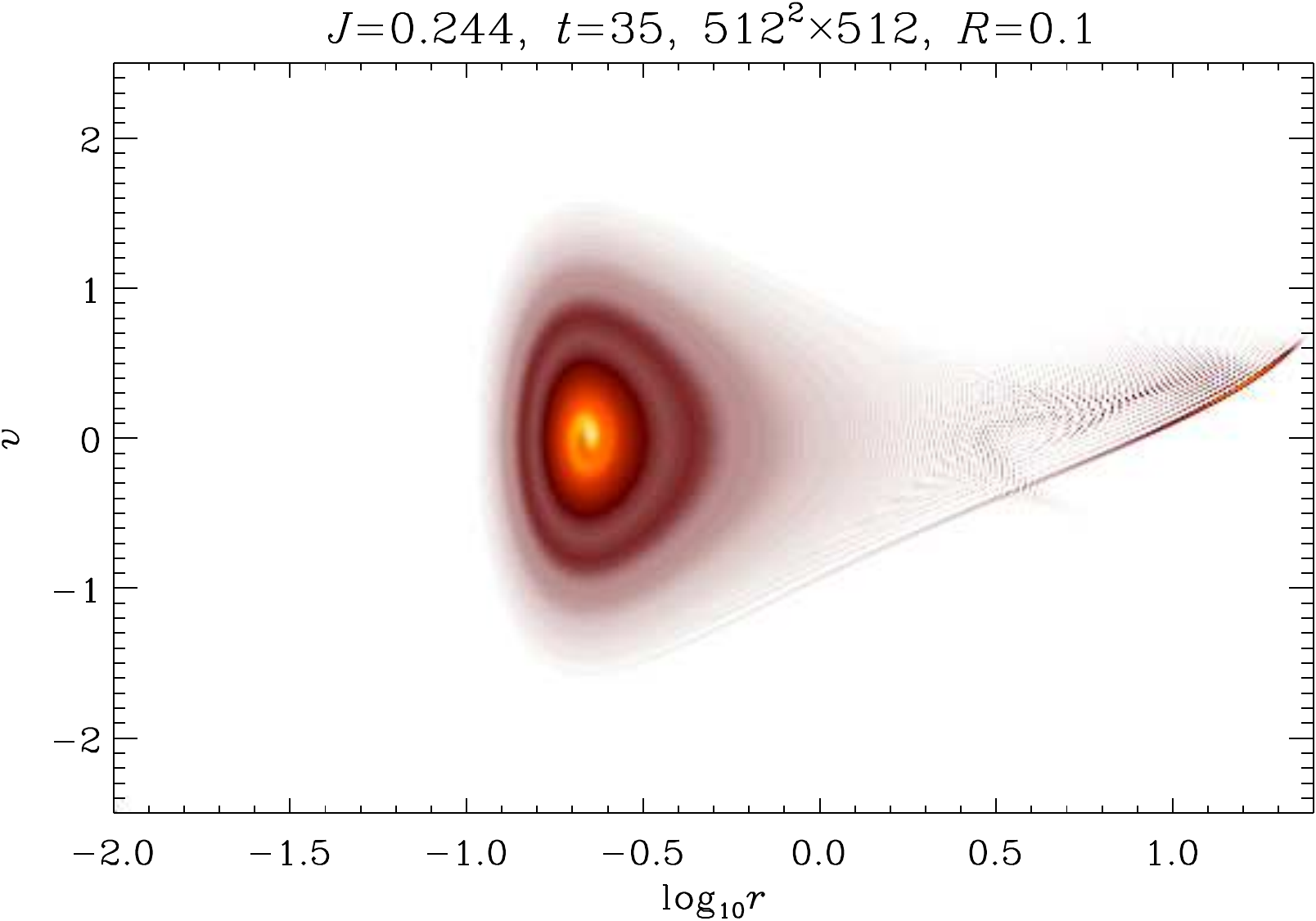}
}
\caption[]{Effect of resolution in the Vlasov code: phase-space density for $R=0.1$ and $j =0.244$. Each column of panels corresponds to a given value of time $t$, increasing from left to right, while each line correspond to a given resolution, $(N_r,N_v,N_j)=(2048,2048,32)$, $(1024,1024,32)$, $(1024,1024,512)$ and $(512,512,512)$ from top to bottom, as indicated on each panel.  The pictures show only the $f \geq 0$ part of the phase-space density, while it can actually become negative because of aliasing. However, this choice of representation does not hide aliased regions. The prominent one corresponds to the textured zone above the large $r$ tail of the system on the right panels.}
\label{fig:testvla}
\end{figure*}
\begin{figure*}
\hbox{
\includegraphics[width=4.25cm]{myfig/snapshot_5.0000.ND.0.1_2047_2047_32.gridALLJ-crop.pdf}
\includegraphics[width=4.25cm]{myfig/snapshot_10.0000.ND.0.1_2047_2047_32.gridALLJ-crop.pdf}
\includegraphics[width=4.25cm]{myfig/snapshot_15.0000.ND.0.1_2047_2047_32.gridALLJ-crop.pdf}
\includegraphics[width=4.25cm]{myfig/snapshot_35.0000.ND.0.1_2047_2047_32.gridALLJ-crop.pdf}
}
\hbox{
\includegraphics[width=4.25cm]{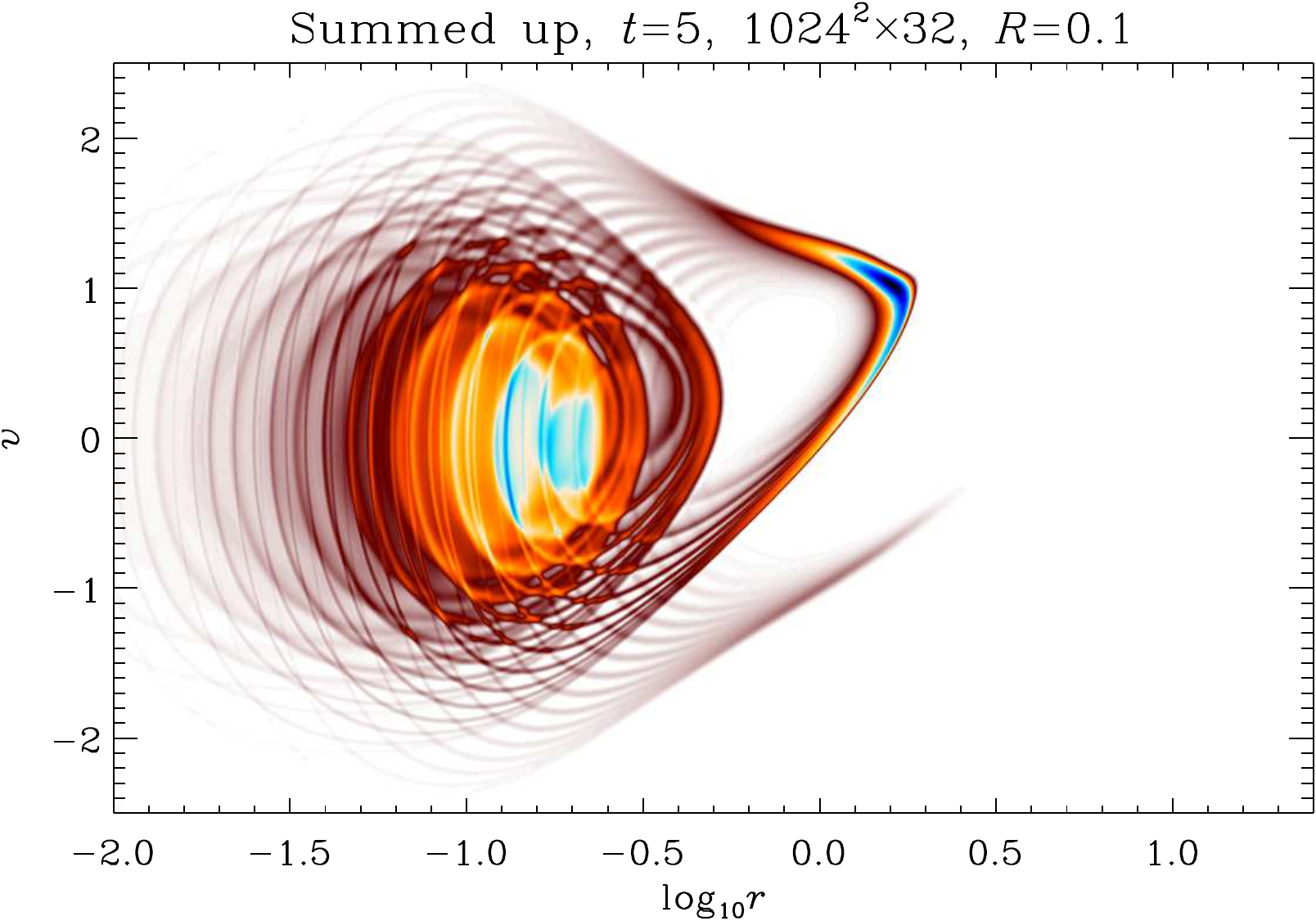}
\includegraphics[width=4.25cm]{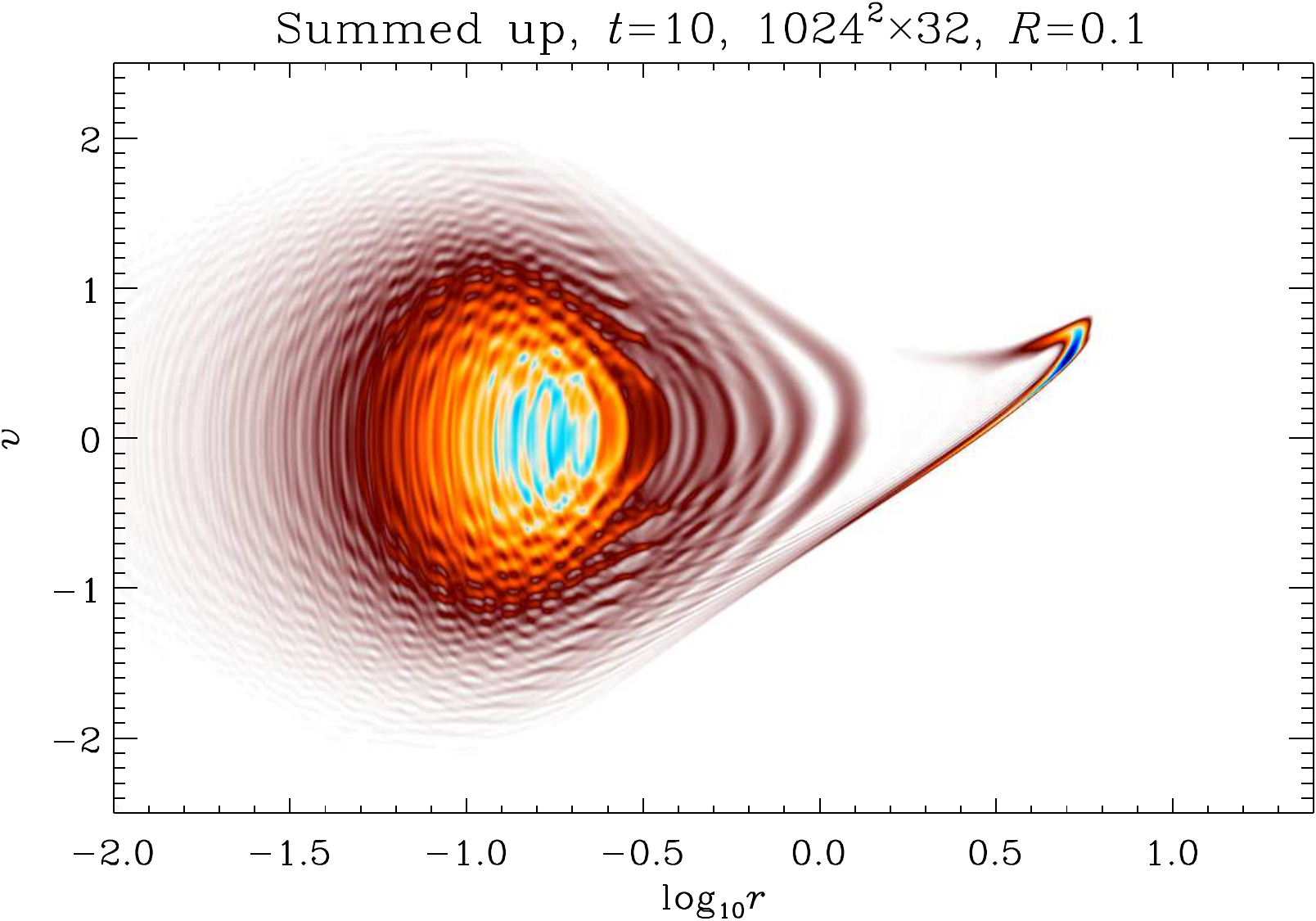}
\includegraphics[width=4.25cm]{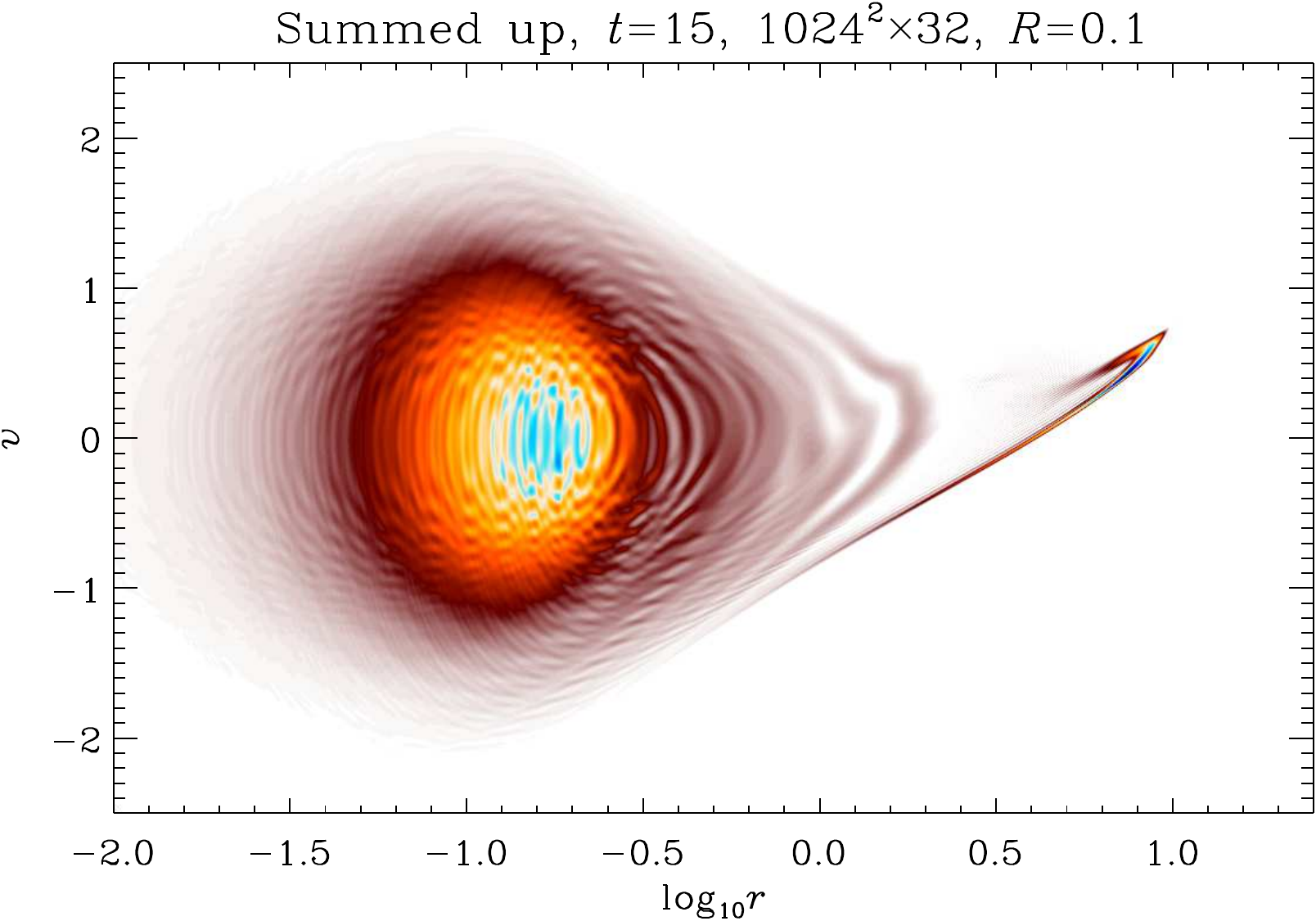}
\includegraphics[width=4.25cm]{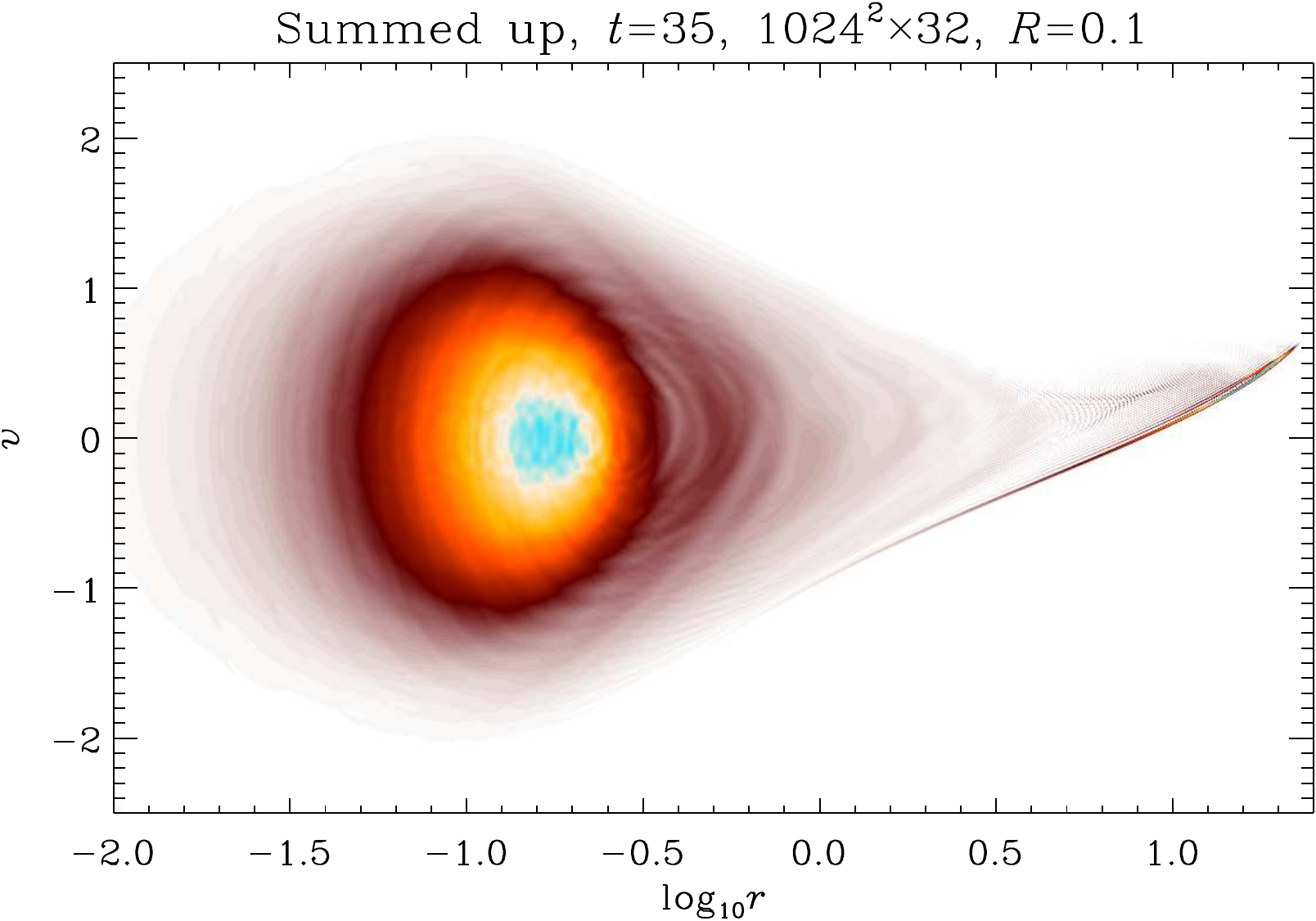}
}
\hbox{
\includegraphics[width=4.25cm]{myfig/snapshot_5.0000.ND.0.1_1023_1023_512.gridALLJ-crop.pdf}
\includegraphics[width=4.25cm]{myfig/snapshot_10.0000.ND.0.1_1023_1023_512.gridALLJ-crop.pdf}
\includegraphics[width=4.25cm]{myfig/snapshot_15.0000.ND.0.1_1023_1023_512.gridALLJ-crop.pdf}
\includegraphics[width=4.25cm]{myfig/snapshot_35.0000.ND.0.1_1023_1023_512.gridALLJ-crop.pdf}
}
\hbox{
\includegraphics[width=4.25cm]{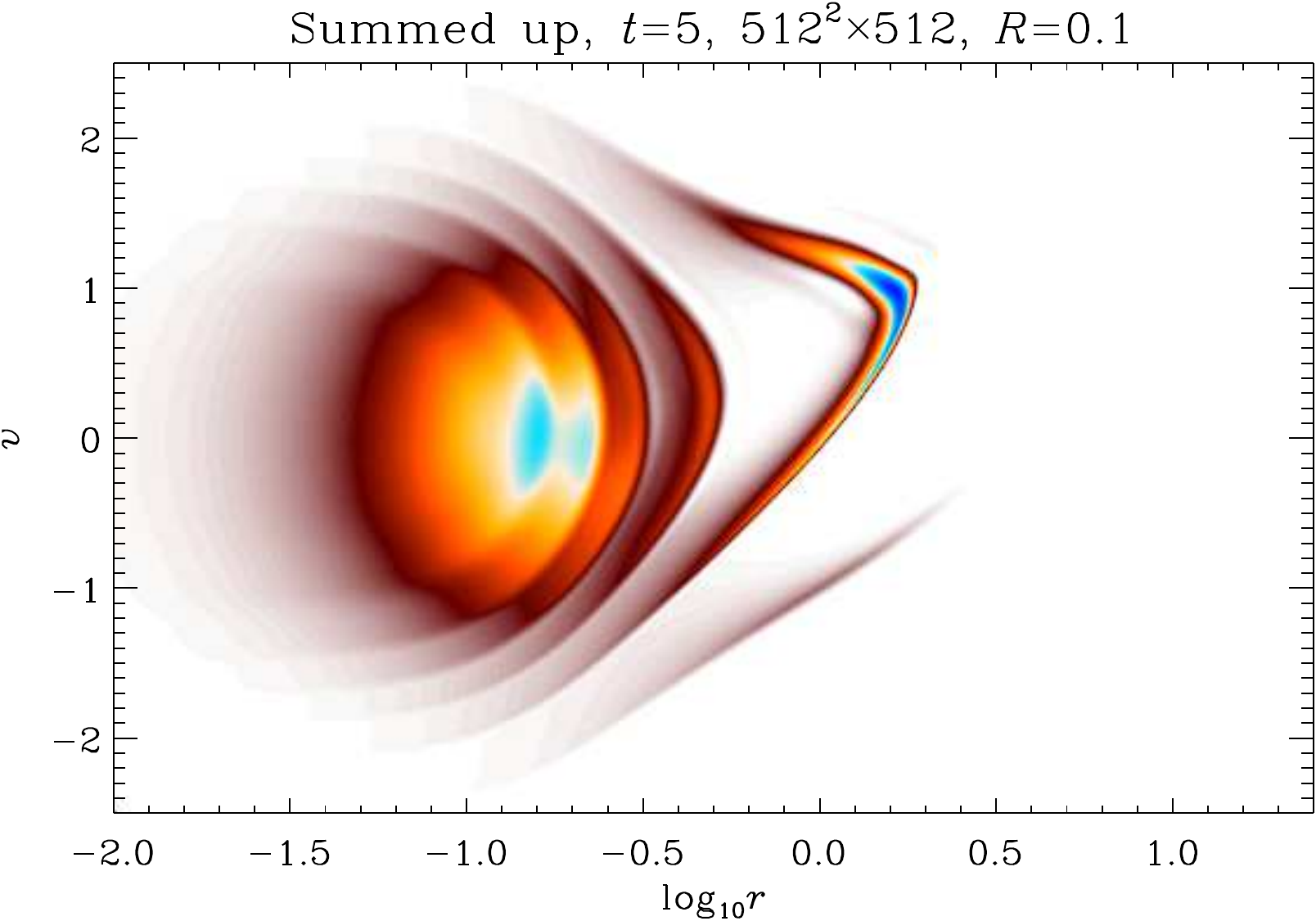}
\includegraphics[width=4.25cm]{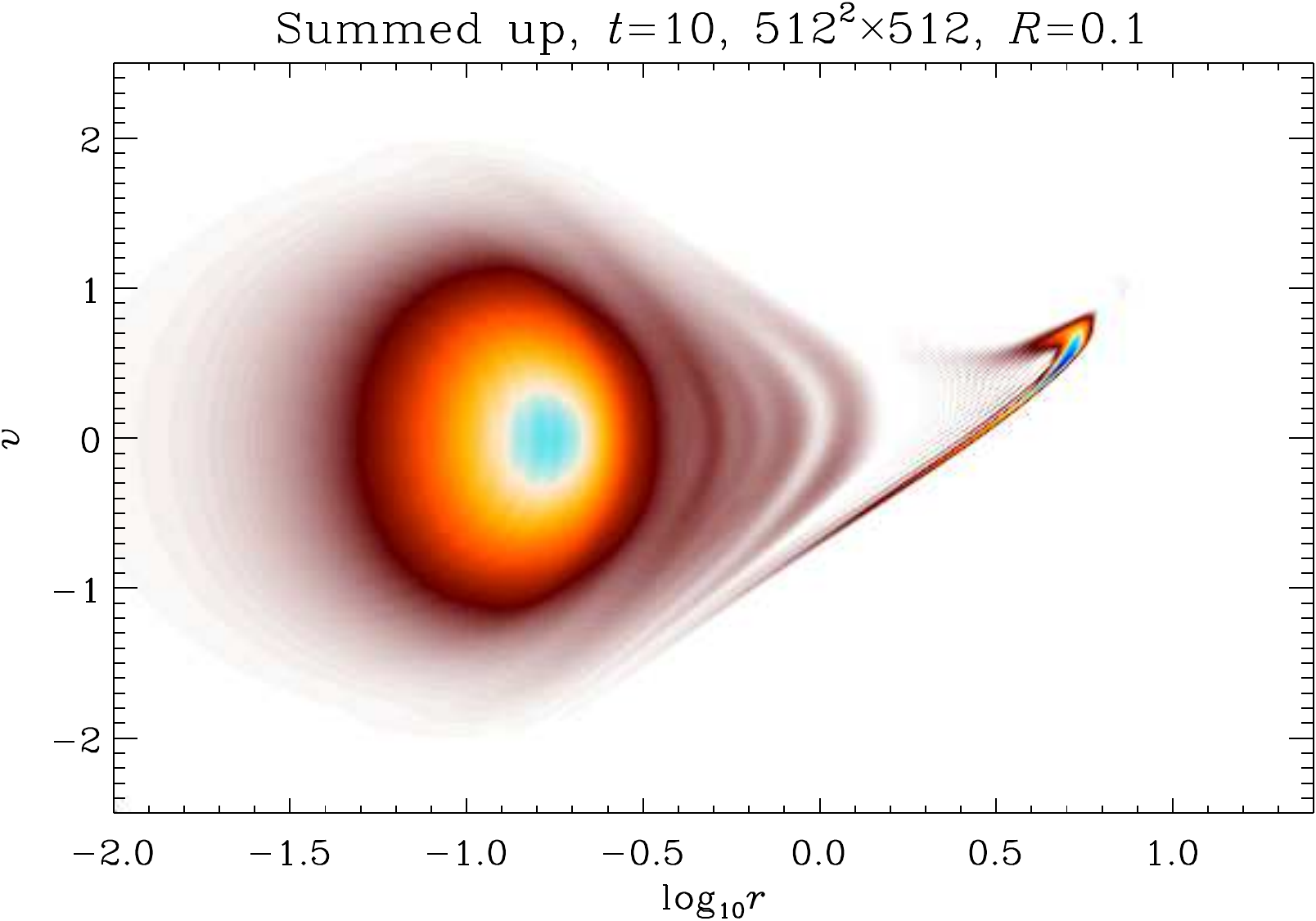}
\includegraphics[width=4.25cm]{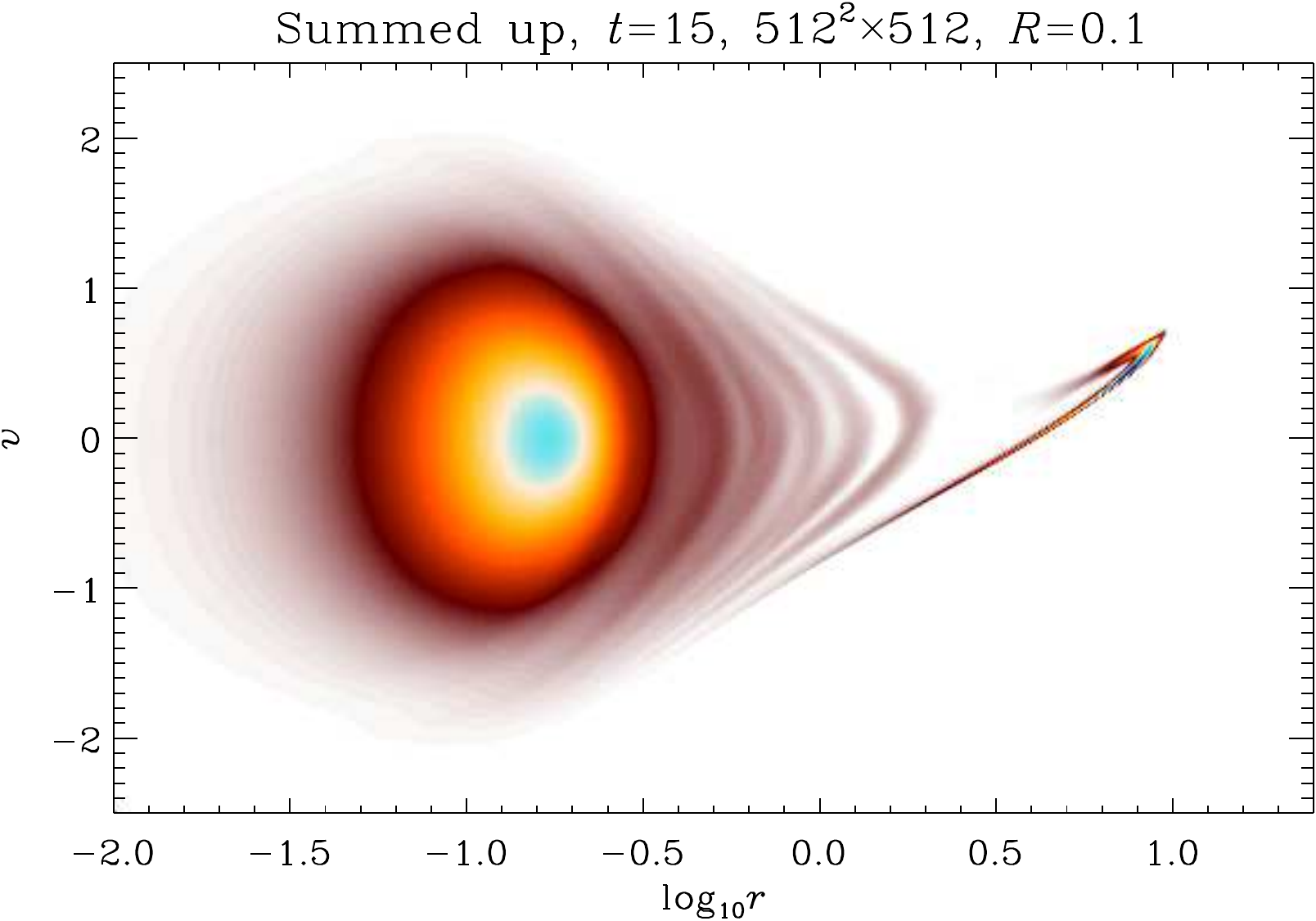}
\includegraphics[width=4.25cm]{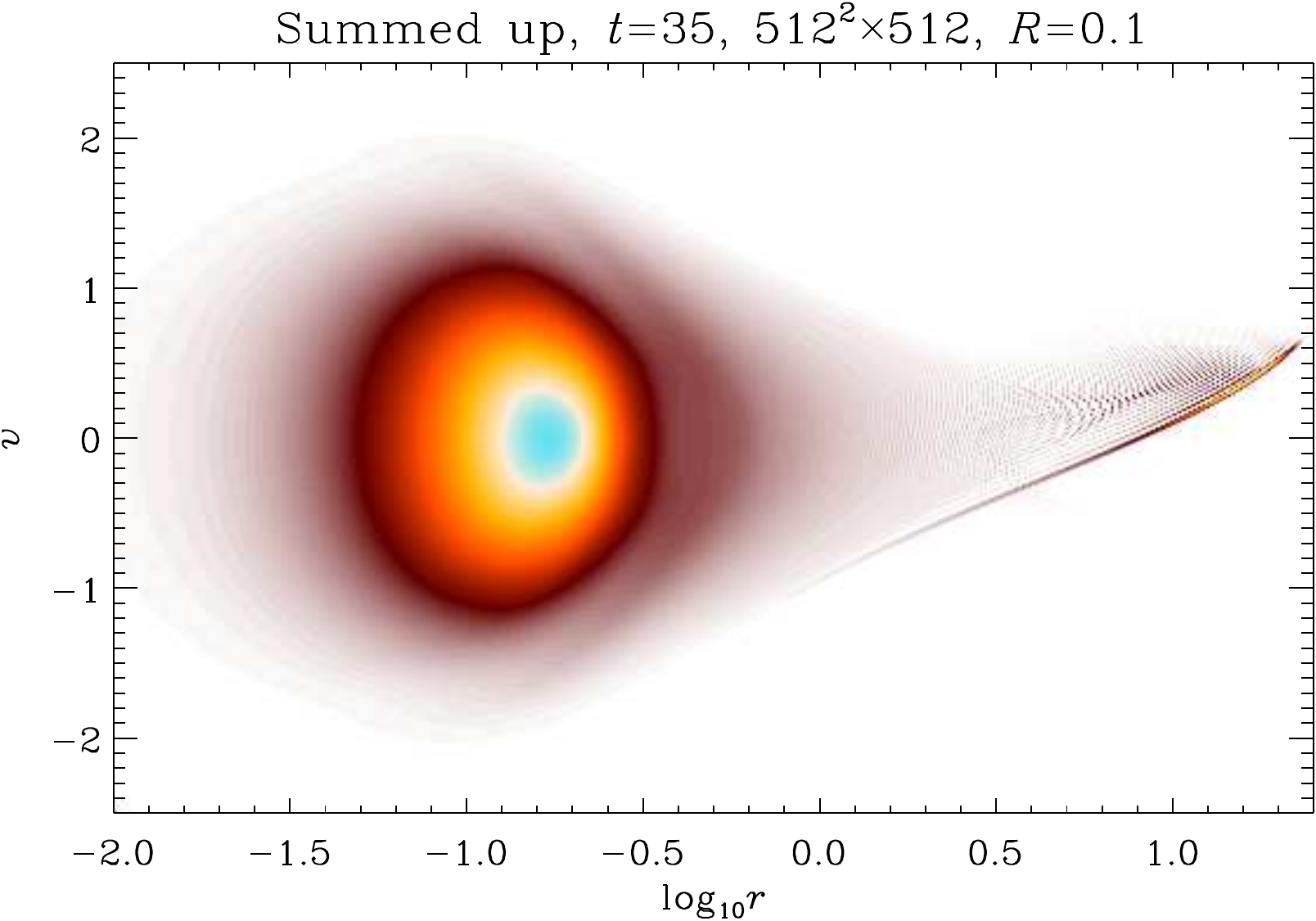}
}
\caption[]{Same as in Fig.~\ref{fig:testvla}, but the phase-space distribution function has now been summed up over the whole available range of values of $j \in [0,J_{\rm max}=1.6]$, where $J_{\rm max}$ is the maximum sampled value of $j$.}
\label{fig:testvla2}
\end{figure*}
 In this appendix, we explain how reflecting boundaries conditions with time delay are implemented in {\tt VlaSolve}. 

If the mass inside the sphere of radius $R_{\rm min}$ is neglected, the trajectories followed by each test particle associated to a grid site that penetrates the sphere are fixed and do not depend on time. This property, combined with the fact that we use a constant time step, allows us to pre-compute these trajectories once and for all. The delayed central sphere method is then implemented by associating a linked list to each grid site whose associate test particle radial position $r$ half a time step backward in time is such that $r \leq R_{\rm min}$. Each linked list contains as many elements as the number of time steps needed for the particle to travel a distance of $2R_{\rm min}$ and the $n^{\rm th}$ element in the list stores the coordinates of the test particle $n$ time steps backward in time. Before starting the simulation, we initialize each element coordinate and the corresponding value of the initial distribution function. For each time step, the value of each element is then simply updated by assigning to it the value of its successor while the last element value, whose coordinates fall inside the computing domain, $r \geq R_{\rm min}$, is interpolated. A comparison of the results obtained with the reflective central sphere to our improved delayed central sphere is shown on figure \ref{fig:delayed}. The improvements are unquestionable. 
\subsection{Parallelization issues}
\label{sec:para}
We implemented a hybrid shared and distributed memory version of {\tt VlaSolve} via the OpenMP and MPI libraries, respectively. 

Shared memory parallelism is relatively straightforward to achieve in the spherically symmetric case, by taking advantage of the fact that the angular momentum $j$ is a conserved quantity. Spline interpolations, which represent the most expensive part of the code, can thus be computed independently for each slice of constant $j$. We therefore easily reach an almost perfect parallelization up to a number of tasks equal to the grid resolution $N_j$ of angular momentum space, which is typically larger than the number of available cores on a shared memory system. 

Distributed memory parallelization via MPI is not as simple. Indeed, spline interpolations are intrinsically non-local, which makes the parallelization along dimensions other than $j$ non trivial. Sticking with the trivial parallelization described above unfortunately limits the maximum total number of processes running in parallel to $N_j$, which is suboptimal.  We overcome this limitation by performing MPI domain decomposition in $(r,v)$ space, following the approach of \cite{Crouseilles09}, who propose to localize the cubic spline interpolation to each domain by using Hermite boundary conditions between the domains with an ad hoc reconstruction of the derivatives. 
\subsection{Effects of resolution}
\label{app:resvlaeff}
Figures~\ref{fig:testvla} and \ref{fig:testvla2} show, respectively for $j=0.244$ and integrated over angular momentum, the phase-space distribution function measured in {\tt VlaSolve} simulations with different resolutions. 
 These simulations have been performed for a H\'enon sphere with initial virial ratio $R=0.1$. Beside the very good global agreement between the various runs, these figures bring out three effects, which increase when the resolution of the phase-space grid is reduced:
\begin{itemize}
\item Diffusion smearing out fine details that build up in phase-space during the course of dynamics, for instance clearly visible when one compares top to bottom middle panels of Fig.~\ref{fig:testvla}. One concern with diffusion is that it might prevent the appearance of unstable modes. However, we did not perform any simulation in this work that would prove this. 
\item Aliasing due to artificial oscillations in the spline interpolation: for the problem studied here, aliasing becomes particularly visible after relaxation in the region above the large $r$ tail, but this does not have significant impact on the dynamics. 
\item Aliasing due to undersampling angular momentum space: it is visible at all times when one examines the phase-space distribution function integrated over angular momentum (top panels of Fig.~\ref{fig:testvla2}) and can have dramatic consequences on the dynamics. The two top lines of panels of Fig.~\ref{fig:testvla} and \ref{fig:testvla2}, corresponding to a sparse sampling in $j$ space with only 32 slices, indeed show the appearance of an instability, which presents, on the third column of these figures, the same pattern whether $(N_r,N_v)=(2048,2048)$ or $(1024,1024)$. This instability is not present in the simulations with higher resolution in $j$, as shown by the two bottom lines of panels. Note that the presence of this instability depends on initial conditions: for $R=0.5$, we did not notice it for the time coverage considered, $t \leq 100$ (upper line of panels of Figs.~\ref{fig:0v5_12} and \ref{fig:0v5_ALLJ}). 
\end{itemize} 
\label{sec:resoeffects}
\section{$N$-body simulations: exploration of the control parameter space}
\label{app:instab}
\begin{figure*}
\begin{center}
\begin{minipage}{0.25\textwidth}
\centerline{\includegraphics[width=4.25cm]{myfig/vlasov_0.1_1m_150.gridALLJ-crop.pdf}}
\end{minipage}
\begin{minipage}{0.25\textwidth}
\centerline{\includegraphics[width=4.25cm]{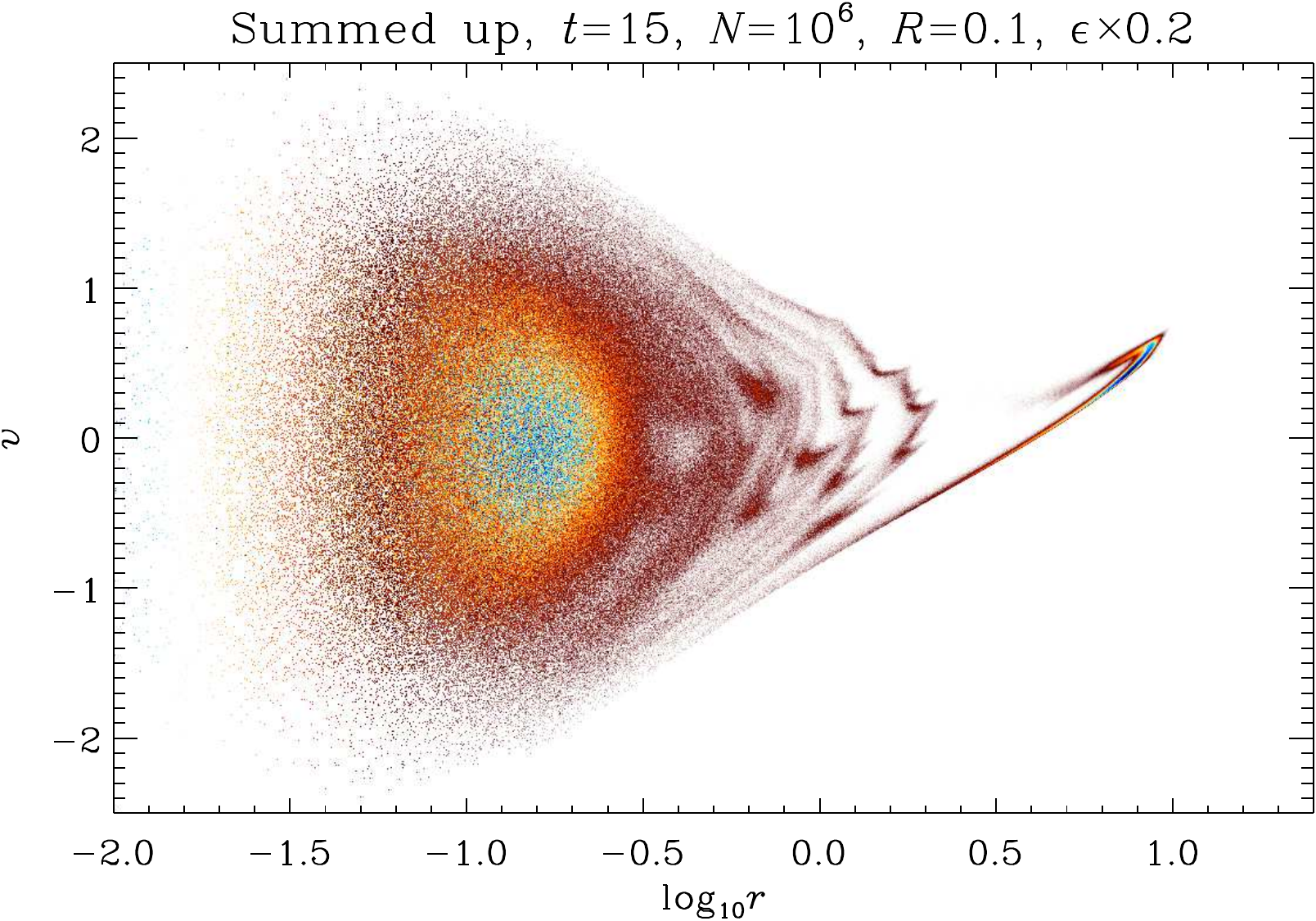}}
\centerline{\includegraphics[width=4.25cm]{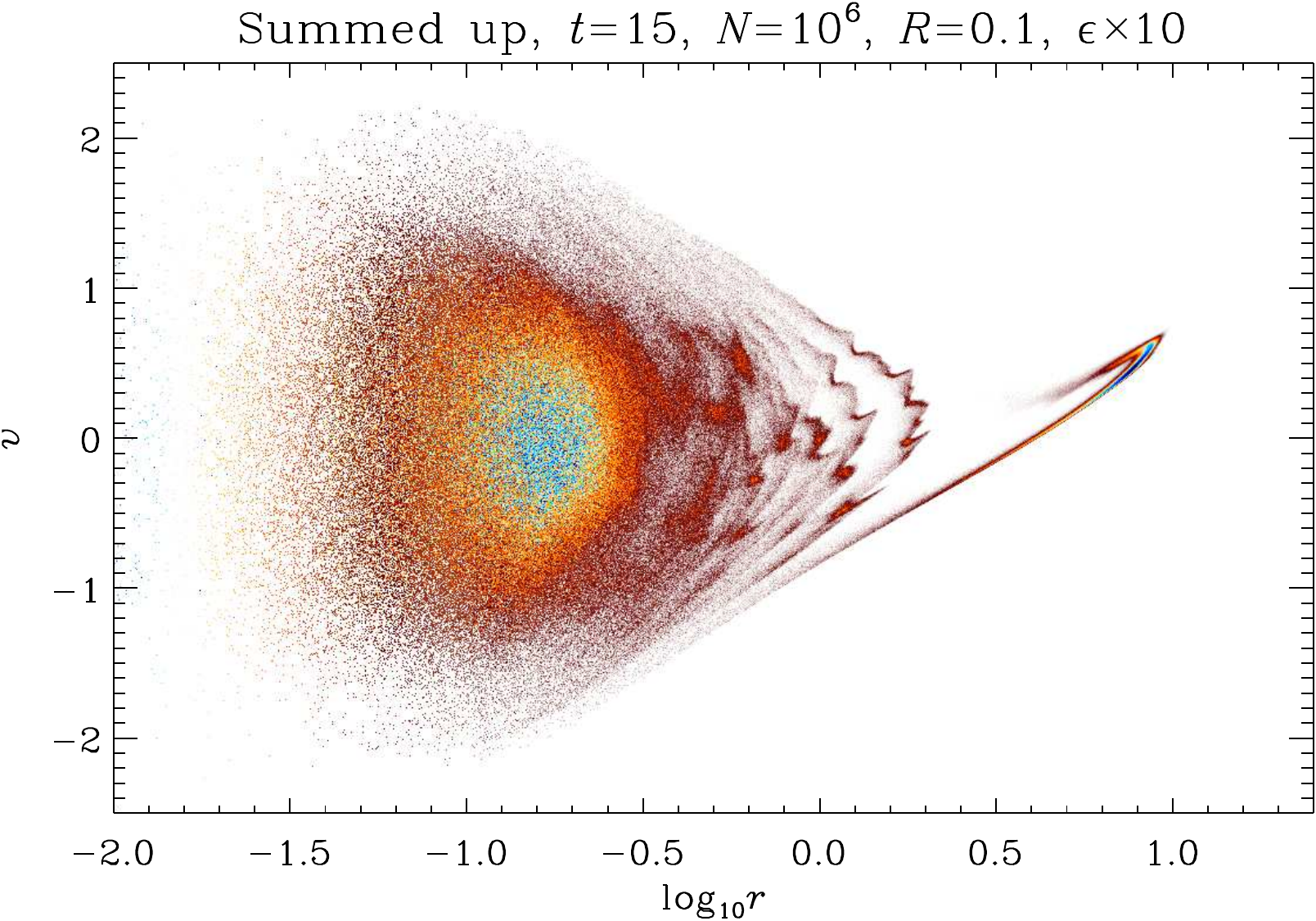}}
\end{minipage}
\begin{minipage}{0.25\textwidth}
\centerline{\includegraphics[width=4.25cm]{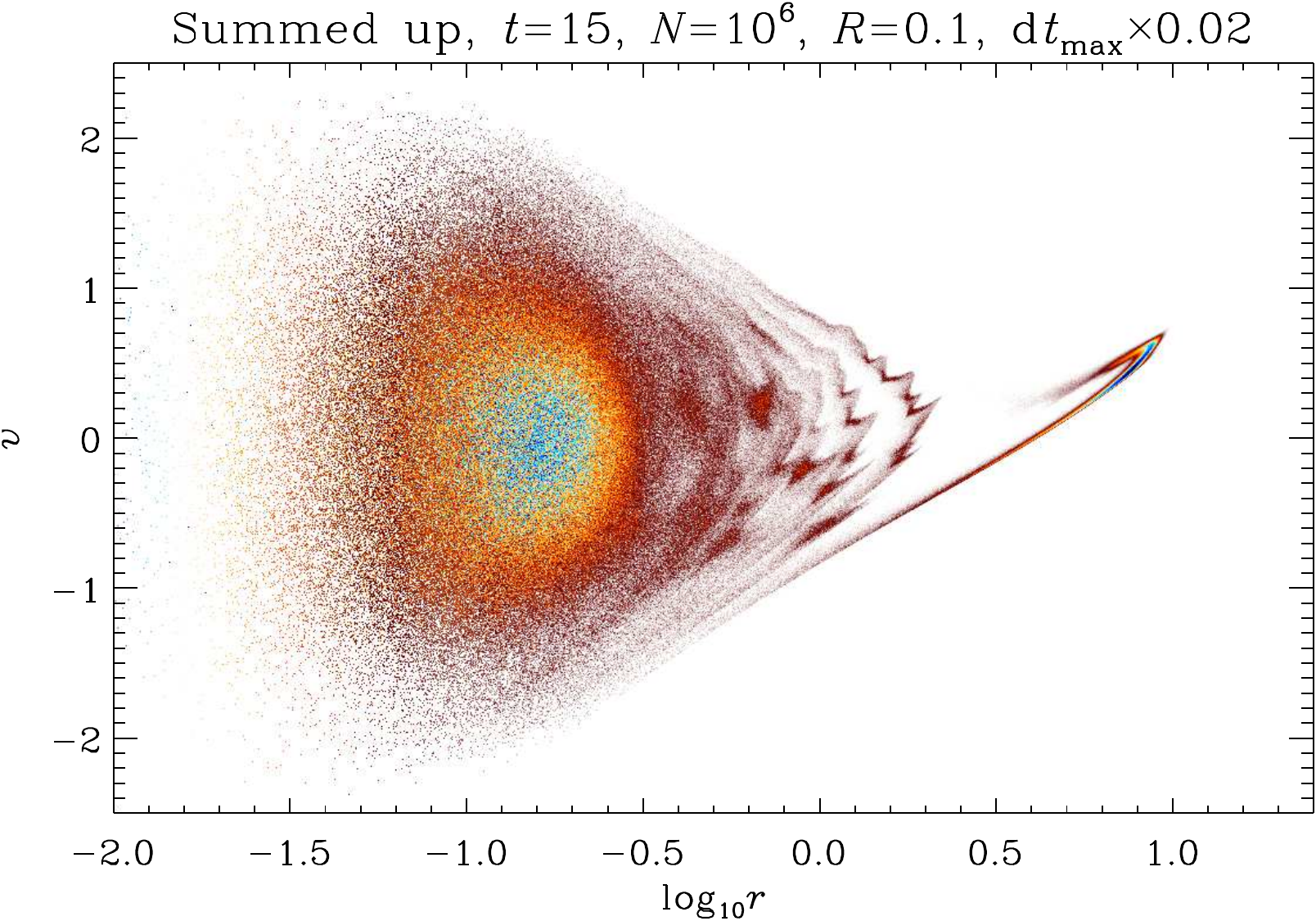}}
\centerline{\includegraphics[width=4.25cm]{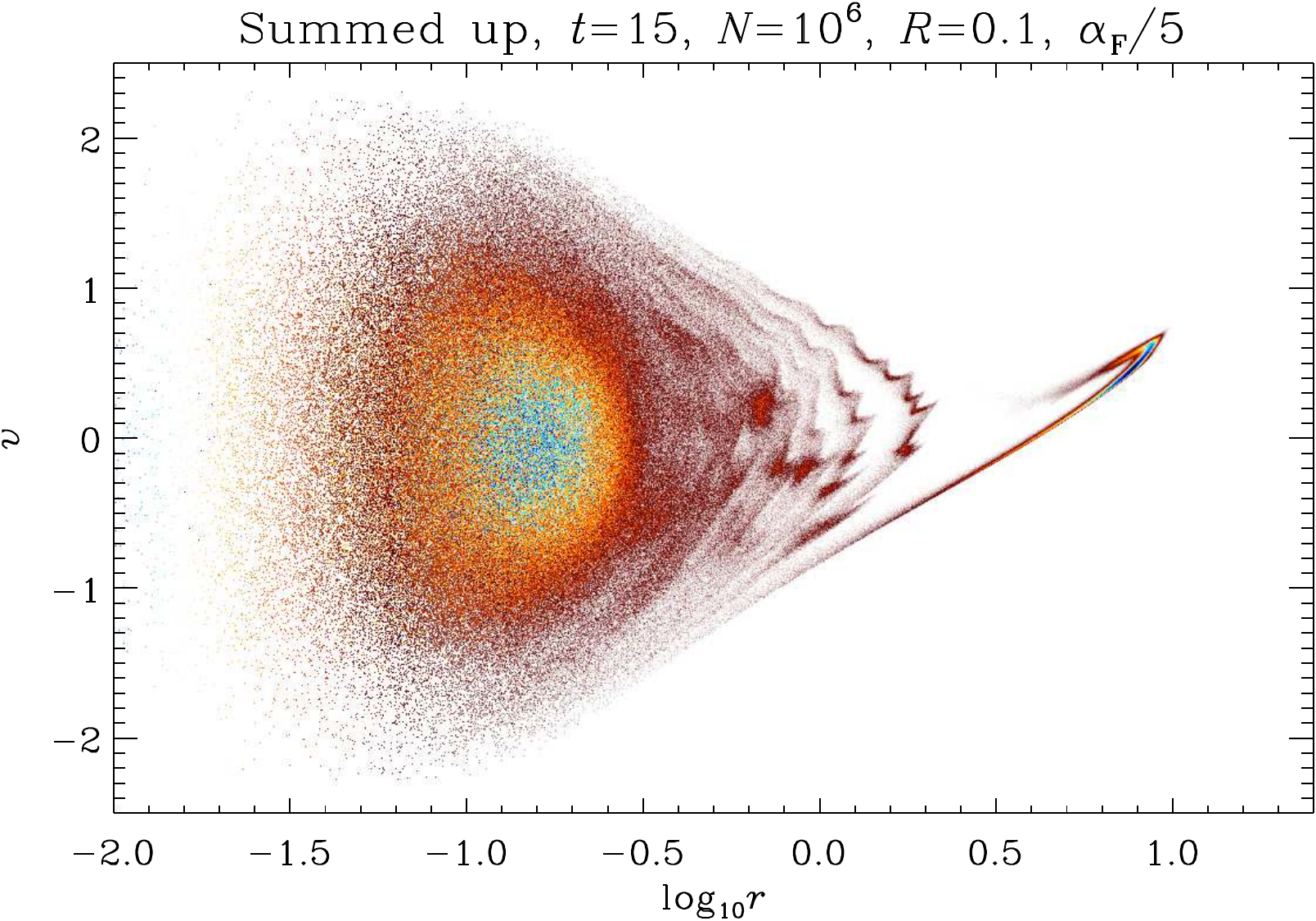}}
\end{minipage}
\end{center}
\caption[]{Effect of changing the important control parameters in {\tt Gadget}. The phase-space density is shown at $t=15$ for {\tt Gadget} simulations with the same initial conditions corresponding to the H\'enon sphere with $R=0.1$ and involving $N=10^6$ particles. In each of the simulations, one control parameter was changed compared to the fiduciary simulation shown on left panel and which uses the settings of \S~\ref{sec:gadpres}. On top and bottom left panels, the softening length of the force was decreased by a factor 5 and increased by a factor 10, respectively.  In top-right panel, the maximum possible time step was divided by a factor 50, while in the bottom-right panel, the tolerance parameter $\alpha_{\rm F}$ defined in \S~\ref{sec:gadpres} was divided by a factor 5. }
\label{fig:testgad}
\end{figure*}
 In \S~\ref{sec:visu} we noticed the presence of an instability in the $R=0.1$ $N$-body simulations. One aim of this appendix is to confirm that this instability is related to the number of particles used in the simulations and not to any other control parameter of the {\tt Gadget} code. In the same time, it is also an opportunity to check that our fiducial choice of the {\tt Gadget} control parameters, given in \S~\ref{sec:gadpres}, is correct. 

Figure~\ref{fig:testgad} illustrates the main results of the tests we performed for simulations with $10^6$ particles. These tests consisted in changing the softening length of the force, the maximum time step value and the tolerance parameter $\alpha_F$ controlling the errors on the force. Improving the accuracy of the force calculation or dividing the maximum time step ${\rm d}t_{\rm max}$ by a factor 50, which corresponds to imposing ${\rm d}t \leq 2\times 10^{-4}$, does not change the results. This is confirmed as well by the measurements of the correlators $C_k$ introduced in \S~\ref{sec:statana}, that we do not show here for simplicity. Only the value of the softening parameter of the force $\epsilon$ has an impact on the dynamics for the tests we did. Reducing $\epsilon$ by a factor 5 seems to slightly blur the phase-space density, although this effect is difficult to decipher, while increasing $\epsilon$ by a factor 10 sharpens the fine structures of the phase-space density. Since $\epsilon$ controls the intensity of close encounters between particles, this is not surprising. Note that increasing $\epsilon$ by a factor 10 is probably an exaggeration, because it {\em worsens dramatically} the match during the mixing phase between the $N$-body simulation and the Vlasov code when examining the correlators $C_k$, a sign that $\epsilon$ is probably getting too close to a physical characteristic scale of the system.\footnote{Increasing $\epsilon$ by a factor ten gives $\epsilon=0.02$, to be compared for example to the size of the core of the system after relaxation, $R_{\rm c} \simeq 0.1$.} We indeed noticed that increasing $\epsilon$ only by a factor 5 does not have much impact, on the other hand, on $C_k$.   However, all these effects do not affect the amplitude of the large scale irregularities on the pattern of $f(r,v,j)$, which are present whatever value of $\epsilon$. This is also a strong indication that close particle encounters are not at the origin of these irregularities. 

We can therefore only conclude that these irregularities and the associated nonlinear instability are the result of non trivial collective effects related to particle shot noise. This argument is also supported by the fact that in addition, the moment of their appearance is particle number dependent, as discussed in \S~\ref{sec:visu}. 
\label{lastpage}
\end{document}